\documentclass[aps,prfluids,preprint,groupedaddress,longbibliography,floatfix]{revtex4-2}

\usepackage{graphicx}
\usepackage{dcolumn}
\usepackage{bm}
\usepackage{amsmath}
\usepackage{physics}

\usepackage{amsfonts}
\usepackage{amssymb}
\usepackage{derivative}
\usepackage{tabularx}
\usepackage{subcaption}
\expandafter\def\csname ver@subfig.sty\endcsname{}
\usepackage{svg}
\usepackage{subfig}
\usepackage{array, makecell}
\usepackage{booktabs}
\usepackage{fancybox}
\usepackage{fancyhdr}
\usepackage{caption}
\usepackage{multirow}
\usepackage{float}
\usepackage{enumitem}
\usepackage{booktabs}
\usepackage{hyperref}

\hypersetup{
    colorlinks = True,
    citecolor = {cyan},
    linkcolor = {cyan},
    urlcolor = {cyan}
}
\usepackage{cleveref}
\usepackage{accents}

\makeatother
\begin{document}
\makeatletter
\renewcommand{\subsection}[1]{%
  \par\addvspace{2.0ex}
  \noindent\textbf{#1}\par\nobreak
  \addvspace{1.0ex}
}
\makeatother

\title{Numerical investigation of buoyancy-aided mixed convective flow past a square cylinder inclined at 45 degrees}

\author{\textbf{Kavin Kabilan}}
\email[]{kavinkabilan.211me323@nitk.edu.in}
\affiliation{Department of Mechanical Engineering, National Institute of Technology Karnataka, Surathkal-575025, India
}
\author{\textbf{Swapnil Sen}}
\email[]{f20211756@pilani.bits-pilani.ac.in}
\affiliation{Department of Mechanical Engineering, Birla Institute of Technology and Science, Pilani, Vidya Vihar-333031, India}
\author{\textbf{Arun K Saha}}
\email[]{aksaha@iitk.ac.in (Corresponding author)}
\affiliation{Department of Mechanical Engineering, Indian Institute of Technology Kanpur, Kanpur-208016, India}

\date{\today}
\preprint{The following article has been accepted for publication in \textit{Physical Review Fluids}.}

\begin{abstract}
The present study numerically investigates the two-dimensional mixed convective flow of air past a square cylinder placed at an angle of incidence $\alpha = 45^{\circ}$ to the free-stream. We perform direct numerical simulations (DNS) for a Reynolds number (Re) of 100, a range of Richardson numbers (Ri) between 0.0 and 1.0, and a Prandtl number (Pr) of 0.7. The critical Richardson number at which the near-field becomes a steady flow from an unsteady one lies between $0.65$ and $0.7$, along with a simultaneous emergence of the far-field unsteadiness. There is no range of Ri for which the entire flow field is seen to be steady. At a relatively moderate Ri, the flow field reveals the presence of vorticity inversion through the momentum addition in the downstream region. We discuss the dual wake-plume nature of the flow behind the cylinder. The wake exhibits characteristics similar to those of a plume, revealing a self-similar behavior in the far-field at increased buoyancy. We explore the cause of the far-field unsteadiness and discuss the mechanism of the observed flow physics using instantaneous and time-averaged flow fields. The important flow quantities, such as force coefficients, vortex shedding frequency, and Nusselt number, are discussed at various Richardson numbers.
\end{abstract}

\maketitle

\section{\label{introduction}Introduction}
\pagestyle{plain}
Flow past bluff bodies is of significant academic and practical interest. Offshore structures, aircraft flows, tall buildings, and heat exchangers are a few situations where understanding of the flow past obstacles is crucial. Many phenomena, such as B\'{e}nard-von K\'{a}rm\'{a}n vortex street formation, oscillation of the shear layer, and its eventual detachment, have been observed in such flows. The wake dynamics are complex due to their dependence on multiple parameters, including the aspect ratio, blockage ratio, and boundary conditions. Such flows demonstrate the traits of nonlinear dynamical systems because the unsteady Navier-Stokes equations are highly nonlinear, even at low Reynolds numbers (Re).

It is now well-known that bluff body flows reveal a series of transitions. At Re $\approx$ 5, the flow past a square cylinder at an angle of incidence $\alpha = 0^{\circ}$ separates at the front sharp edges, and the two separating shear layers merge downstream of the cylinder forming a symmetric steady recirculation region, whose length increases with an increase in Re. The flow undergoes a Hopf bifurcation (steady to unsteady flow) at Re $\approx$ 45 \cite{sen,gerrardmech}. Depending on Re, these two shear layers sometimes reattach on the side walls before finally separating at the rear edges of the cylinder. For Re in the range (150-175), a second (spatial) transition takes the two-dimensional flow towards three-dimensionality. Further, an increase in Re takes the flow to turbulence along the Ruelle-Takens-Newhouse route to chaos, as shown by Robichaux \textit{et al.} \cite{robichaux,sahatransitionchaos}. Remarkably, despite geometrical differences, the wake characteristics of square and circular cylinders exhibit striking similarities \cite{noack,karniadakis,williamson,sahaphd}.

In the case of $\alpha = 45^{\circ}$, the Hopf bifurcation occurs at Re $=42$ due to the lower bluffness compared to the $\alpha = 0^{\circ}$ case \cite{sohankarblockage}. This is the lowest critical Reynolds number among all angles of incidence in the range $0^{\circ} < \alpha < 45^{\circ}$. Upto an Re of 100, there exist five distinct regimes of the two-dimensional flow with the formation of the main wake along with sub-wakes around the trailing downstream corner of the cylinder \cite{senpaper}. Due to the symmetric shape of the obstacle about the centerline, the wake possesses spatio-temporal symmetry, wherein the flow reflects about the centerline after each half shedding period \cite{jiangjfm}, similar to the $\alpha = 0^{\circ}$ case. The second transition occurs around Re $\approx$ 120 at which the flow becomes three-dimensional, as reported by Jiang \cite{jiangjfm} and Yoon \textit{et al.} \cite{yoon}. Once again, this critical value is the lowest among all angles of incidence in the range $0^{\circ} < \alpha < 45^{\circ}$. While the sequence of flow transitions remains identical between $0^{\circ}$ and $45^{\circ}$, the critical Re for each transition is significantly lower in the $\alpha = 45^{\circ}$ case. All computations in this study are two-dimensional, as the transition to three-dimensionality occurs around Re $=120$, while our study is for Re $=100$.

In the computational modeling of any free shear flow, a significant challenge is resolving a physically infinite domain into a computationally finite one. Due to computational limitations, it is not possible to use an infinite domain. As a result, we must impose artificial boundaries, which raises an issue with the domain "finiteness," quantified by blockage ratio ($\beta$). The blockage ratio, defined as $\beta$ = $d/H$, is the ratio of the projected length of the cylinder ($d$) to the transverse length of the domain ($H$). Sohankar \textit{et al.} \cite{sohankarblockage} reported that a reduction in blockage ratio causes drag coefficient ($\overline{C}_D$), pressure drag coefficient ($\overline{C}_P$), and Strouhal number (St) to decrease. Many authors have conducted experimental studies \cite{tritton,okajima,sahaexp}, and the effect of flow confinement has been studied in detail \cite{davisetal,camarri,suzukietal,suzuki2}.

The heating of the obstacle in such flows leads to numerous new phenomena. In mixed convective flows, the heated cylinder generates density differences in the fluid, adding momentum to the near-field that would otherwise have a momentum deficit due to the presence of the cylinder, which "suppresses" vortex shedding \cite{dushe,samtaney,leblond}. Sharma and Eswaran reported that the flow field is steady beyond a critical Richardson number (Ri) of 0.15 \cite{sharmaeswaran,dushe,changsa}. They have also studied the effect of buoyancy-aided (Ri $>0.0$) and buoyancy-opposed (Ri $<0.0$) configurations, where the cylinder is at a higher or lower temperature than the free-stream, respectively. Studies have also investigated flow in the regimes of forced and natural convection \cite{ranjan,sahafree} as well as natural convection flows with other geometries, which reveal a similar route to chaos as that of forced flows past cylinders \cite{qiao1,qiao2,jiang}. The effect of angle of incidence has been studied for isothermal \cite{sohankarblockage,yoon} and non-isothermal flows \cite{arifhasan, dulhani, hasaninclination, vssnobhasan}. The effect of heating level, free-stream inclination \cite{fsinclinationhasan}, and Prandtl number \cite{prandtleffectali} has been studied for square cylinders, and similar studies exist for elliptical cylinders \cite{ellipticalhasan}. Many works have concentrated on the mixed convective flow in the cross-buoyancy configuration with emphasis on the effect of baroclinic vorticity on the wake structures \cite{kieftjfm, baroclinichasan, alisanghi}. There is an effect of the downstream length of the domain on the flow physics, as shown by Dushe \cite{dushe}, who reported that the increased strength of the buoyancy far downstream causes previously unexamined physical phenomena to take place. Most studies have not considered a domain longer than $50d$ ($d$ is the projected length of the cylinder), leading to the loss of far-field flow physics. Motivated by the work of Dushe, we have considered an exceptionally long domain of $110d$ downstream of the cylinder. For ease of further discussion, we have divided the domain beyond the cylinder into three distinct regions: (i) near-field (Y $< 15$), (ii) intermediate-field ($15 \leq$ Y $\leq 60$), and (iii) far-field (Y $> 60$).

Natural convective plumes occur in a variety of industrial and environmental situations \cite{woodhousejfm}. Studying the plume behaviour, which occurs far downstream of the source, is key in order to characterize such phenomena. Most studies on mixed convective flows past bluff bodies have dealt with the near-field \cite{kakade,arifhasan,dulhani,changsa, hasaninclination,vssnobhasan,fsinclinationhasan}. There has been no numerical investigation on the far-field characteristics of mixed convective flows past cylinders except that of Dushe \cite{dushe}, who has performed a detailed far-field investigation for the $\alpha = 0^{\circ}$ case and found further transitions in the far-field at higher Ri. Very few studies exist for the $\alpha = 45^{\circ}$ case in the buoyancy-aided regime \cite{arifhasan, hasaninclination, vssnobhasan}, where only the near-field flow has been discussed. An experimental study by Kimura and Bejan \cite{kimurabejan} provides qualitative evidence of unsteadiness in the far-field of a natural convection plume resulting from a point heat source. The heat is increased progressively, and changes in the plume structure in the far-field are qualitatively examined. For $\alpha = 45^{\circ}$, no work exists that discusses the physics of both the near-field and far-field flow in the buoyancy-aided flow regime and, thus, is the motivation to perform this study to observe the far-field characteristics as well as examine the effect of the cylinder orientation on the nature of the downstream flow.

In forced flows, the far-field becomes steady due to the diffusion of the shed vortices. Since we are considering buoyancy-aided flow, the natural convective part of the flow may dominate over the forced convective part in the far-field. The reduced strength of the wake, combined with the increased strength of the plume arising from natural convection, may lead to unsteadiness in the far-field, which is supported by our results and those of Dushe \cite{dushe}. Therefore, our objective is to investigate the mixed convective flow past an inclined square cylinder at $\alpha = 45^{\circ}$, with the aim of investigating the near-field and far-field dynamics in detail.

To achieve the objectives outlined above, we have conducted direct numerical simulations (DNS) of two-dimensional, unconfined, buoyancy-aided (vertically upward flow with gravity acting downward) mixed convective flow of air past a heated square cylinder (heated to a constant wall temperature higher than the free-stream temperature) at a fixed incidence angle, $\alpha = 45^{\circ}$ to the free-stream. The present investigation considers flow in the buoyancy-aided regime as the buoyancy force assists the flow in the upward direction. The flow parameters used are Re $=100$ and $0.0 \leq Ri \leq 1.0$. Air has been considered as the fluid, the Prandtl number (Pr) of which is taken as 0.7.

The remaining manuscript is structured as follows: In \Cref{num}, we have discussed the numerical method, governing equations and boundary conditions, details of the validation, grid independence test, and domain independence test. In \Cref{results}, we have presented the results of our investigation, which include integral parameters, instantaneous, and time-averaged results. These results are followed by a discussion on the inversion of vorticity, suppression of vortex shedding, and far-field unsteadiness. Finally, the main findings are summarized in \Cref{conclusion}.

\section{\label{num}Numerical Method}
\subsection{Numerical Domain}

\begin{figure}[t]
    \centering
    \includegraphics[width=\linewidth]{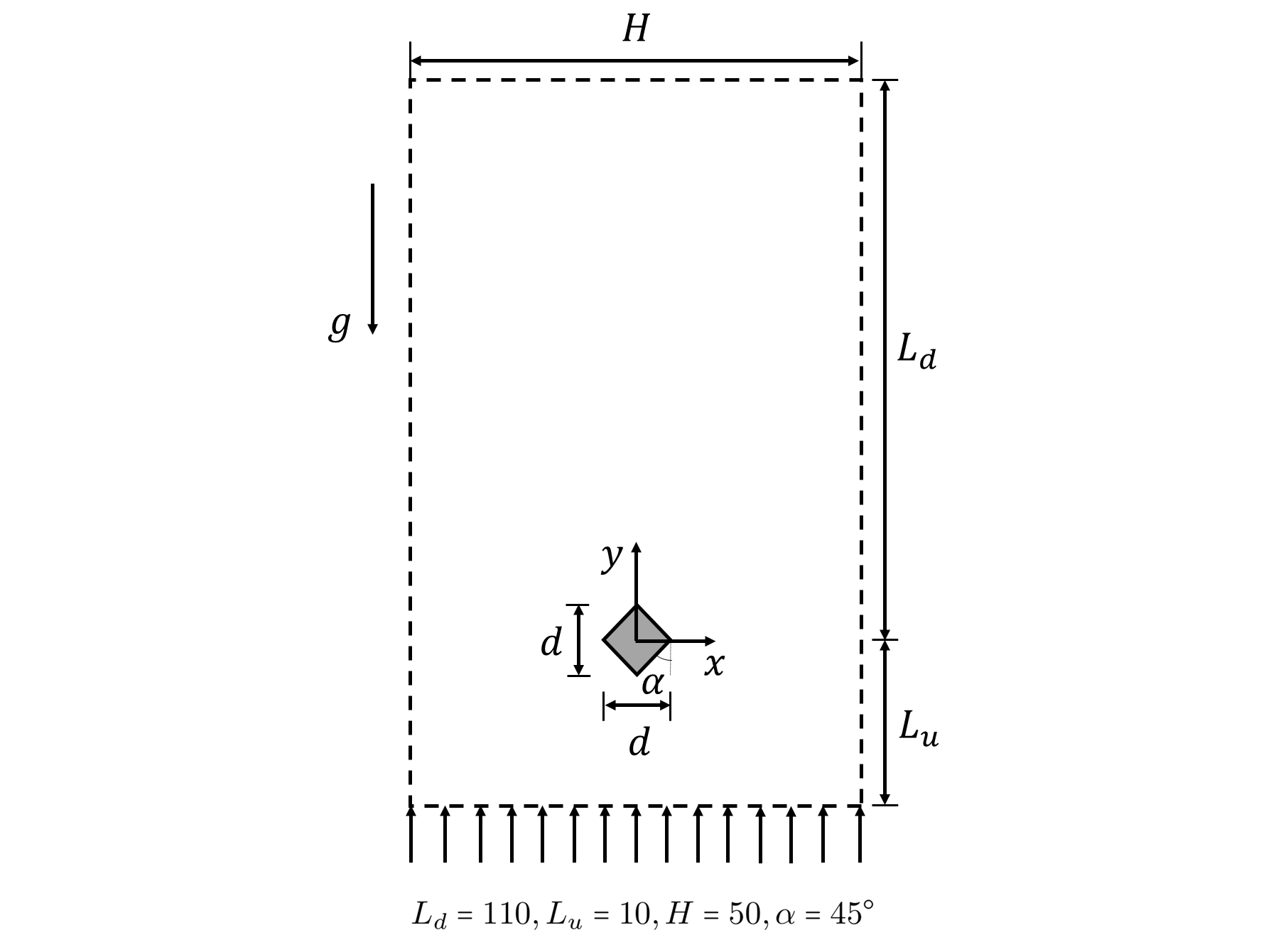}
    \caption{Two-dimensional computational domain.}
    \label{domain1}
\end{figure}

\Cref{domain1} illustrates the numerical domain with its dimensions provided alongside. The $Y$ direction is the streamwise flow direction, and the $X$ direction is the transverse flow direction. Gravity acts in the negative $Y$ direction. The solid cylinder is heated to a constant temperature greater than the free-stream temperature. The square cylinder has a projected length of $d$, which is the length scale used in all calculations. It is placed at an angle of incidence $\alpha = 45^{\circ}$ with the free-stream. The streamwise length of the domain ($L$) is $120d$, and the transverse length ($H$) is $50d$. The upstream length of the domain ($L_u$) is $10d$. The blockage ratio in the current investigation is 2$\%$, which is reasonably low for the solution to be free from dependence on blockage. 

\subsection{Governing Differential Equations and Boundary Conditions}
Two-dimensional, incompressible flow past a square cylinder at an angle of incidence $\alpha = 45^{\circ}$ has been calculated. The governing differential equations, namely the continuity equation, the unsteady Navier-Stokes equations, and the energy equation, given in \Cref{eq1,eq2,eq3}, respectively, are solved. In the dimensionless form, they are expressed as:

\begin{equation} \label{eq1}
	\frac{\partial u_i}{\partial x_i} = 0
\end{equation}	

\begin{equation} \label{eq2}
	\frac{\partial u_i}{\partial \tau} + \frac{\partial (u_j u_i)}{\partial x_j}  = -\frac{\partial p}{\partial x_i} + \frac{1}{Re}\left[\frac{\partial^2 u_i}{\partial x_i^2}\right] + Ri \theta \delta_{i2}
\end{equation}

\begin{equation} \label{eq3}
	\frac{\partial \theta}{\partial \tau} + \frac{\partial (u_i \theta)}{\partial x_i}  = \frac{1}{RePr} \left[\frac{\partial^2 \theta}{\partial x_i^2}\right]
\end{equation}

Where, 
\begin{equation*}
Gr = \dfrac{g\beta(T_w - T_{\infty})d^3}{\nu^2}, \hspace{5pt} Re = \dfrac{\rho V_{\infty}d}{\mu}, \hspace{5pt} Ri = \dfrac{Gr}{Re^2}, \hspace{5pt} Pr= \dfrac{\mu C_{ph}}{K}
\end{equation*}

$Gr$ is the Grashof number, $g$ is the acceleration due to gravity, $\beta$ is the coefficient of volumetric thermal expansion, $Pr$ is the Prandtl number of the fluid, $\nu$ is the kinematic viscosity of the fluid, $\rho$ is the density of the fluid, $\mu$ is the dynamic viscosity of the fluid, $C_{ph}$ is the specific heat of the fluid, and $K$ is the thermal conductivity of the fluid.  $\delta_{i,j}$ is the Kronecker delta.

Air is the fluid under consideration, and the Prandtl number is taken as 0.7. The effect of viscous dissipation has been neglected. The Boussinesq approximation has been utilized to convert density differences to corresponding temperature differences, and the buoyancy force term has been included in the $y$-momentum equation. In \Cref{eq1,eq2,eq3}, all velocities and lengths are non-dimensionalized with the free-stream velocity $V_{\infty}$ and the projected length of the cylinder in the streamwise direction $d$, respectively. Similarly, time and pressure are non-dimensionalized with $d/V_{\infty}$ and $\rho V_\infty^2$, respectively.  
Temperature is non-dimensionalized as follows:
    \begin{equation}
    \theta = \dfrac{T-T_{\infty}}{T_w - T_{\infty}}
    \end{equation}
    
    Where $T_{\infty}$ is the free-stream temperature, and $T_w$ is the temperature at the cylinder wall. The boundary conditions are as follows:

    At the inlet ($Y=0$), constant streamwise velocity and temperature are used. Transverse velocity is set to zero.
    \begin{equation*}
        v=1.0, \hspace{5pt} u=0
    \end{equation*}
    \begin{equation*}
      \ \theta=0
    \end{equation*}

    Transverse confining surfaces ($X = \pm H/2$) are unbounded, and subsequently, they are modeled as free-slip, adiabatic surfaces.
    \begin{equation*}
    \pdv{v}{x}=0,\hspace{5pt} u=0
    \end{equation*}
    \begin{equation*}
    \pdv{\theta}{x}=0
    \end{equation*}

     At the outlet ($Y=L$), selecting an appropriate outflow boundary condition is challenging because there is no unique prescription for the outflow boundary of external flows. It has been shown that the selection of the outflow boundary condition greatly affects the upstream flow, particularly in subsonic flow. Therefore, it should be chosen to have minimal effect on upstream flow. It has been shown in the past that the convective boundary condition (CBC) is superior to other boundary conditions such as zero Neumann boundary condition (NBC) or vanishing second derivatives (VSD) by Sohankar \textit{et al.} \cite{sohankarblockage,sohankar2} and others \cite{abbassi}. In particular, Sohankar \textit{et al.} \cite{sohankarblockage} have shown that the CBC reduces computational time, iterations for convergence, and the time taken to reach the periodic flow. Moreover, the CBC allows for the smooth passage of vortices out of the domain, while the NBC and VSD conditions distort and smear vortices exiting the domain. Accordingly, the convective boundary condition, first prescribed by Orlanski \cite{orlanski}, has been employed for both velocity and temperature:
    \begin{equation*}
        \pdv{u_i}{t}+v_c \pdv{u_i}{y} = 0
    \end{equation*}
    \begin{equation*}
        \pdv{\theta}{t}+v_c \pdv{\theta}{y} = 0
    \end{equation*}
    
    Where, $v_c$ is the average convective velocity of vortices leaving the domain. An optimum value of $v_c$ = 0.8, which provides a faster convergence without any loss of accuracy, has been used for all calculations performed in the present study.

    For the cylinder, no-slip and constant temperature boundary conditions are used for all surfaces.
    \begin{equation*}
        u=v=0
    \end{equation*}
    \begin{equation*}
        \theta=1.0
    \end{equation*}

\begin{figure}
    \centering
     \begin{subfigure}[t]{0.3\textwidth}
         \centering
         \includegraphics[width=\textwidth]{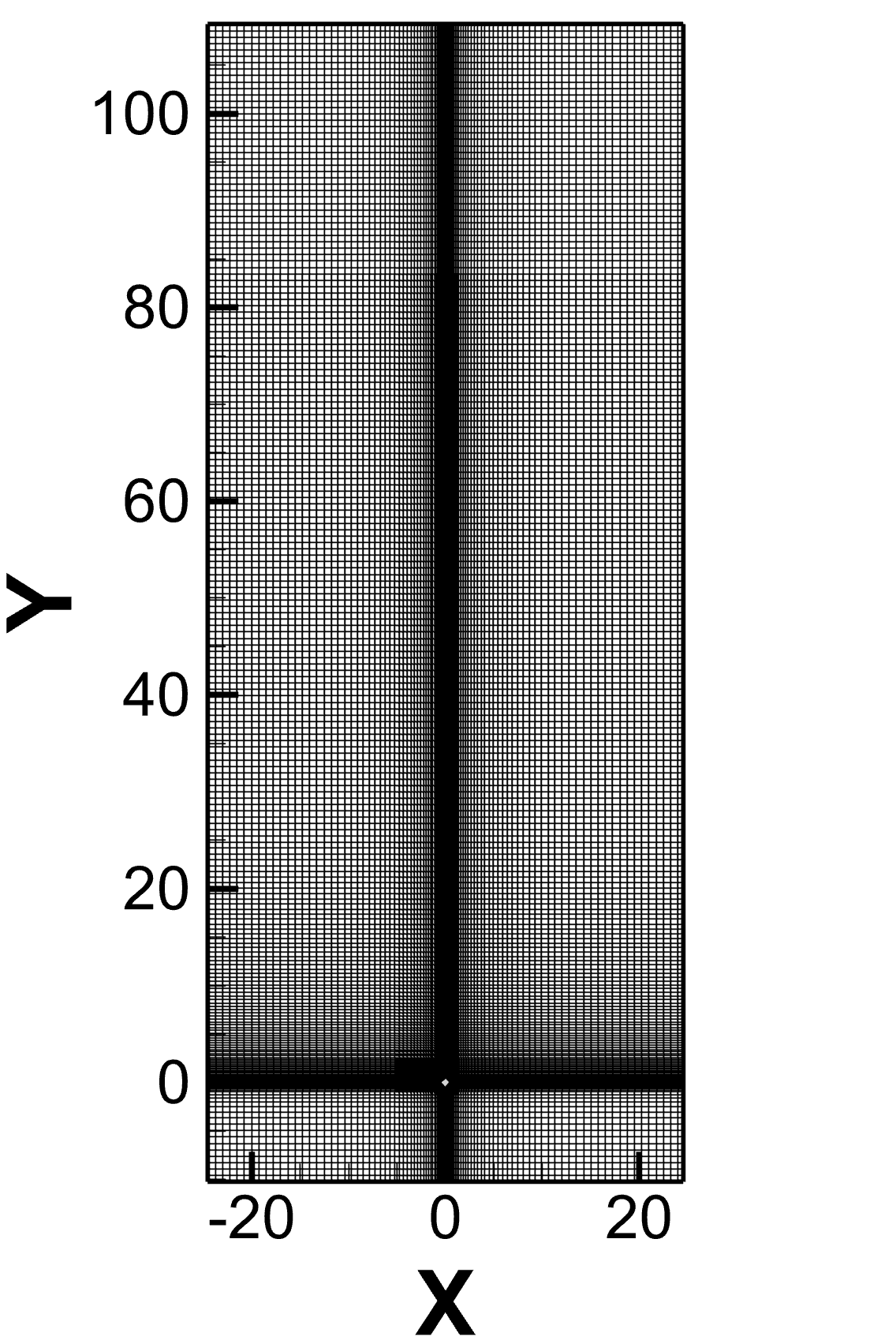}
         \caption{Complete domain ($-25 \leq$ X $\leq 25$, $-10 \leq$ Y $\leq 110$)}
         \label{griddetail}
     \end{subfigure}
     \hspace{25pt}
     \begin{subfigure}[t]{0.3\textwidth}
         \centering
         \includegraphics[width=\textwidth]{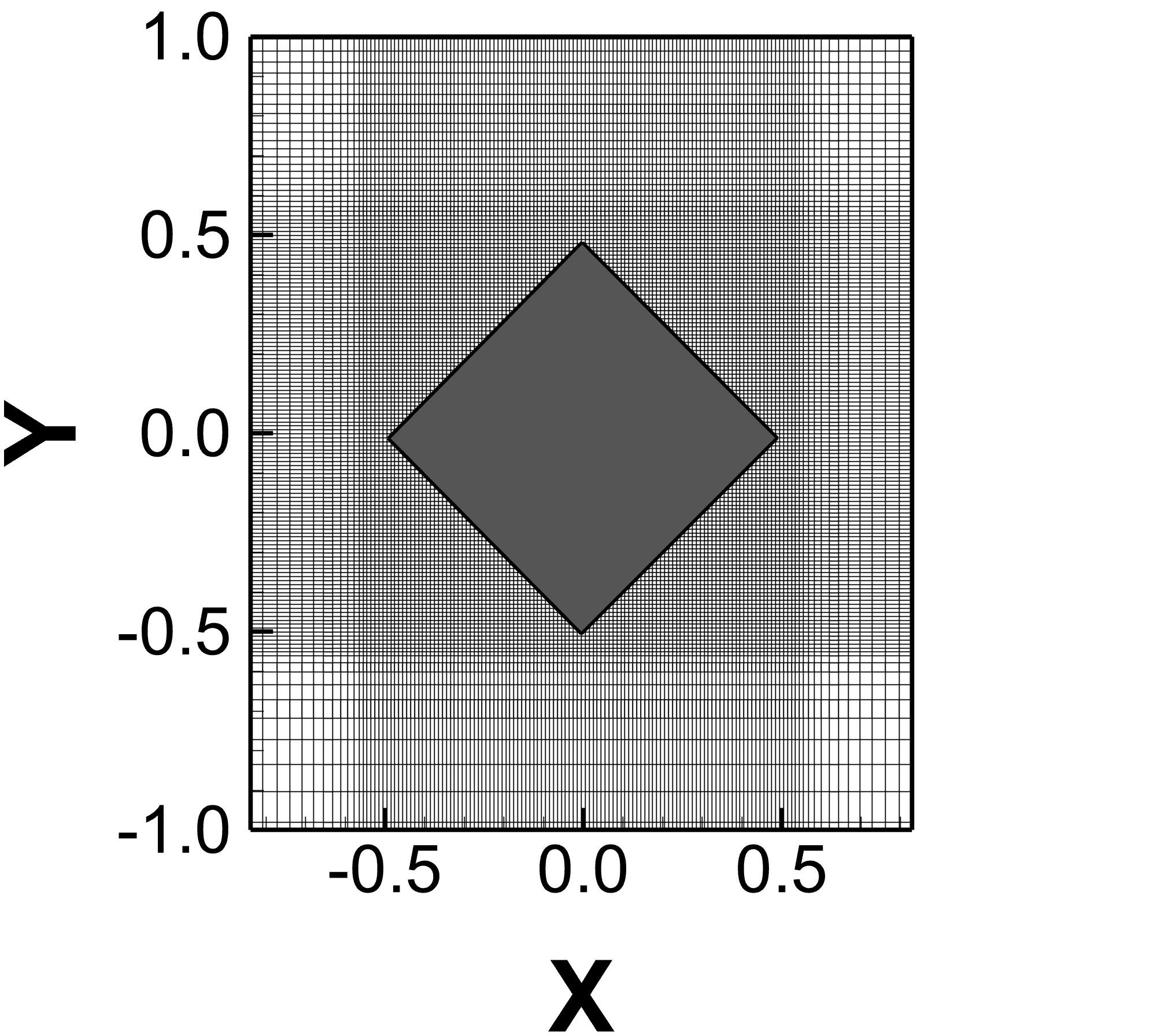}
         \caption{Near the cylinder vicinity ($-0.85 \leq$ X $\leq 0.85$, $-1 \leq$ Y $\leq 1$)}
         \label{gridnearwall}
     \end{subfigure}
     \caption{Details of grid used in the (a) full domain, and (b) in the vicinity of the cylinder.}
     \label{griddetails}
\end{figure}
	
\subsection{Solution Methodology}
The numerical method used in the present work is an improved version of the finite-difference Marker and Cell algorithm proposed by Harlow and Welch \cite{harlow}. The governing equations have been solved on a staggered grid that helps in the coupling of pressure and velocity fields. Velocity components are located on the cell face to which they are normal and pressure is located at the cell center. Velocity components are interpolated to the cell center when required. Convective and diffusive terms are discretized using a second-order central difference scheme. Time advancement of both convective and diffusive terms is done using a second-order time-accurate explicit Adams-Bashforth method. 

A two-step predictor-corrector method is used to obtain the velocity and pressure fields. In the first step, provisional values of velocity are calculated explicitly using the results of the previous time step. In the subsequent step, pressure and velocity are corrected through the solution of \Cref{eq7,eq8,eq9} such that the resulting velocity field is divergence-free. This iterative procedure is continued till the velocity error between consecutive time steps is below a defined threshold.

The momentum equation can be written using a space operator G (containing convective and diffusive terms) as: 
\begin{equation} \label{eq5}
    \pdv{u_i}{t} = G(u_i,u_j) - \pdv{p}{x_i}
\end{equation}
Then, the predictor step for marching forward in time takes the form:
\begin{equation} \label{eq6}
    \frac{u_i^* - u_i^n}{\Delta t} = \left[\frac{3}{2} G(u_i,u_j)^n - \frac{1}{2} G(u_i,u_j)^{n-1} \right] - \pdv{p^n}{x_i}
\end{equation}
Subsequently, the pressure correction step involves the solution of the pressure correction equation: 
\begin{equation} \label{eq7}
    p' = \dfrac{-\omega_0 \pdv{u_i^*}{x_i}}{2 \Delta t \left[ \dfrac{1}{(\Delta x)^2} + \dfrac{1}{(\Delta y)^2}\right]  } 
\end{equation}
The final solutions for pressure and velocity, as part of the correction step, are given as:
\begin{equation} \label{eq8}
    p^{n+1} \leftarrow p^n + p'
\end{equation}
\begin{equation} \label{eq9}
    u_i^{n+1} \leftarrow u_i^* + \dfrac{\Delta t}{\Delta x_i}p'
\end{equation}

The corrector step (\Cref{eq7}) is solved using the point Gauss-Seidel method with an over-relaxation parameter $ \omega_0 $, which accelerates pressure correction. An optimum value of $ \omega_0 $ = 1.8 has been used for all calculations in the present work.

The grid utilized is shown in \Cref{griddetails}. It is a non-uniform, structured Cartesian grid with fine meshes in the near-wall region. In the vicinity of the cylinder, the grid is uniform with a cell size of $\Delta x = \Delta y = 0.01$. Beyond a certain distance from the obstacle in all directions, the grid spacing is increased in a proper ratio to minimize computational cost.

The characteristics of flow and heat transfer past a bluff object are studied by calculating the different integral parameters. The vortex shedding frequency $f_s$ is calculated using the Fast Fourier transform (FFT) of any time signal in the wake. Subsequently, \textbf{Strouhal number} (St) is calculated as:
\begin{equation*}
St = \dfrac{ f_s d}{V_{\infty}}
\end{equation*} 

The \textbf{drag coefficient} ($C_D$) and \textbf{lift coefficient} ($C_L$) are calculated as:
\begin{equation*}
C_D = \dfrac{F_D}{\dfrac{1}{2}\rho V_{\infty}^2 A_x}
\end{equation*} 

\begin{equation*}
C_L = \dfrac{F_L}{\dfrac{1}{2}\rho V_{\infty}^2 A_y}
\end{equation*}

The heat transfer balance at the solid-liquid interface is given by

\begin{equation*}
\left.-K \pdv{T}{n}\right\vert_{n=0} = h(T_w - T_{\infty})
\end{equation*}

Using the above relation, \textbf{Nusselt number} (Nu) can be written as:
\begin{equation*}
\implies Nu = \dfrac{h d }{K} = \left.-\pdv{\theta}{n}\right\vert_{n=0}
\end{equation*} 

Where $f_s$ is the vortex shedding frequency, $F_D$ is the streamwise or drag force, $F_L$ is the transverse or lift force, $h$ is the wall heat transfer coefficient, $n$ is the wall normal direction to the surface of the cylinder, $A_x$ and $A_y$ are the streamwise and the transverse projected areas of the cylinder, respectively.

Rigorous validation has been performed for two different angles of incidence ($\alpha = 0^{\circ}$ and $\alpha = 45^{\circ}$) by comparing the time-averaged values of drag coefficient, the Strouhal number, and the Nusselt number. Validation has been carried out against Sohankar \textit{et al.} \cite{sohankarblockage}, Yoon \textit{et al}. \cite{yoon}, Sharma and Eswaran \cite{sharmaeswaran}, Ranjan \textit{et al}. \cite{ranjan}, and Yadav \textit{et al.} \cite{senpaper}. The results of the validation are presented in \Cref{valsq1,valsq2,valslant1,valslant2}. For the $\alpha = 0^{\circ}$ case, the mean drag coefficient reveals a maximum error of 3.09$\%$, while the corresponding errors in the case of Strouhal number and Nusselt number are 1.85$\%$ and 4.64 $\%$, respectively. For the $\alpha = 45^{\circ}$ case, the errors in the mean drag coefficient, Strouhal number, and Nusselt number are 5.16$\%$, 3.37$\%$, and 3.45$\%$, respectively. Therefore, the current simulated results are in good agreement with those available in the literature.

\begin{table}[h!]
\centering
\caption{Comparison of time-averaged drag coefficient and Strouhal number (isothermal flow, $\alpha = 0^{\circ}$).}
\setlength{\tabcolsep}{0.5em}
\begin{tabular}{c ccc cc}
\toprule
\toprule
\multirow{2}{*}{\begin{tabular}[c]{@{}c@{}}Reynolds \\ number (Re)\end{tabular}} & \multicolumn{3}{c}{\begin{tabular}[c]{@{}c@{}}Mean drag coefficient \\ ($\overline{C}_D$)\end{tabular}} & \multicolumn{2}{c}{\begin{tabular}[c]{@{}c@{}}Strouhal number \\ (St)\end{tabular}} \\ 
\cmidrule(lr){2-4} \cmidrule(lr){5-6} 
& Present & Sohankar \textit{et al}. \cite{sohankarblockage} & Yoon \textit{et al}. \cite{yoon} & Present & Sohankar \textit{et al}. \cite{sohankarblockage} \\ 
\midrule
50 & 1.57 & 1.62 & 1.57 & 0.114 & - \\ 
100 & 1.48 & 1.46 & 1.44 & 0.147 & 0.147 \\ 
150 & 1.41 & 1.41 & 1.40 & 0.159 & 0.162 \\ 
\bottomrule
\bottomrule
\end{tabular}
\label{valsq1}
\end{table}

\begin{table}[h!]
\centering
\caption{Comparison of mean Nusselt number and drag coefficient at Re $=100$ ($\alpha = 0^{\circ}$).}
\setlength{\tabcolsep}{0.75em}
\begin{tabular}{c ccc cc}
\toprule
\toprule
\multirow{2}{*}{\begin{tabular}[c]{@{}c@{}}Richardson \\ number (Ri)\end{tabular}} 
& \multicolumn{2}{c}{\begin{tabular}[c]{@{}c@{}}Mean Nusselt number \\ ($\overline{Nu}$)\end{tabular}} 
& \multicolumn{2}{c}{\begin{tabular}[c]{@{}c@{}}Mean drag coefficient \\ ($\overline{C}_D$)\end{tabular}} \\ 
\cmidrule(lr){2-3} \cmidrule(lr){4-5}
& Present & Sharma and Eswaran \cite{sharmaeswaran} 
& Present & Sharma and Eswaran \cite{sharmaeswaran} \\ 
\midrule
0.0  & 3.90 & 4.09 & 1.48 & 1.57 \\ 
0.25 & 4.12 & 4.30 & 1.66 & 1.76 \\ 
0.50 & 4.37 & 4.55 & 1.98 & 2.07 \\ 
0.75 & 4.55 & 4.73 & 2.28 & 2.36 \\ 
1.0  & 4.70 & 4.87 & 2.57 & 2.63 \\ 
\bottomrule
\bottomrule
\end{tabular}
\label{valsq2}
\end{table}

\begin{table}[h!]
\centering
\caption{Comparison of time-averaged drag coefficient and Strouhal number (isothermal flow, $\alpha = 45^{\circ}$).}
\small
\setlength{\tabcolsep}{0.4em}
\begin{tabular}{c cccc ccc}
\toprule
\toprule
\multirow{2}{*}{\begin{tabular}[c]{@{}c@{}}Reynolds  \\ number (Re)\end{tabular}} & \multicolumn{4}{c}{\begin{tabular}[c]{@{}c@{}}Mean drag coefficient \\ ($\overline{C}_D$)\end{tabular}} & \multicolumn{3}{c}{\begin{tabular}[c]{@{}c@{}}Strouhal number \\ (St)\end{tabular}} \\ 
\cmidrule(lr){2-5} \cmidrule(lr){6-8} 
& Present 
& \begin{tabular}[c]{@{}c@{}}Sohankar \\ \textit{et al.} \cite{sohankarblockage}\end{tabular} 
& \begin{tabular}[c]{@{}c@{}}Yoon \\ \textit{et al.} \cite{yoon}\end{tabular} 
& \begin{tabular}[c]{@{}c@{}}Yadav \\ \textit{et al.} \cite{senpaper}\end{tabular} 
& Present 
& \begin{tabular}[c]{@{}c@{}}Sohankar \\ \textit{et al.} \cite{sohankarblockage}\end{tabular} 
& \begin{tabular}[c]{@{}c@{}}Yadav \\ \textit{et al.} \cite{senpaper}\end{tabular} \\ 
\midrule
50  & 1.63 & 1.55 & 1.59 & 1.64 & 0.144 & 0.140 & 0.141 \\ 
100 & 1.77 & 1.71 & 1.72 & 1.78 & 0.180 & 0.178 & 0.180 \\ 
150 & 1.86 & 1.85 & 1.79 & 1.94 & 0.190 & 0.194 & 0.200 \\ 
\bottomrule
\bottomrule
\end{tabular}
\label{valslant1}
\end{table}

\begin{table}[h!]
\centering
\caption{Comparison of mean Nusselt number at Re $=100$ and Ri $=0.0$ ($\alpha = 45^{\circ}$).}
\setlength{\tabcolsep}{0.75em}
\begin{tabular}{c cc}
\toprule
\toprule
\multirow{2}{*}{\begin{tabular}[c]{@{}c@{}}Reynolds  \\ number (Re)\end{tabular}} & \multicolumn{2}{c}{\begin{tabular}[c]{@{}c@{}}Mean Nusselt number \\ ($\overline{Nu}$)\end{tabular}} \\ 
\cmidrule(lr){2-3} 
& Present & Ranjan \textit{et al}. \cite{ranjan} \\ 
\midrule
60  & 4.00 & 4.08 \\ 
100 & 5.20 & 5.35 \\ 
150 & 6.43 & 6.66 \\ 
\bottomrule
\bottomrule
\end{tabular}
\label{valslant2}
\end{table}

The accuracy of the computed solution depends on the discretization error, which is controlled by the grid resolution. To ensure that the grid is sufficiently fine to reduce the discretization error, a grid independence test has been carried out using three different grid sizes, namely, G1 (619 $\times$ 320), G2 (777 $\times$ 396), and G3 (883 $\times$ 504) at Re $=100$ and Ri $=1.0$ by comparing the same parameters used in the validation. The parameters computed using the three grid sizes, along with the minimum grid spacing in the vicinity of the cylinder, are presented in \Cref{gridtest}. The domain size is kept constant for the grid independence test (120 $\times$ 50). We observe that G2 produces results with minimal deviation from G3. The mean drag coefficient reveals an error of $2.82\%$ and $1.23\%$ for G1 and G2, respectively, with respect to G3. The corresponding error in the case of the Strouhal number is $2.09 \%$ and $5.88\%$, and for the mean Nusselt number, it is $1.09\%$ and $0.55\%$, respectively. Given the moderate computational cost and closeness of results with G3, the grid G2 is utilized for all subsequent calculations.
\begin{table}[ht!]
\centering
\caption{Mean drag coefficient, mean Nusselt number and Strouhal number for different grid resolutions at Re $=100$ and Ri $=1.0$.}
\setlength{\tabcolsep}{0.6em}
\begin{tabular}{c c c c c} 
\toprule
\toprule
Grid & Minimum spacing & \begin{tabular}[c]{@{}c@{}}Mean drag coefficient \\ ($\overline{C}_D$)\end{tabular} & \begin{tabular}[c]{@{}c@{}}Mean Nusselt number \\ ($\overline{Nu}$)\end{tabular} & \begin{tabular}[c]{@{}c@{}}Strouhal number \\ (St)\end{tabular} \\ 
\midrule
G1 & 0.025 & 2.48 & 5.50 & 0.215 \\ 
G2 & 0.01  & 2.44 & 5.47 & 0.221 \\ 
G3 & 0.0075 & 2.41 & 5.44 & 0.234 \\ 
\bottomrule
\bottomrule
\end{tabular}
\label{gridtest}
\end{table}
As mentioned in \Cref{introduction}, integral parameters depend on the blockage ratio. To ensure that the calculated integral parameters are free from dependence on the blockage ratio, a domain independence test has been carried out using two different domains, namely D1 ($120 \times 50$, $\beta = 2\%$) and D2 ($120 \times 67$, $\beta = 1.5\%$) at Re $=100$ and Ri $=1.0$ by comparing the same parameters that are used in the validation. Both domains have the same streamwise extent. As the original domain is already very long, it is expected that an increase in its length downstream of the cylinder will not change the flow physics. A detailed investigation on the effect of the domain length has been conducted by Dushe \cite{dushe}. The mean drag coefficient ($\overline{C}_D$), mean Nusselt number ($\overline{Nu}$), and Strouhal number (St) for the domain D1 are 2.44, 5.47, and 0.221. The corresponding values for the domain D2 are 2.45, 5.44, and 0.237. The mean drag coefficient reveals an error of $0.4\%$ for D1 with respect to D2, whereas for the mean Nusselt number and Strouhal number, it is $0.6\%$, and $6.8\%$, respectively. The domain independence test reveals that the results do not depend on blockage. Therefore, domain D1 is used for all subsequent calculations.

In order to prevent numerical instabilities caused by the use of an explicit scheme for temporal discretization in the present study, the time increment is determined by subjecting it to two restrictions, namely, the Courant–Friedrichs–Lewy (CFL) condition, and the grid Fourier number (Fo) restriction corresponding to convection and diffusion dominated regions respectively. We have calculated the minimum time step throughout the domain as prescribed by the above two criteria. Further, the lower value between the two conditions is multiplied by a factor less than unity (between 0.2 and 0.4) to ensure stability for the second-order accurate spatially and temporally discretized equations used in the present study. The typical value of the time step is found to be 0.0005, which was chosen after performing a time-step independence study.

\section{\label{results}Results and Discussion}
In this section, we have used the results of DNS to provide insight into the effect of buoyancy on the flow field behind the cylinder. Based on the instantaneous and time-averaged results, we have discussed the physical mechanisms for the suppression of vortex shedding, vorticity inversion, and far-field unsteadiness. We have also discussed the effect of buoyancy on the integral parameters of the flow field. It is to be noted that the results presented in this study are only applicable to the streamwise length of the domain used in the current study. The choice of domain length influences the observed transitions, and a longer domain could reveal additional flow physics. We set the total averaging time (T) to two different values for different calculations to achieve statistical independence. The values taken for T are 200 and 300, which correspond to 36 and 54 shedding cycles at Ri = 0.0, respectively (which has the highest shedding time period due to having the lowest Strouhal number as seen in \Cref{stvsri}).  It is found that the results obtained do not vary more than $1\%$ between two different averaging times. Therefore, it is expected that the presented results are statistically independent.

\subsection{Instantaneous Flow}
\begin{figure}[htbp]
	\centering
	\begin{subfigure}[t]{0.3\textwidth}
		\includegraphics[width=\linewidth]{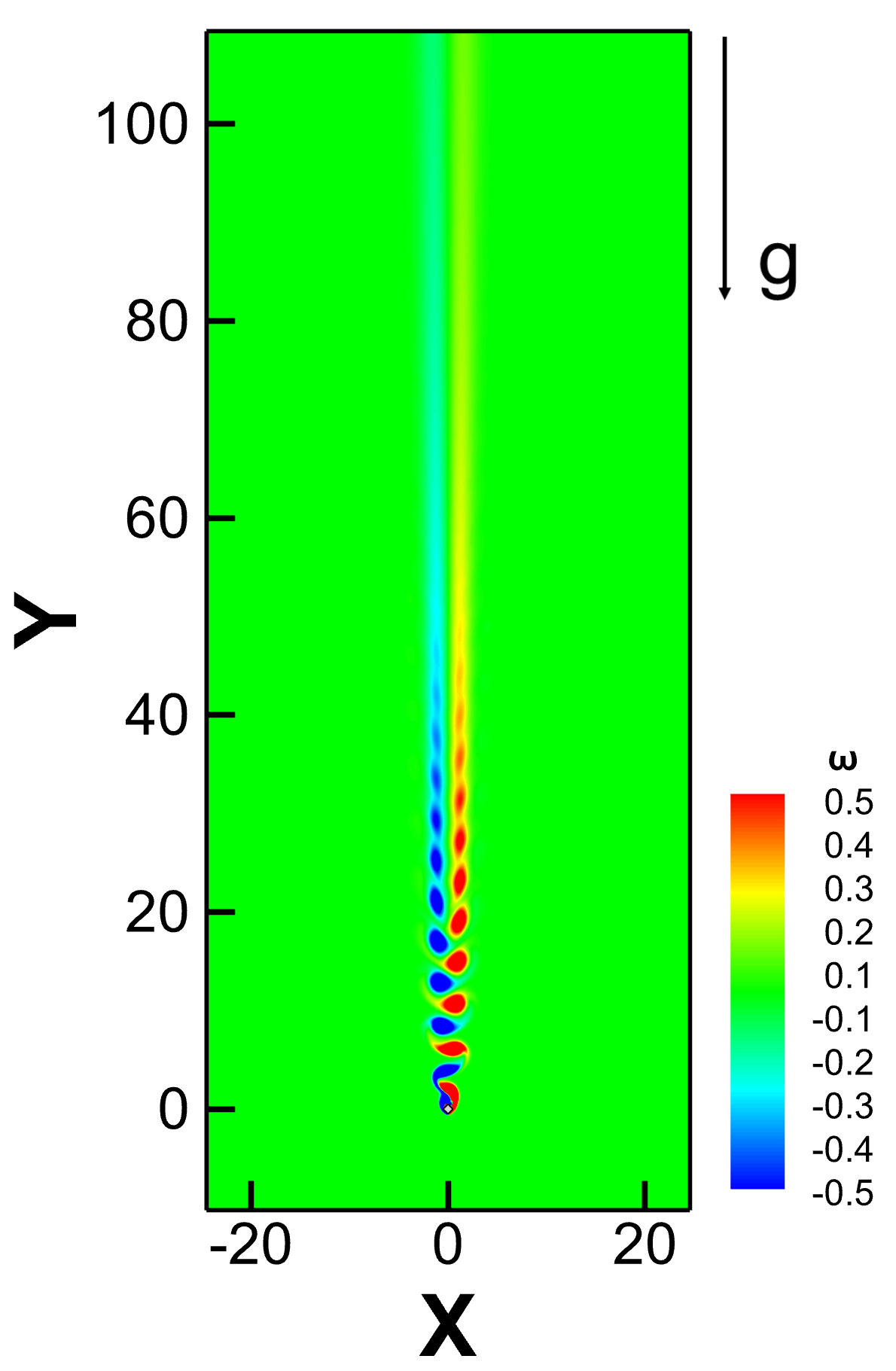}
		\caption{Ri $=0.0$}
		\label{insvort0}
	\end{subfigure}\hfill
	\begin{subfigure}[t]{0.3\textwidth}
		\includegraphics[width=\linewidth]{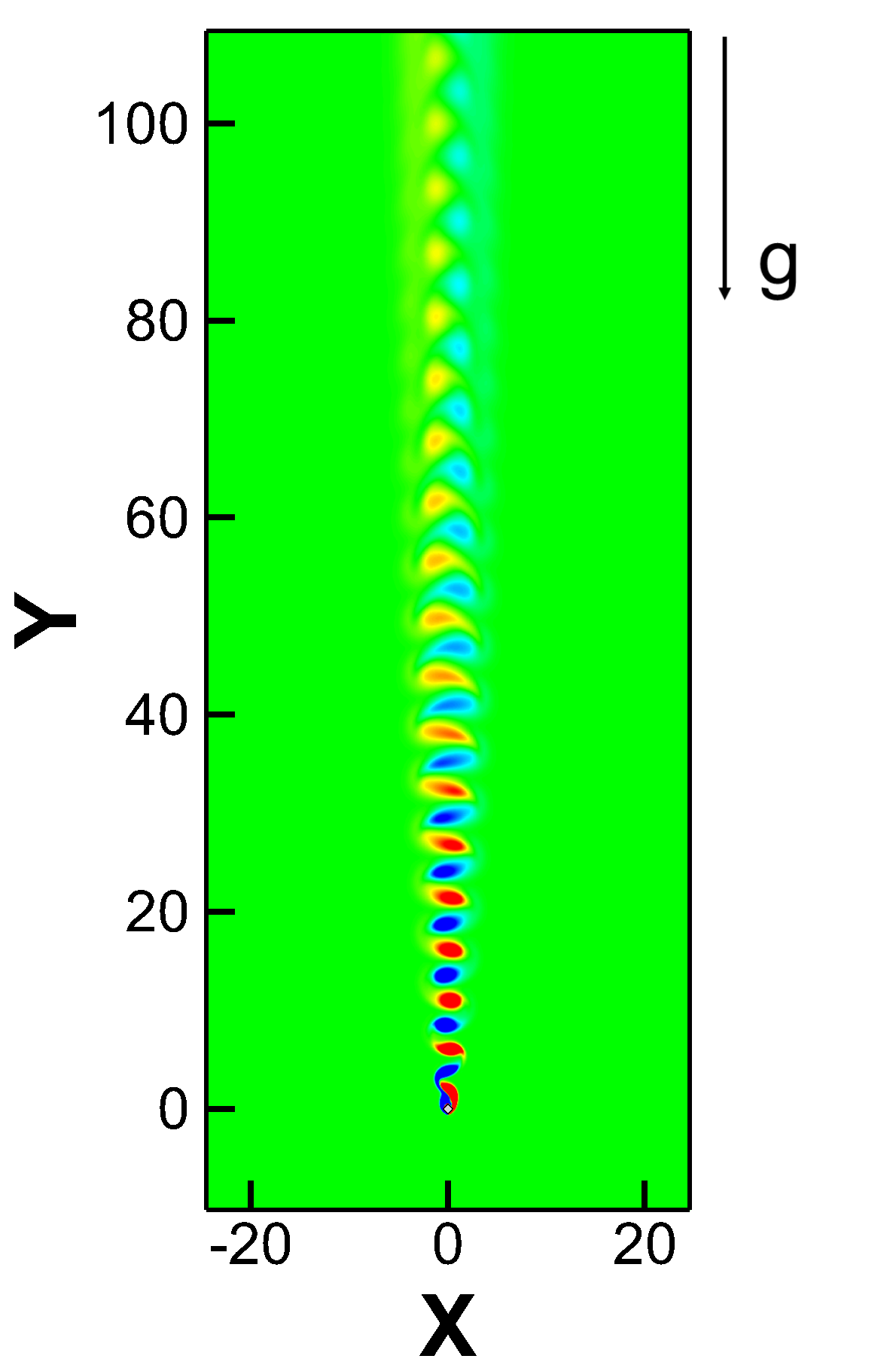}
		\caption{Ri $=0.2$}
		\label{insvort02}
	\end{subfigure}\hfill
	\begin{subfigure}[t]{0.3\textwidth}
		\includegraphics[width=\linewidth]{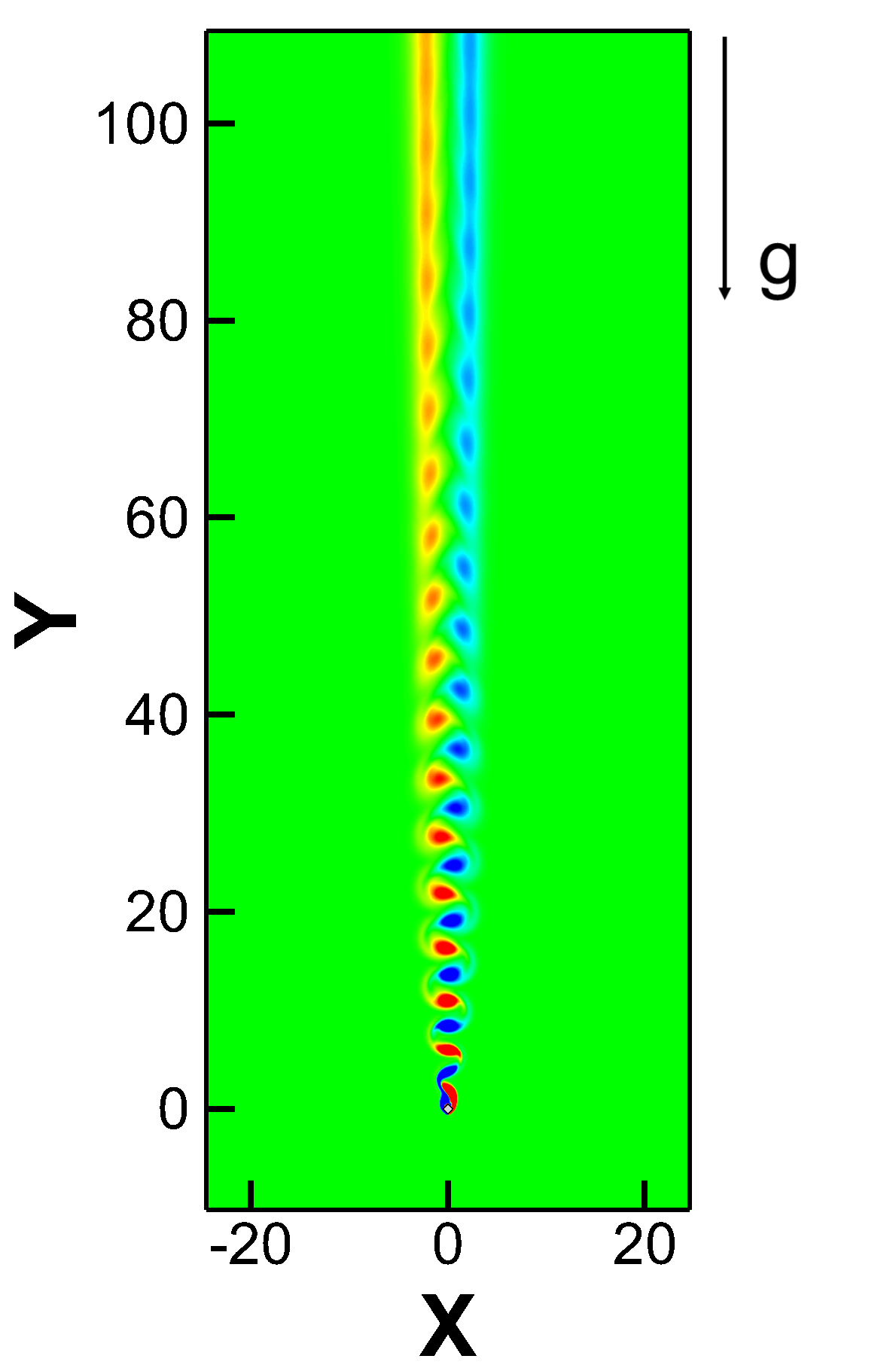}
		\caption{Ri $=0.4$}
		\label{insvort04}
	\end{subfigure}\hfill

	\begin{subfigure}[t]{0.3\textwidth}
		\includegraphics[width=\textwidth]{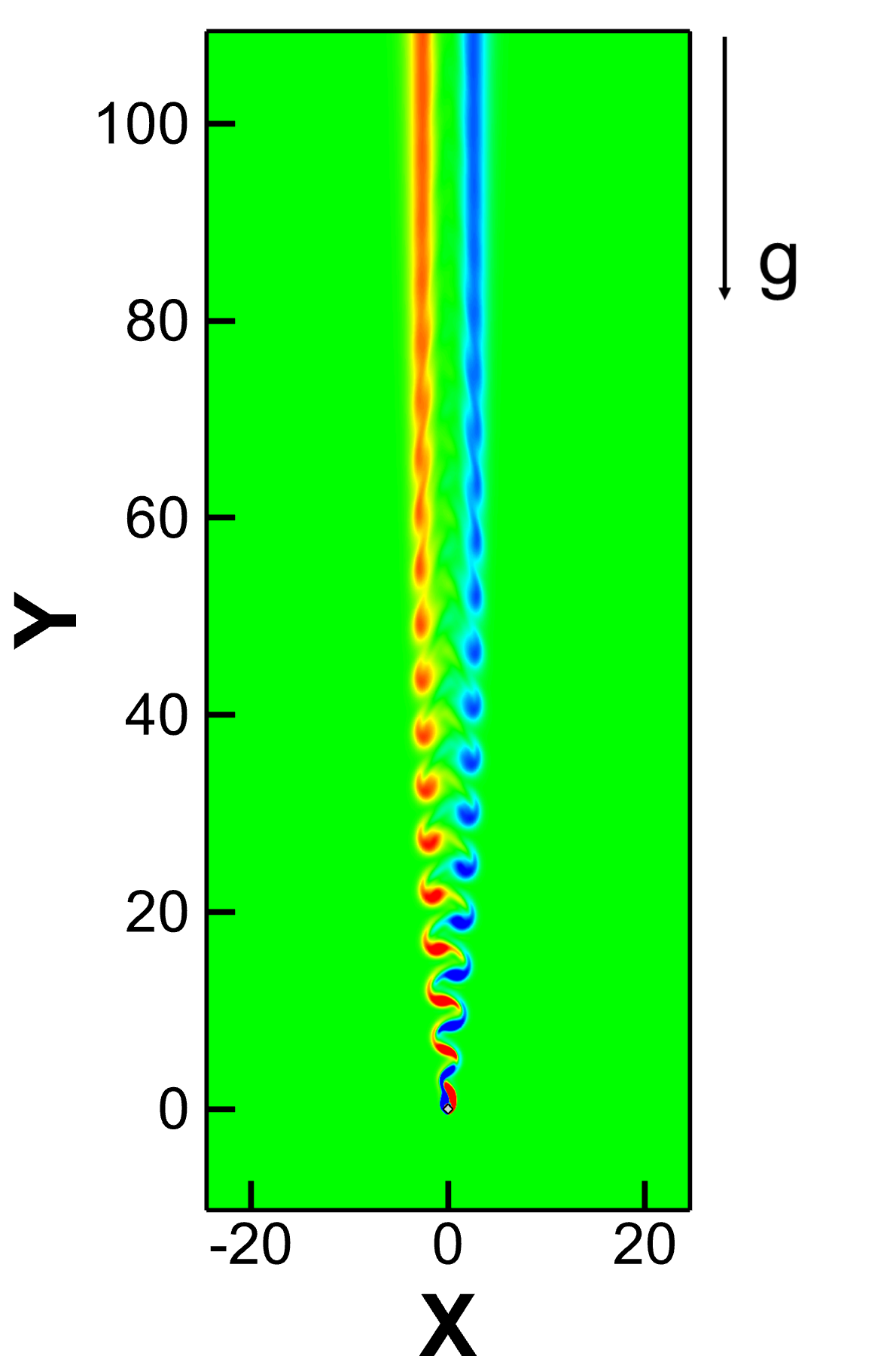}
		\caption{Ri $=0.6$}
		\label{insvort06}
	\end{subfigure}\hfill
	\begin{subfigure}[t]{0.3\textwidth}
		\includegraphics[width=\linewidth]{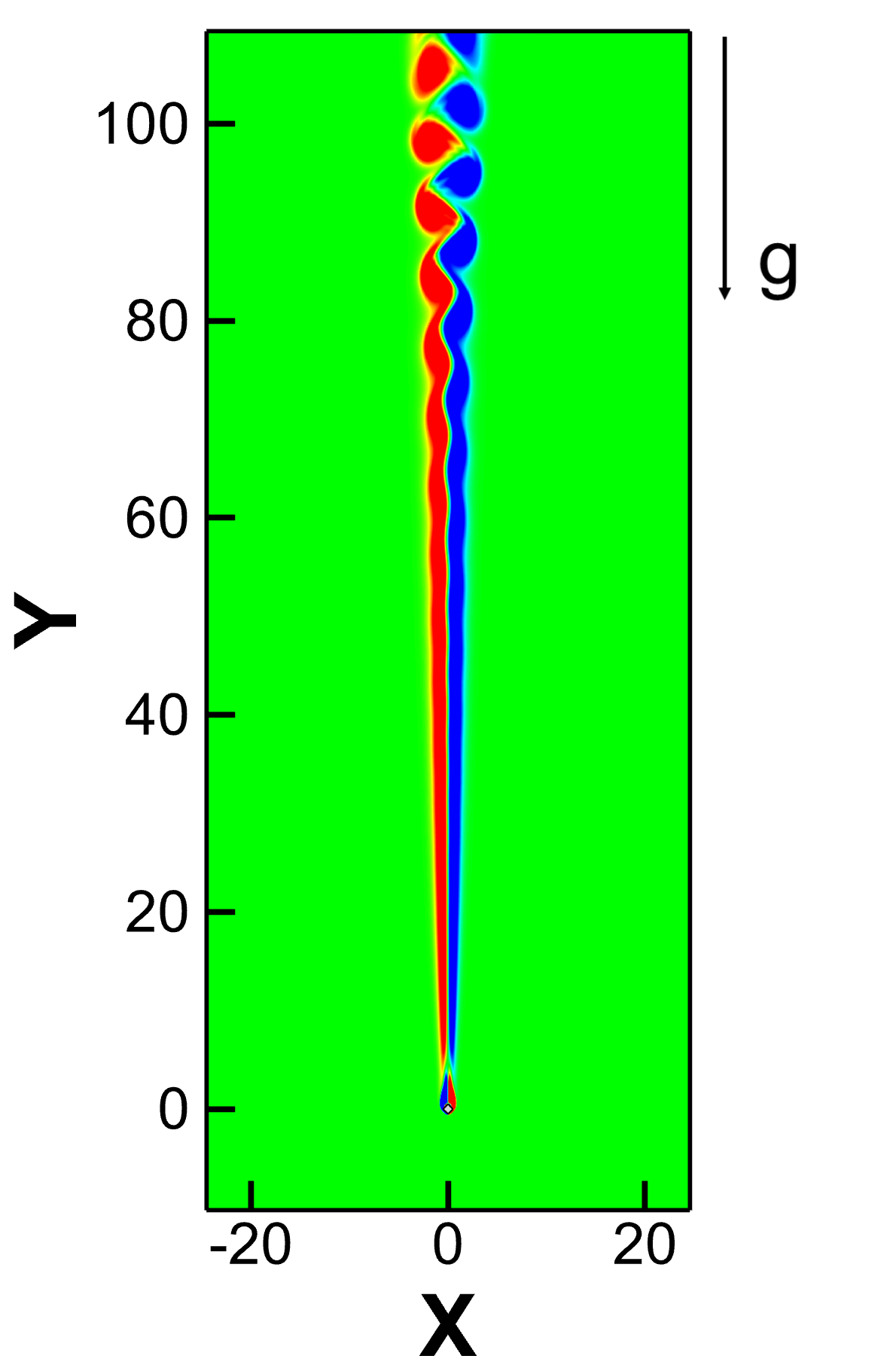}
		\caption{Ri $=0.7$}
		\label{insvort07}
	\end{subfigure}\hfill
	\begin{subfigure}[t]{0.3\textwidth}
		\includegraphics[width=\linewidth]{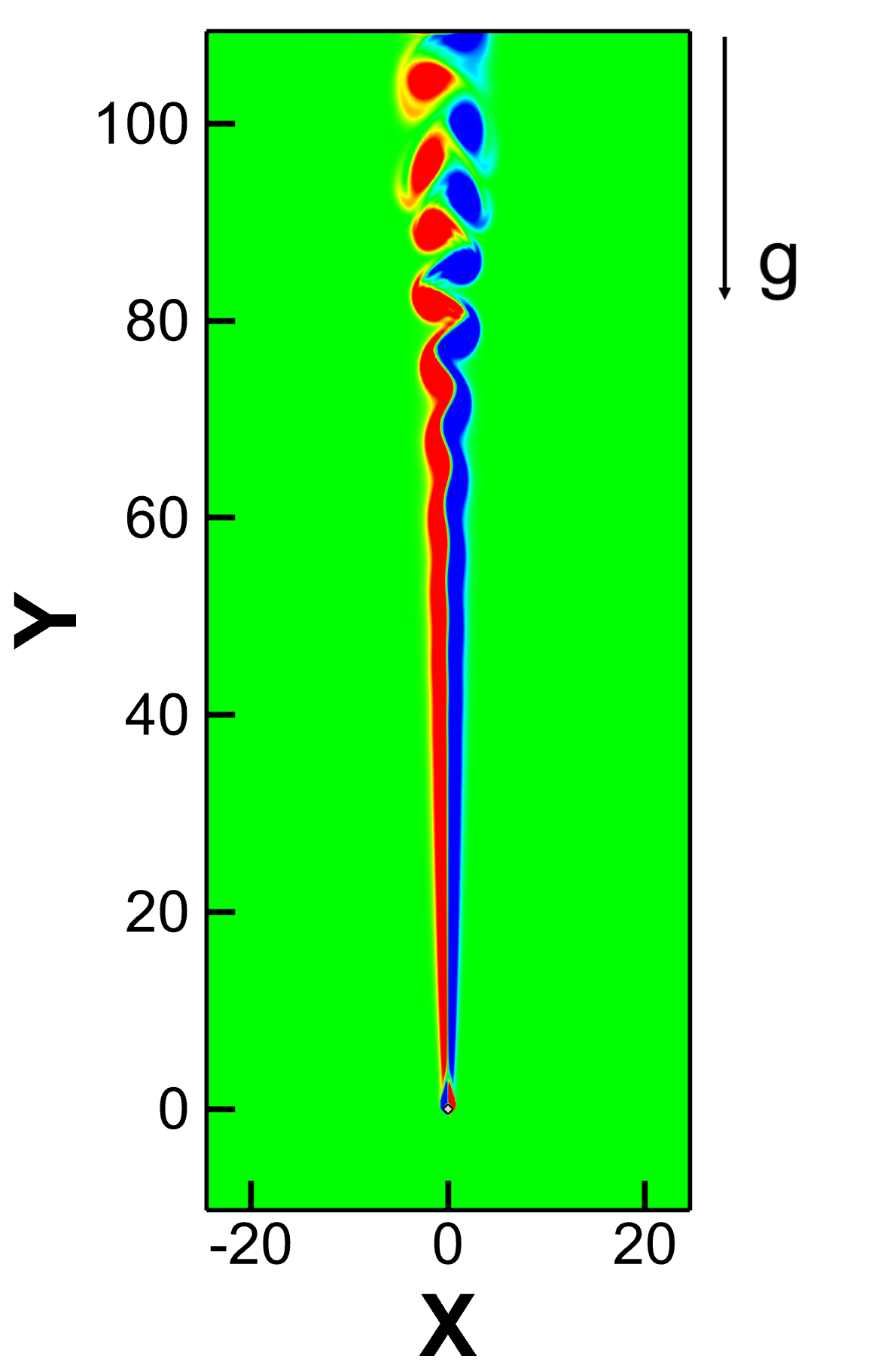}
		\caption{Ri $=1.0$}
		\label{insvort10}
	\end{subfigure}\hfill
	
	\caption{Instantaneous vorticity contours in the complete domain for Re $=100$ and $0.0\leq$ Ri $\leq1.0$: (a) Ri $=0.0$, (b) Ri $=0.2$, (c) Ri $=0.4$, (d) Ri $=0.6$, (e) Ri $=0.7$, (f) Ri $=1.0$; ($\omega_{min},\omega_{max},\Delta\omega_{}) \equiv (-0.5,0.5,0.01)$.}
	\label{insvort}
\end{figure}

\begin{figure}[htbp]
	\centering
	\begin{subfigure}[t]{0.3\textwidth}
		\includegraphics[width=\linewidth]{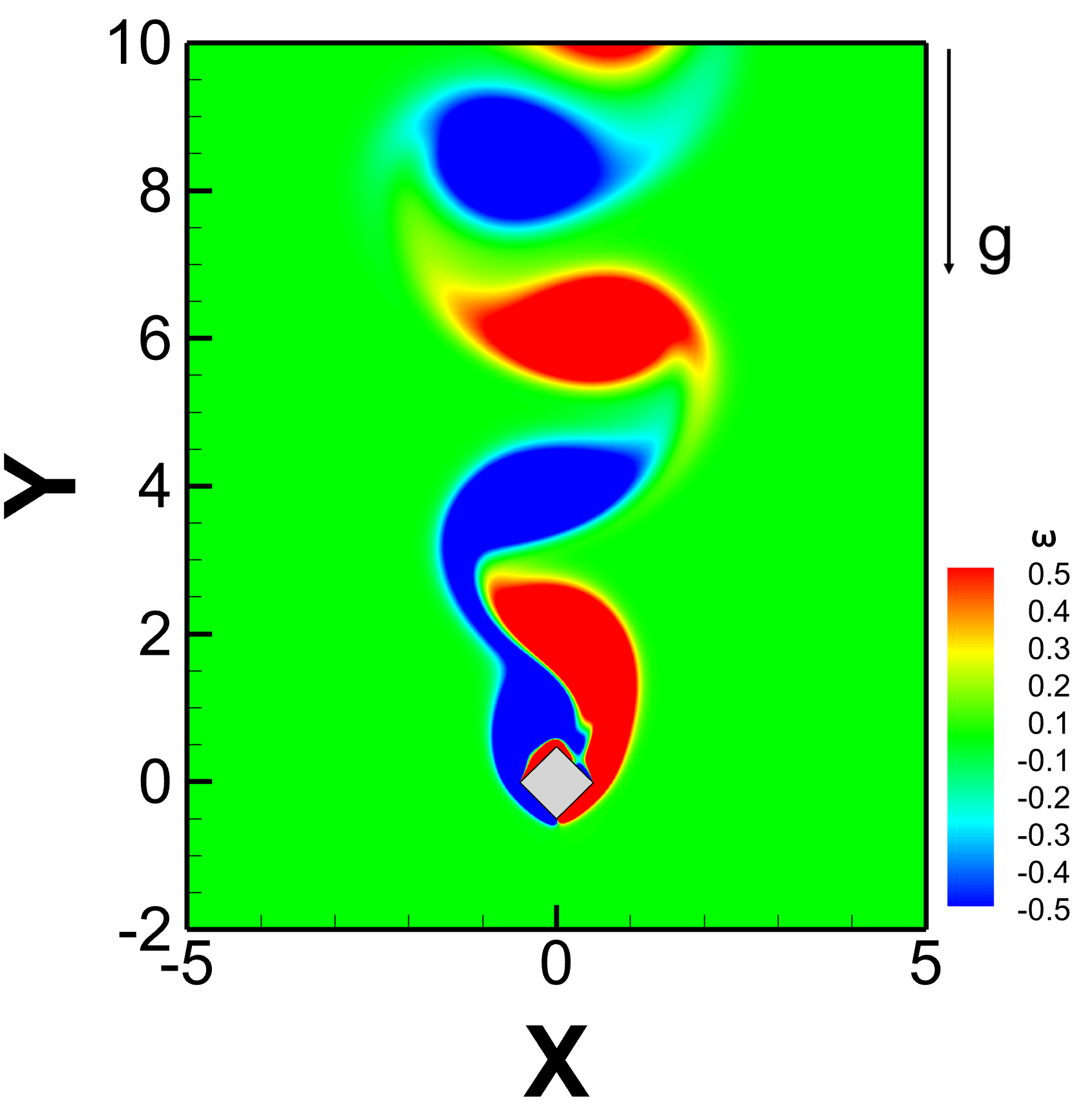}
		\caption{Ri $=0.0$}
		\label{insnearvort0}
	\end{subfigure}\hfill
	\begin{subfigure}[t]{0.3\textwidth}
		\includegraphics[width=\linewidth]{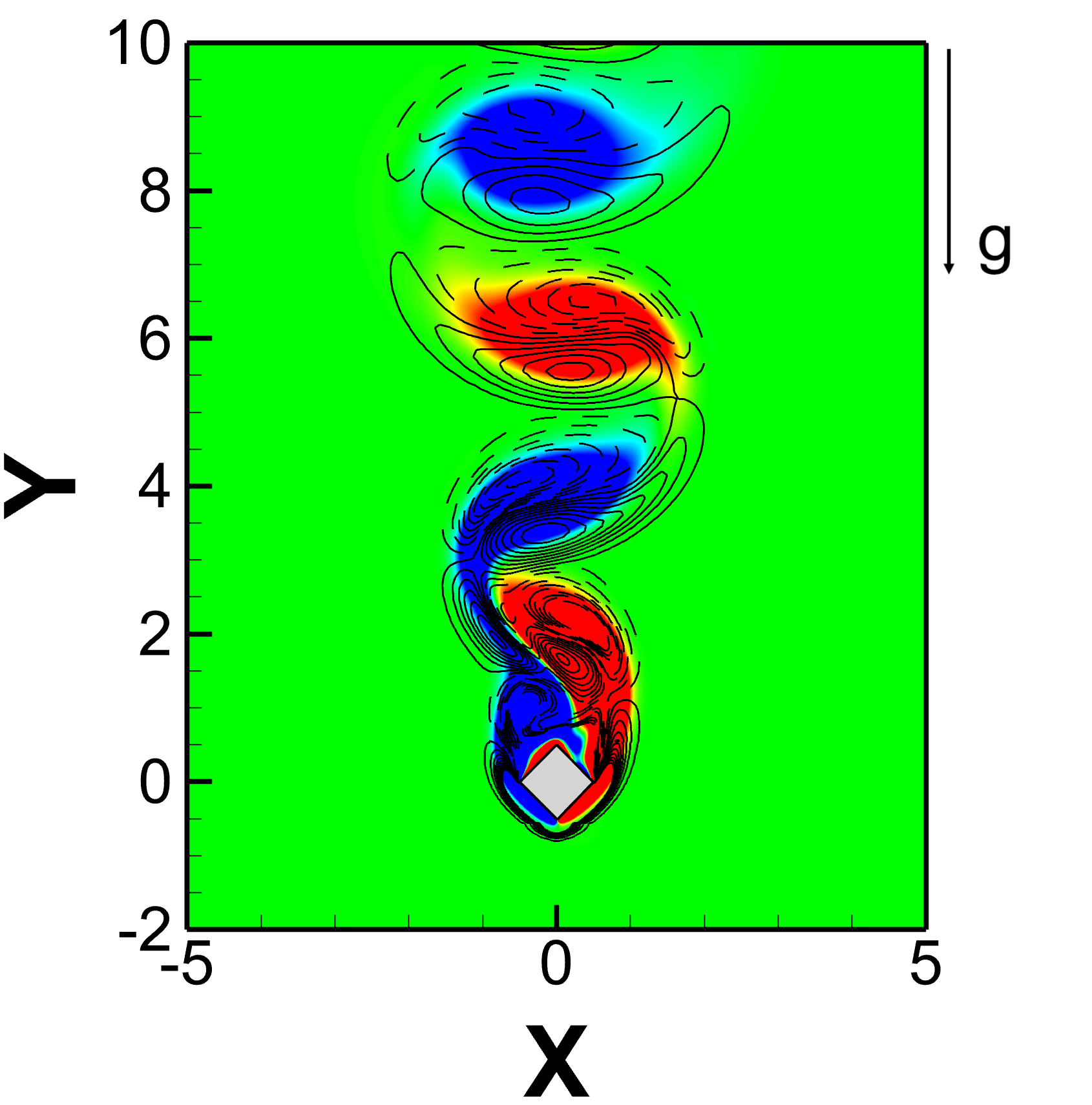}
		\caption{Ri $=0.2$}
		\label{insnearvort02}
	\end{subfigure}\hfill
	\begin{subfigure}[t]{0.3\textwidth}
		\includegraphics[width=\linewidth]{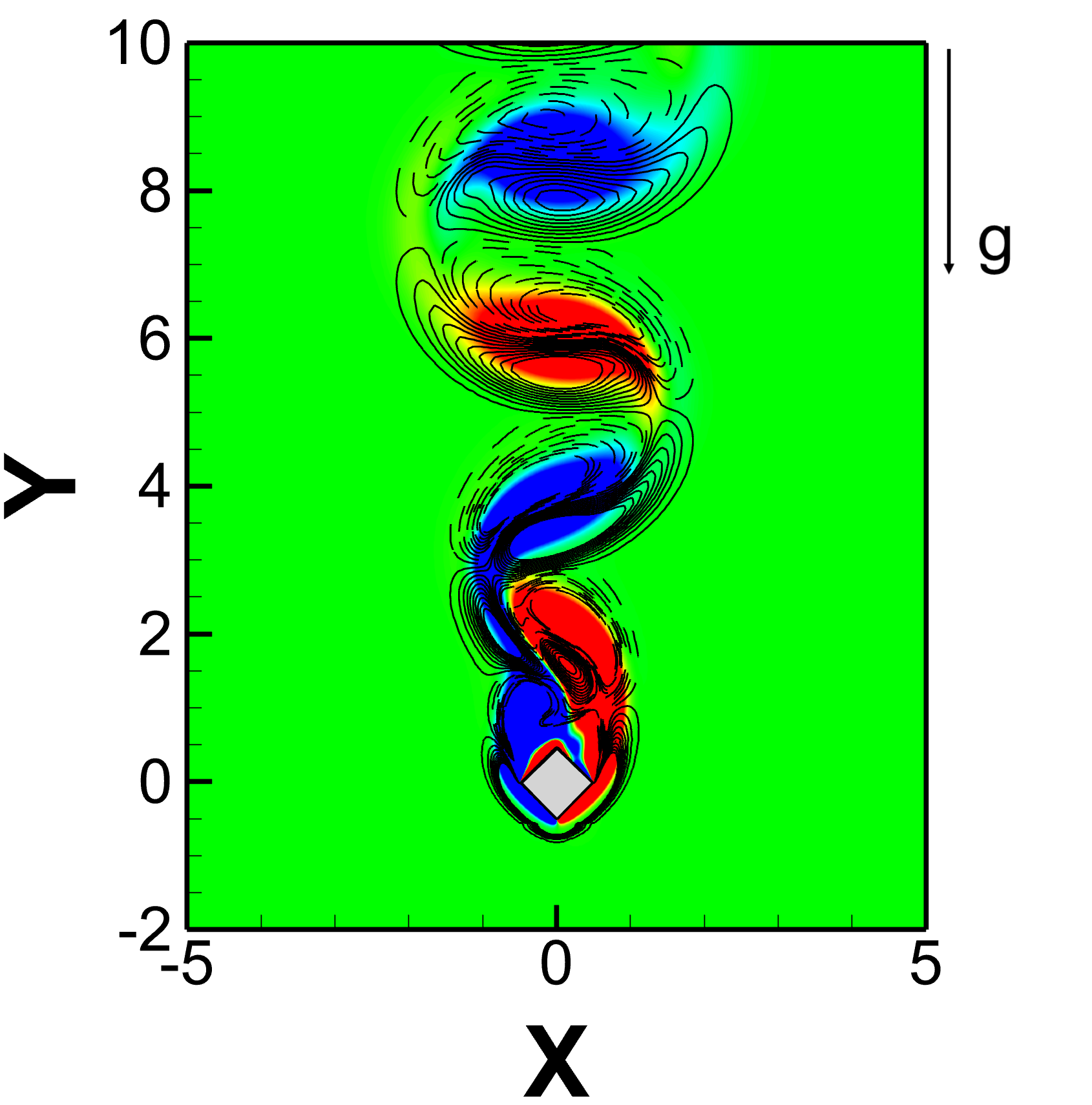}
		\caption{Ri $=0.4$}
		\label{insnearvort04}
	\end{subfigure}\hfill

	\begin{subfigure}[t]{0.3\textwidth}
		\includegraphics[width=\textwidth]{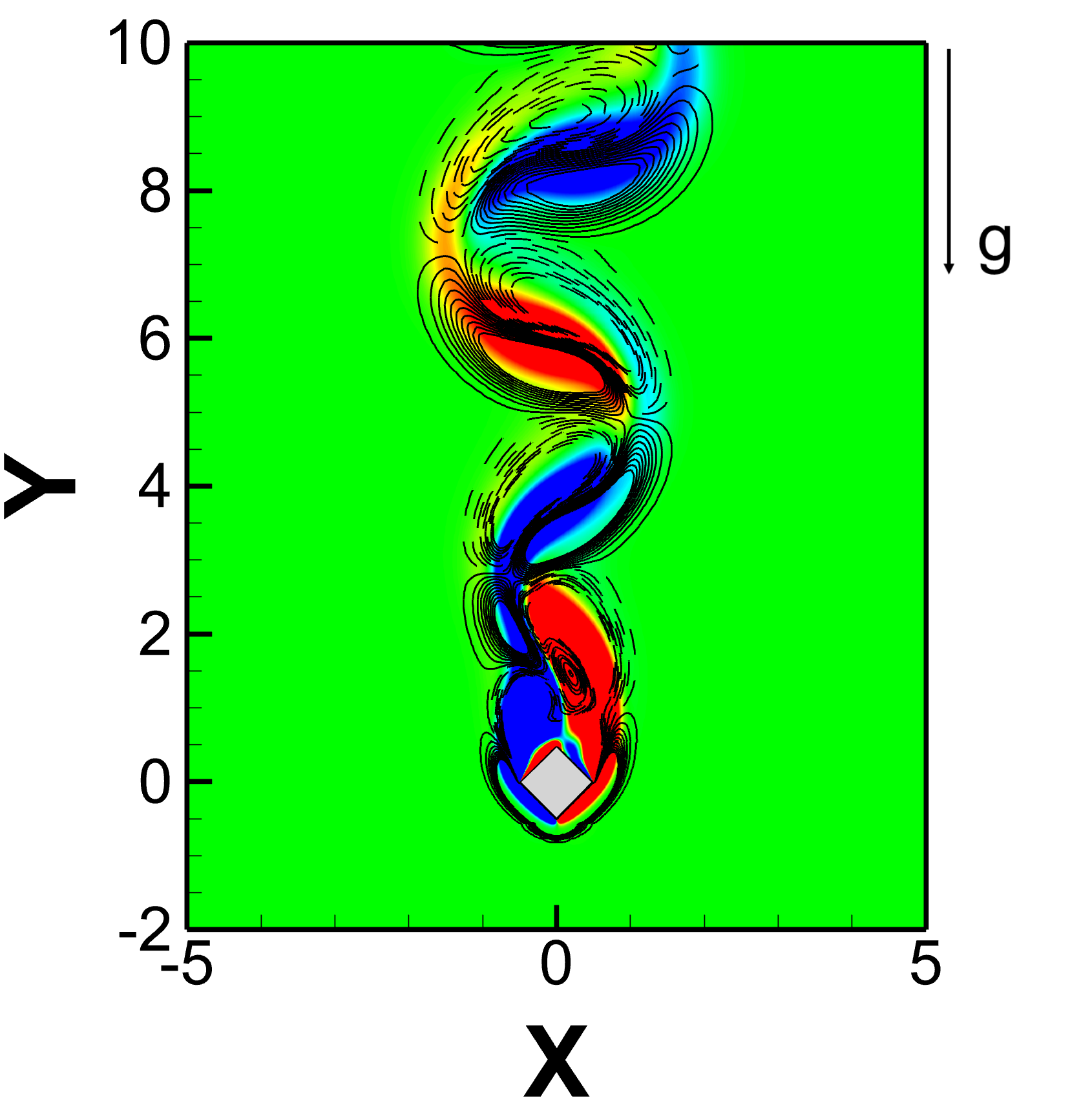}
		\caption{Ri $=0.6$}
		\label{insnearvort06}
	\end{subfigure}\hfill
	\begin{subfigure}[t]{0.3\textwidth}
		\includegraphics[width=\linewidth]{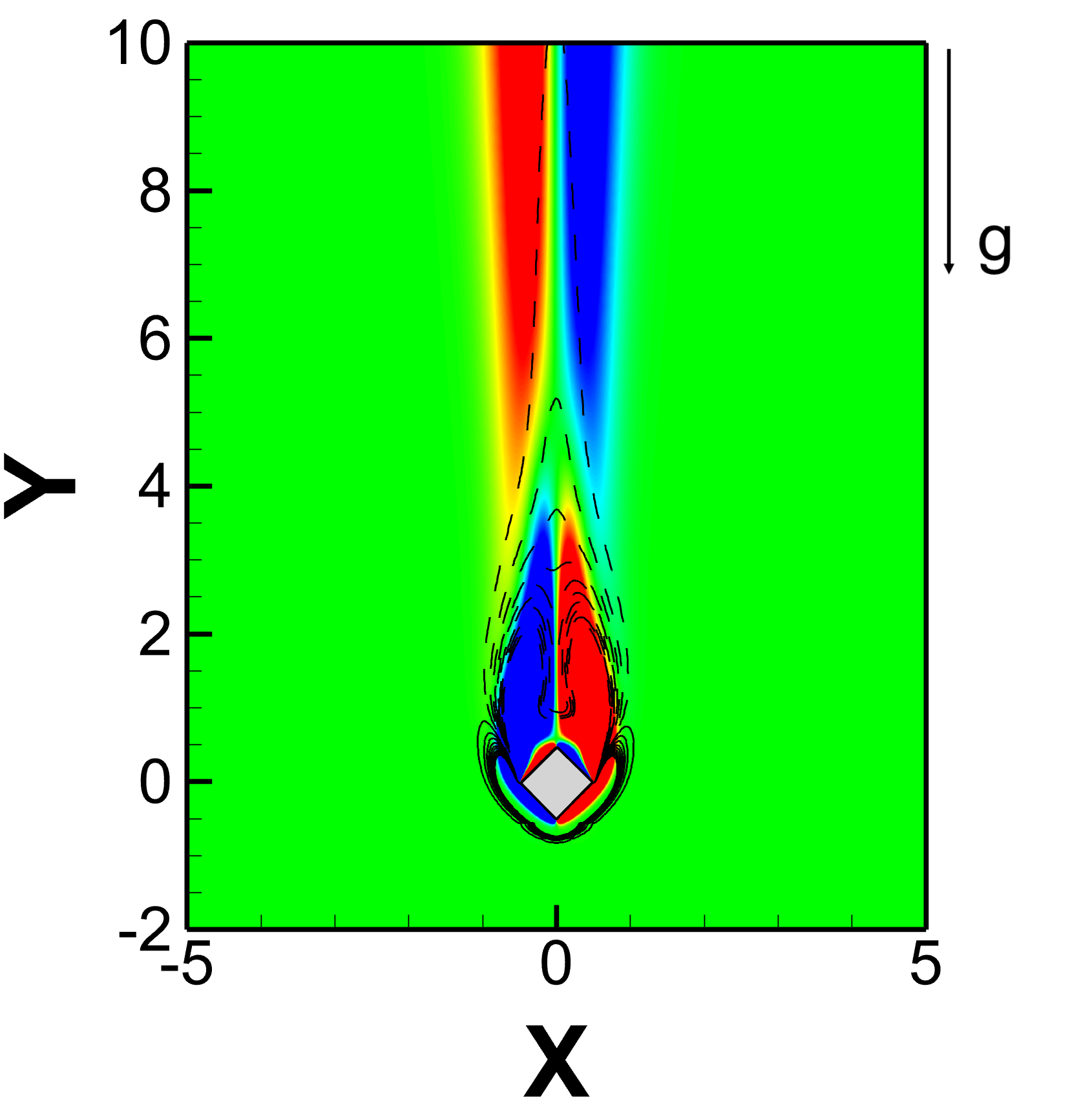}
		\caption{Ri $=0.7$}
		\label{insnearvort07}
	\end{subfigure}\hfill
	\begin{subfigure}[t]{0.3\textwidth}
		\includegraphics[width=\linewidth]{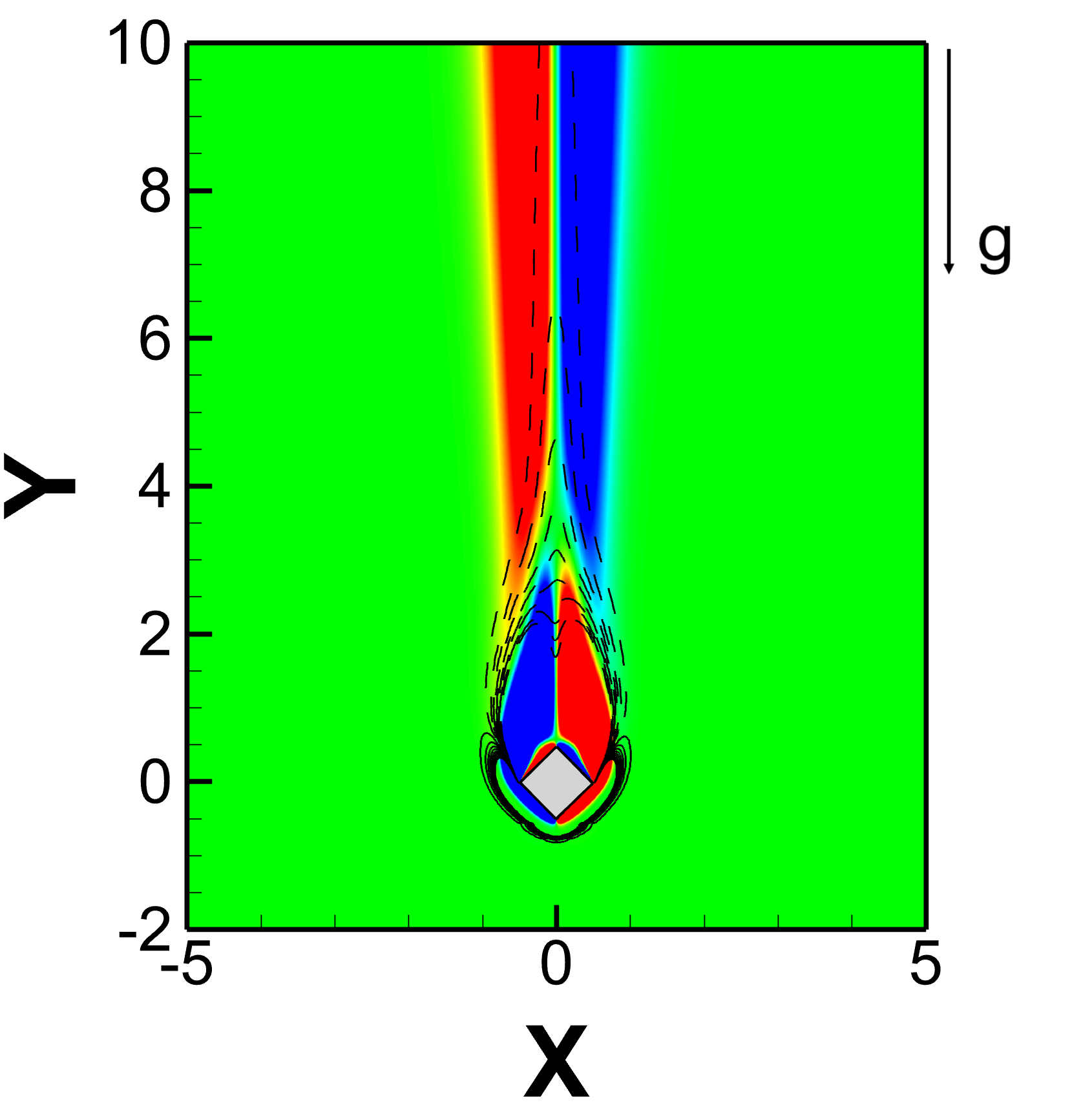}
		\caption{Ri $=1.0$}
		\label{insnearvort10}
	\end{subfigure}\hfill
	
	\caption{Instantaneous vorticity contours in the near-field for Re $=100$ and $0.0\leq$ Ri $\leq1.0$: (a) Ri $=0.0$, (b) Ri $=0.2$, (c) Ri $=0.4$, (d) Ri $=0.6$, (e) Ri $=0.7$, (f) Ri $=1.0$; ($\omega_{min},\omega_{max},\Delta\omega_{}) \equiv (-0.5,0.5,0.01)$, ($\omega_{baro,min},\omega_{baro,max},\Delta\omega_{baro}) \equiv (-0.1,0.1,0.01)$. The flood represents total vorticity, while the lines represent baroclinic vorticity.}
	\label{insnearvort}
\end{figure}

\begin{figure}[htbp]
\centering
\begin{subfigure}[t]{0.3\textwidth}
    \includegraphics[width=\linewidth]{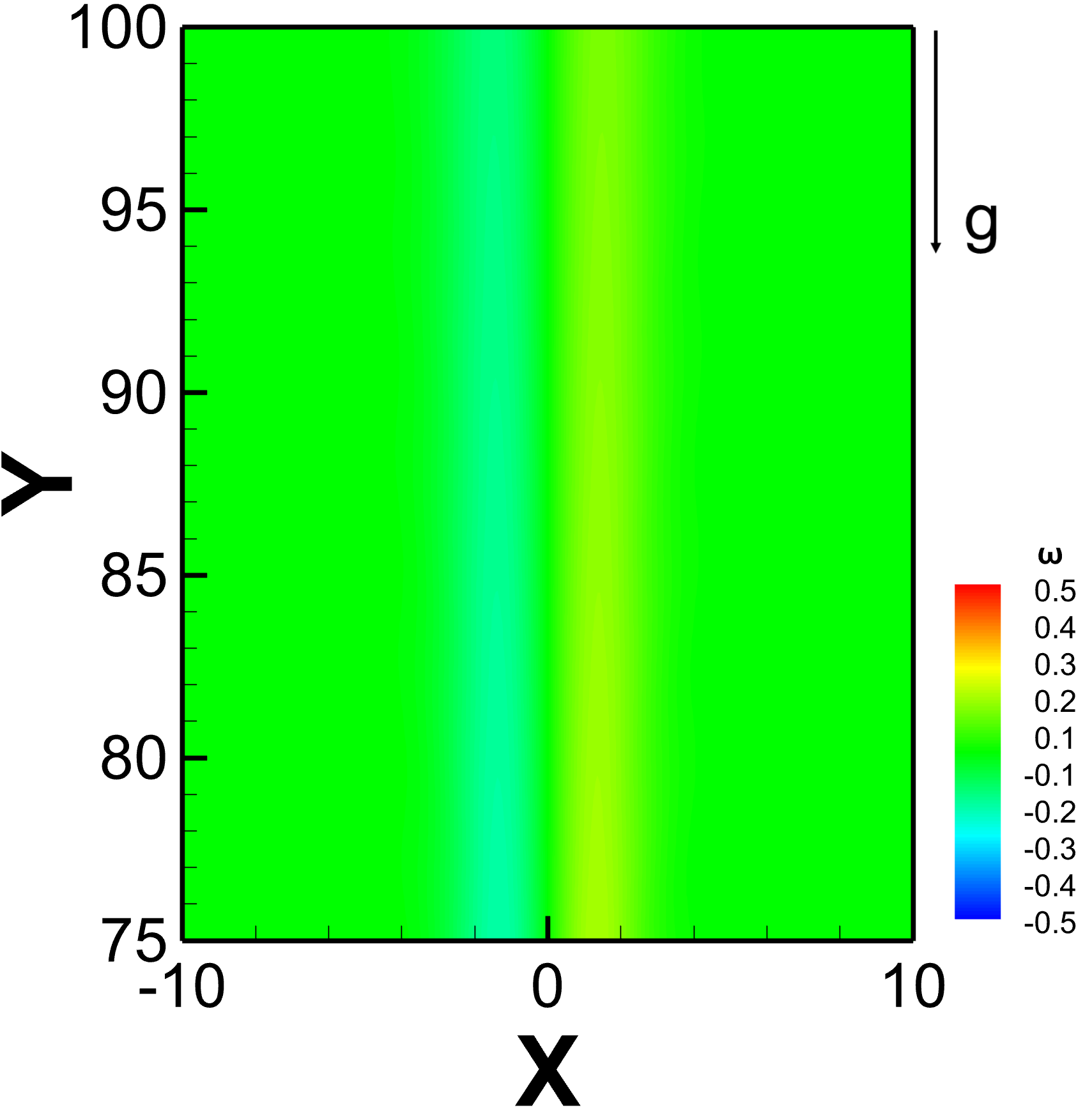}
\caption{Ri $=0.0$}
\label{insfarvort0}
\end{subfigure}\hfill
\begin{subfigure}[t]{0.3\textwidth}
  \includegraphics[width=\linewidth]{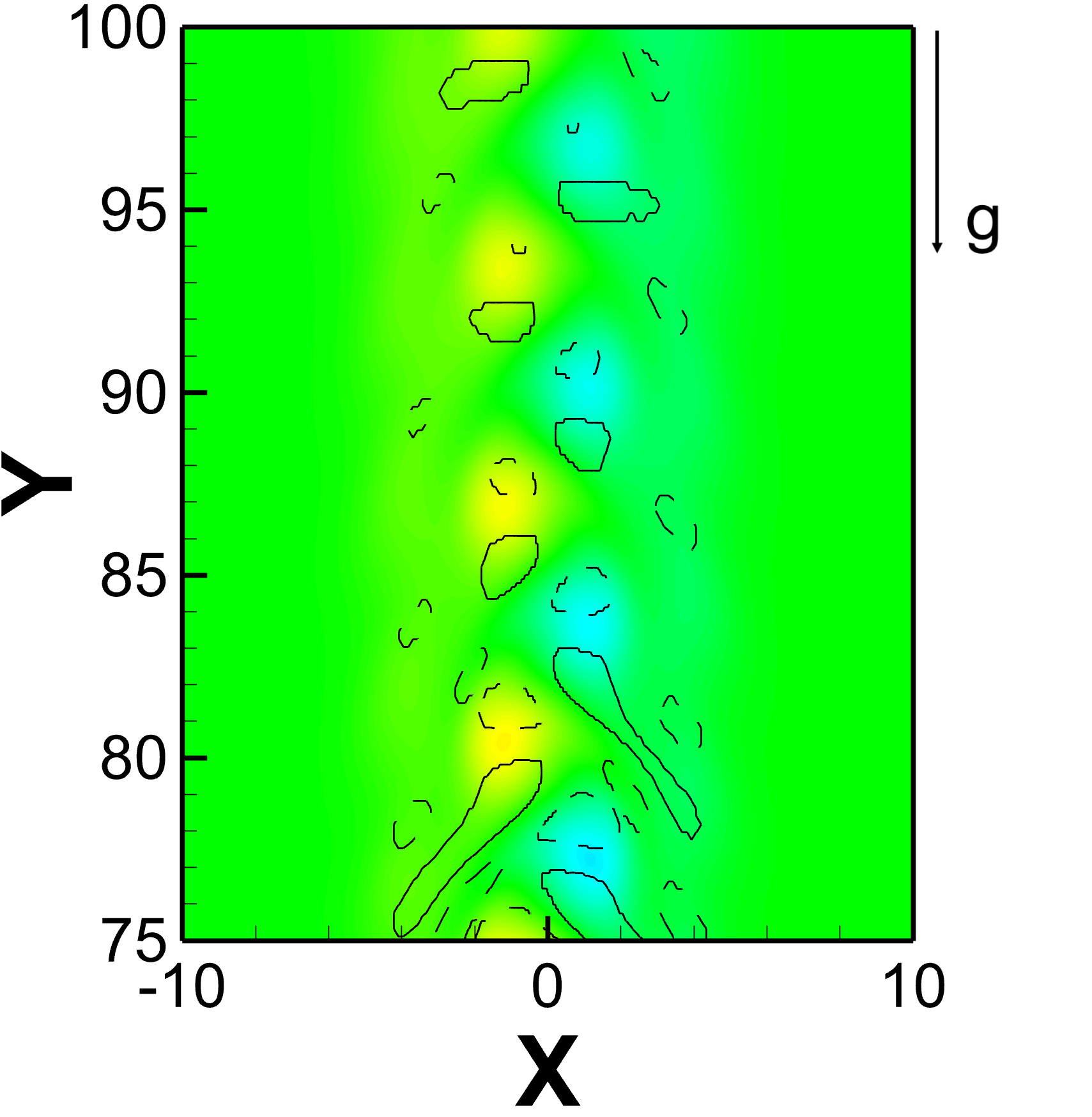}
\caption{Ri $=0.2$}
\label{insfarvort02}
\end{subfigure}\hfill
\begin{subfigure}[t]{0.3\textwidth}
    \includegraphics[width=\linewidth]{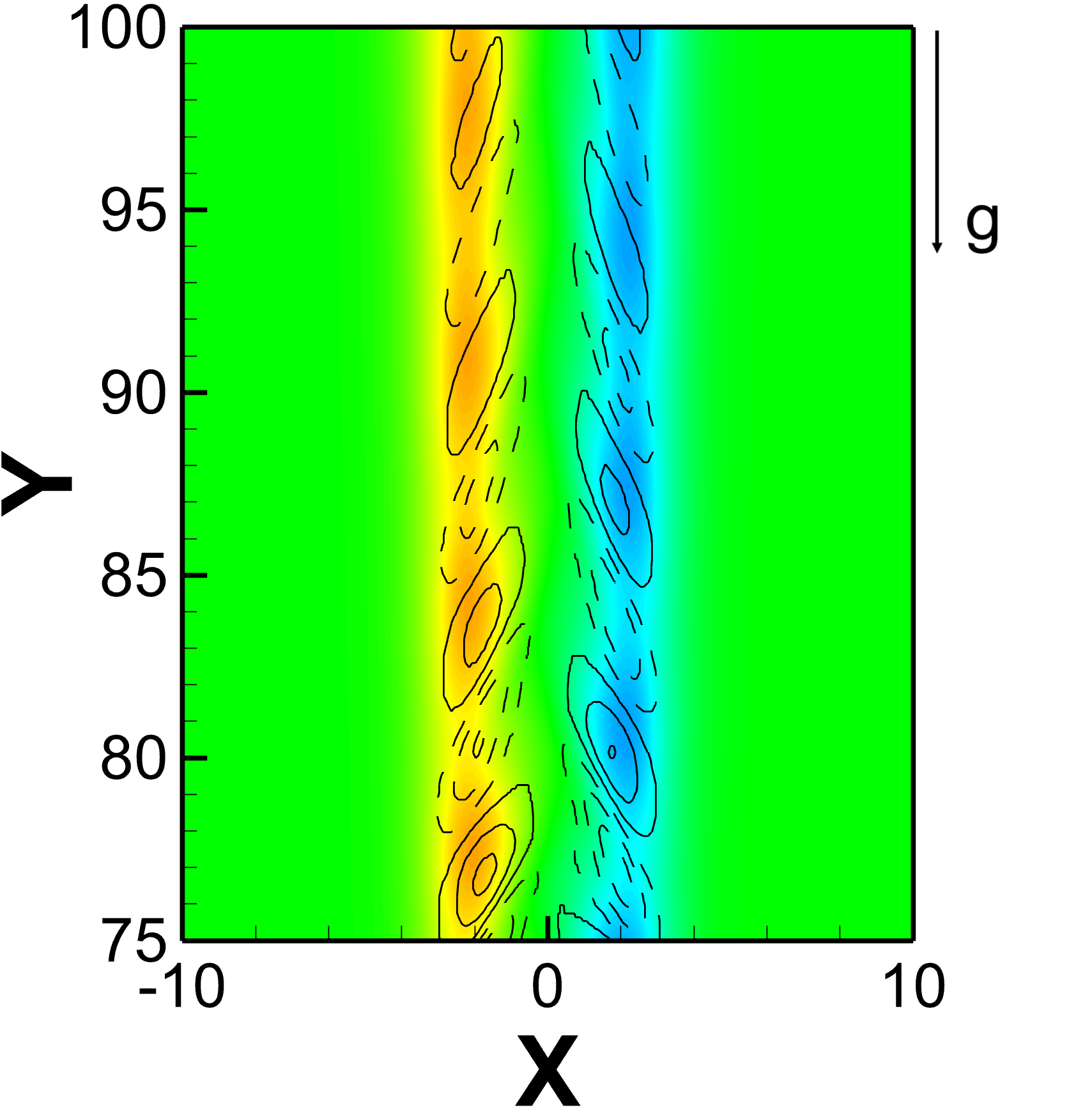}
\caption{Ri $=0.4$}
\label{insfarvort04}
\end{subfigure}\hfill

\begin{subfigure}[t]{0.3\textwidth}
    \includegraphics[width=\textwidth]{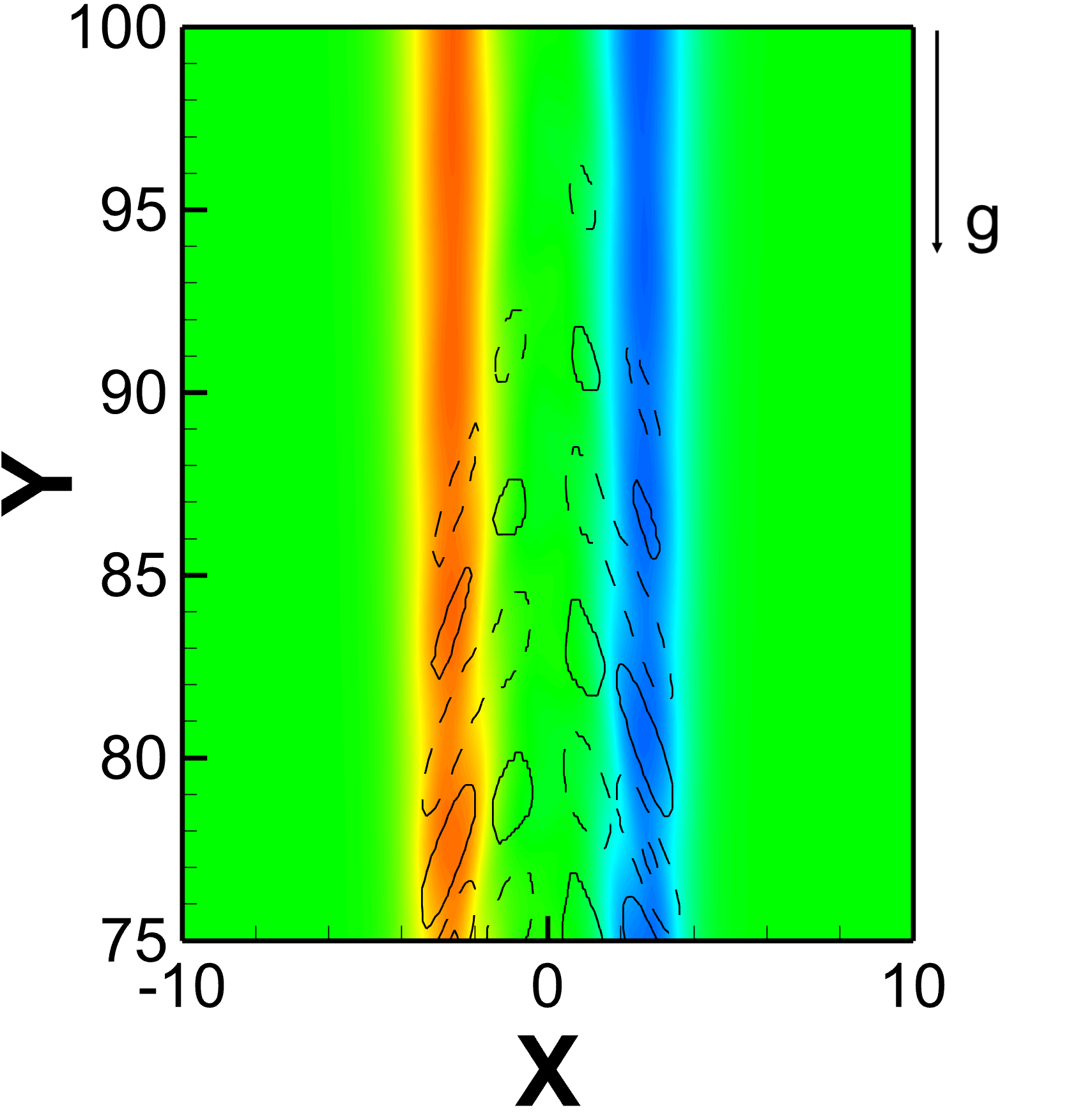}
\caption{Ri $=0.6$}
\label{insfarvort06}
\end{subfigure}\hfill
\begin{subfigure}[t]{0.3\textwidth}
    \includegraphics[width=\linewidth]{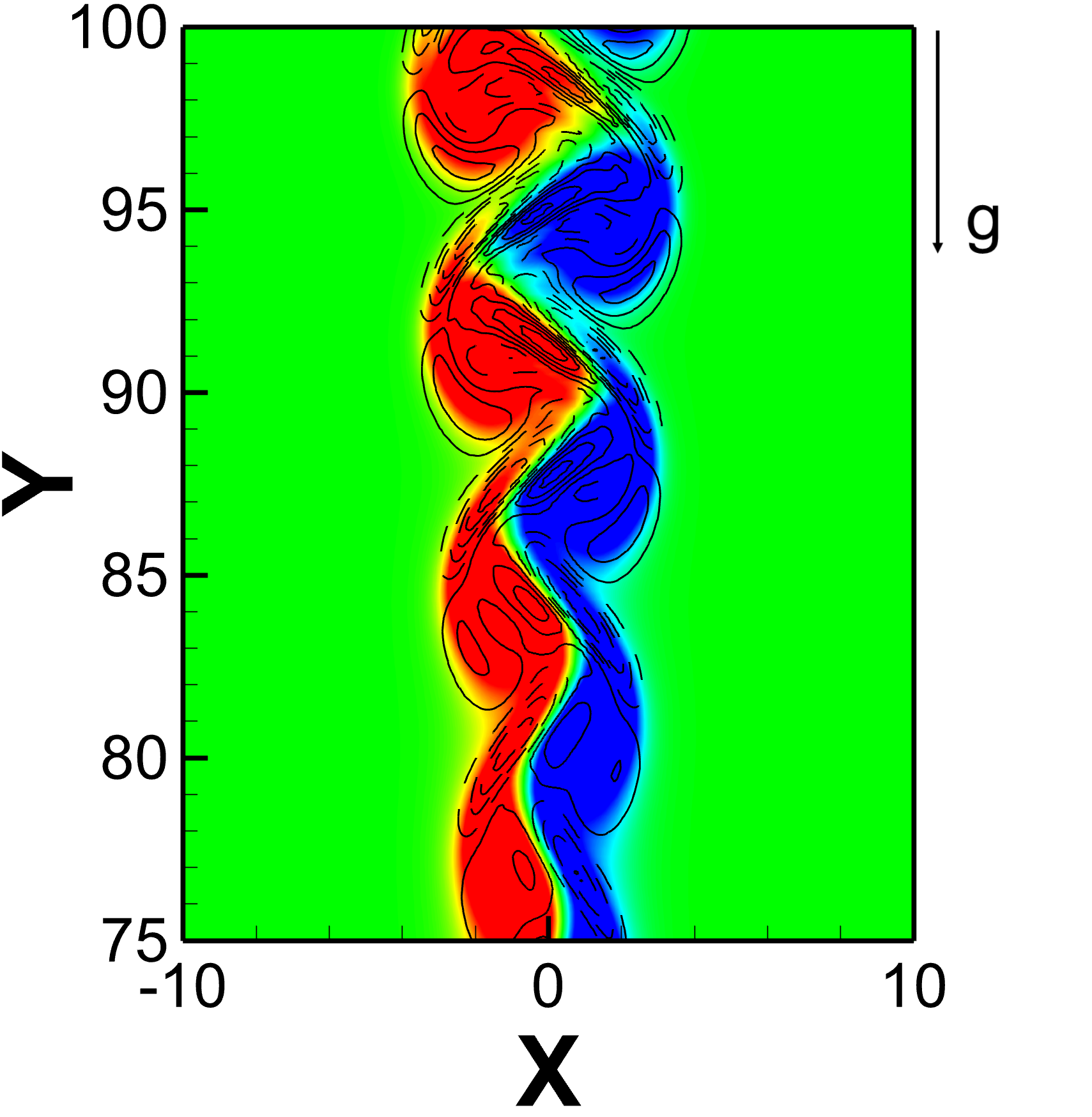}
\caption{Ri $=0.7$}
\label{insfarvort07}
\end{subfigure}\hfill
\begin{subfigure}[t]{0.3\textwidth}
    \includegraphics[width=\linewidth]{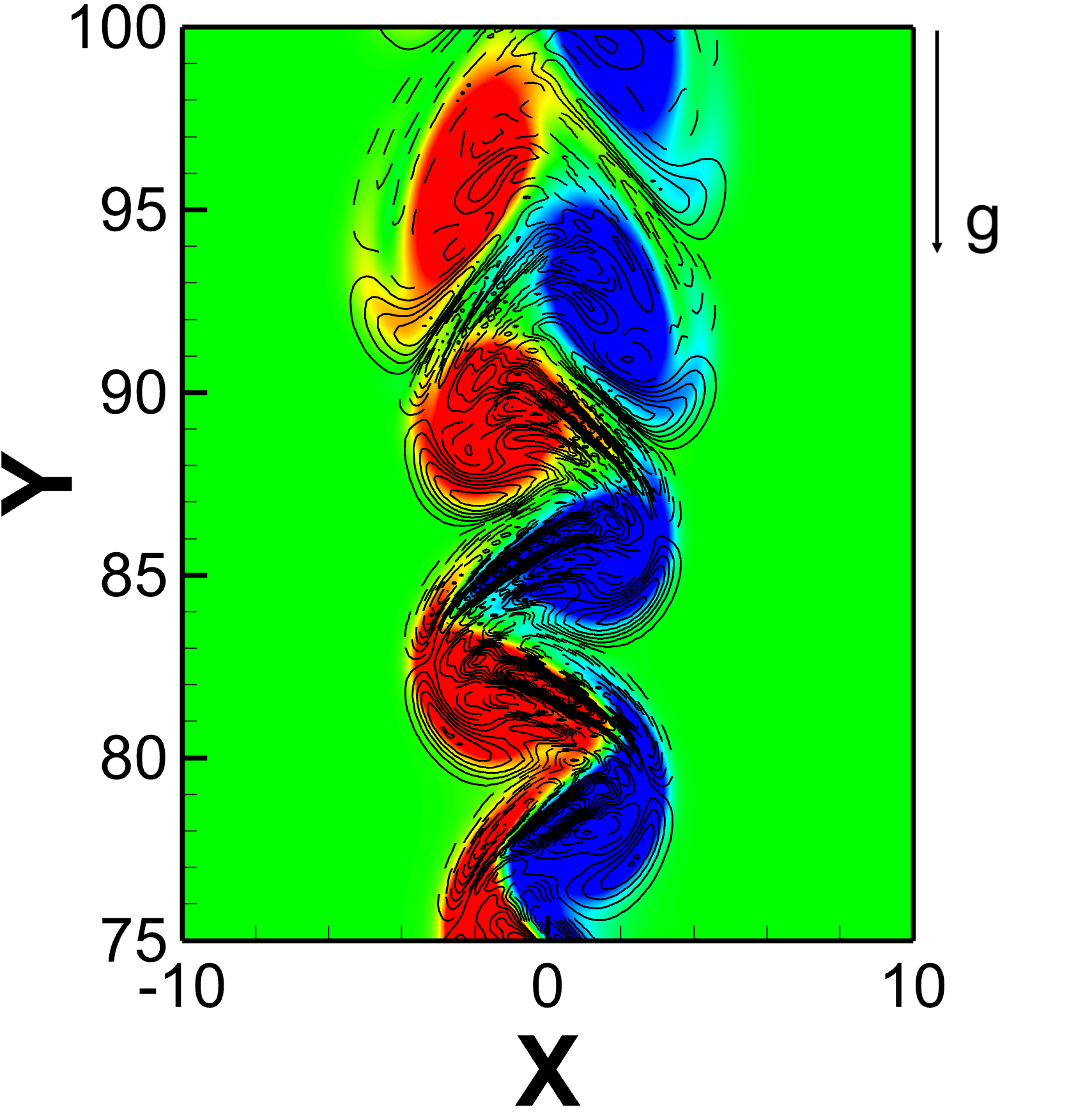}
\caption{Ri $=1.0$}
\label{insfarvort10}
\end{subfigure}\hfill
\caption{Instantaneous vorticity contours in the far-field for Re $=100$ and $0.0\leq$ Ri $\leq1.0$: (a) Ri $=0.0$, (b) Ri $=0.2$, (c) Ri $=0.4$, (d) Ri $=0.6$, (e) Ri $=0.7$, (f) Ri $=1.0$; ($\omega_{min},\omega_{max},\Delta\omega_{}) \equiv (-0.5,0.5,0.01)$, ($\omega_{baro,min},\omega_{baro,max},\Delta\omega_{baro}) \equiv (-0.05,0.05,0.005)$. The flood represents total vorticity while the lines represent baroclinic vorticity.}
\label{insfarvort}
\end{figure}

\begin{figure}[htbp]
\centering
\begin{subfigure}[t]{0.3\textwidth}
    \includegraphics[width=\linewidth]{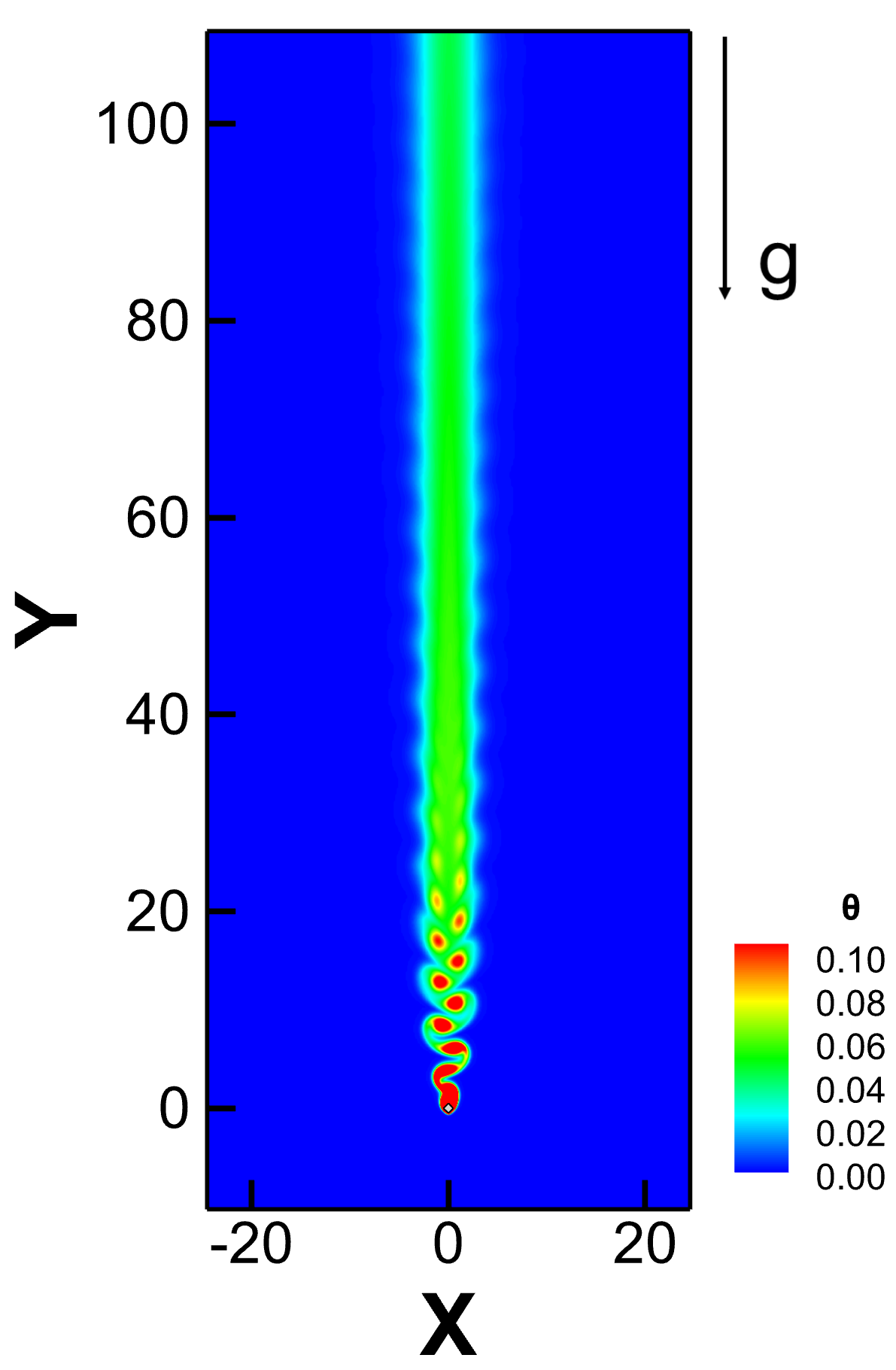}
\caption{Ri $=0.0$}
\label{instemp0}
\end{subfigure}\hfill
\begin{subfigure}[t]{0.3\textwidth}
  \includegraphics[width=\linewidth]{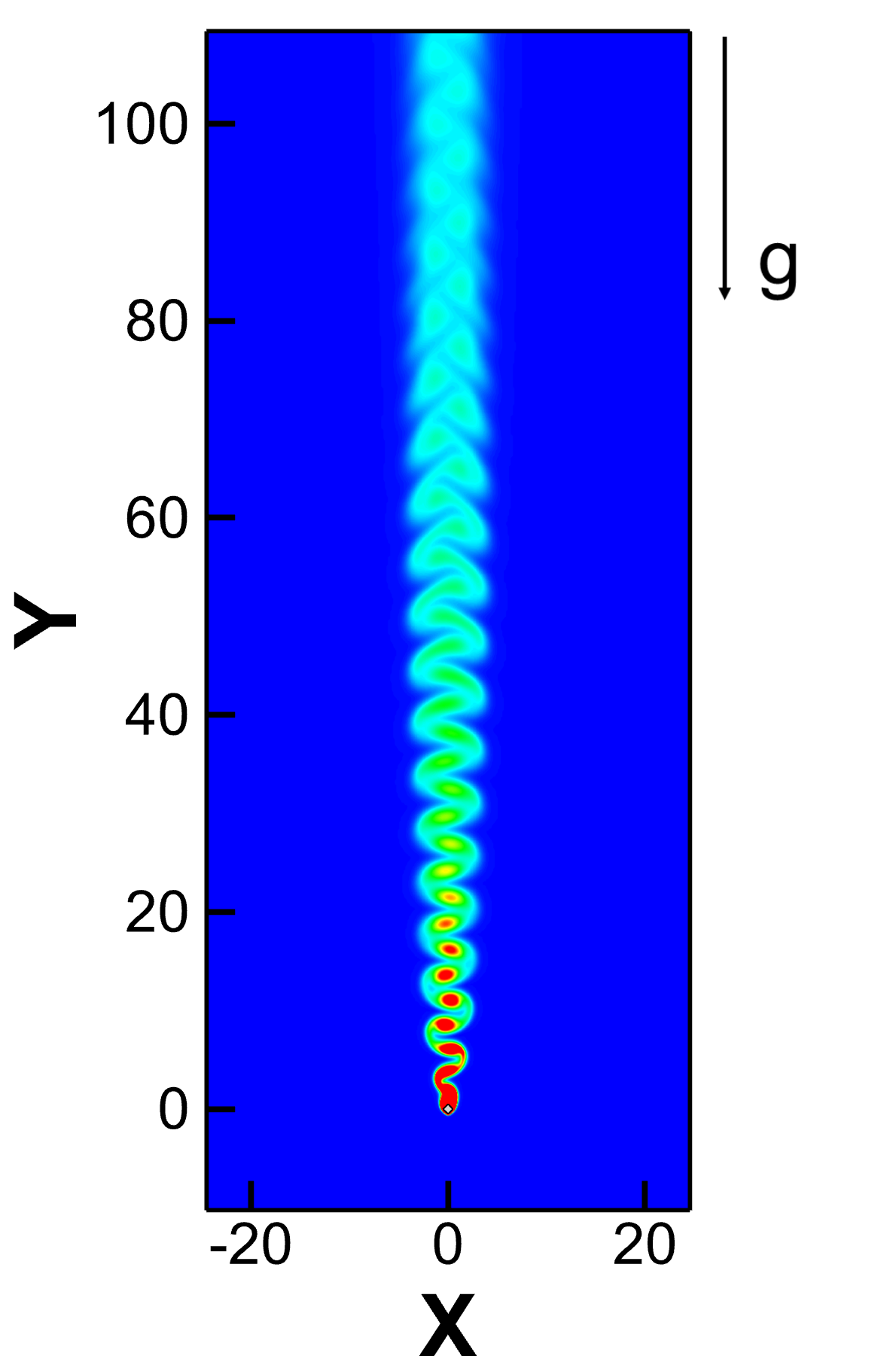}
\caption{Ri $=0.2$}
\label{instemp2}
\end{subfigure}\hfill
\begin{subfigure}[t]{0.3\textwidth}
    \includegraphics[width=\linewidth]{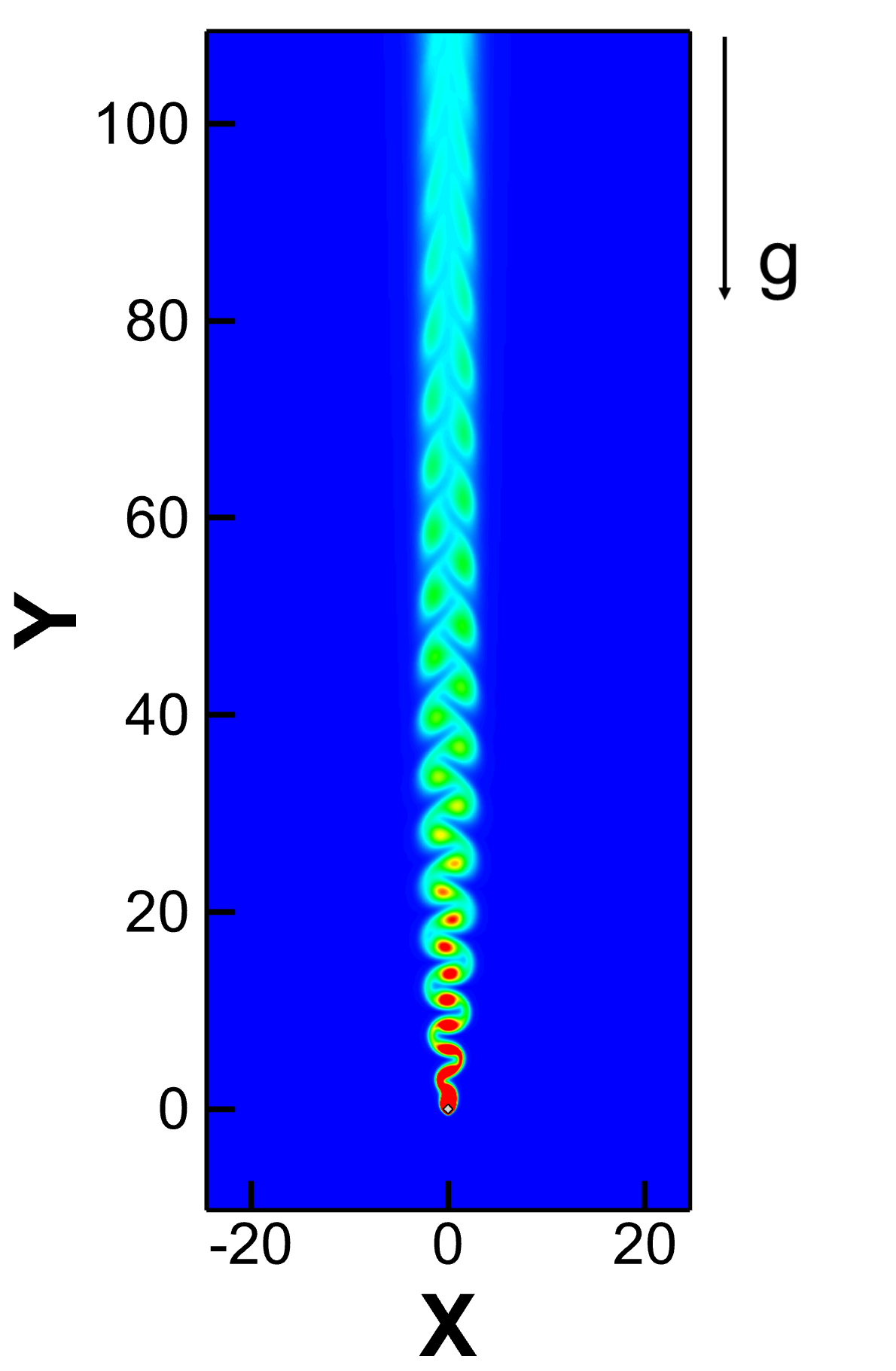}
\caption{Ri $=0.4$}
\label{instemp4}
\end{subfigure}\hfill

\begin{subfigure}[t]{0.3\textwidth}
    \includegraphics[width=\textwidth]{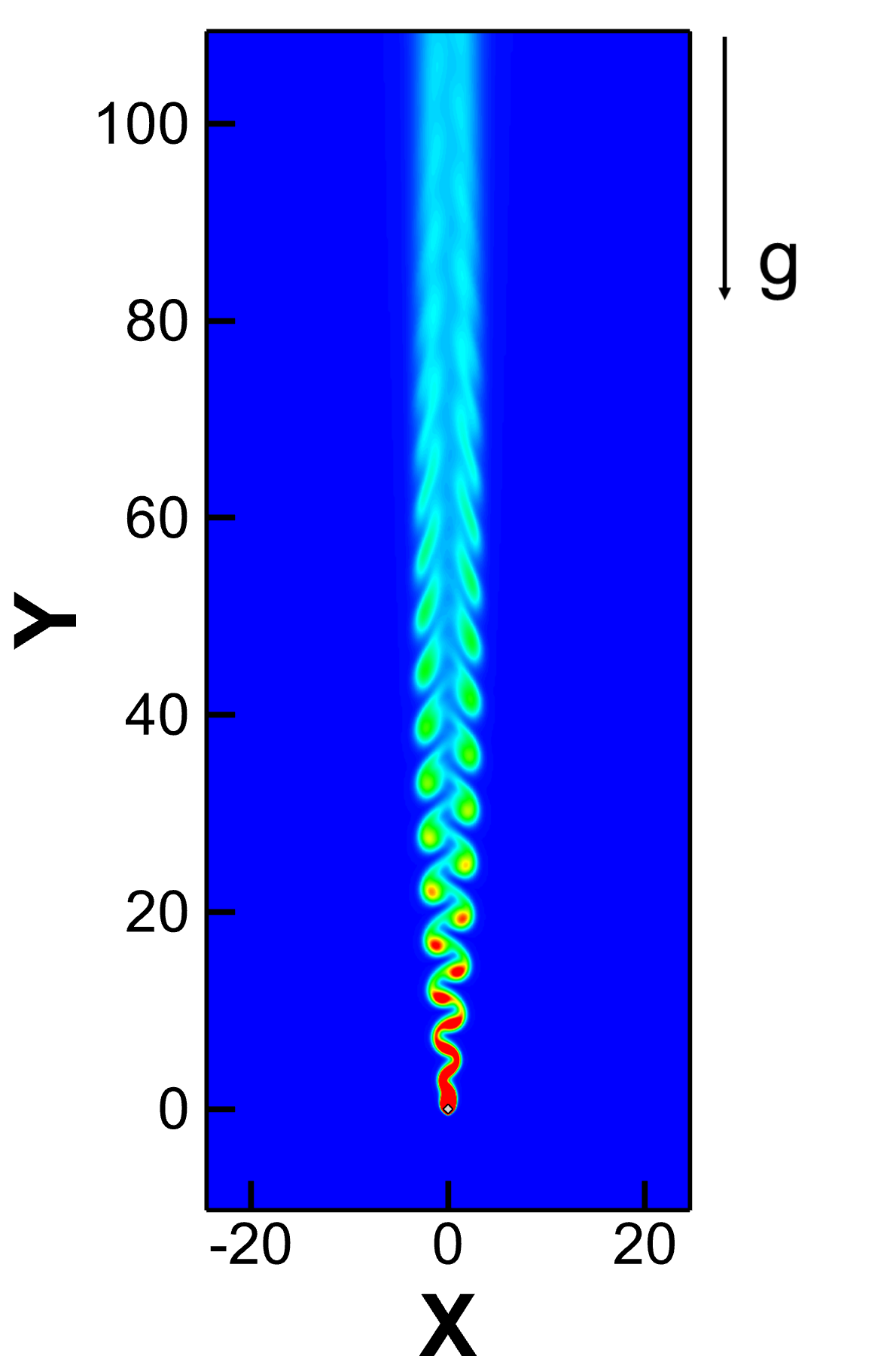}
\caption{Ri $=0.6$}
\label{instemp6}
\end{subfigure}\hfill
\begin{subfigure}[t]{0.3\textwidth}
    \includegraphics[width=\linewidth]{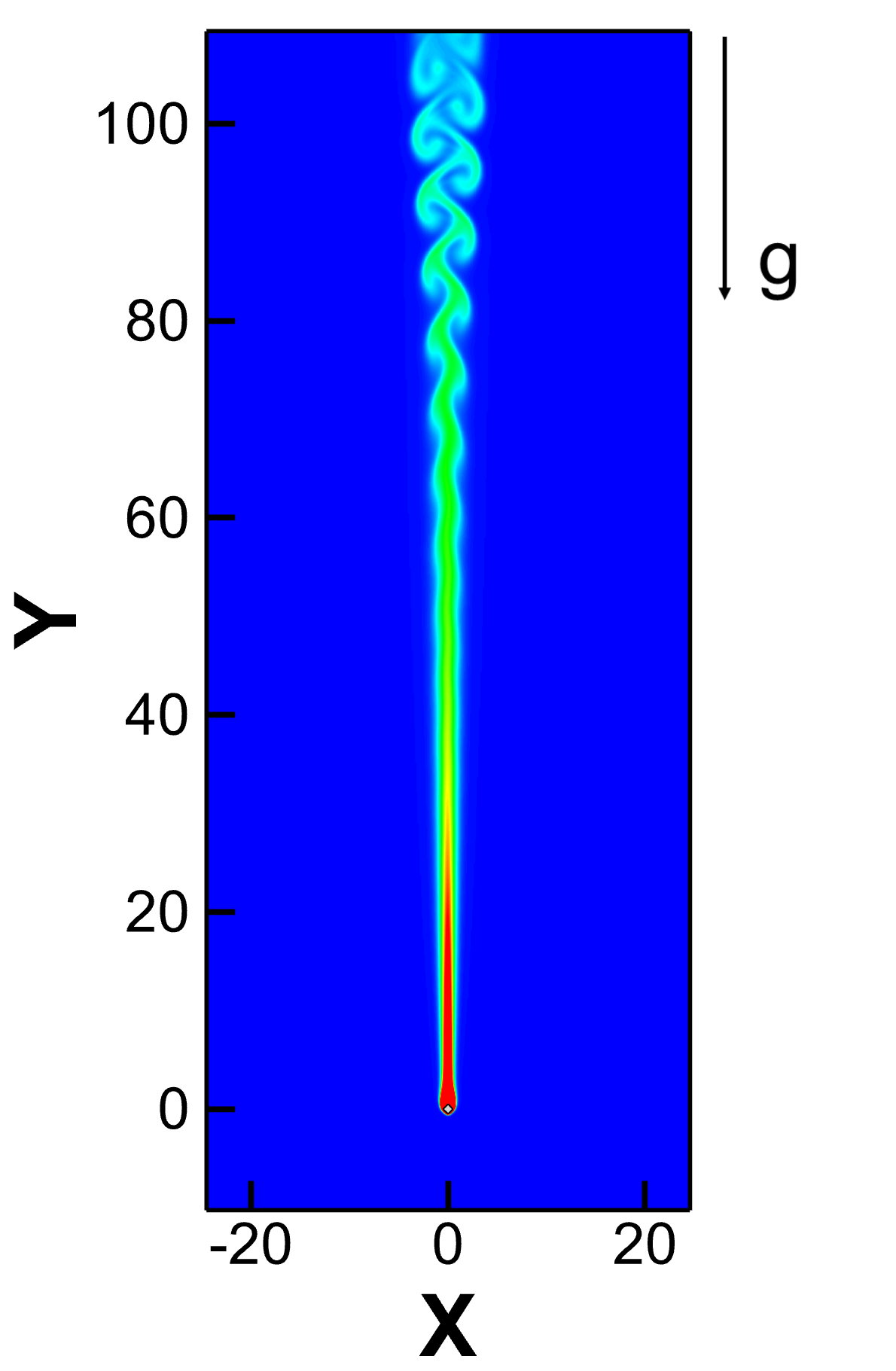}
\caption{Ri $=0.7$}
\label{instemp7}
\end{subfigure}\hfill
\begin{subfigure}[t]{0.3\textwidth}
    \includegraphics[width=\linewidth]{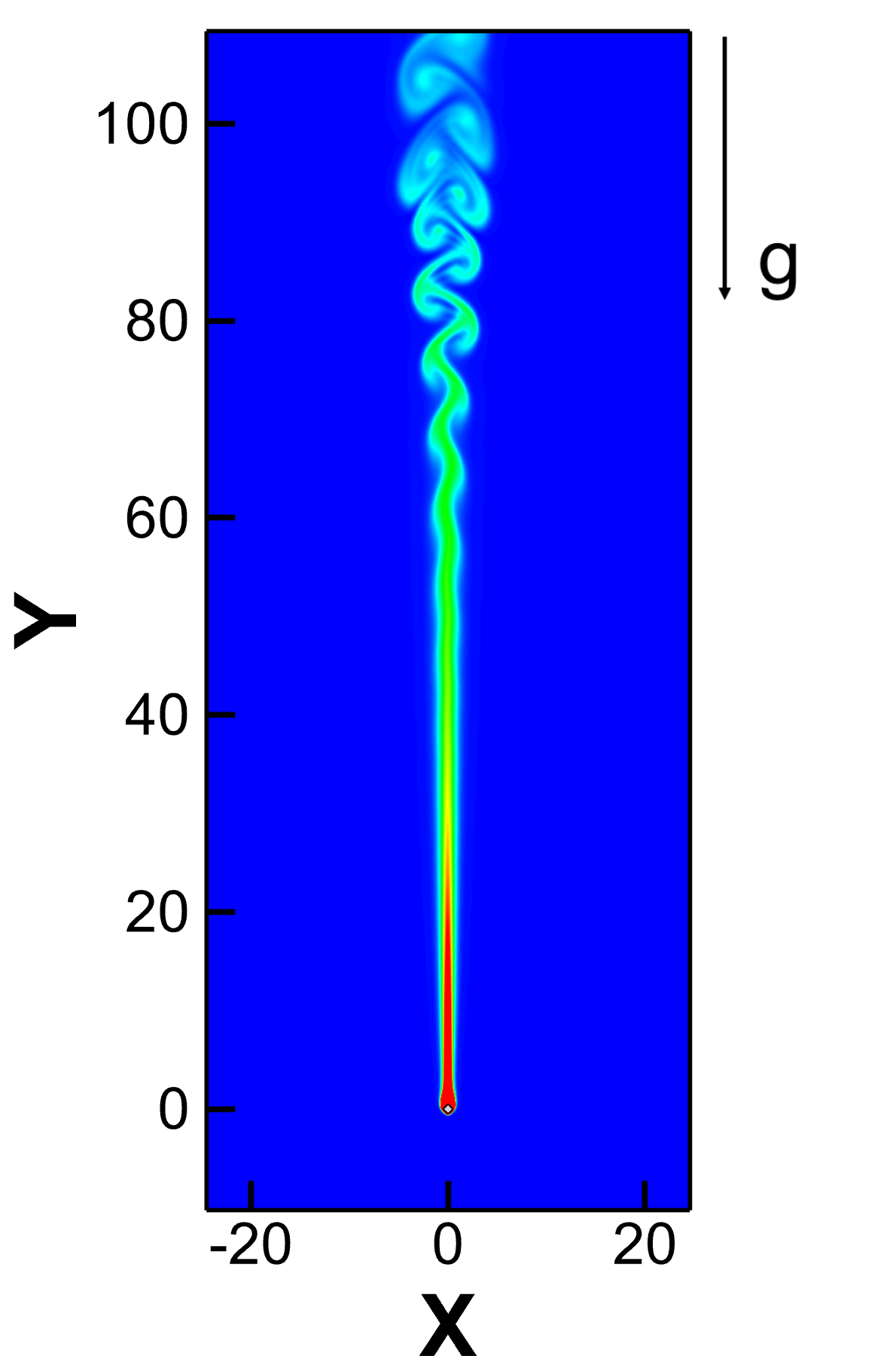}
\caption{Ri $=1.0$}
\label{instemp10}
\end{subfigure}\hfill

\caption{Instantaneous temperature contours for Re $=100$ and $0.0\leq$ Ri $\leq1.0$: (a) Ri $=0.0$, (b) Ri $=0.2$, (c) Ri $=0.4$, (d) Ri $=0.6$, (e) Ri $=0.7$, (f) Ri $=1.0$; ($\theta_{min},\theta_{max},\Delta\theta) \equiv (0,0.1,0.001)$.}
\label{instemp}
\end{figure}

\begin{figure}[htbp]
\centering
\begin{subfigure}[t]{0.3\textwidth}
    \includegraphics[width=\linewidth]{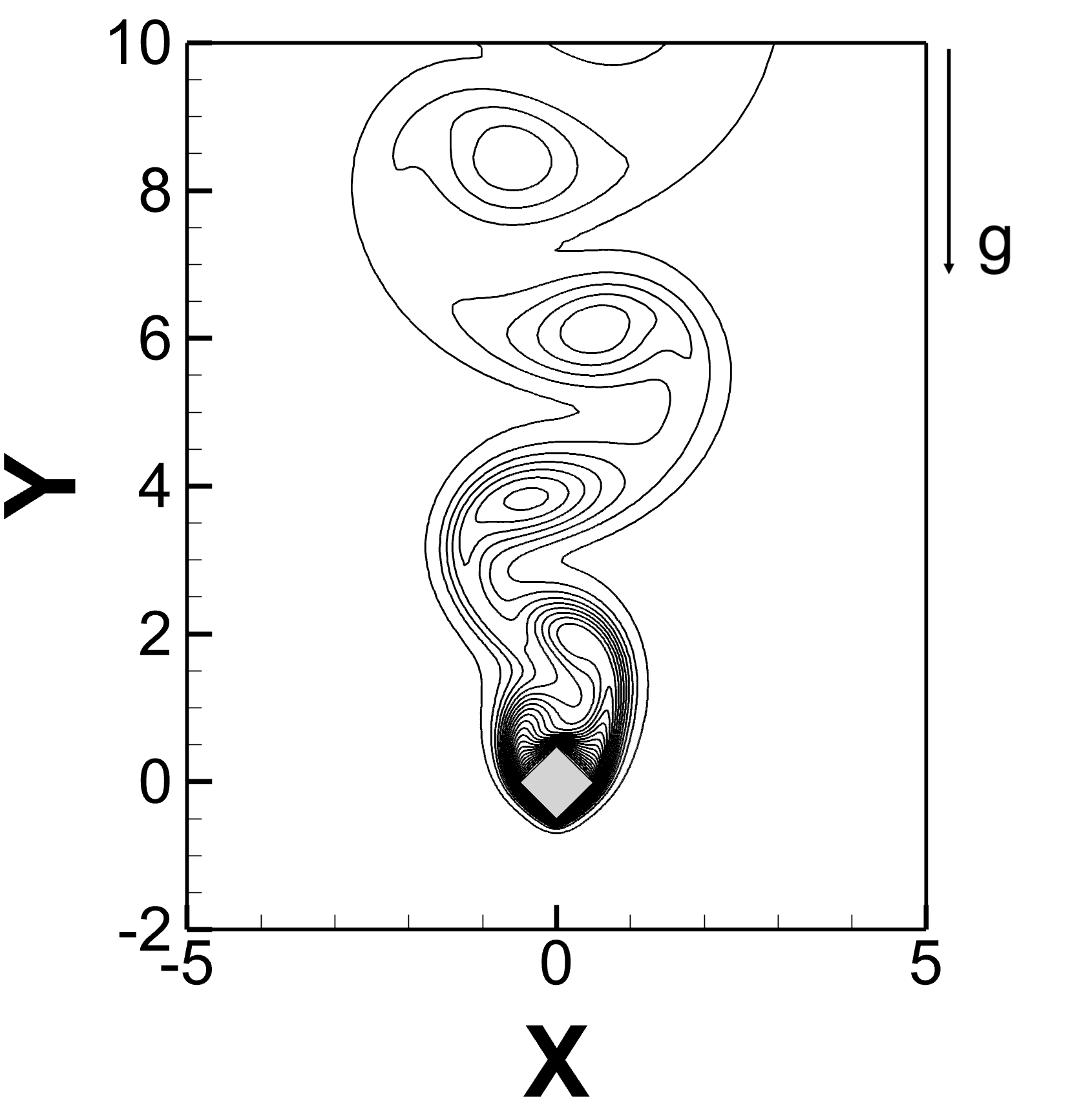}
\caption{Ri $=0.0$}
\label{insneartemp0}
\end{subfigure}\hfill
\begin{subfigure}[t]{0.3\textwidth}
  \includegraphics[width=\linewidth]{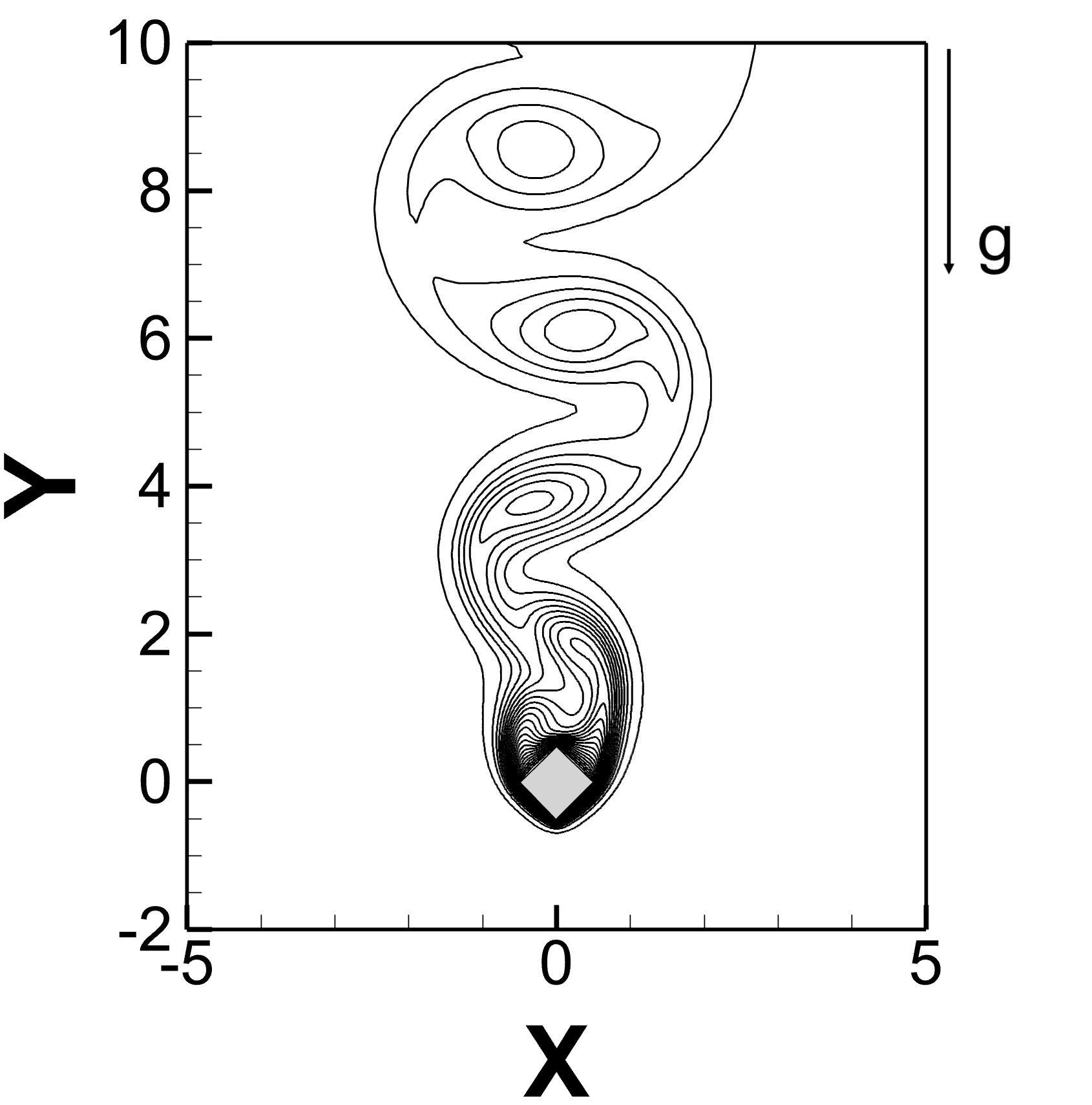}
\caption{Ri $=0.2$}
\label{insneartemp02}
\end{subfigure}\hfill
\begin{subfigure}[t]{0.3\textwidth}
    \includegraphics[width=\linewidth]{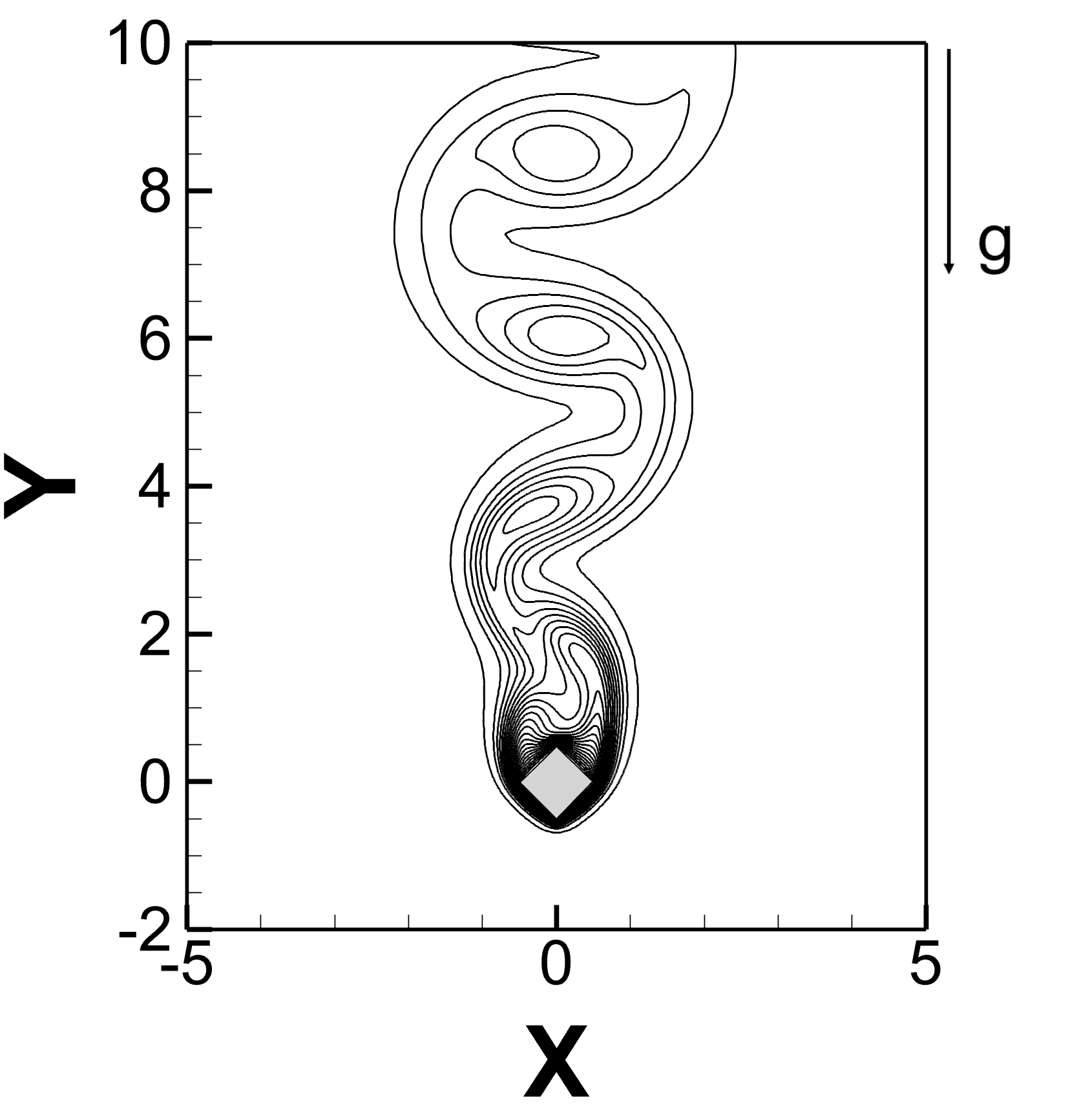}
\caption{Ri $=0.4$}
\label{insneartemp04}
\end{subfigure}\hfill

\begin{subfigure}[t]{0.3\textwidth}
    \includegraphics[width=\textwidth]{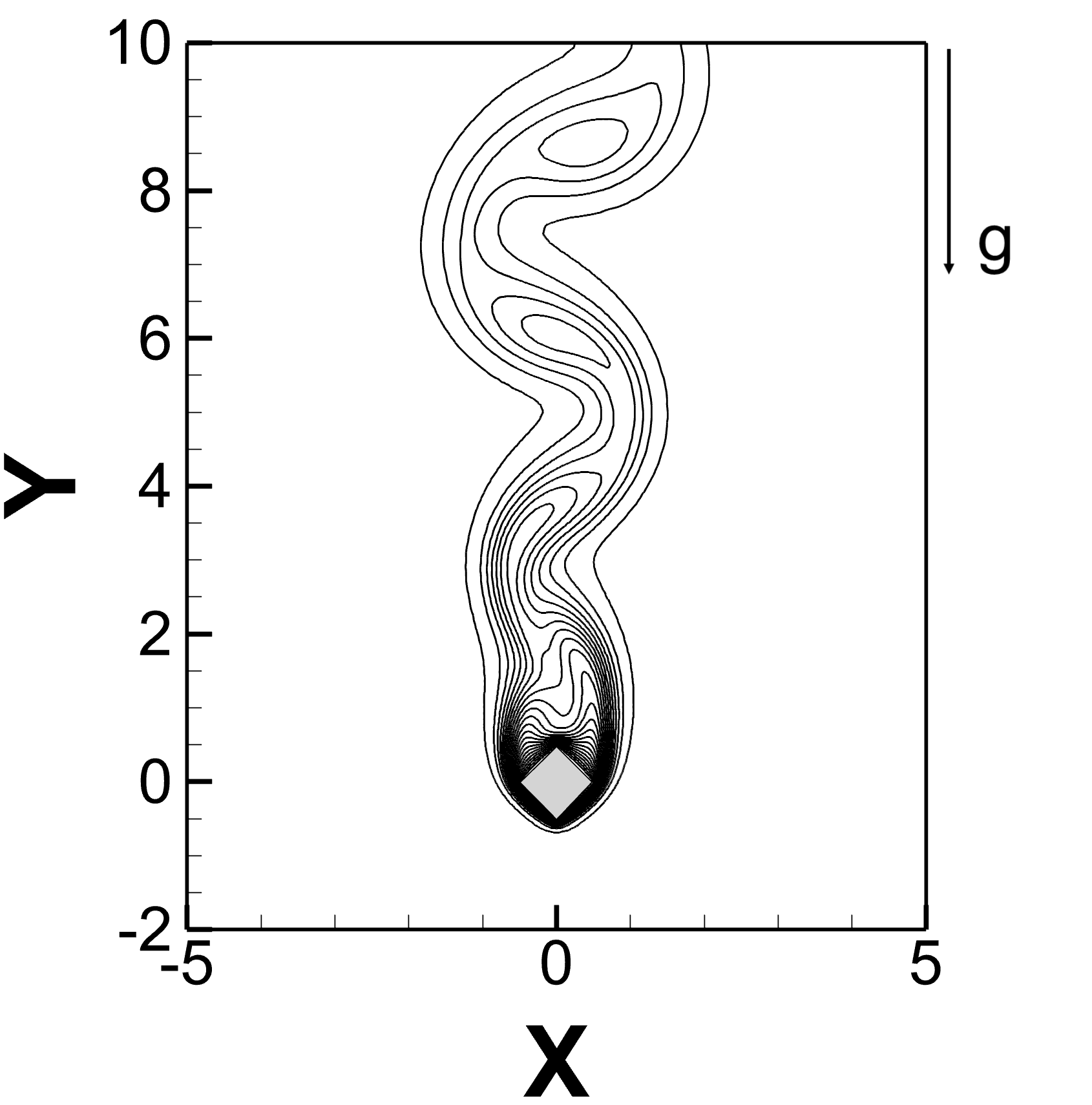}
\caption{Ri $=0.6$}
\label{insneartemp06}
\end{subfigure}\hfill
\begin{subfigure}[t]{0.3\textwidth}
    \includegraphics[width=\linewidth]{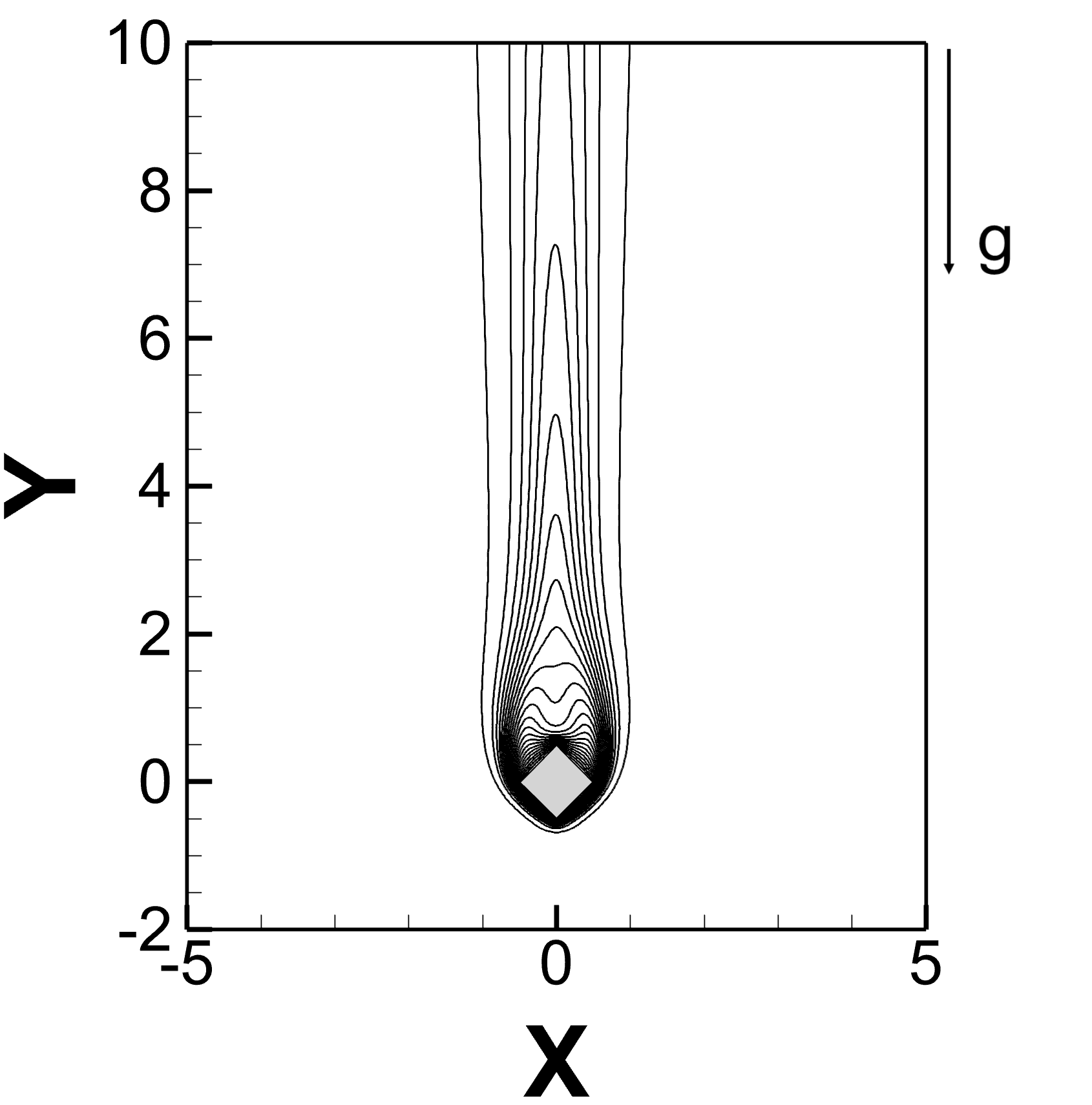}
\caption{Ri $=0.7$}
\label{insneartemp07}
\end{subfigure}\hfill
\begin{subfigure}[t]{0.3\textwidth}
    \includegraphics[width=\linewidth]{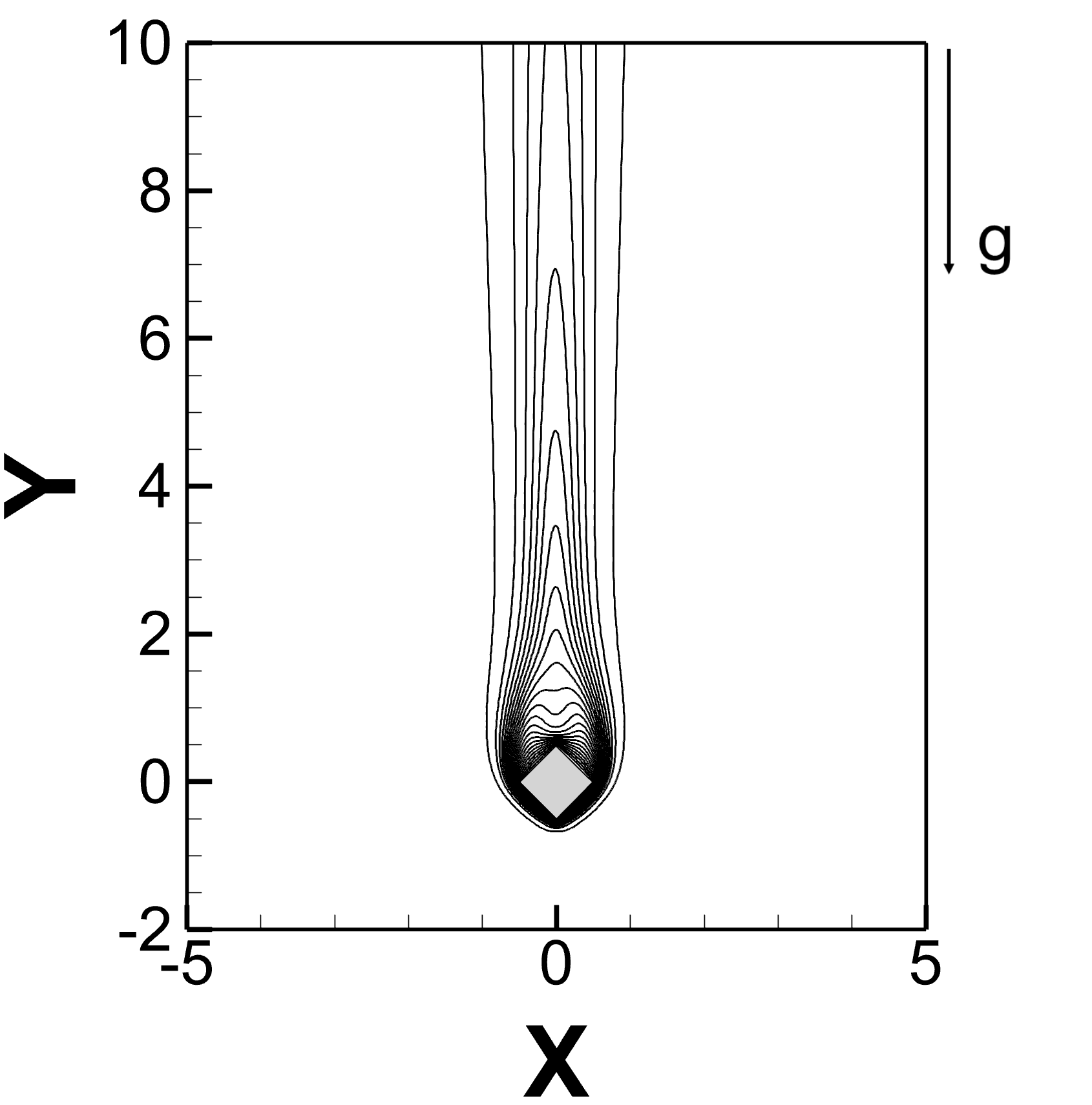}
\caption{Ri $=1.0$}
\label{insneartemp10}
\end{subfigure}\hfill

\caption{Instantaneous temperature contours (in the cylinder vicinity) for Re $=100$ and $0.0\leq$ Ri $\leq1.0$: (a) Ri $=0.0$, (b) Ri $=0.2$, (c) Ri $=0.4$, (d) Ri $=0.6$, (e) Ri $=0.7$, (f) Ri $=1.0$; ($\theta_{min},\theta_{max},\Delta\theta) \equiv (0,0.1,0.001)$.}
\label{insneartemp}
\end{figure}

Instantaneous vorticity contours for Re $=100$ and $0.0\leq$ Ri $\leq1.0$ in the full domain, near-field, and far-field are presented in \Cref{insvort,insnearvort,insfarvort} respectively. Instantaneous temperature contours in the full domain and near-field are shown in \Cref{instemp,insneartemp} respectively. All instantaneous contours are phase-synchronized (phase in the shedding cycle when the lift coefficient is minimum) to ensure uniformity in flow visualization. 

The instantaneous vorticity contours (see \Cref{insvort}) reveal the presence of the B\'{e}nard-von K\'{a}rm\'{a}n vortex street, signifying the shedding of vortices in the near-field for Ri $=0.0$ (forced convection). There is negative vorticity in the left and positive vorticity in the right half, which spans the entire domain downstream. The two oppositely signed shear layers attached to the cylinder interact, leading to the shedding of vortices. The flow becomes steady at the far-field due to viscous diffusion. When Ri is increased to a non-zero positive value, it is seen that the sign of vorticity starts as that of a forced flow but slowly inverts beyond a critical distance downstream and persists with this inverted sign for the rest of the domain beyond the critical distance. Additionally, the far-field flow is no longer steady, which was the case for Ri $=0.0$. At higher Ri (0.4-0.7), the inversion of vorticity takes place closer to the cylinder, and the width between the two shear layers increases downstream. The shape of the shed vortices also differs greatly with increasing Ri, which requires further investigation. For Ri $\geq$ 0.7, the near-field flow becomes steady with the suppression of vortex shedding due to the effect of buoyancy, and a large-scale undulation of the shear layers is seen in the far-field region. The shear layer becomes discontinuous, forming isolated patches of vorticity due to the instability developing within the stronger shear layer. As Ri increases, the inception point of the far-field unsteadiness moves closer to the cylinder. The instantaneous vorticity in the cylinder vicinity (see \Cref{insnearvort}) shows regular vortex shedding at Ri $=0.0$ with vortices of the opposite sign being shed alternatively into the wake. With increasing Ri, the transverse extent of the interaction of the positively-signed shear layer with the oppositely-signed shear layer reduces gradually, appearing as though their mutual interaction is being inhibited. For Ri $>$ 0.7, there is minimal interaction between the oppositely-signed shear layers, and they form a separation bubble, similar to forced flows of a sub-critical Re at which the transition to unsteadiness occurs. Vortex shedding from the cylinder does not happen for Ri $>$ 0.7, and the point of inversion of vorticity comes so close to the cylinder that it can be visualized in the near-field plots of vorticity as well (see \Cref{insnearvort07,insnearvort10}). 

The inversion of vorticity has been reported in the literature for isothermal confined flows \cite{davisetal, camarri, suzukietal, suzuki2}, where the interplay between several sources of vorticity in the flow causes the reversal of vorticity in the wake of the cylinder. In unconfined flows, a reversal of the B\'{e}nard-von K\'{a}rm\'{a}n vortex street takes place when the obstacle (bluff body or aerofoil) is imparted a periodic motion. Under appropriate conditions of the shape of the undulating body or the control parameter, the shed vortices cross the centerline and switch their position after travelling downstream, which manifests as a jet profile with a momentum excess in the time-averaged velocity plot, compared to the original wake profile with a momentum deficit \cite{godoydiana,schnipper,jones}. The vorticity inversion seen in our case resembles such structures. However, the shed vortices do not physically convect to the opposite side in this case. Rather, there is an inversion of vorticity due to the interplay between the vorticity fields due to forced convection and natural convection, which will be discussed in a later section.

The forced convective component of the flow causes shear layer separation at the transverse corners of the cylinder, causing a deficit of momentum in the wake. On the other hand, the natural convective component causes jet-like behavior downstream, competing with the momentum deficit caused by the wake, with the respective vorticities signed oppositely to each other. The opposing effects of these two flow components lead to a vorticity inversion. This inversion does not occur for Ri $=0.0$ due to the absence of buoyancy at this Ri. For positive Ri, vorticity inversion occurs downstream at an inception point, which comes further upstream as Ri is increased (see \Cref{recircinversion}) due to the increased buoyancy (see \Cref{insvort,meanvort}).

Isotherms also reveal the presence of vortex shedding in the near-field for Ri $=0.0$ and the steady far-field, as confirmed by examining temperature signals. At Ri $=0.2$, the undulation of shear layers extends to Y $=65$, and the complete flow field is unsteady up to Ri $=0.6$. The shape of the isotherm structures continues to change when Ri is increased due to the increased strength of the buoyancy force and the constant strength of the inertial force. With a further increase in Ri, the temperature in the near-field becomes steady, while the far-field reveals the presence of structures similar to those seen in plumes, as shown by Kimura and Bejan \cite{kimurabejan}. The near-field plots of temperature show similar behavior as the vorticity contours (see \Cref{insneartemp}). The undulation of the isotherms becomes significantly lower with increasing Ri, and for Ri $\geq$ 0.7, the structures reveal a steady flow and thermal field in the near-field.

\begin{figure}[htbp]
			\centering
			\begin{subfigure}[t]{0.45\textwidth}
				\includegraphics[width=\linewidth]{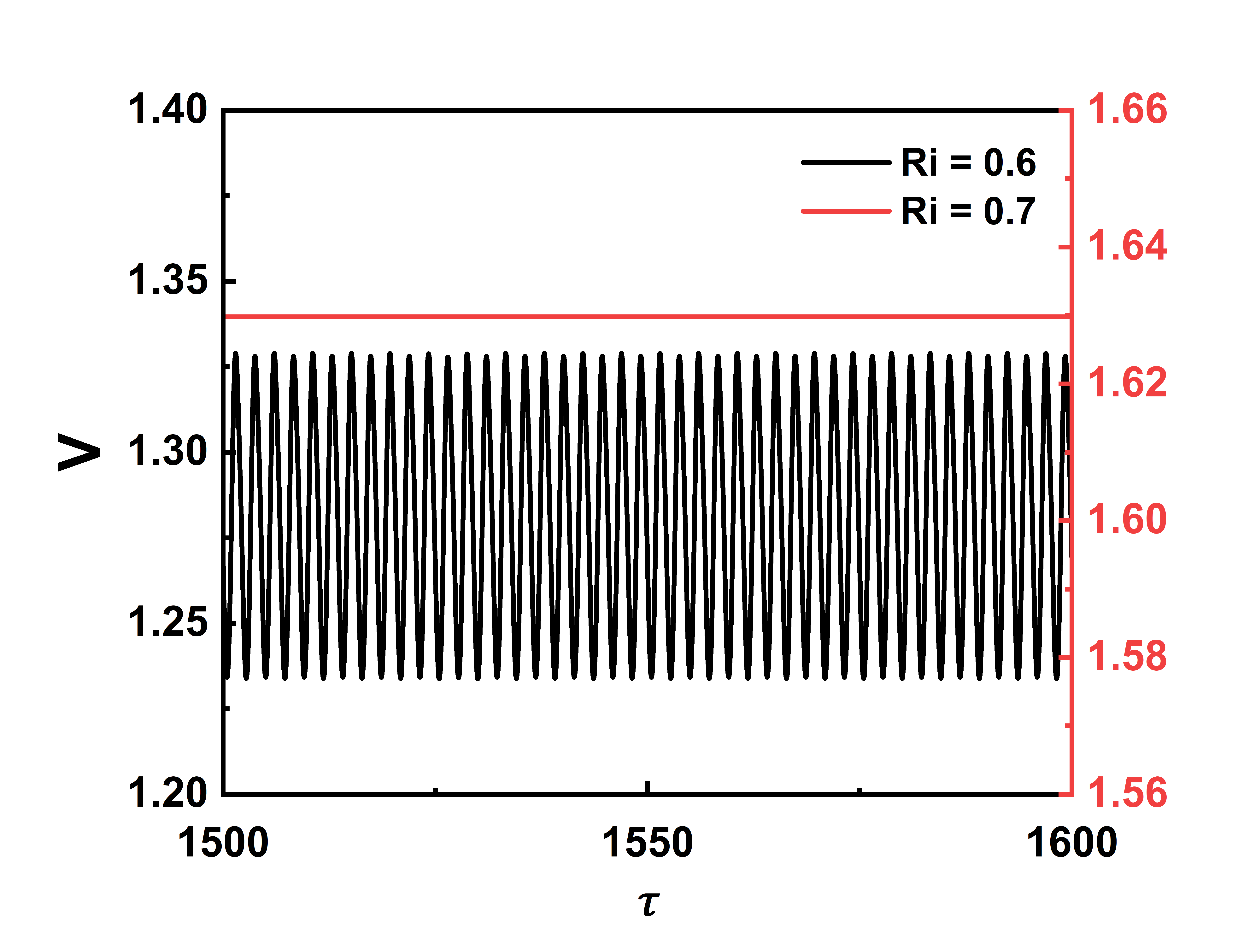}
				\caption{Y $=11.5$}
				\label{strsignear}
			\end{subfigure}\hspace{30pt}
			\begin{subfigure}[t]{0.45\textwidth}
				\includegraphics[width=\linewidth]{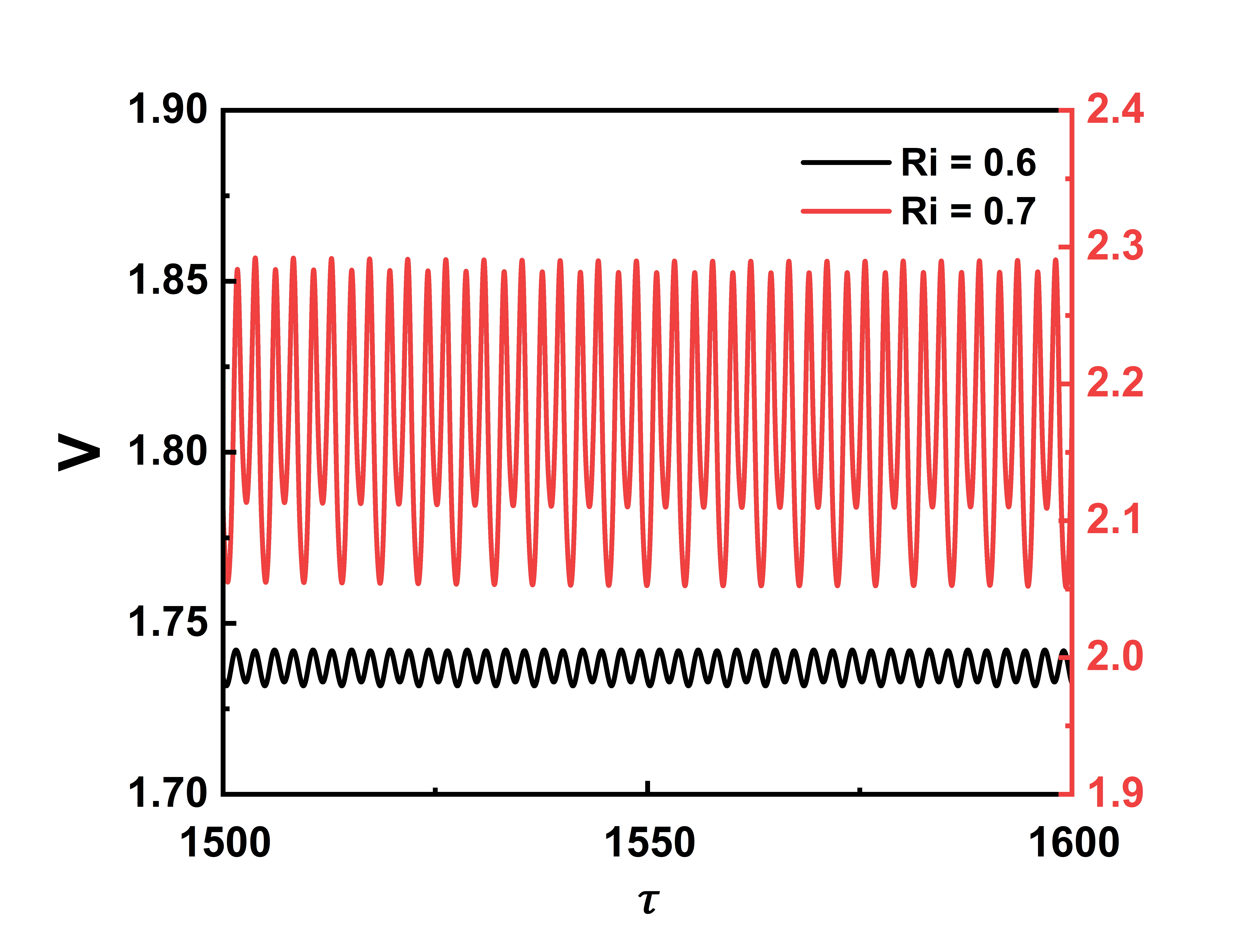}
				\caption{Y $=73.5$}
				\label{strsigfar}
			\end{subfigure}
			
			\begin{subfigure}[t]{0.45\textwidth}
				\includegraphics[width=\linewidth]{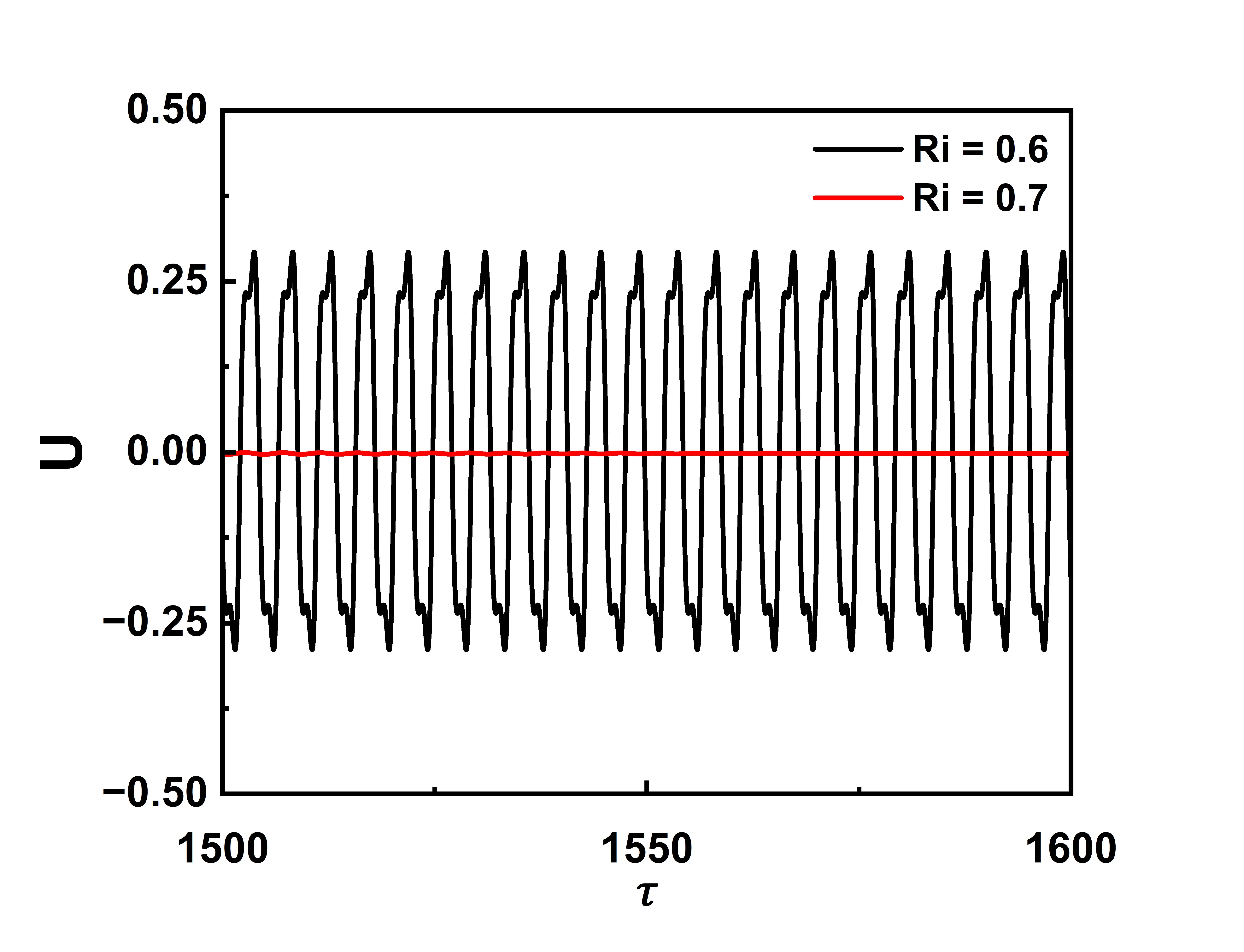}
				\caption{Y $=11.5$}
				\label{trasignear}
			\end{subfigure}\hspace{30pt}
			\begin{subfigure}[t]{0.45\textwidth}
				\includegraphics[width=\linewidth]{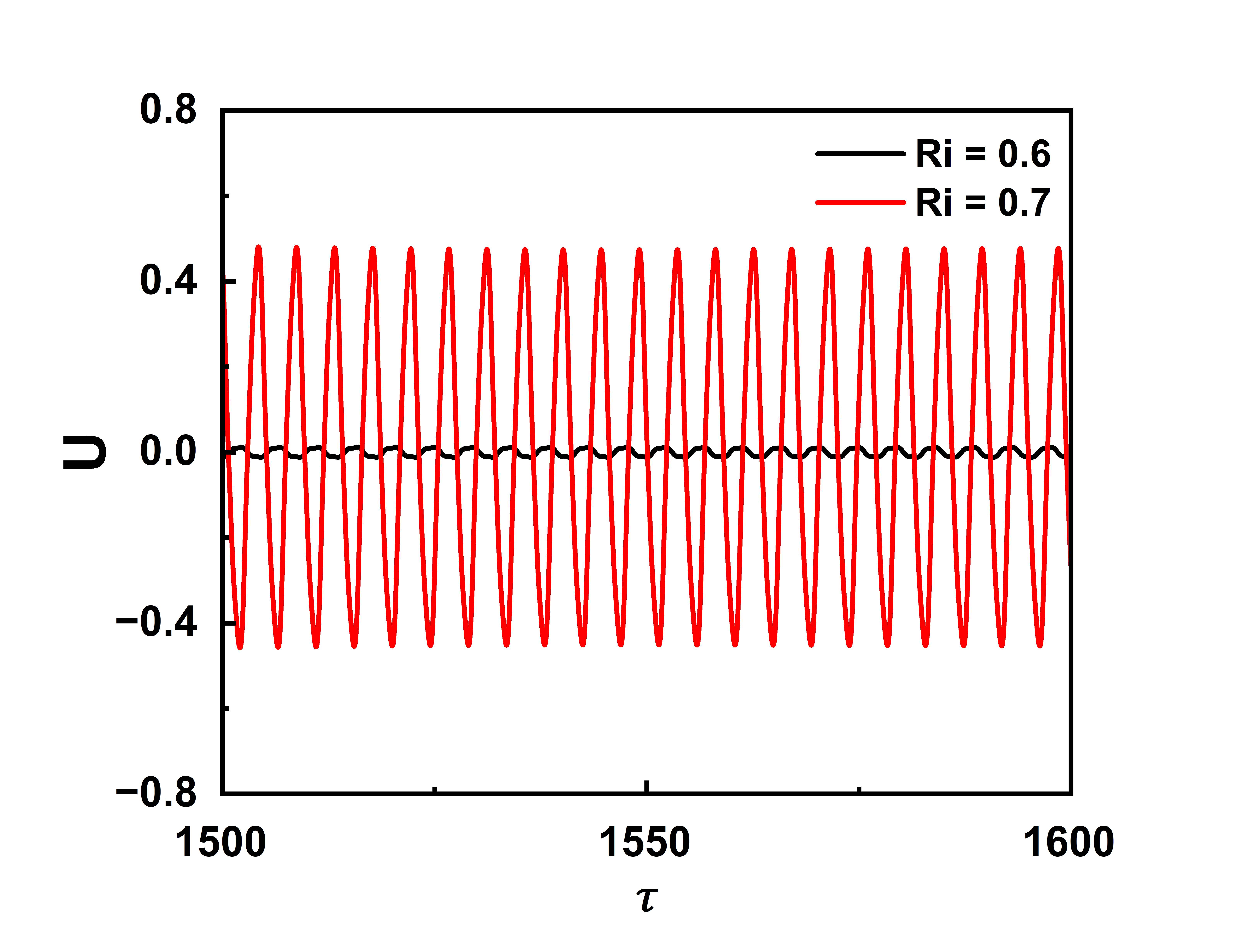}
				\caption{Y $=73.5$}
				\label{trasigfar}
			\end{subfigure}\hspace{30pt}
			
			\begin{subfigure}[t]{0.45\textwidth}
				\centering
				\includegraphics[width=\linewidth]{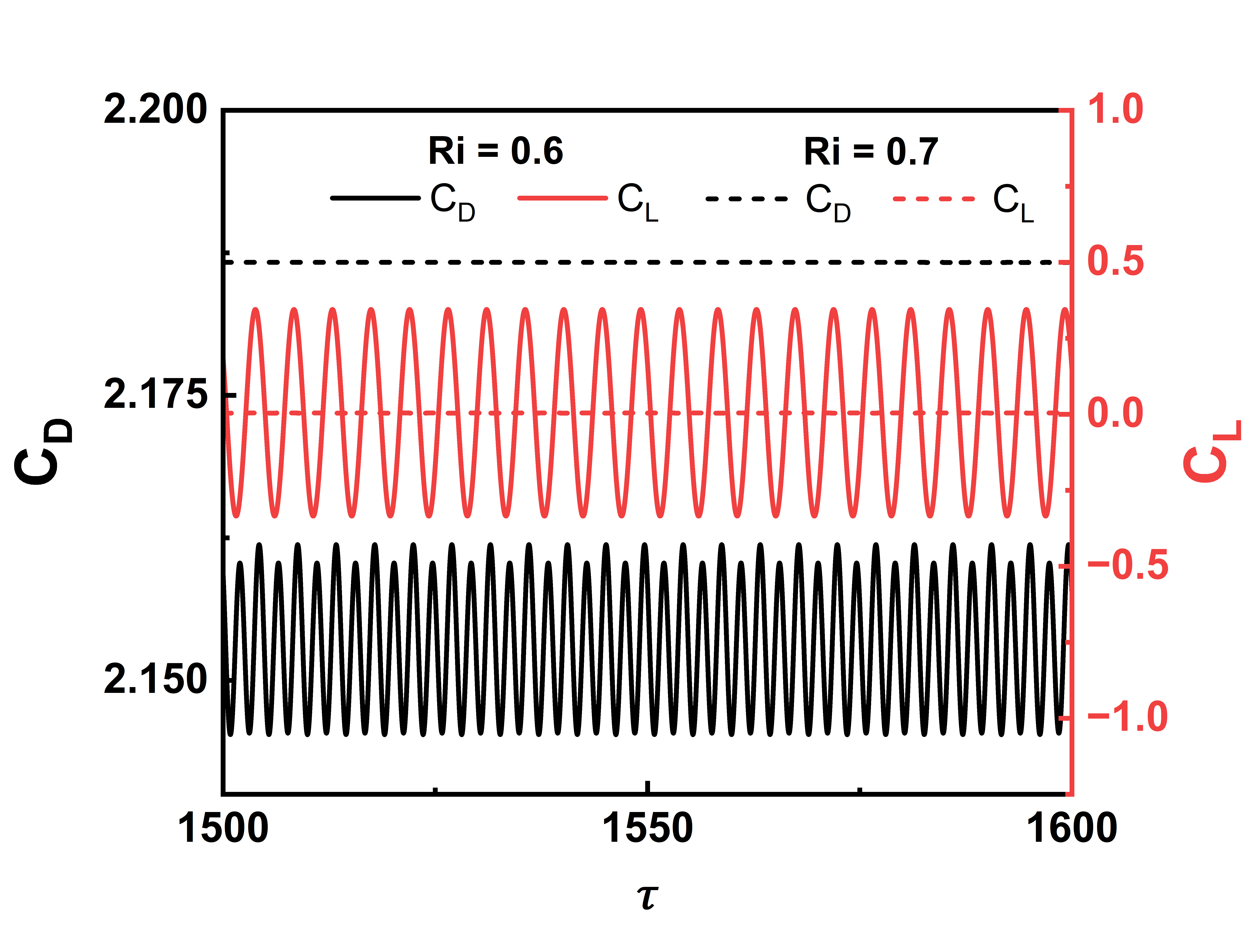}
				\caption{Lift and drag coefficient signals}
				\label{cdclsignal}
			\end{subfigure}
					
			\caption{Streamwise (a, b) and transverse (c, d) velocity signals in the near- and far-field for Ri $=0.6$ and Ri $=0.7$, along with (e) drag and lift coefficient signals. The near-field signals reveal the suppression of vortex shedding, and the far-field signals show the far-field unsteadiness. In contrast, the force coefficient signals characterize the near-field flow excellently.}
			\label{vsig}
		\end{figure}
		
		To confirm the behavior shown by the flow field in the contours, we have plotted time signals of the streamwise and transverse velocity in the near-field (Y $=11.5$) and far-field (Y $=73.5$), which are shown in \Cref{vsig} for Ri $=0.6$ and Ri $=0.7$. For Ri $=0.6$, streamwise and transverse velocity plots in the near-field reveal unsteady, periodic behavior, and for Ri $=0.7$, a steady behavior without any periodicity is observed (see \Cref{strsignear,trasignear}). Far-field streamwise and transverse velocity plots show the opposite behavior, with negligible unsteadiness in the case of Ri $=0.6$ but a large-scale unsteadiness for Ri $=0.7$ associated with the onset of the far-field plume-like unsteadiness (see \Cref{strsigfar,trasigfar}). 

The plots of lift and drag coefficient (see \Cref{cdclsignal}) show similarities with the near-field velocity signals, suggesting 
that the near-field flow is unsteady for Ri $=0.6$ and steady for Ri $=0.7$, supporting the suppression of vortex shedding in the near-field. This makes the force coefficients excellent predictors of the nature of the near-field flow.

\begin{table}[h!]
\centering
\caption{Variation of Strouhal number (St) with Richardson number (Ri) in the near-field (Y $=11.5$) and far-field (Y $=73.5$). The transverse velocity signal along the centerline (X $=0$) has been used to calculate the Strouhal number.}
\setlength{\tabcolsep}{0.75em}
\begin{tabular}{c cc}
\toprule
\toprule
\multirow{2}{*}{\begin{tabular}[c]{@{}c@{}}Richardson \\ number (Ri)\end{tabular}} & \multicolumn{2}{c}{Strouhal number (St)} \\ 
\cmidrule(l){2-3}
& Near-field (Y $=11.5$, X $=0$) & Far-field (Y $=73.5$, X $=0$) \\ 
\midrule
0.0 & 0.180 & 0.0 \\ 
0.2 & 0.197 & 0.194 \\ 
0.4 & 0.207 & 0.204 \\ 
0.6 & 0.220 & 0.220 \\ 
0.7 & - & 0.227 \\ 
1.0 & - & 0.221 \\ 
\bottomrule
\bottomrule
\end{tabular}
\label{stvsri}
\end{table}

	\Cref{stvsri} shows the variation of Strouhal number (St) with Ri. The transverse velocity signal along the centerline (X $=0$) is used to compute the Strouhal number at two different streamwise locations, Y $=11.5$ and Y $=73.5$, which fall in the near-field and far-field, respectively. St shows negligible variation between the near-field and far-field for Ri $<$ 0.7. Above the critical Ri, the near-field flow has no periodicity, and the corresponding St is zero. The flow in the wake is accelerated due to the buoyancy force, which is more significant at elevated Richardson numbers, resulting in increased Strouhal number with increasing Ri.

\newpage

\subsection{Time-Averaged Flow} \label{timeavgflow}
\begin{figure}[htbp]
	\centering
	\begin{subfigure}[t]{0.3\textwidth}
		\includegraphics[width=\linewidth]{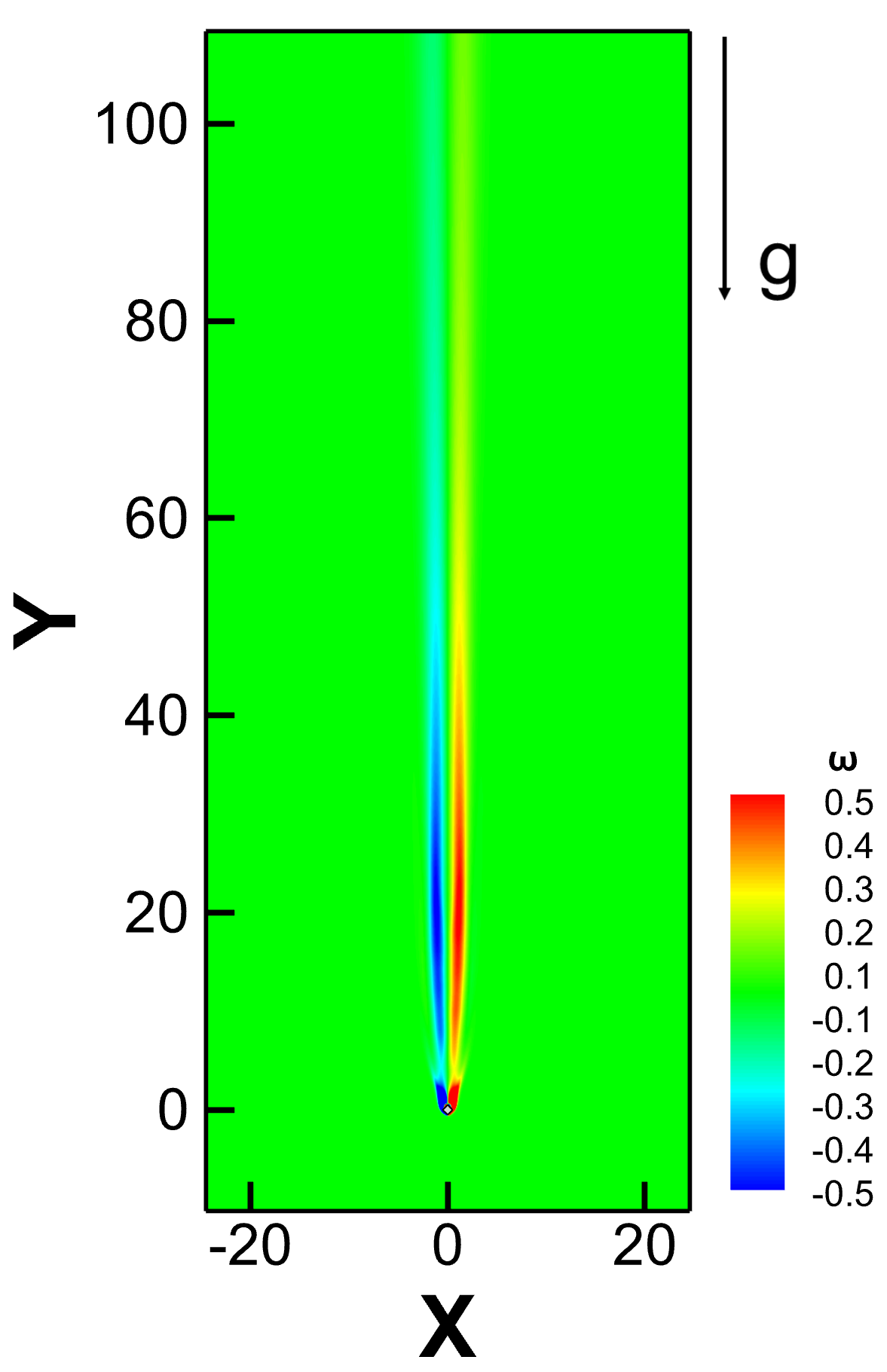}
		\caption{Ri $=0.0$}
		\label{meanvort0}
	\end{subfigure}\hfill
	\begin{subfigure}[t]{0.3\textwidth}
		\includegraphics[width=\linewidth]{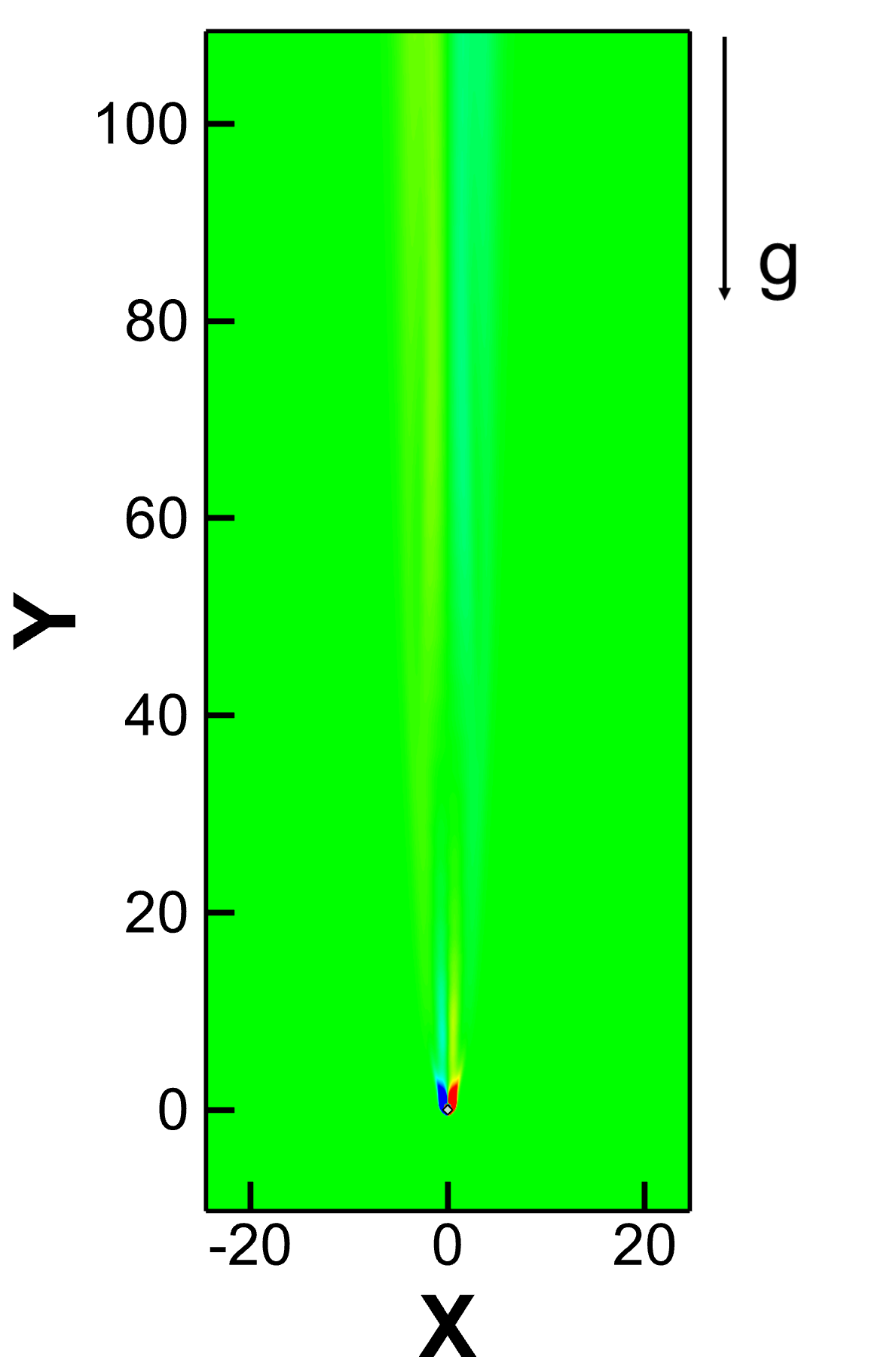}
		\caption{Ri $=0.2$}
		\label{meanvort02}
	\end{subfigure}\hfill
	\begin{subfigure}[t]{0.3\textwidth}
		\includegraphics[width=\linewidth]{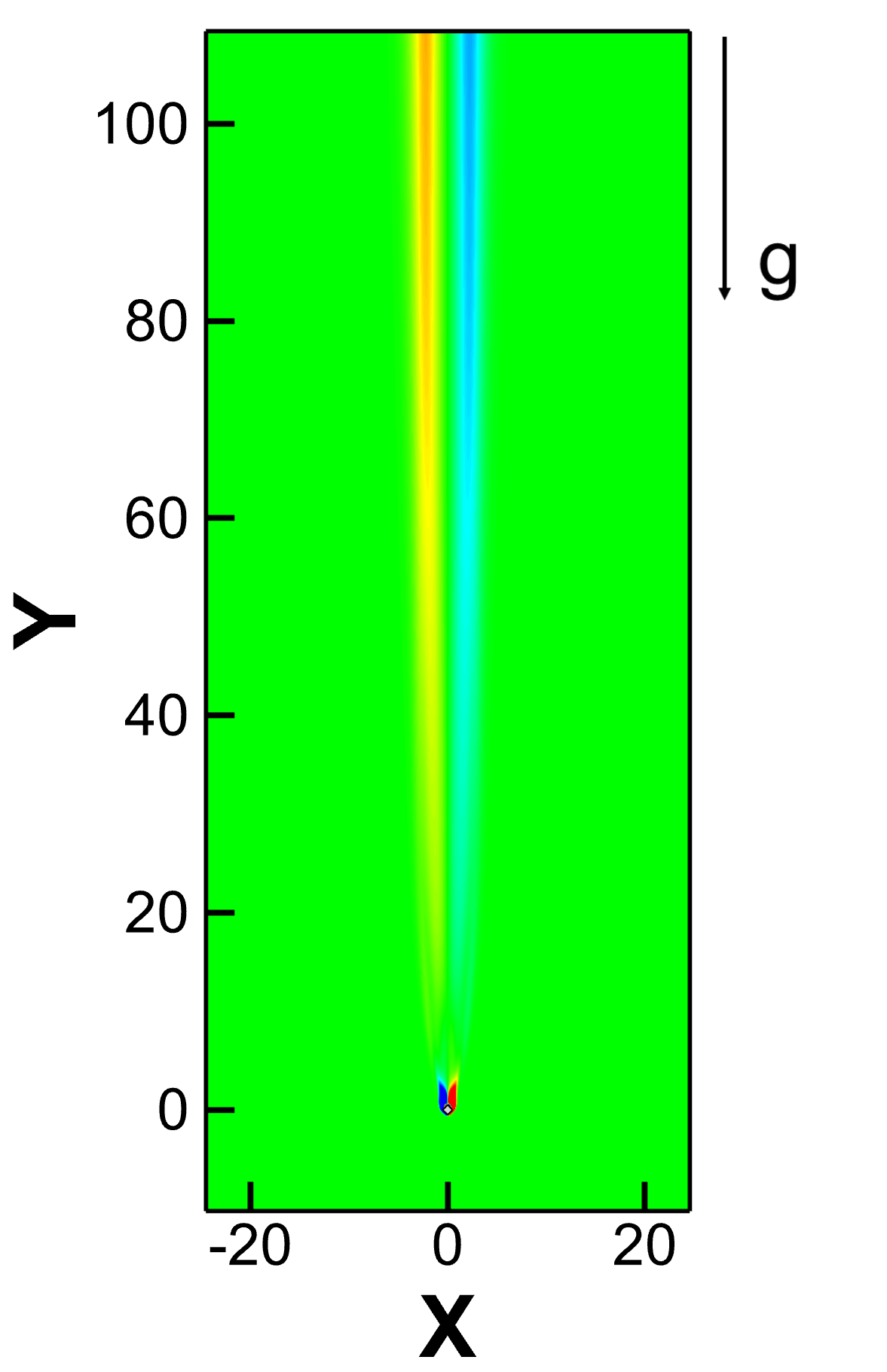}
		\caption{Ri $=0.4$}
		\label{meanvort04}
	\end{subfigure}\hfill

	\begin{subfigure}[t]{0.3\textwidth}
		\includegraphics[width=\textwidth]{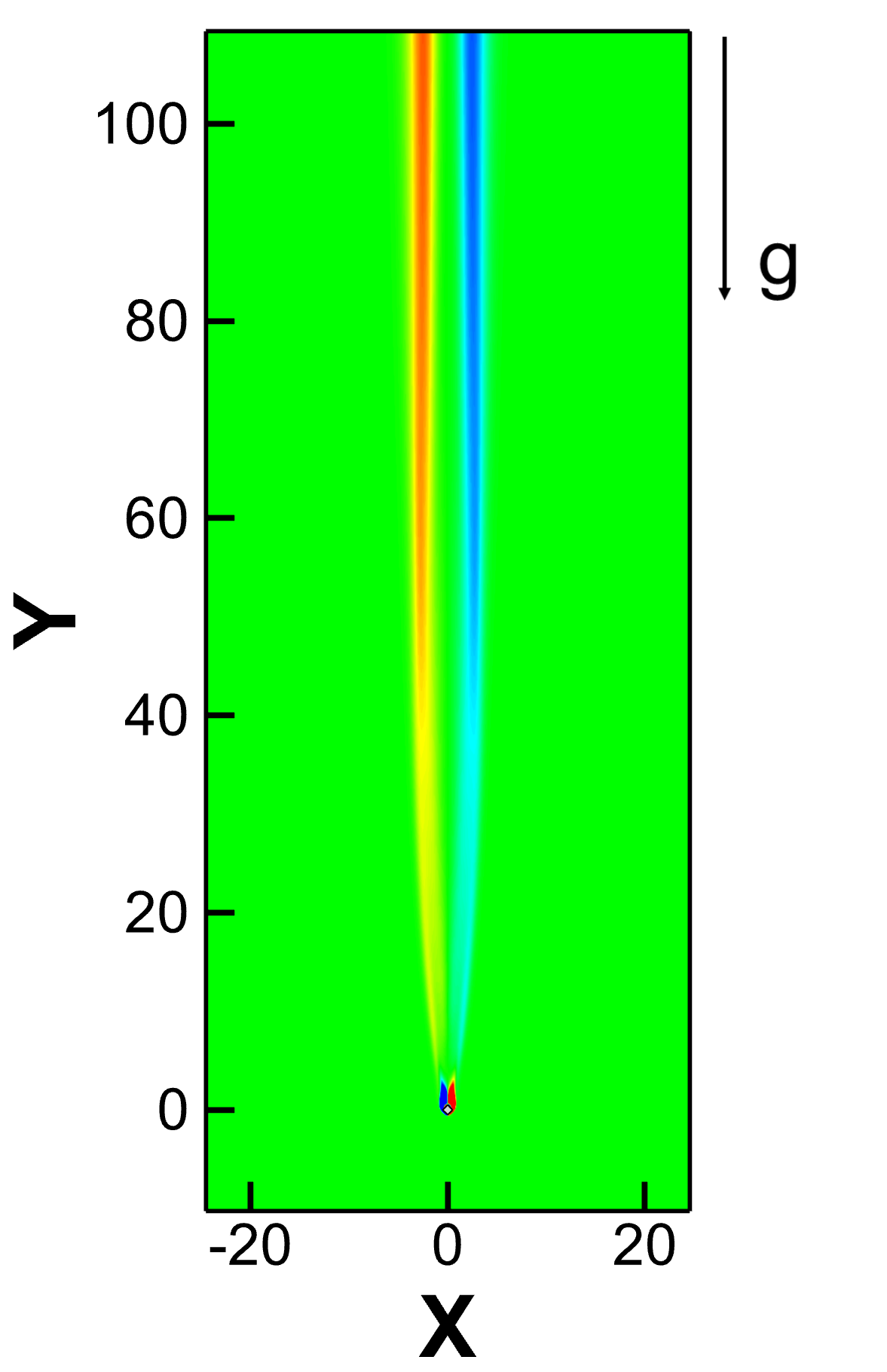}
		\caption{Ri $=0.6$}
		\label{meanvort06}
	\end{subfigure}\hfill
	\begin{subfigure}[t]{0.3\textwidth}
		\includegraphics[width=\linewidth]{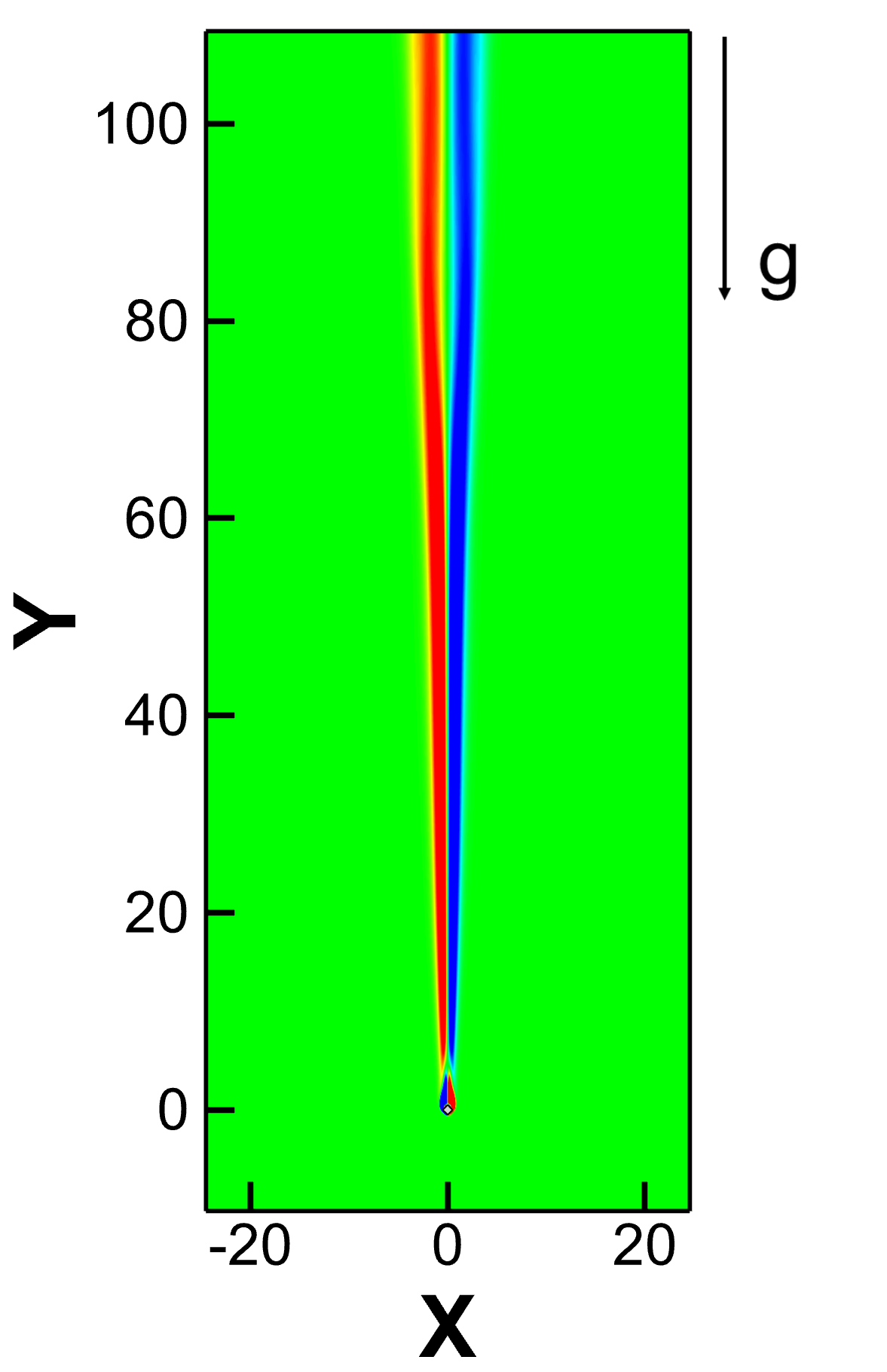}
		\caption{Ri $=0.7$}
		\label{meanvort07}
	\end{subfigure}\hfill
	\begin{subfigure}[t]{0.3\textwidth}
		\includegraphics[width=\linewidth]{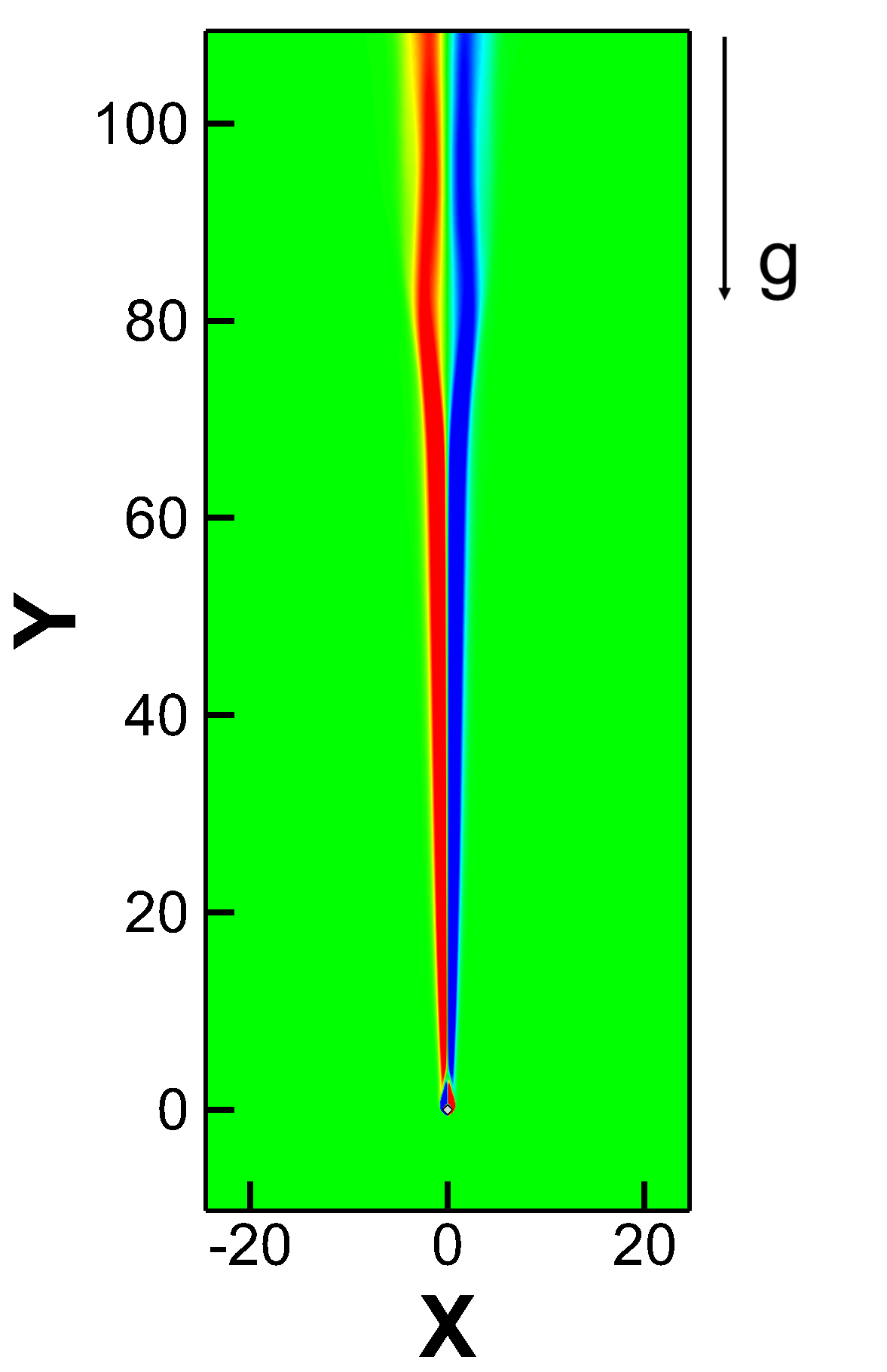}
		\caption{Ri $=1.0$}
		\label{meanvort10}
	\end{subfigure}\hfill
	
	\caption{Time-averaged vorticity contours for Re $=100$ and $0.0\leq$ Ri $\leq1.0$: (a) Ri $=0.0$, (b) Ri $=0.2$, (c) Ri $=0.4$, (d) Ri $=0.6$, (e) Ri $=0.7$, (f) Ri $=1.0$; ($\overline{\omega}_{min},\overline{\omega}_{max},\Delta\overline{\omega}_{}) \equiv (-0.5,0.5,0.01)$.}
	\label{meanvort}
\end{figure}

\begin{figure}[htbp]
	\centering
	\begin{subfigure}[t]{0.3\textwidth}
		\includegraphics[width=\linewidth]{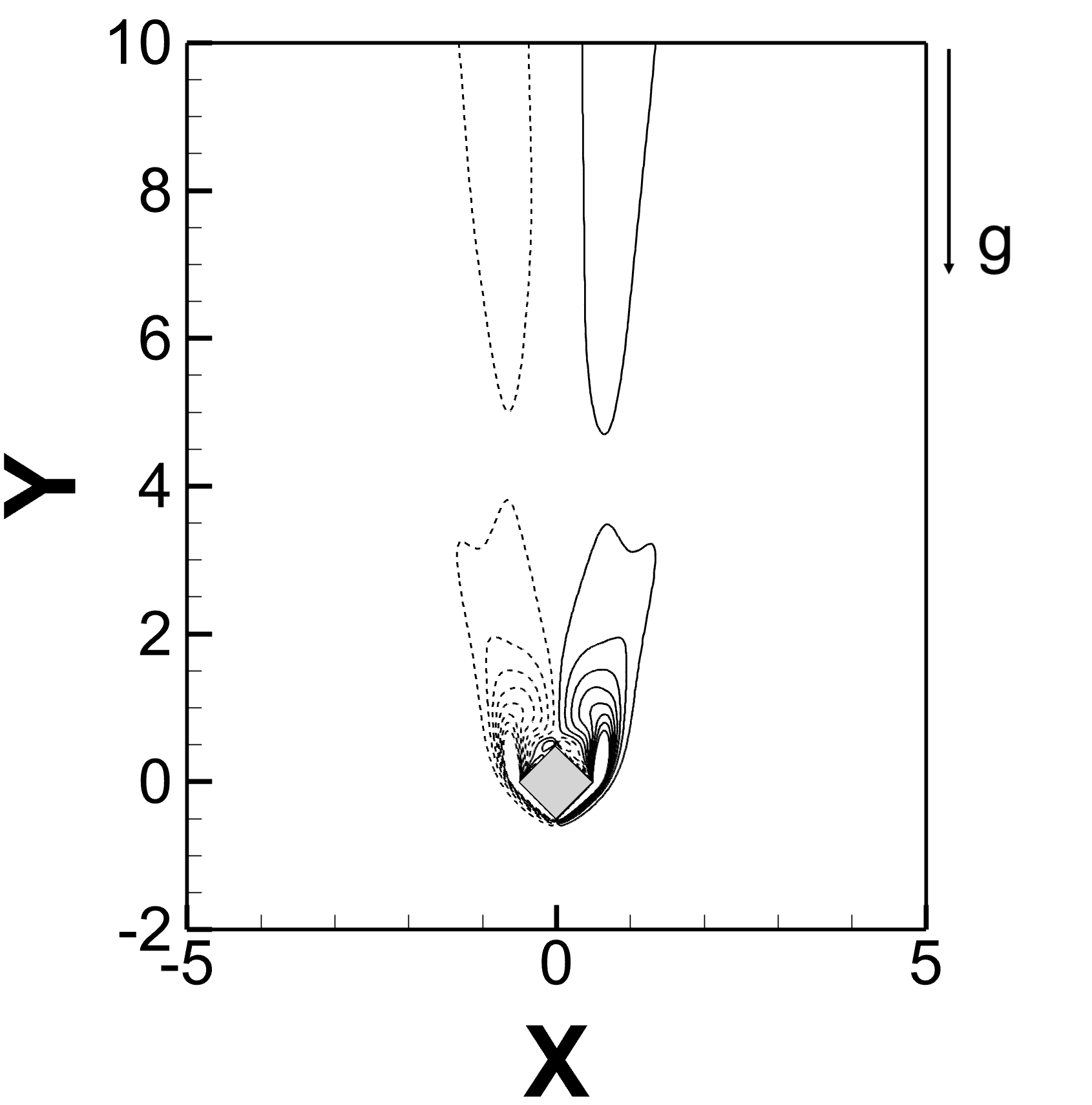}
		\caption{Ri $=0.0$}
		\label{meannearvort0}
	\end{subfigure}\hfill
	\begin{subfigure}[t]{0.3\textwidth}
		\includegraphics[width=\linewidth]{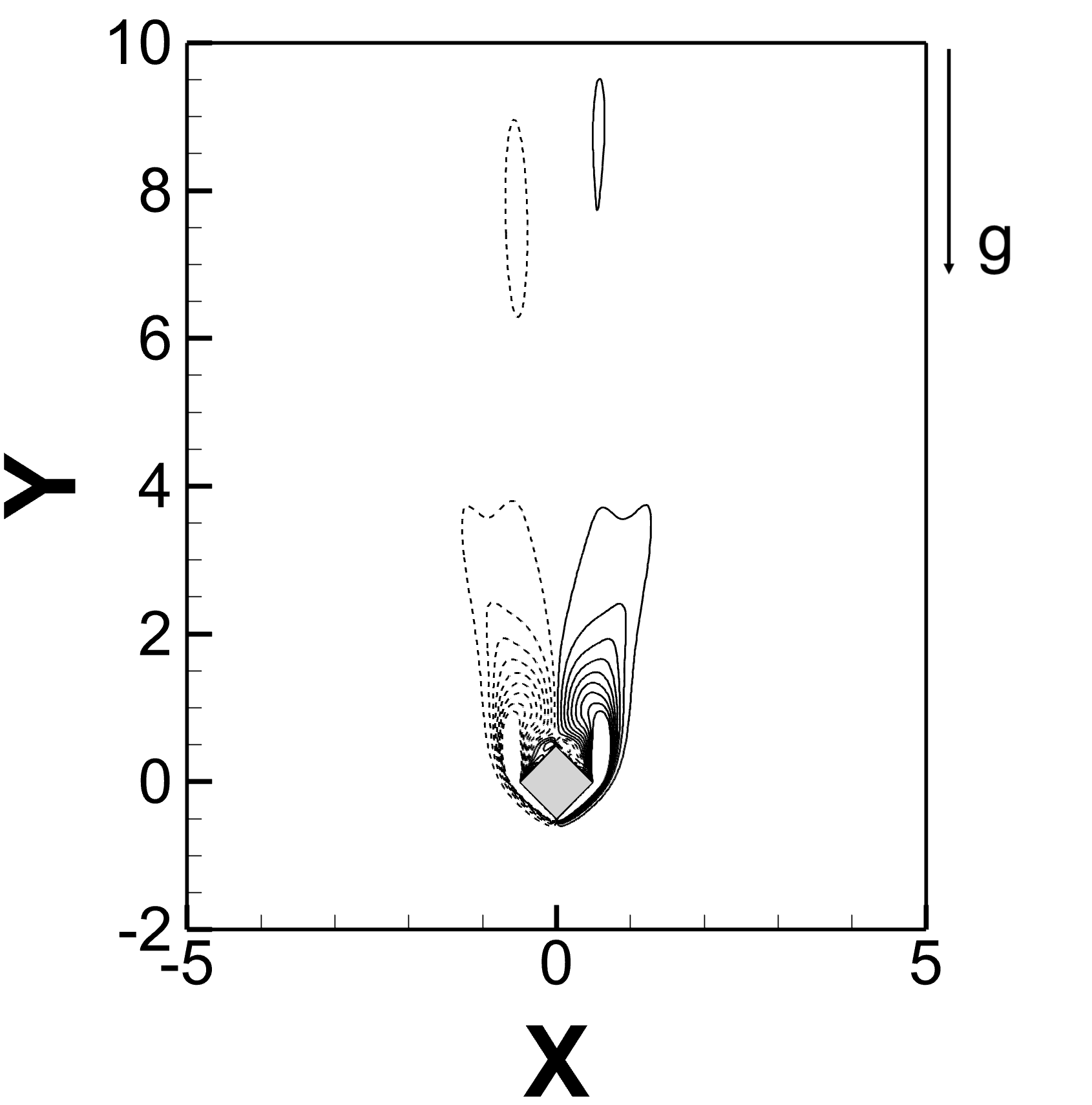}
		\caption{Ri $=0.2$}
		\label{meannearvort02}
	\end{subfigure}\hfill
	\begin{subfigure}[t]{0.3\textwidth}
		\includegraphics[width=\linewidth]{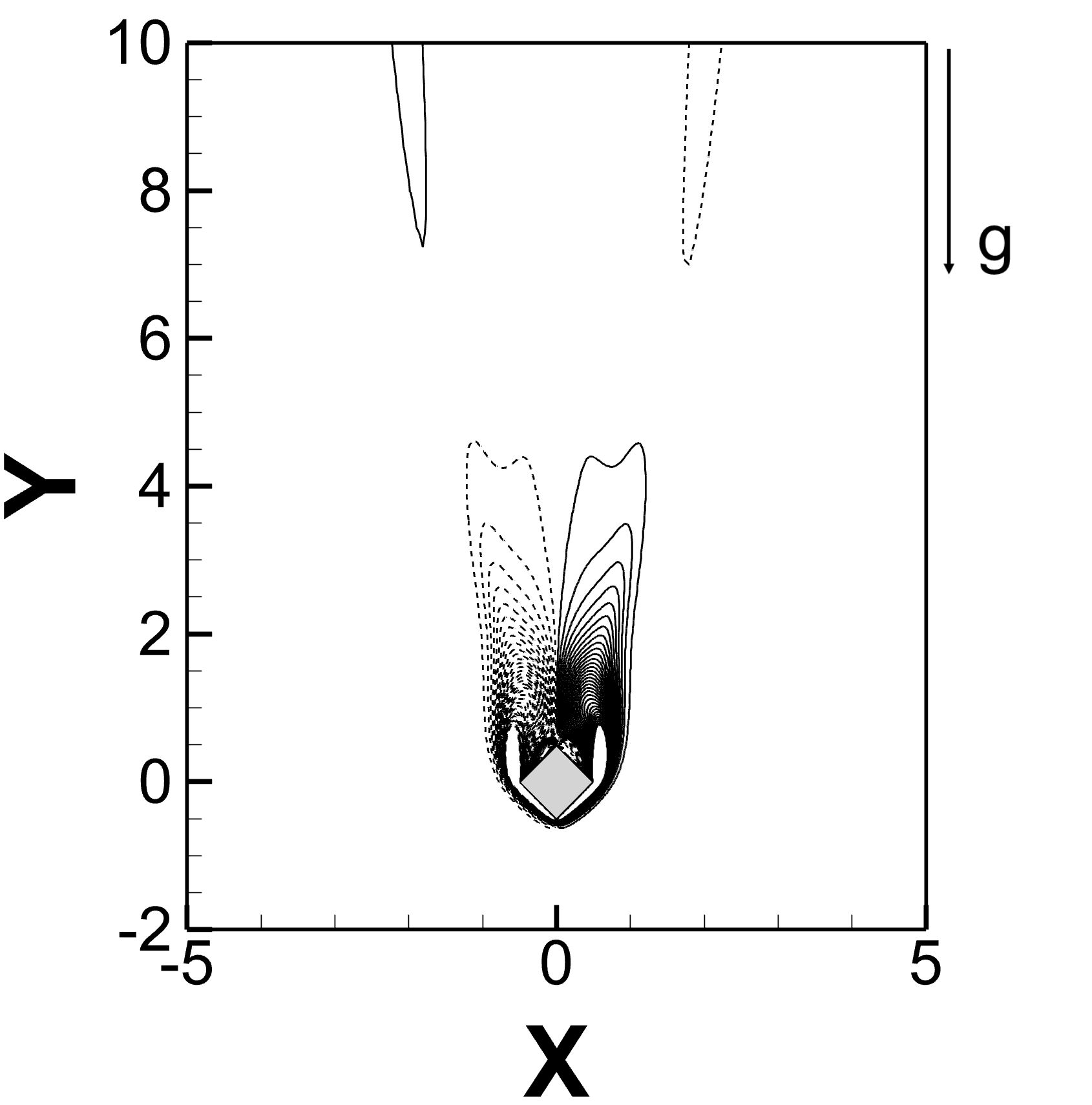}
		\caption{Ri $=0.4$}
		\label{meannearvort04}
	\end{subfigure}\hfill

	\begin{subfigure}[t]{0.3\textwidth}
		\includegraphics[width=\textwidth]{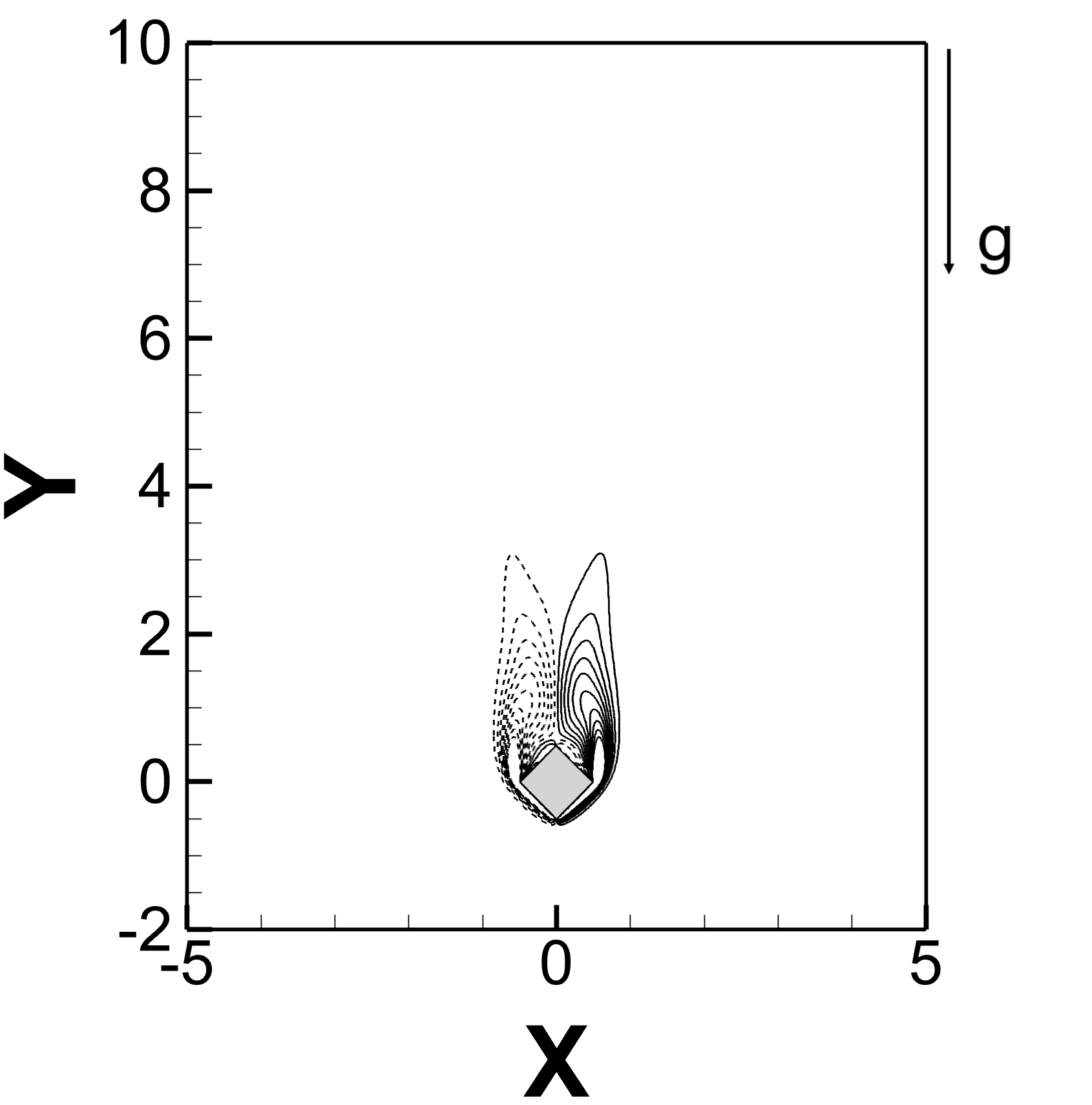}
		\caption{Ri $=0.6$}
		\label{meannearvort06}
	\end{subfigure}\hfill
	\begin{subfigure}[t]{0.3\textwidth}
		\includegraphics[width=\linewidth]{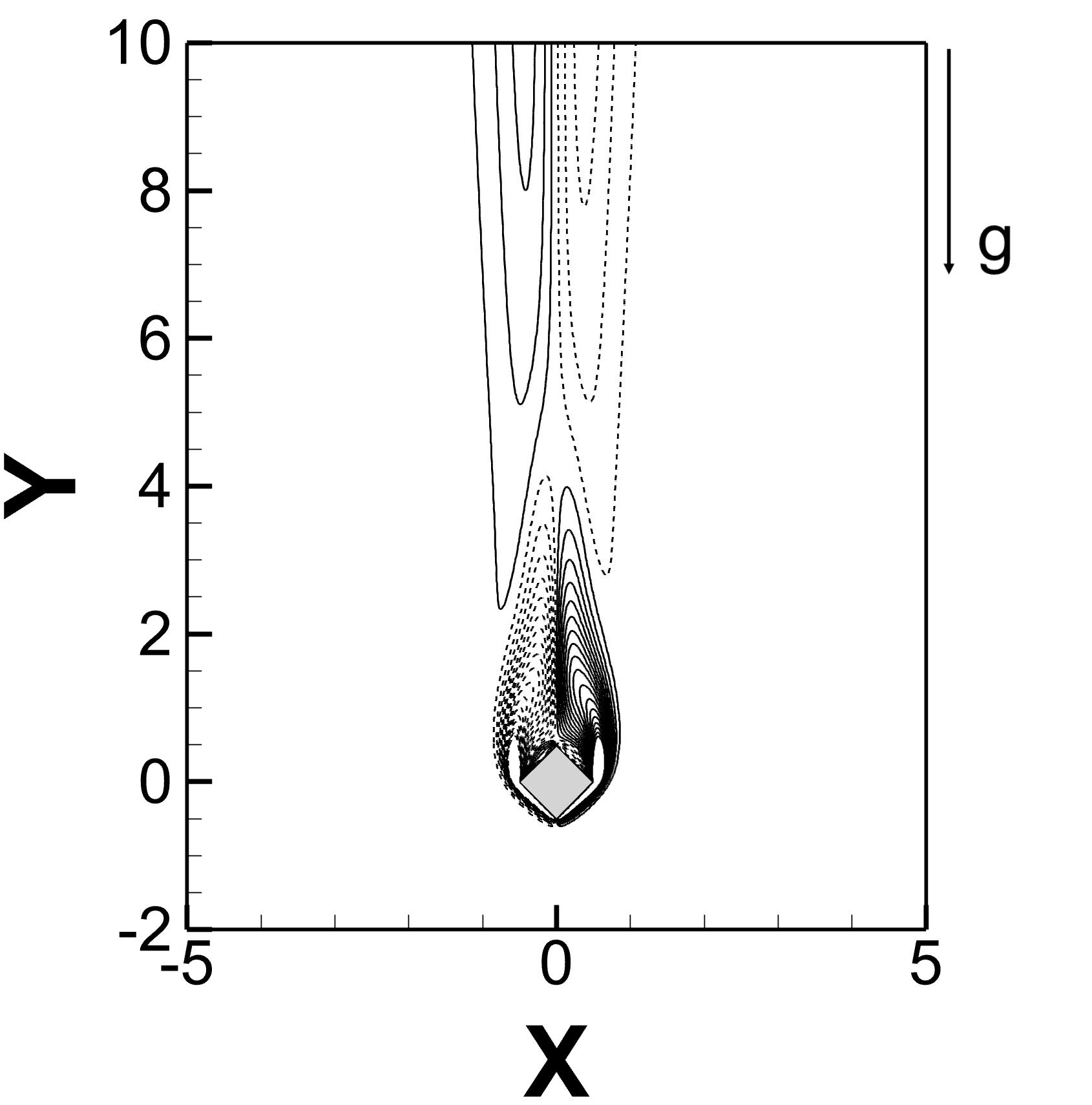}
		\caption{Ri $=0.7$}
		\label{meannearvort07}
	\end{subfigure}\hfill
	\begin{subfigure}[t]{0.3\textwidth}
		\includegraphics[width=\linewidth]{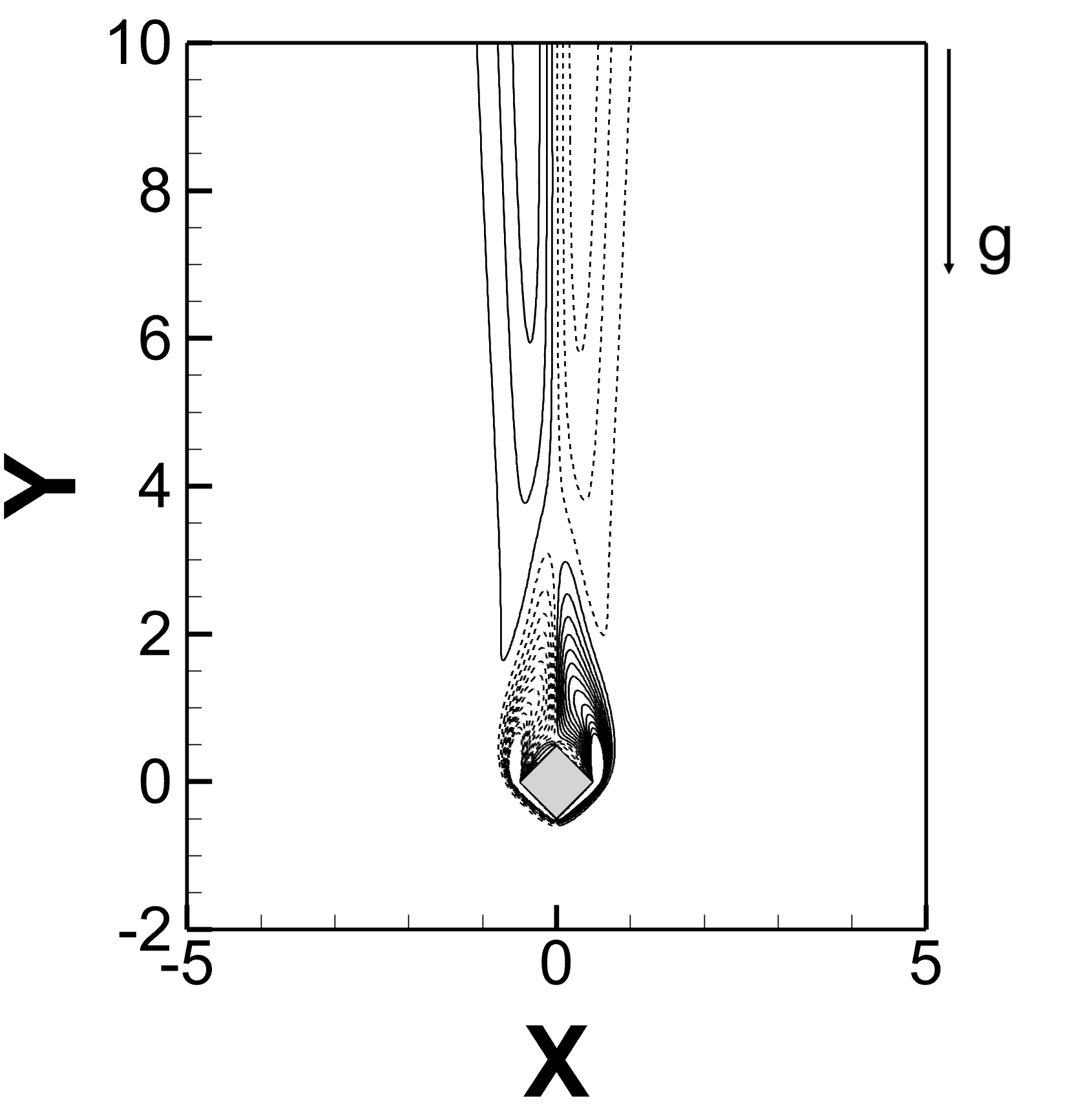}
		\caption{Ri $=1.0$}
		\label{meannearvort10}
	\end{subfigure}\hfill
	
	\caption{Time-averaged vorticity contours (in the cylinder vicinity) for Re $=100$ and $0.0\leq$ Ri $\leq1.0$: (a) Ri $=0.0$, (b) Ri $=0.2$, (c) Ri $=0.4$, (d) Ri $=0.6$, (e) Ri $=0.7$, (f) Ri $=1.0$; ($\overline{\omega}_{min},\overline{\omega}_{max},\Delta\overline{\omega}_{}) \equiv (-4,4,0.12)$.}
	\label{meannearvort}
\end{figure}

\begin{figure}[htbp]
	\centering
	\begin{subfigure}[t]{0.3\textwidth}
		\includegraphics[width=\linewidth]{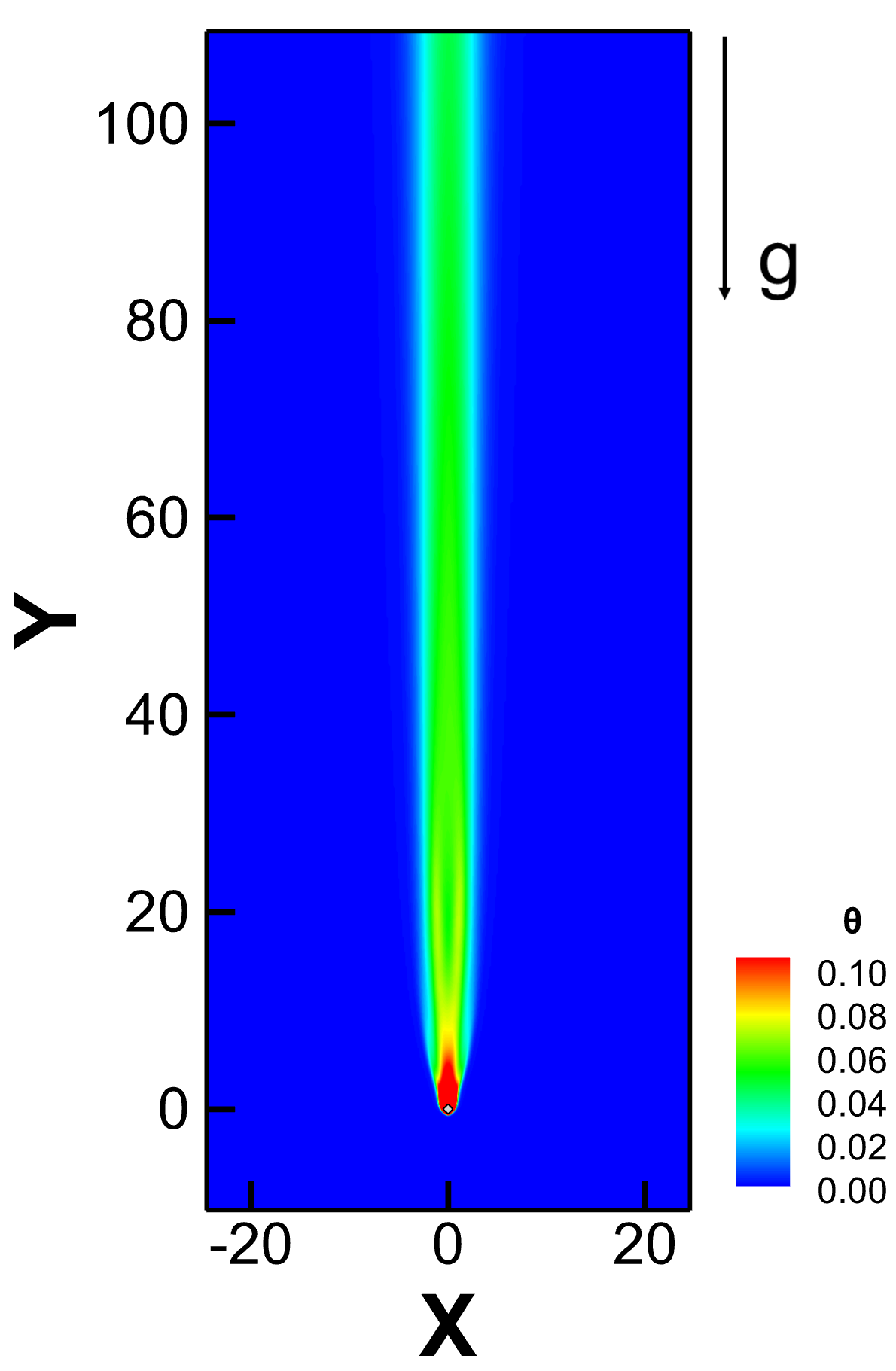}
		\caption{Ri $=0.0$}
		\label{meantemp0}
	\end{subfigure}\hfill
	\begin{subfigure}[t]{0.3\textwidth}
		\includegraphics[width=\linewidth]{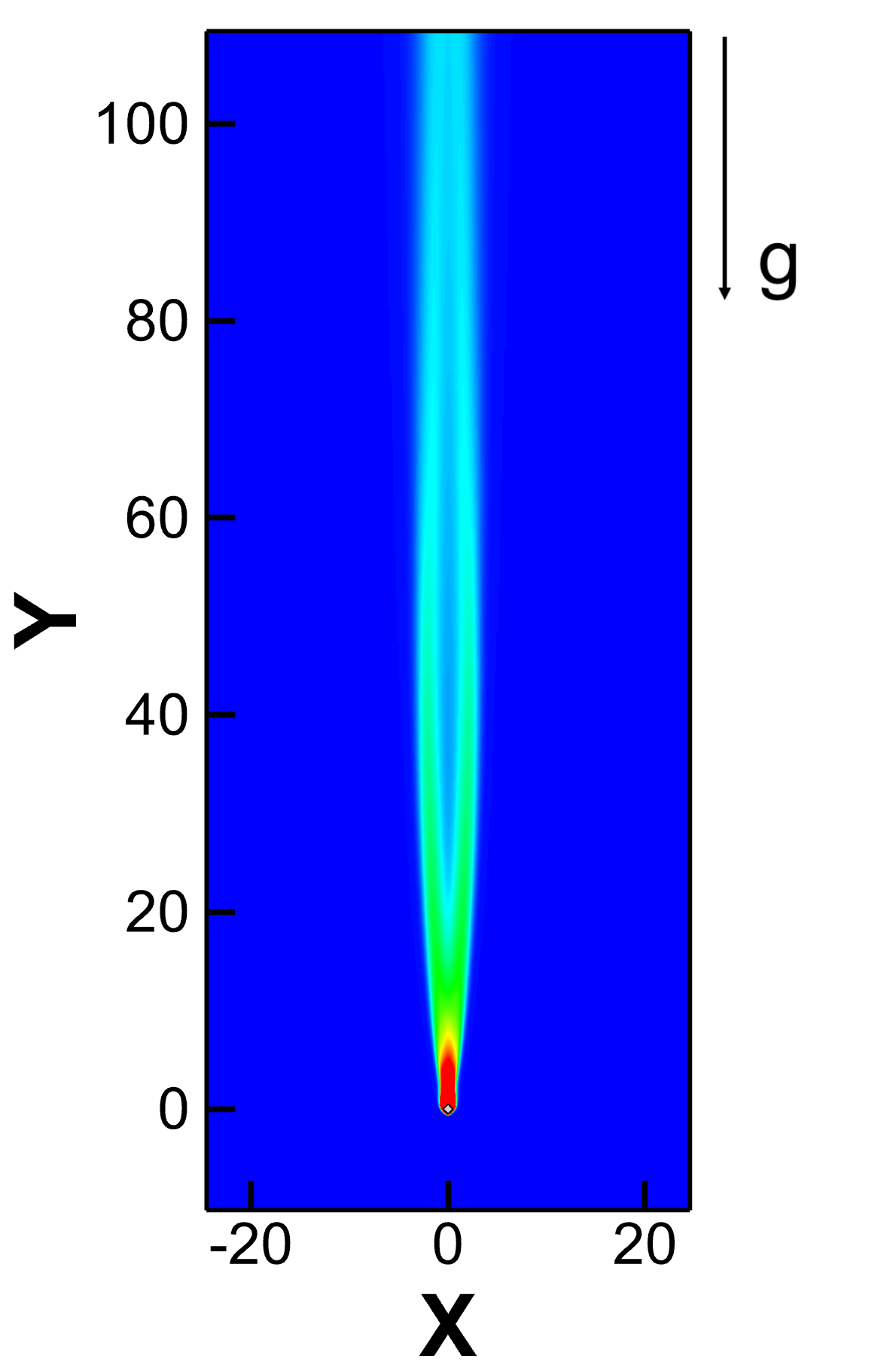}
		\caption{Ri $=0.2$}
		\label{meantemp02}
	\end{subfigure}\hfill
	\begin{subfigure}[t]{0.3\textwidth}
		\includegraphics[width=\linewidth]{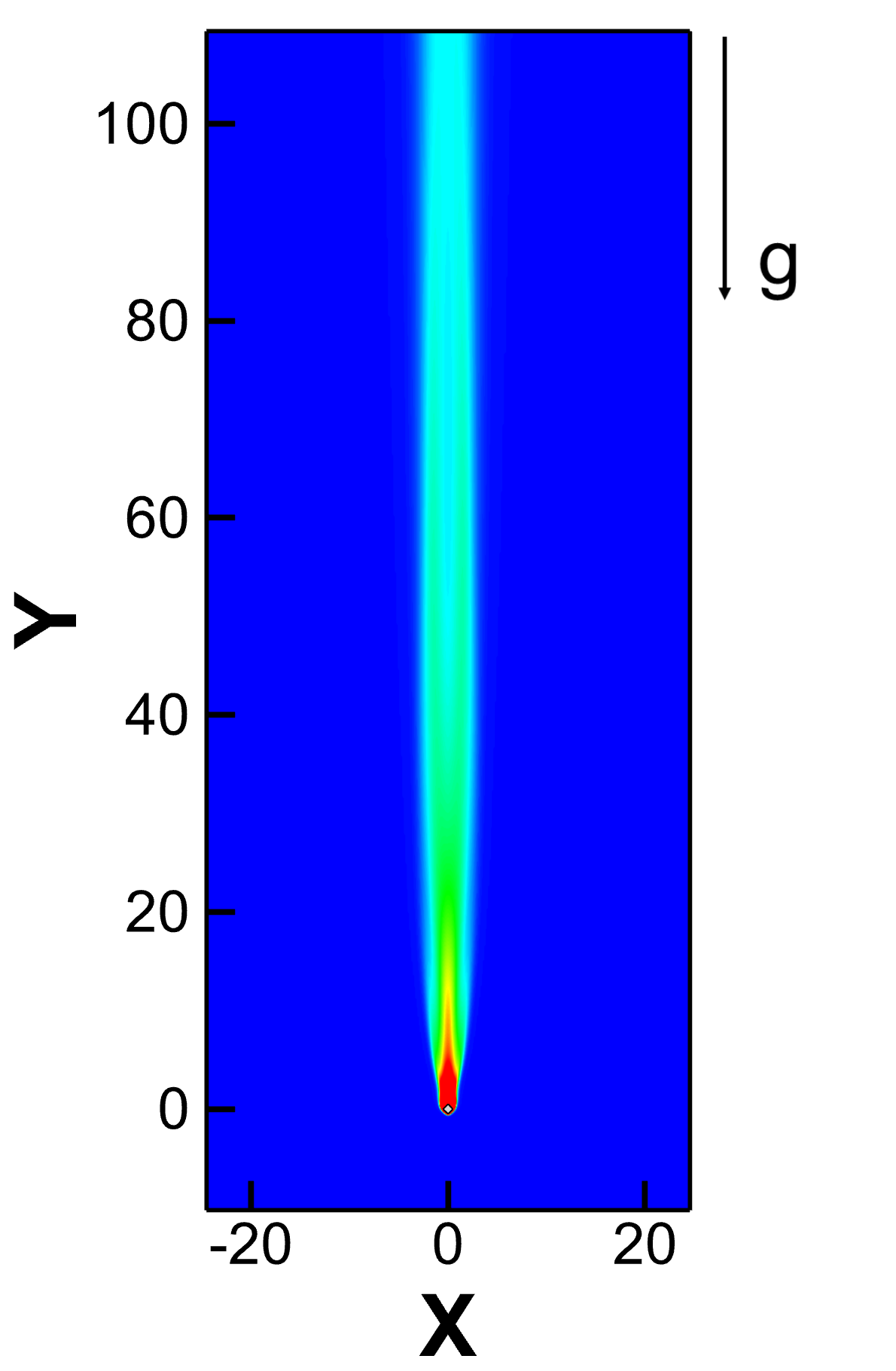}
		\caption{Ri $=0.4$}
		\label{meantemp04}
	\end{subfigure}\hfill

	\begin{subfigure}[t]{0.3\textwidth}
		\includegraphics[width=\textwidth]{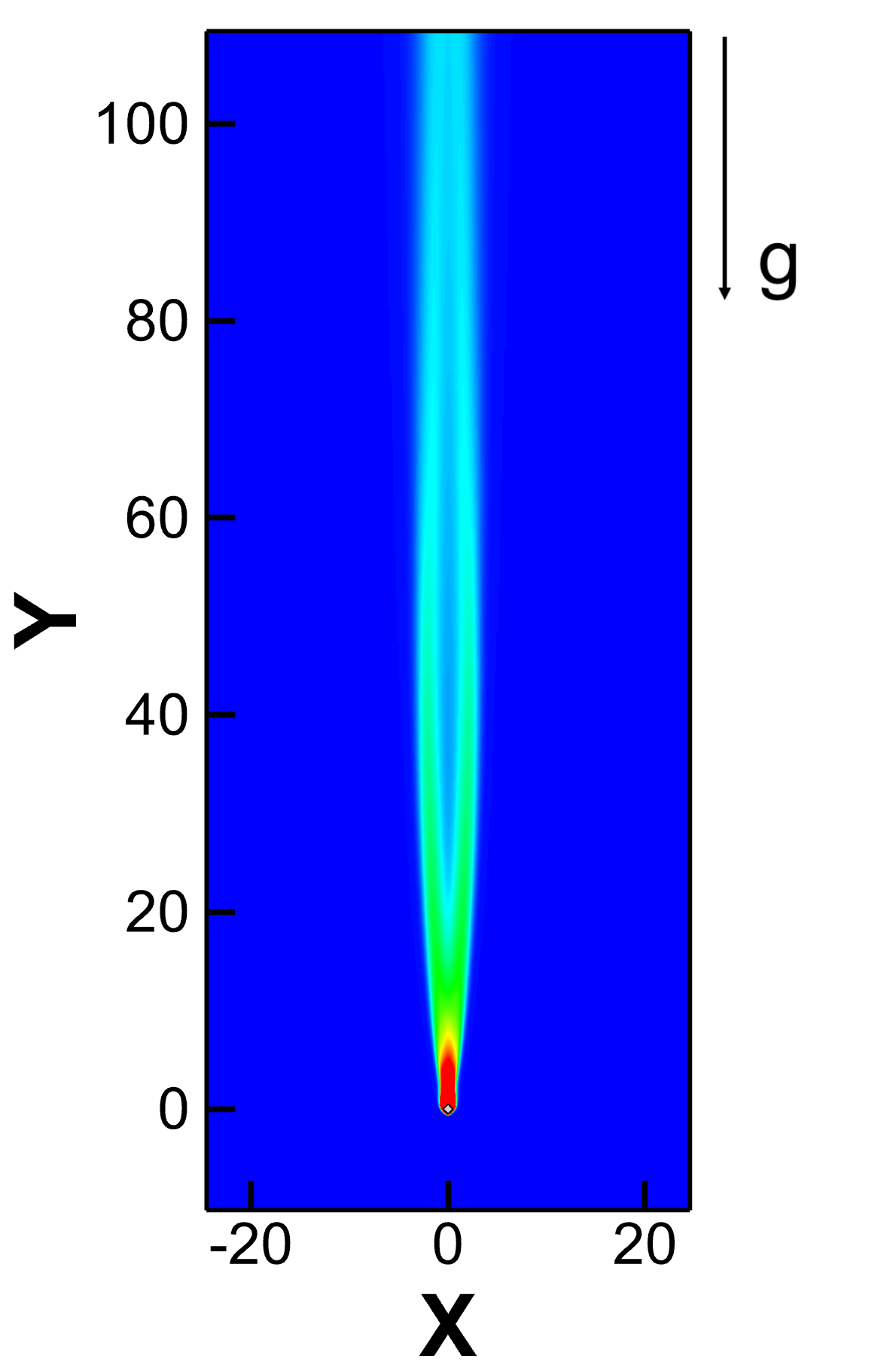}
		\caption{Ri $=0.6$}
		\label{meantemp06}
	\end{subfigure}\hfill
	\begin{subfigure}[t]{0.3\textwidth}
		\includegraphics[width=\linewidth]{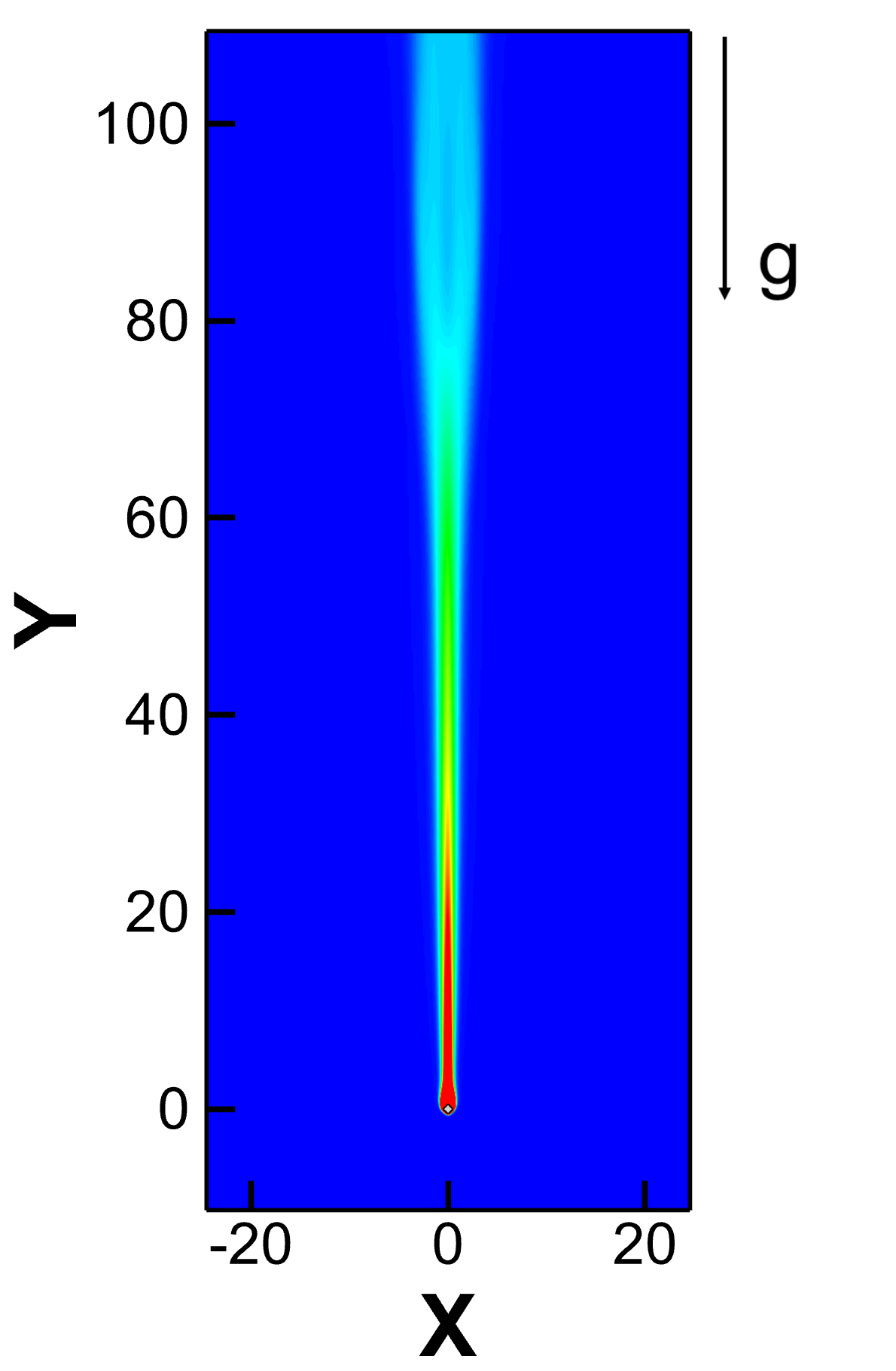}
		\caption{Ri $=0.7$}
		\label{meantemp07}
	\end{subfigure}\hfill
	\begin{subfigure}[t]{0.3\textwidth}
		\includegraphics[width=\linewidth]{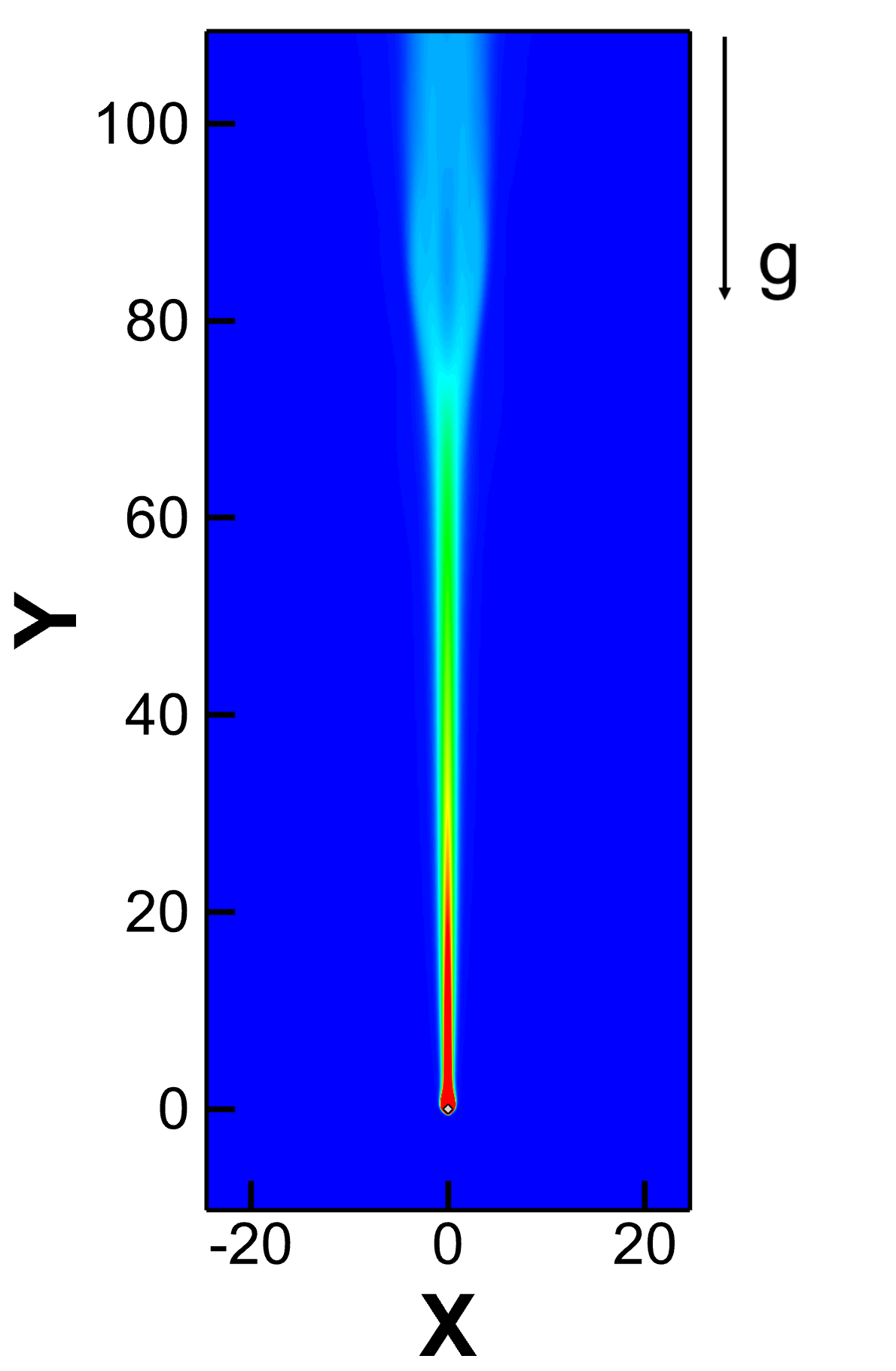}
		\caption{Ri $=1.0$}
		\label{meantemp10}
	\end{subfigure}\hfill
	
	\caption{Time-averaged temperature contours for Re $=100$ and $0.0\leq$ Ri $\leq1.0$: (a) Ri $=0.0$, (b) Ri $=0.2$, (c) Ri $=0.4$, (d) Ri $=0.6$, (e) Ri $=0.7$, (f) Ri $=1.0$; ($\overline{\theta}_{min},\overline{\theta}_{max},\Delta\overline{\theta}_{}) \equiv (0,0.1,0.001)$.}
	\label{meantemp}
\end{figure}

\begin{figure}[htbp]
	\centering
	\begin{subfigure}[t]{0.3\textwidth}
		\includegraphics[width=\linewidth]{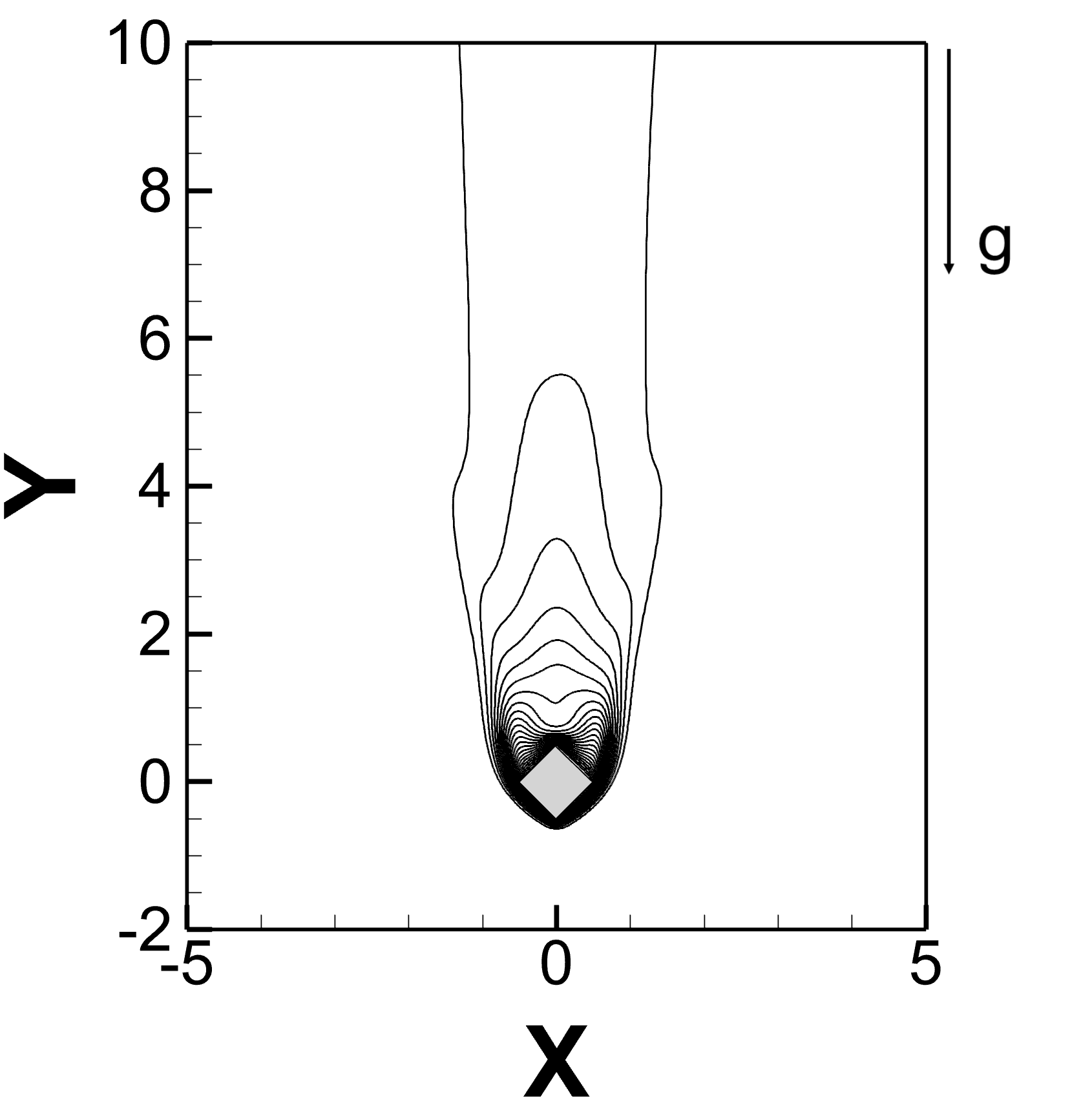}
		\caption{Ri $=0.0$}
		\label{meanneartemp0}
	\end{subfigure}\hfill
	\begin{subfigure}[t]{0.3\textwidth}
		\includegraphics[width=\linewidth]{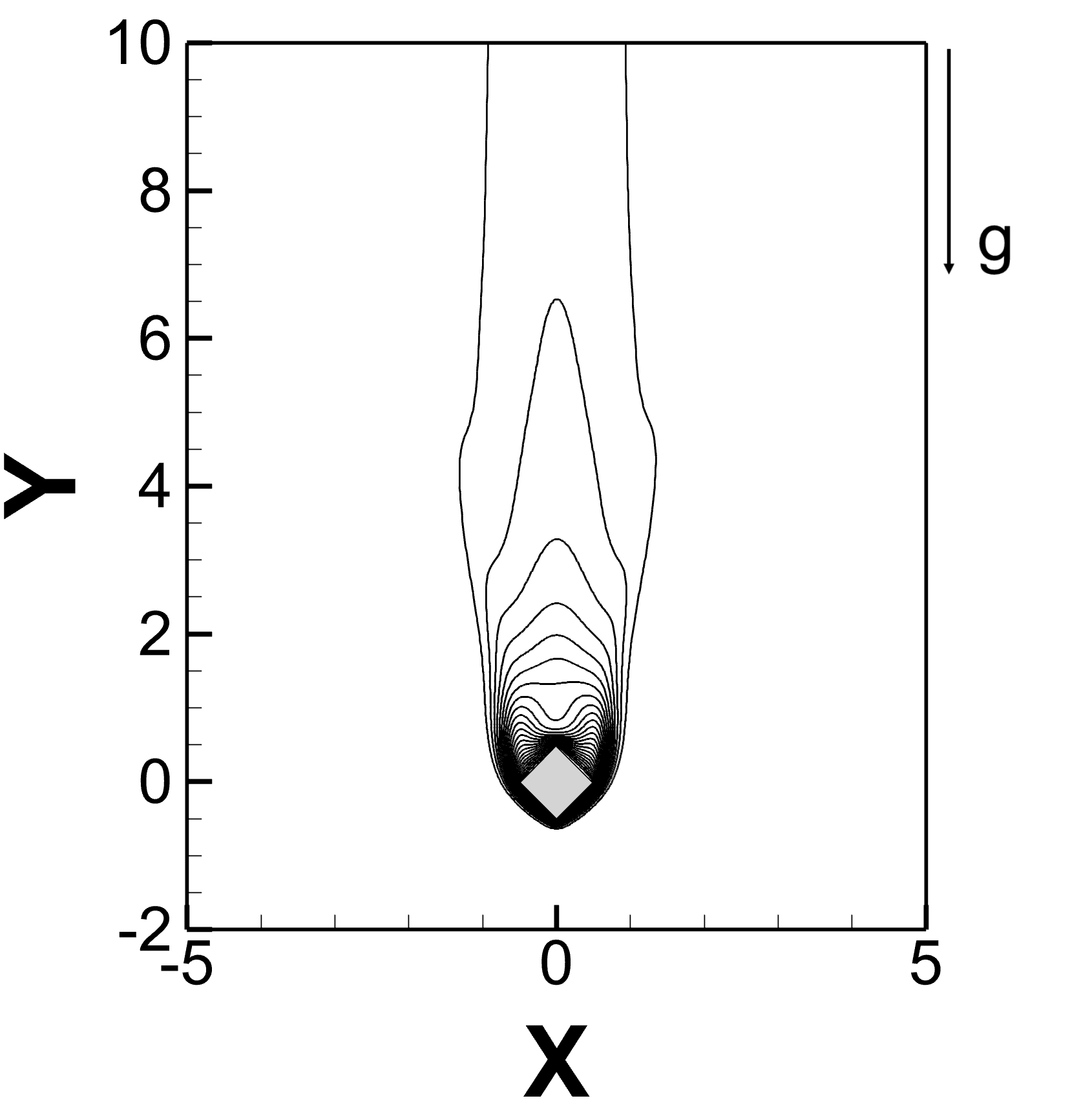}
		\caption{Ri $=0.2$}
		\label{meanneartemp02}
	\end{subfigure}\hfill
	\begin{subfigure}[t]{0.3\textwidth}
		\includegraphics[width=\linewidth]{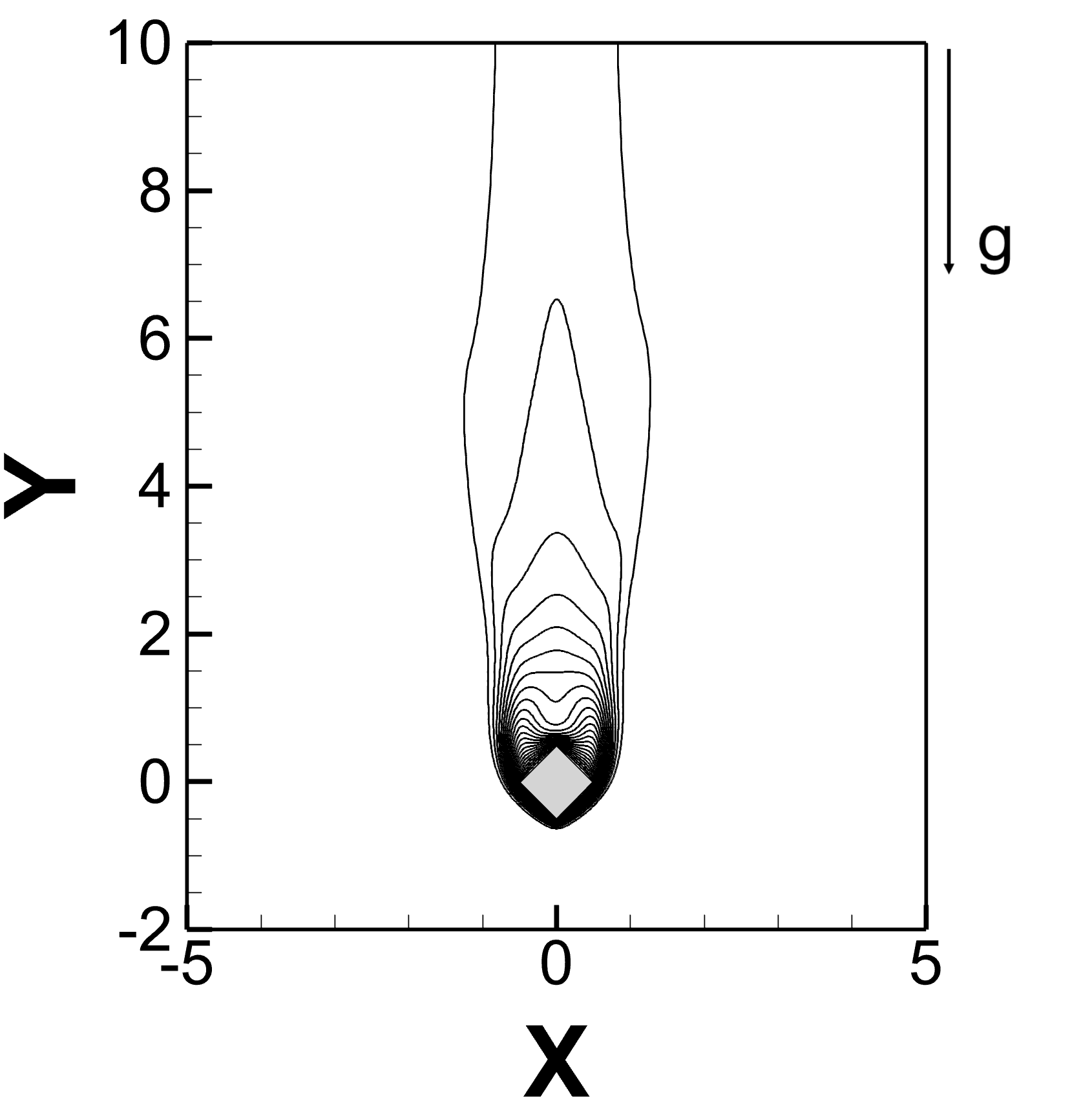}
		\caption{Ri $=0.4$}
		\label{meanneartemp04}
	\end{subfigure}\hfill

	\begin{subfigure}[t]{0.3\textwidth}
		\includegraphics[width=\textwidth]{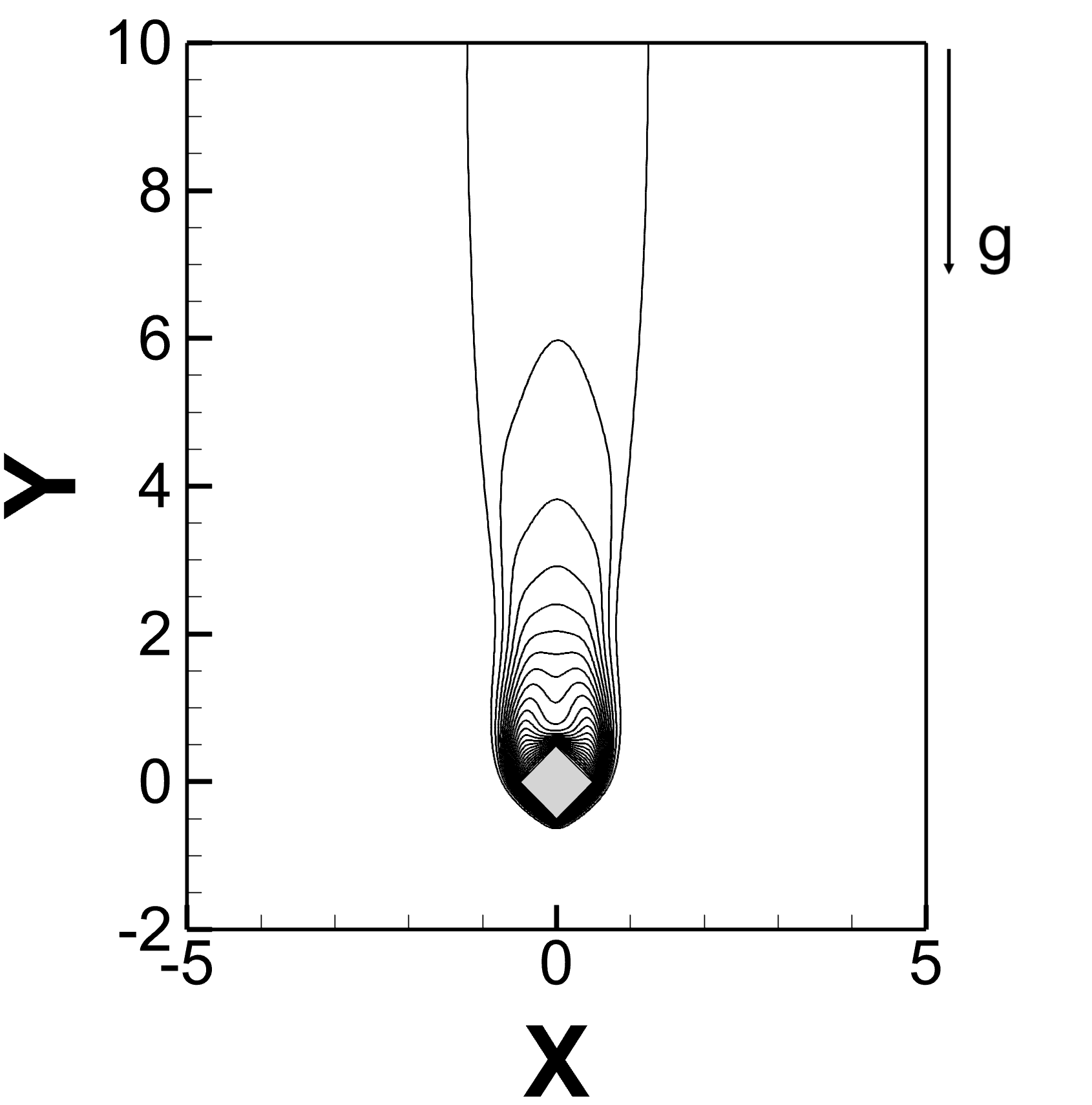}
		\caption{Ri $=0.6$}
		\label{meanneartemp06}
	\end{subfigure}\hfill
	\begin{subfigure}[t]{0.3\textwidth}
		\includegraphics[width=\linewidth]{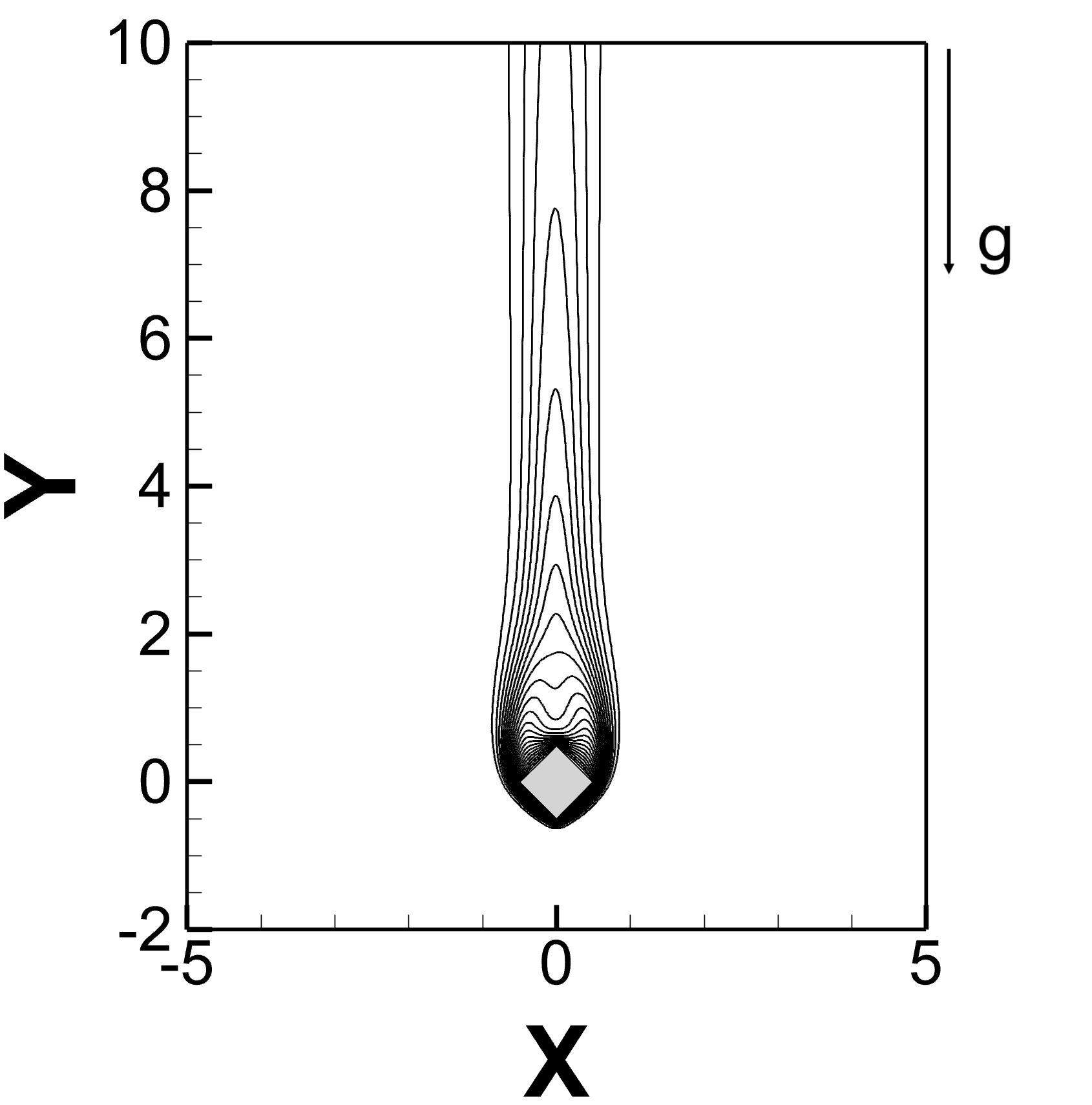}
		\caption{Ri $=0.7$}
		\label{meanneartemp07}
	\end{subfigure}\hfill
	\begin{subfigure}[t]{0.3\textwidth}
		\includegraphics[width=\linewidth]{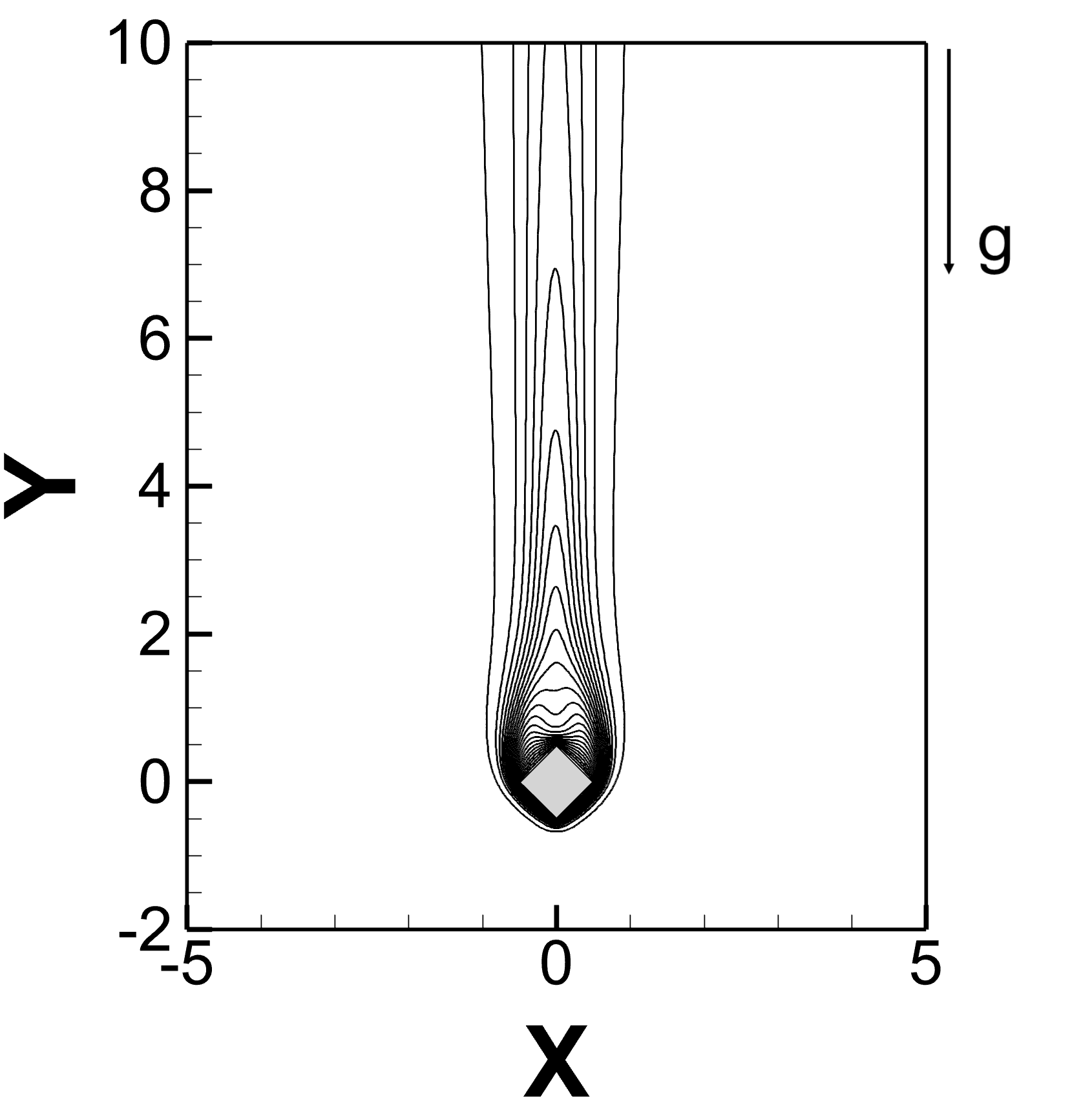}
		\caption{Ri $=1.0$}
		\label{meanneartemp10}
	\end{subfigure}\hfill
	
	\caption{Time-averaged temperature contours (in the cylinder vicinity) for Re $=100$ and $0.0\leq$ Ri $\leq1.0$: (a) Ri $=0.0$, (b) Ri $=0.2$, (c) Ri $=0.4$, (d) Ri $=0.6$, (e) Ri $=0.7$, (f) Ri $=1.0$; ($\overline{\theta}_{min},\overline{\theta}_{max},\Delta\overline{\theta}_{}) \equiv (0.05,1,0.04)$.}
	\label{meanneartemp}
\end{figure}

Time-averaged vorticity and temperature contours for Re $=100$ and $0.0\leq$ Ri $\leq1.0$ are presented in \Cref{meanvort,meantemp} in the full domain and \Cref{meannearvort,meanneartemp} in the vicinity of the cylinder, respectively. The time-averaged vorticity contours in the complete domain (see \Cref{meanvort}) reveal that at Ri $=0.0$, the strength of the vorticity is high in the near-field but diminishes in the far-field. In contrast, for Ri $>$ 0.0, the strength of the vorticity is weak in the near-field but grows on moving downstream, showing the opposite nature of the forced convective vorticity and the natural convective vorticity. The forced convective vorticity is introduced due to the shear layer separation at the transverse/lateral corners of the cylinder, while the natural convective vorticity is introduced due to buoyancy, which causes a jet-like effect in the wake. The interplay of these opposing vorticities is the cause of many physical phenomena that we have observed in this study and will be discussed in detail later. Beyond Ri $=0.7$, the strength of the vorticity is significantly higher than that of the cases below Ri $=0.7$, and the oppositely-signed vortices remain attached up to the far-field, which was not the case at lower Ri. Further, the two vortices are found to separate suddenly in the far-field at a point (around Y $=65$) which also coincides with the point of onset of the far-field unsteadiness. All positive Ri cases display vorticity inversion, as seen in the instantaneous contours earlier. In the near-field plot of vorticity, the width between the shear layers is seen to reduce with increasing Ri, similar to the instantaneous case. Beyond Ri $=0.7$, the inversion can be seen in the near-field, and the vortices form a steady recirculation bubble, whose length decreases with increasing Ri. These cases resemble the instantaneous plots as the flow is steady in the near-field. Time-averaged temperature contours (see \Cref{meantemp}) show a widening of the plume in the near and intermediate fields for Ri $<$ 0.7. Beyond Ri=0.7, the plume is narrow in the near-field, and there is a sharp increase in the plume width in the far-field, associated with the onset of the far-field unsteadiness. Such behavior is also observed in the near-field temperature contours (see \Cref{meanneartemp}), where the plume is wide in the near-field for low Ri, and beyond Ri $=0.7$, the plume is narrow.

\begin{figure}[htbp]
    \centering
  
    \begin{subfigure}[t]{0.45\textwidth}
        \centering
        \includegraphics[width=\linewidth]{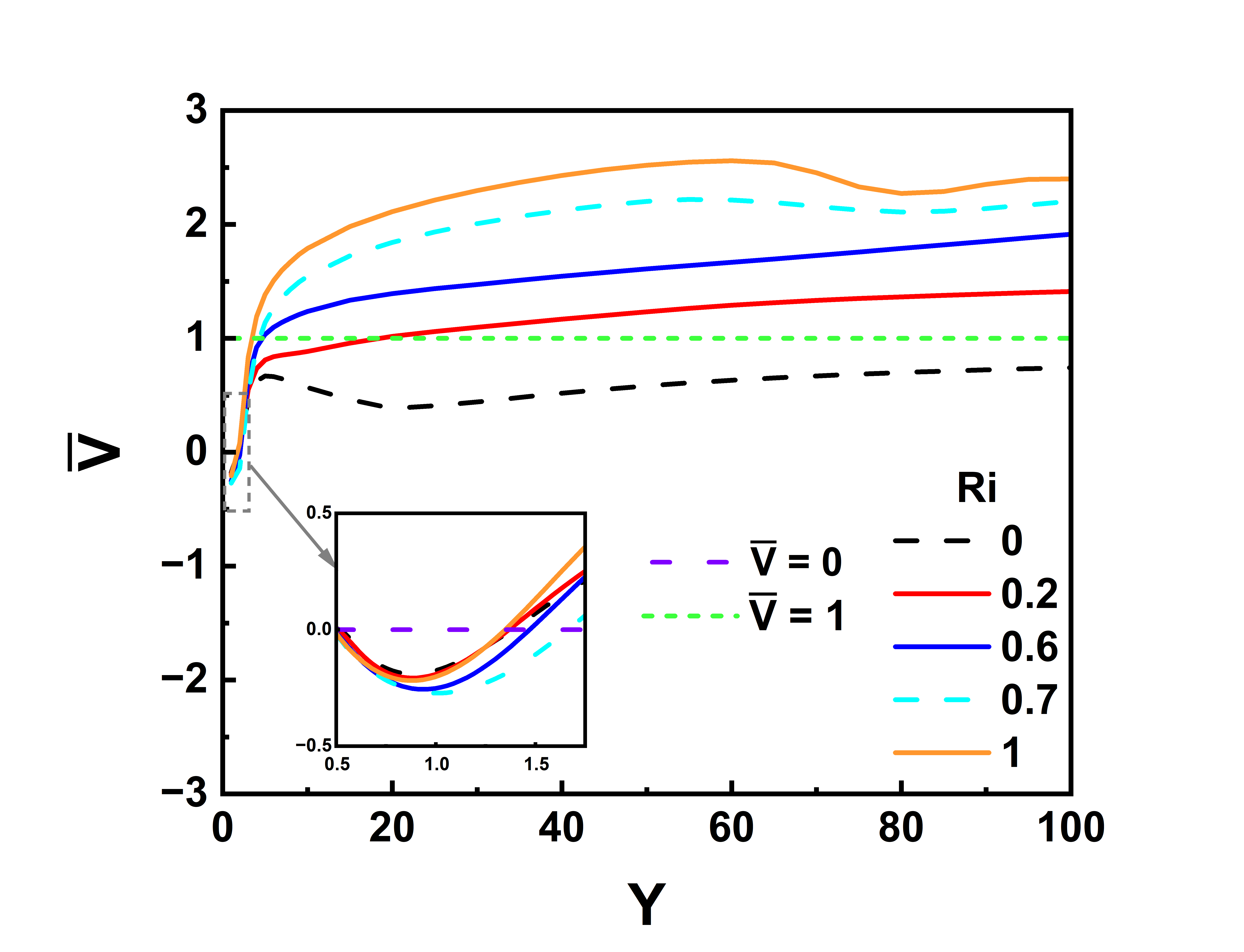}
        \caption{}
        \label{axialvel}
    \end{subfigure}
    \hspace{0.05\textwidth}
    \begin{subfigure}[t]{0.45\textwidth}
        \centering
        \includegraphics[width=\linewidth]{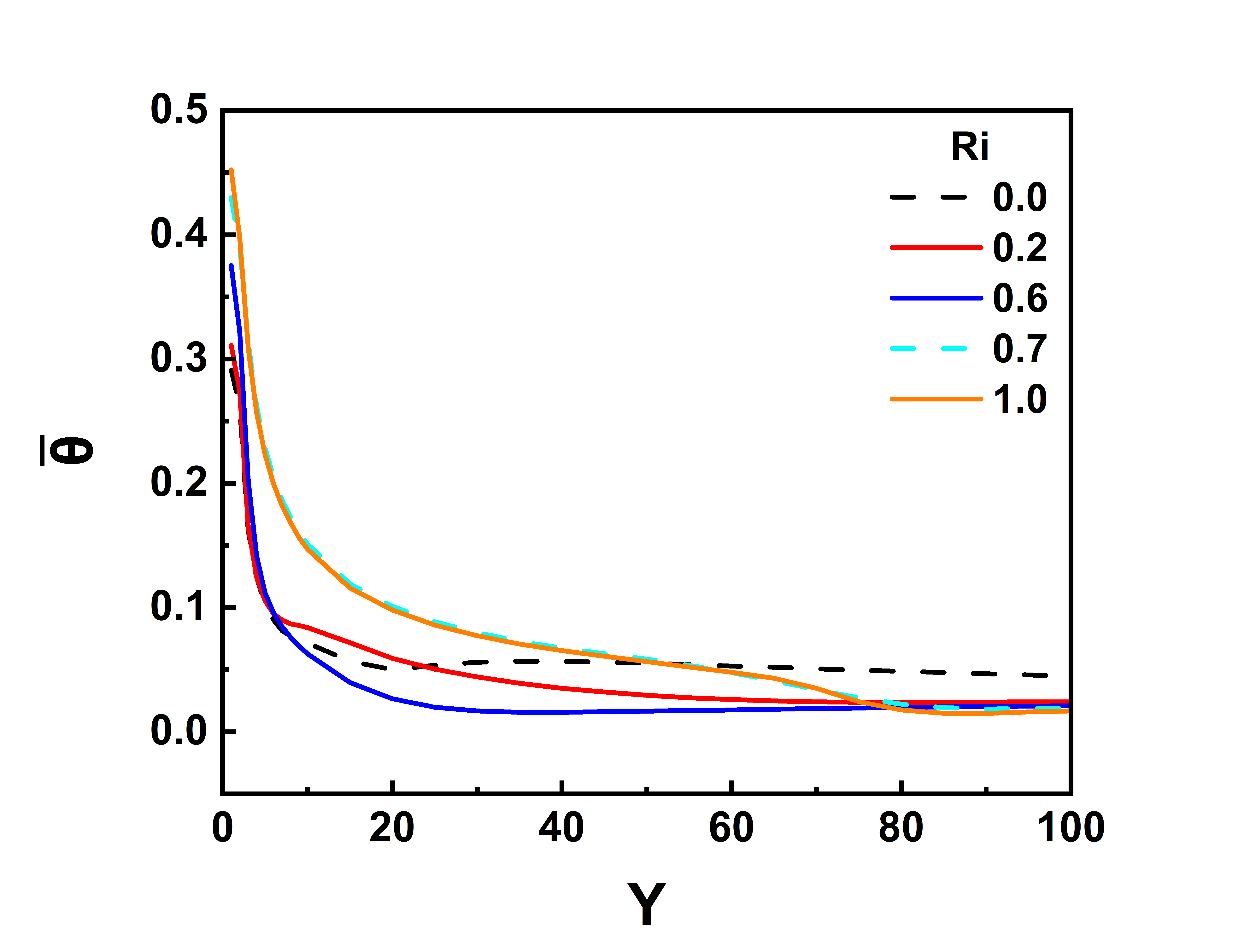}
        \caption{}
        \label{axialtemp}
    \end{subfigure}
    \caption{Variation of (a) centerline streamwise velocity and (b) centerline temperature with streamwise distance.}
    \label{axial}
\end{figure}

\Cref{axial} shows the centerline streamwise velocity and centerline temperature along the streamwise length of the domain for different Ri. \Cref{axialvel} reveals the buoyancy-driven jetting effect, which becomes more pronounced as one moves further downstream. The average streamwise velocity ($V_{avg}$) at any particular downstream location Y is equal to the free-stream velocity ($V_{\infty}$ = 1.0, in the present case). It is to be noted that the wake has a momentum deficit when the centerline streamwise velocity is lower than $V_{avg}$ and becomes a wake with an excess momentum if the centerline velocity exceeds $V_{avg}$ at that location. There is a deficit of momentum in the immediate wake of the cylinder due to shear layer separation at the front lateral corners. However, the presence of the natural convective flow counteracts this deficit and accelerates the flow behind the cylinder, leading to a faster recovery and, eventually, an excess of momentum beyond the cylinder exhibiting jet-like characteristics far downstream, while the flow immediately downstream of the cylinder shows a wake-like behavior in the near-field. For the purpose of visualizing the location of momentum recovery and subsequent excess, $\overline{V}$ = 1.0 has been plotted in \Cref{axialvel}. The centerline velocity variation reveals that the momentum deficit persists in the forced convection case throughout the domain despite the recovery of some momentum later downstream. For positive Ri, the momentum recovery is aided by the buoyancy-driven jet, and the recovery distance decreases with increasing Ri, occurring at Y $=20$ for Ri $=0.2$ and around Y $=5$ for Ri $\geq$ 0.6. Further downstream, for Ri $\geq$ 0.7, there is a decrease in velocity in the far-field, which occurs around the same distance as the onset of the far-field unsteadiness. This phenomenon is attributed to the widening of the wake caused by the unsteadiness of the flow beyond this location. The inset in the same graph shows the centerline velocity in the immediate wake of the cylinder (0.5 $\leq$ Y $\leq$ 1.75), providing an insight into the recirculation length. Starting from the base of the cylinder (Y $=0.5$), there is a negative streamwise velocity due to the recirculation of flow as it turns towards the cylinder (in the time-averaged sense). However, the velocity quickly becomes positive after the recirculation region. The recirculation length shows a nonlinear variation, with Ri $=1$ and Ri $=0.2$ having the shortest recirculation length, followed by Ri $=0.0$, 0.6 and 0.7, respectively. \Cref{axialtemp} shows the centreline variation of temperature with streamwise distance, revealing a reducing trend of temperature with downstream distance. The centerline temperature is highest for higher Ri close to the cylinder; however, in the far-field, the behavior is the opposite, and higher Ri have lower temperatures, albeit the difference is marginal in the far-field.

\begin{figure}[htbp]
	\centering
	\begin{subfigure}[t]{0.45\textwidth}
		\includegraphics[width=\linewidth]{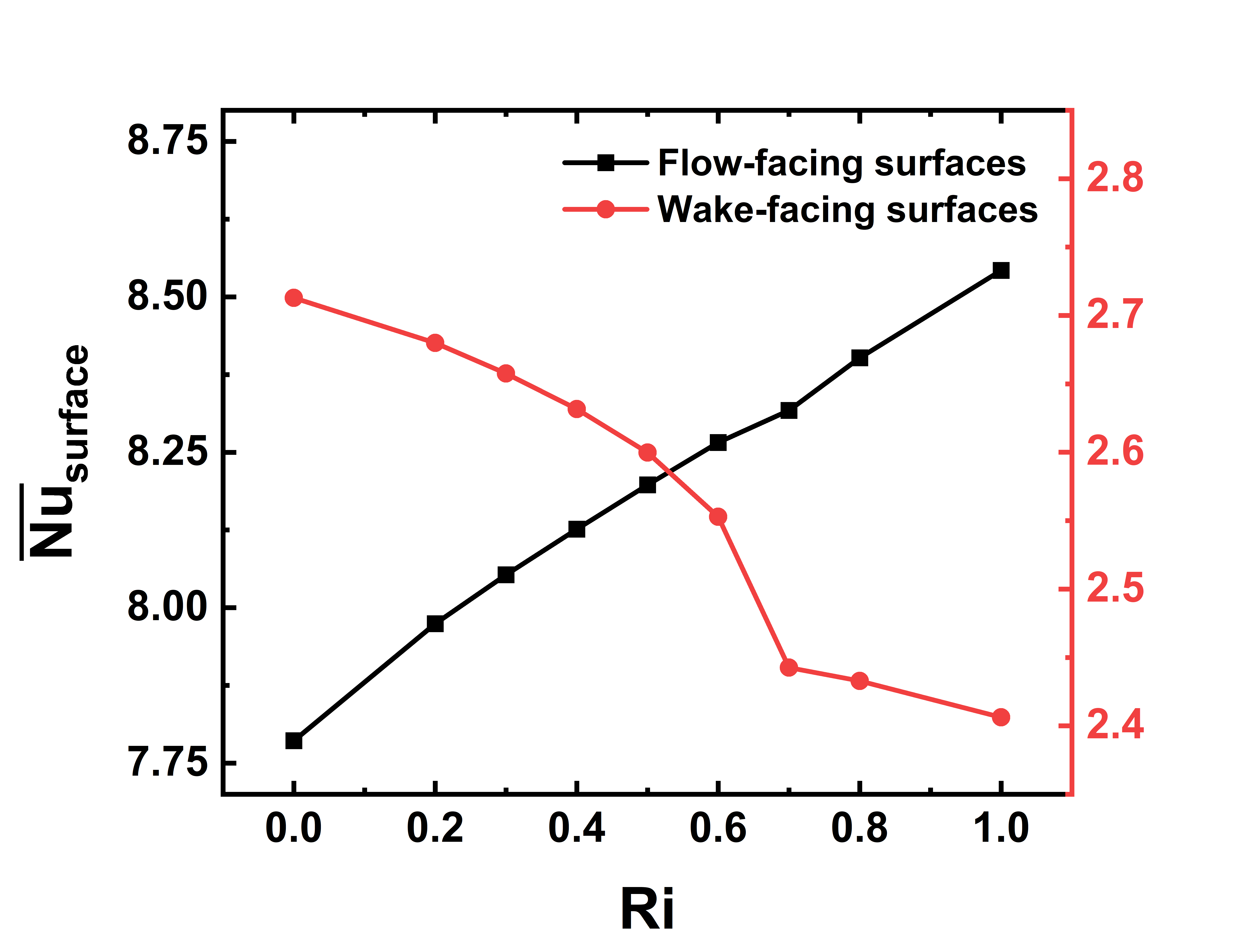}
		\caption{}
		\label{surfacenusselt}
	\end{subfigure}\hfill
	\begin{subfigure}[t]{0.45\textwidth}
		\includegraphics[width=\linewidth]{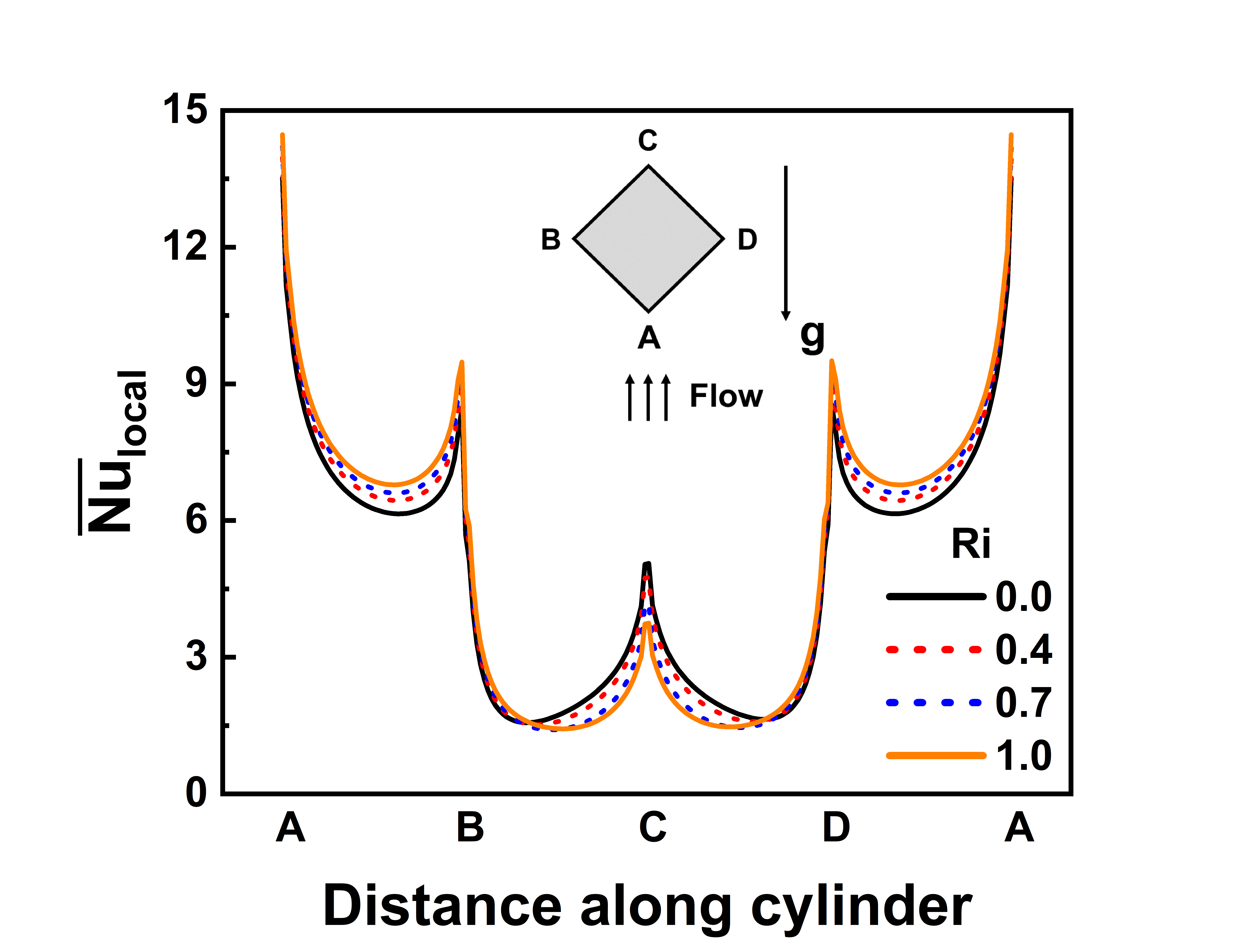}
		\caption{}
		\label{localnusselt}
	\end{subfigure}
	\caption{Variation of (a) time-averaged Nusselt number, and (b) local Nusselt number along the cylinder surface, for various Ri.}
	\label{nusseltvariation}
\end{figure}

Nusselt number is a function of the spatial coordinates as well as time. Therefore, one can average the Nusselt number in space, time, or both. $\overline{Nu}_{local}$ refers to the time-averaged local Nusselt number at a particular location on the cylinder, while $\overline{Nu}_{surface}$ refers to the temporally and spatially averaged Nusselt number for a particular surface on the cylinder, and $\overline{Nu}$ refers to the temporally and spatially averaged Nusselt number for the entire cylinder. \Cref{nusseltvariation} shows the variation of time-averaged Nusselt number for different cylinder surfaces and local Nusselt number along the cylinder surface for various Ri, along with the nomenclature of the cylinder surfaces (AB, BC, CD, and DA). For flow-facing surfaces (AB and DA), the mean Nusselt number increases with increasing Ri, whereas for wake-facing surfaces (BC and CD), the mean Nusselt number decreases with increasing Ri. There is a sharp decrease in the Nusselt number at Ri $=0.7$ for the wake-facing surfaces and, subsequently, in the Nusselt number for the full cylinder. The shedding phenomenon helps in higher heat transfer due to increasing momentum exchange between the thermal boundary layers at various walls and the separating shear layers. When vortex shedding is suppressed, the momentum transport is inhibited as the mutual interaction is no longer active. Therefore, the Nusselt number sharply decreases for the wake-facing surfaces at Ri $=0.7$. The Nusselt number for the surfaces AB and DA, as well as that for surfaces BC and CD, are equal due to the symmetric nature of the time-averaged flow about the centerline. The local Nusselt number is the highest for the vertex A, which directly faces the forced flow, and it is considerably lower for the wake-facing surfaces compared to the flow-facing surfaces due to the recirculation region having very weak streamwise velocities, causing decreased heat transfer from the wake-facing surfaces. The Nusselt number reaches a local maxima at the lateral corners of the cylinder (B and D), due to the acceleration of flow at these corners.

\begin{table}[htbp]
\centering
\caption{Variation of force coefficients and Nusselt number with Richardson number (Ri).}
\setlength{\tabcolsep}{0.4em}
\begin{tabular}{c cc c c}
\toprule
\toprule
\multirow{2}{*}{\begin{tabular}[c]{@{}c@{}}Richardson \\ number (Ri)\end{tabular}} & \multicolumn{2}{c}{Drag coefficient ($C_D$)} & \multirow{2}{*}{\begin{tabular}[c]{@{}c@{}}RMS lift coefficient \\ ($C_{L,\text{rms}}$)\end{tabular}} & \multirow{2}{*}{\begin{tabular}[c]{@{}c@{}}Mean Nusselt number \\ ($\overline{Nu}$)\end{tabular}} \\ 
\cmidrule(l){2-3}
& Mean ($\overline{C}_{D}$) & RMS ($C_{D,\text{rms}}$) &  &  \\ 
\midrule
0.0 & 1.81 & 0.032 & 0.500 & 5.25 \\ 
0.2 & 1.94 & 0.024 & 0.449 & 5.33 \\ 
0.4 & 2.05 & 0.015 & 0.373 & 5.38 \\ 
0.6 & 2.15 & 0.006 & 0.241 & 5.41 \\ 
0.7 & 2.19 & 0.0   & 0.0   & 5.38 \\ 
1.0 & 2.44 & 0.0   & 0.0   & 5.47 \\ 
\bottomrule
\bottomrule
\end{tabular}
\label{integparamvsri}
\end{table}

The variation of the integral parameters with Ri is listed in \Cref{integparamvsri}. $\overline{C}_D$ and $\overline{Nu}$ show an increase with increasing Ri. The buoyancy in the vicinity of the cylinder increases both pressure drag, and skin friction drag on the cylinder. On the other hand, as Ri increases, RMS quantities decrease as expected. All quantities presented in \Cref{integparamvsri} are near-field quantities, and hence, when the near-field is stabilized, these values become constant with no periodicity. Hence, RMS quantities fall to zero beyond the critical Ri. The RMS Nusselt number has not been reported as it is extremely small for all Ri. The mean Nusselt number presents an interesting variation. Below Ri $=0.7$, the Nusselt number increases monotonically, decreasing to a local minimum value at Ri $=0.7$ and subsequently increasing for Ri $>$ 0.7. Below 0.7, flow acceleration due to buoyancy forces causes thinning of the thermal boundary layer, leading to higher Nusselt numbers for the entire cylinder. There is a greater increase in the Nusselt number of the flow-facing surfaces compared to the decrease in the Nusselt number of the wake-facing surfaces, leading to an overall increase in the Nusselt number of the cylinder. However, at Ri $=0.7$, as discussed earlier, the Nu of the wake-facing surfaces sharply decreases due to the inhibition of vortex shedding, and this decrease is greater than the increase in the Nusselt number of flow-facing surfaces, leading to an overall decrease in the Nusselt number of the cylinder. Beyond Ri $=0.7$, an increase in cylinder Nu is seen, similar to the trend below Ri $=0.7$.

 The critical Ri for the transition to steady flow for both square and circular cylinders is 0.15, as reported by Sharma and Eswaran \cite{sharmaeswaran}, and Chang and Sa \cite{changsa}. As reported by \cite{ellipticalhasan}, the corresponding value for an elliptical cylinder with an aspect ratio (AR) of 2 is between 0.75 and 1.0, while for an AR = 4, no vortex shedding suppression is observed for the range of Ri (0.0-1.25) considered in their study. However, the critical Ri for our inclined square cylinder in the present study is between 0.6 and 0.7, which is significantly greater than that reported for square and circular cylinder cases. This large variation may be attributed to the (i) thickness and vorticity content of the separating shear layer and/or (ii) afterbody effect, that is, the part of the obstacle beyond the separation of the flow from it. The shape of the obstacle and the corresponding afterbody are expected to affect the transition sequence in mixed convective flows past obstacles. The exact details of the effect require further investigation.

A comparison of integral parameters among obstacles of different shapes (square \cite{sharmaeswaran}, circular \cite{changsa}, and elliptical cylinders \cite{ellipticalhasan}) against the diamond-shaped cylinder of the present study is given in \Cref{shapecomp}. It is seen that the drag coefficient at Ri $=0.0$ is highest in the case of an elliptical cylinder with AR = 4, followed by a diamond cylinder, an elliptical cylinder with AR = 2, a square cylinder, and finally, a circular cylinder. A similar trend is observed at higher Ri with minor changes. Similarly, the elliptical cylinders reveal the highest Nusselt number, followed by diamond and square cylinders at all Ri. The Strouhal number is approximately the same for elliptical and diamond-shaped cylinders, while it is considerably lower for square and circular cylinders.

\begin{table}[htbp]
\centering
\caption{ Variation of drag coefficient, Nusselt number, and Strouhal number with Ri for different cylinder shapes. }
\setlength{\tabcolsep}{0.6em}
\begin{tabular}{l c c c c}
\toprule
\toprule
\begin{tabular}[c]{@{}c@{}}Cylinder \\ shape\end{tabular} & \begin{tabular}[c]{@{}c@{}}Richardson \\ number (Ri)\end{tabular} & \begin{tabular}[c]{@{}c@{}}Mean drag \\ coefficient ($\overline{C}_D$)\end{tabular} & \begin{tabular}[c]{@{}c@{}}Mean Nusselt \\ number ($\overline{Nu}$)\end{tabular} & \begin{tabular}[c]{@{}c@{}}Strouhal \\ number (St)\end{tabular} \\
\midrule
Square (\cite{sharmaeswaran}) & 0.0 & 1.57 & 4.09 & 0.147 \\
Circular (\cite{changsa}) & 0.0 & 1.22 & 5.22 & 0.154 \\
Elliptical (AR = 2) (\cite{ellipticalhasan}) & 0.0 & 1.77 & 5.93 & 0.183 \\
Elliptical (AR = 4) (\cite{ellipticalhasan}) & 0.0 & 2.10 & 6.26 & 0.176 \\
Diamond (present) & 0.0 & 1.81 & 5.25 & 0.180 \\
\midrule
Square (\cite{sharmaeswaran}) & 0.5 & 2.07 & 4.55 & Steady \\
Elliptical (AR = 2) (\cite{ellipticalhasan}) & 0.5 & 2.14 & 6.17 & 0.218 \\
Elliptical (AR = 4) (\cite{ellipticalhasan}) & 0.5 & 2.82 & 6.65 & 0.224 \\
Diamond (present) & 0.5 & 2.11 & 5.40 & 0.212 \\
\midrule
Square (\cite{sharmaeswaran})  & 1.0 & 2.63 & 4.87 & Steady \\
Elliptical (AR = 2) (\cite{ellipticalhasan}) & 1.0 & 2.43 & 6.15 & Steady \\
Elliptical (AR = 4) (\cite{ellipticalhasan}) & 1.0 & 3.25 & 6.84 & Steady \\
Diamond (present) & 1.0 & 2.44 & 5.47 & Steady \\
\bottomrule
\bottomrule
\end{tabular}
\label{shapecomp}
\end{table}

\begin{figure}[htbp]
	\centering  
	\includegraphics[width=0.6\linewidth]{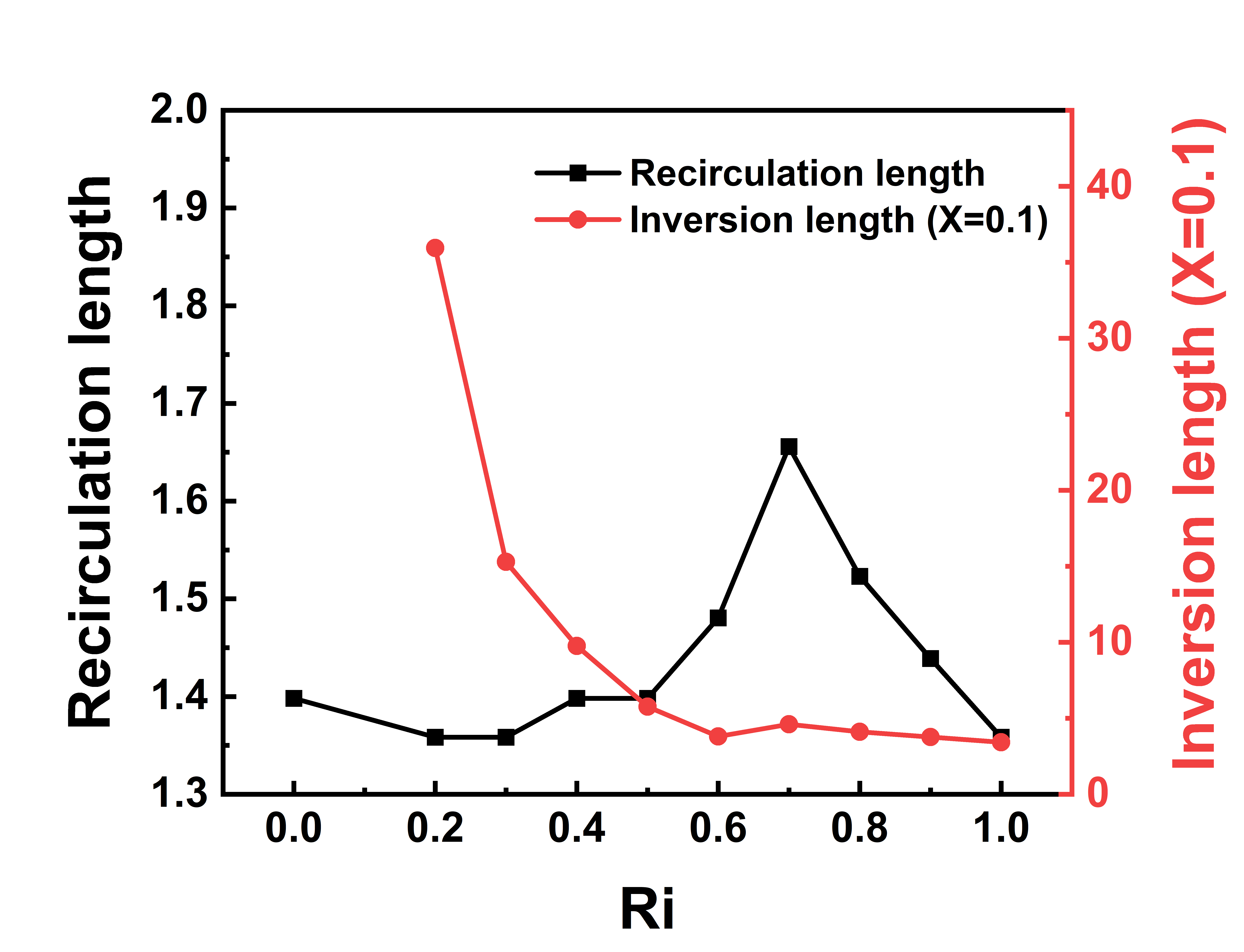}
        \caption{Variation of recirculation and inversion lengths with Ri. Recirculation length has been calculated by considering the time-averaged centerline (X $=0$) velocities, while the inversion length has been calculated by considering time-averaged velocities parallel to the centerline at X $=0.1$.}
        \label{recircinversion}
\end{figure}

\Cref{recircinversion} reveals the variation of recirculation and inversion lengths with Ri. For the calculation of the inversion length, time-averaged vorticity data is taken for all streamwise locations at a transverse location X $=0.1$. The location of the first sign change in the vorticity data yields the inversion length. To ensure independence of the result from the choice of transverse location, we have also calculated inversion length using X $=0.2$, and the results reveal no variation based on the choice of X. For the calculation of recirculation length, time-averaged streamwise velocity data along the centerline are considered. The location of the sign change in this data yields the recirculation length. There is a monotonic decrease in the inversion length up to Ri $=0.6$, followed by a marginal increase at Ri $=0.7$ and a subsequent reduction for higher Ri. The decrease in inversion length is due to the increasing strength of the natural convective vorticity over the strength of the forced convective vorticity with increasing Ri. The forced convective component is strong at low Ri but diminishes with downstream distance. At the same time, the natural convective component is weak in the near-field, but its strength grows with downstream distance. When Ri is increased, the strength of the natural convective vorticity closer to the cylinder also grows. The recirculation length shows an interesting variation, which was discussed briefly with \Cref{axialvel}. It increases up to Ri $=0.7$, where it peaks, followed by a decrease. Prior to Ri $=0.7$, vortex shedding actively occurs in the flow. Due to the jet formed by buoyancy, the interaction between the two shear layers is inhibited, as seen in near-field plots of vorticity and temperature (see \Cref{insnearvort,insneartemp}). This increases the shear layer length and, consequently, the recirculation length, which increases up to Ri $=0.7$. Beyond Ri $=0.7$, there is no vortex shedding, and the near-field flow becomes steady. Since the flow is steady, there is no mutual interaction between the opposite shear layers, which form a steady separation bubble, and the effect of increasing buoyancy beyond the critical suppression value is to add momentum to the separation bubble, reducing its length.

\begin{figure}[htbp]
			\centering
			\begin{subfigure}[t]{0.45\textwidth}
				\includegraphics[width=\linewidth]{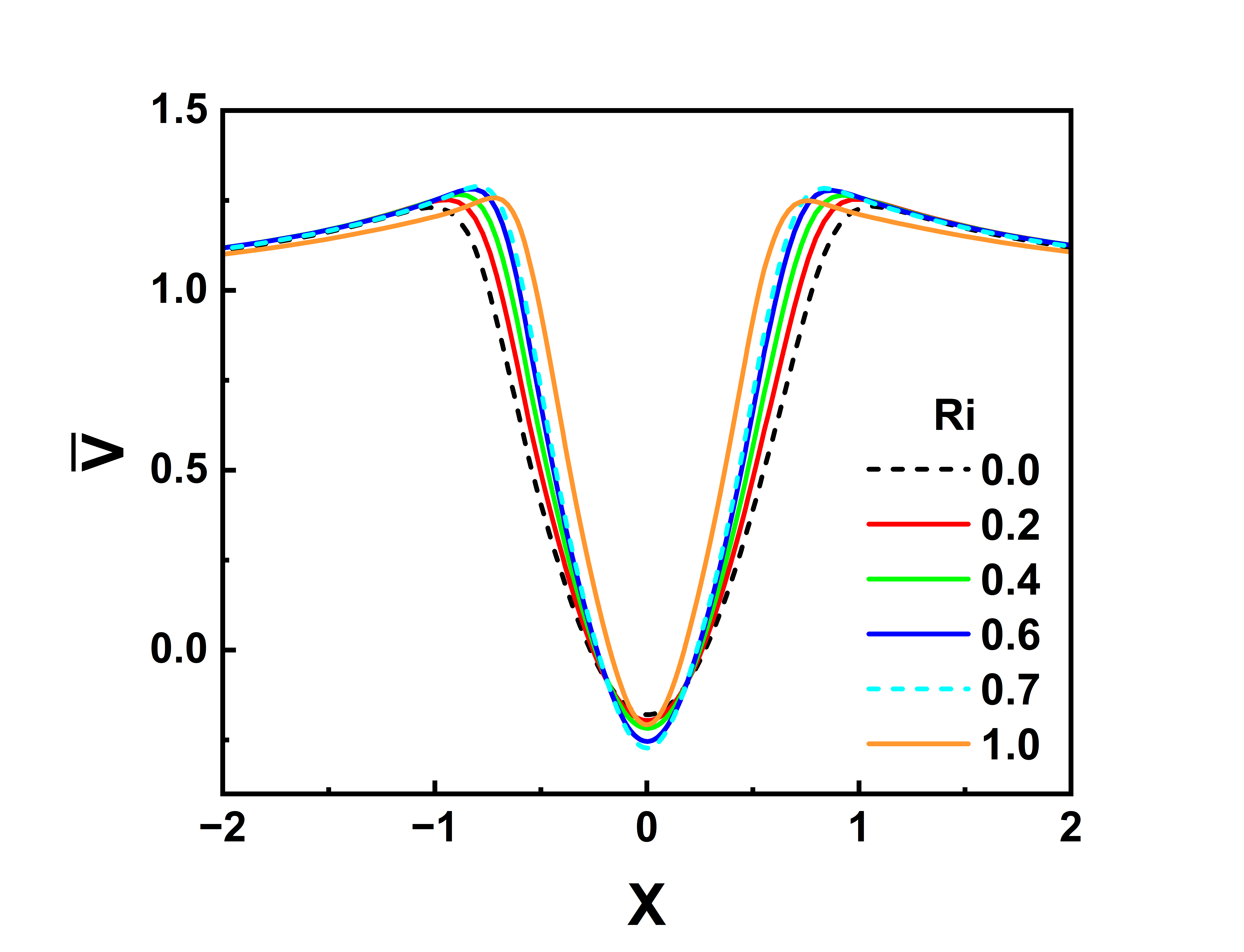}
				\caption{Y $=1$}
				\label{vprof1}
			\end{subfigure}\hspace{30pt}
			\begin{subfigure}[t]{0.45\textwidth}
				\includegraphics[width=\linewidth]{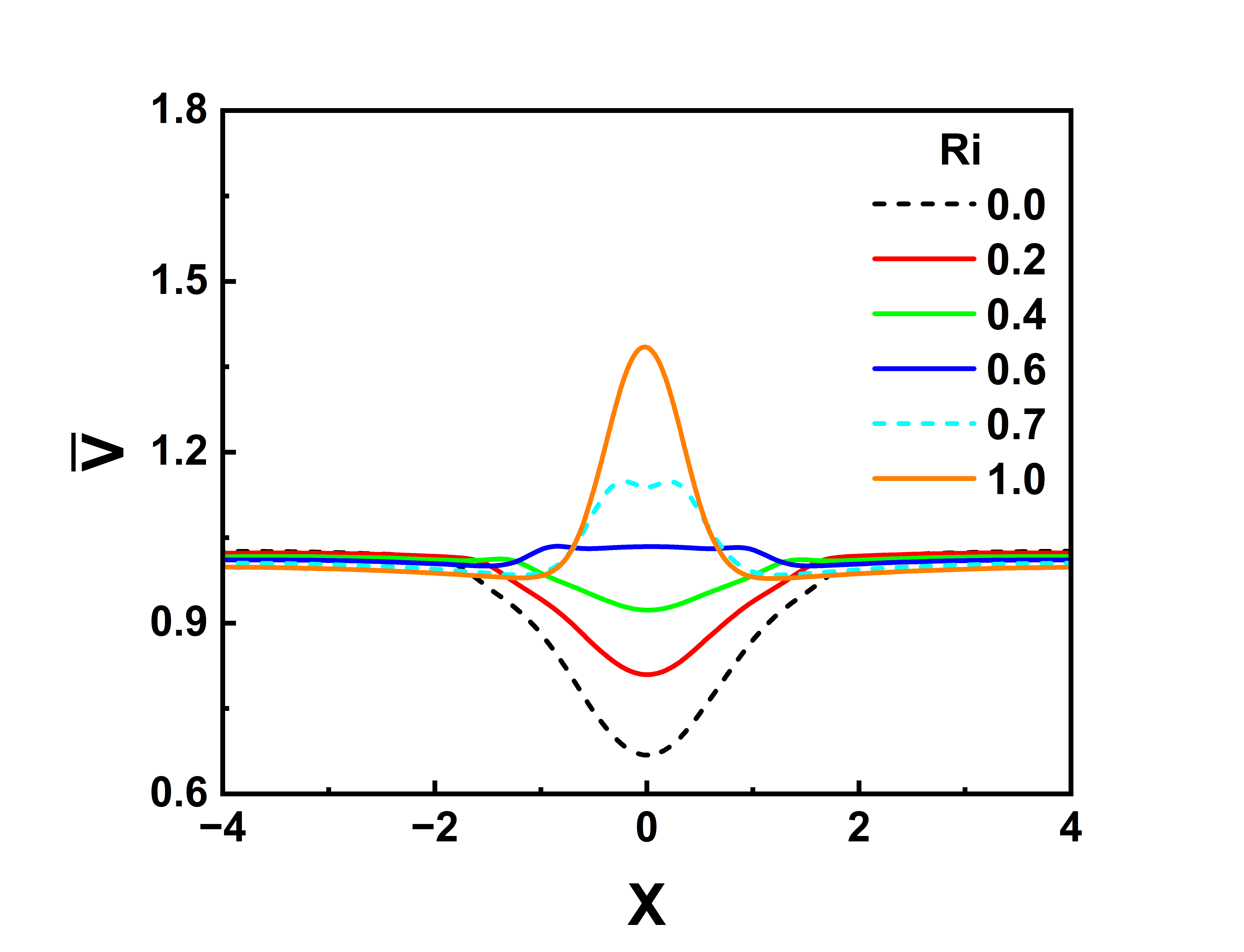}
				\caption{Y $=5$}
				\label{vprof2}
			\end{subfigure}
			
			\begin{subfigure}[t]{0.45\textwidth}
				\includegraphics[width=\linewidth]{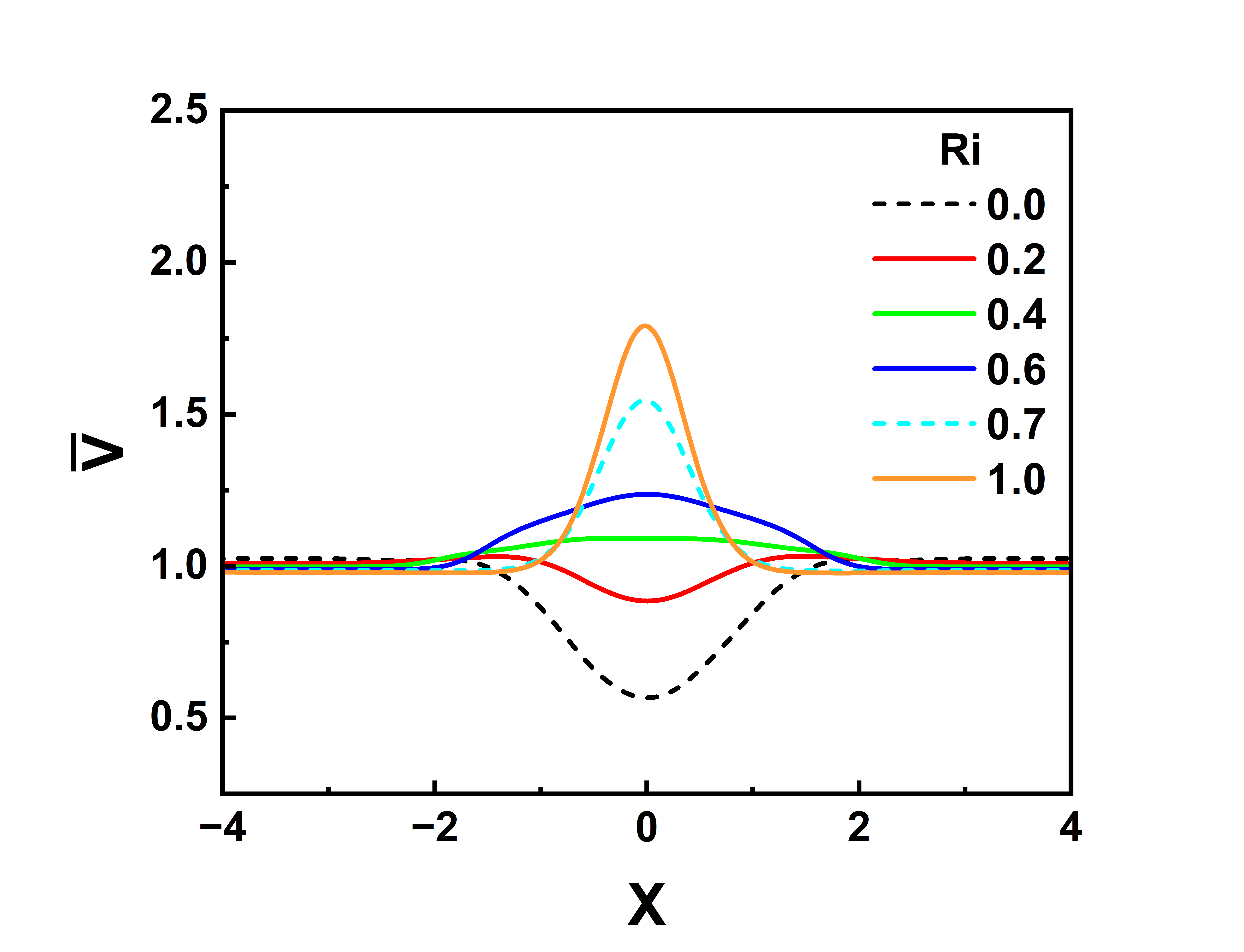}
				\caption{Y $=10$}
				\label{vprof3}
			\end{subfigure}\hspace{30pt}
			\begin{subfigure}[t]{0.45\textwidth}
				\includegraphics[width=\linewidth]{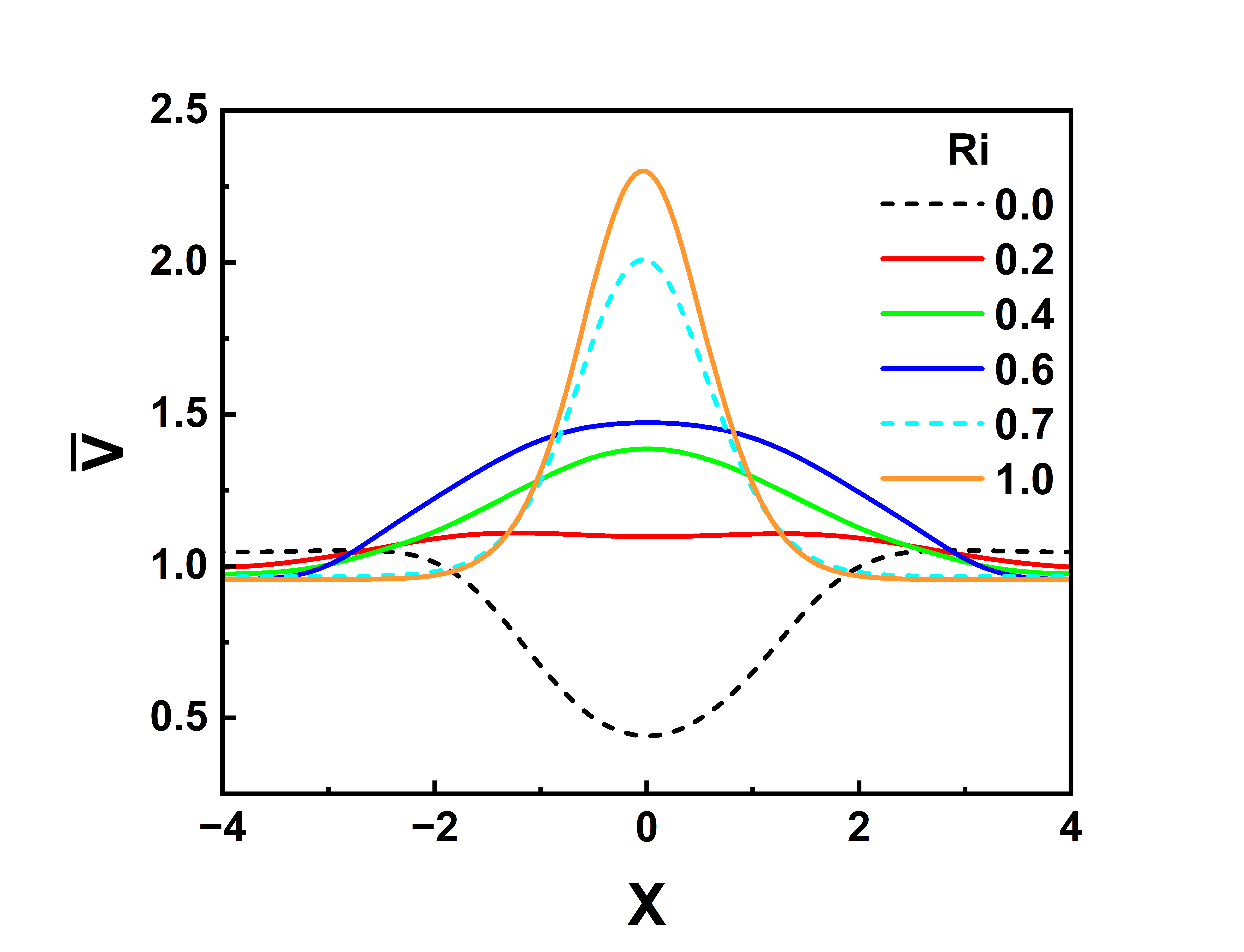}
				\caption{Y $=30$}
				\label{vprof4}
			\end{subfigure}\hspace{30pt}
			
			\begin{subfigure}[t]{0.45\textwidth}
			\includegraphics[width=\linewidth]{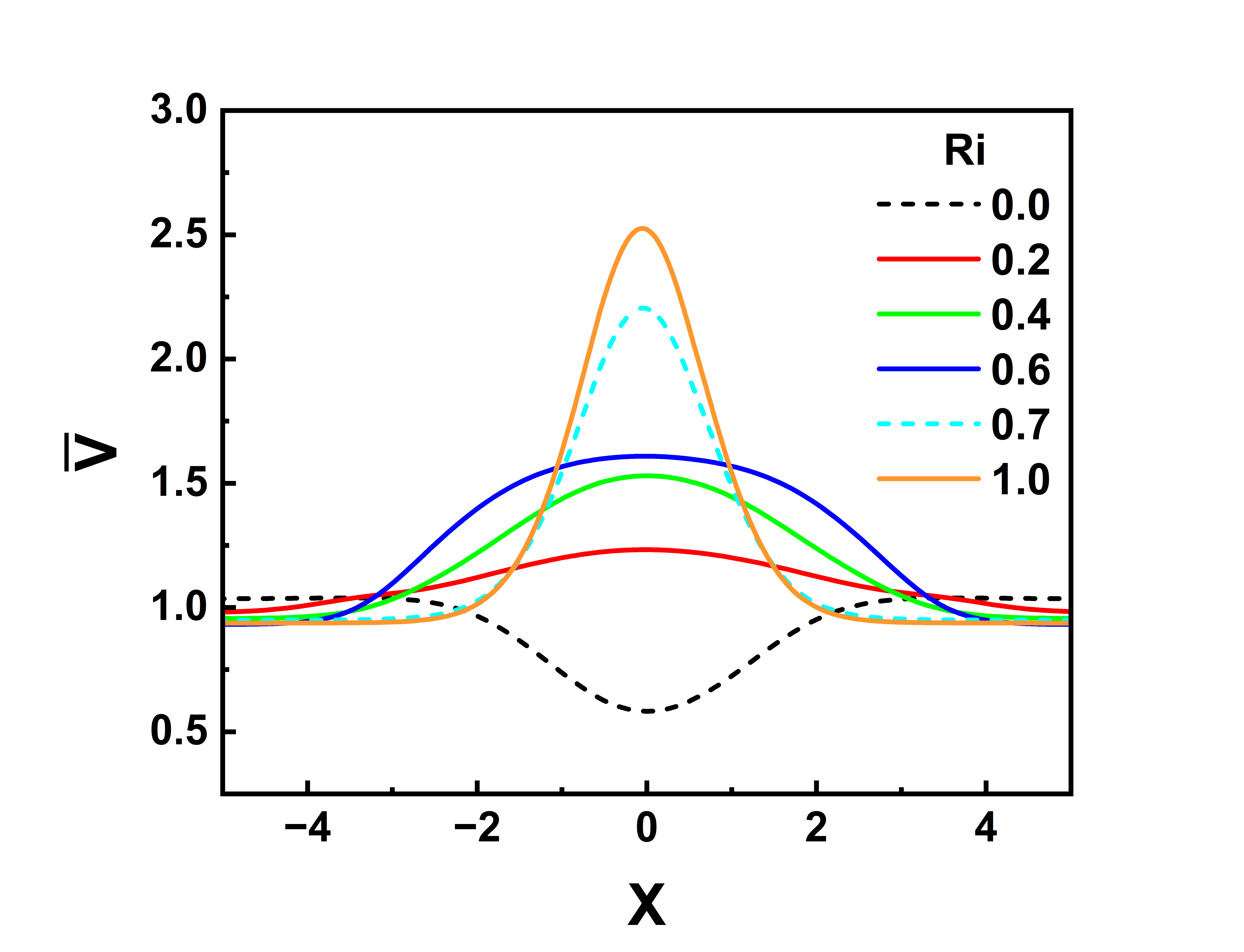}
			\caption{Y $=50$}
			\label{vprof5}
			\end{subfigure}\hspace{30pt}
			\begin{subfigure}[t]{0.45\textwidth}
			\includegraphics[width=\linewidth]{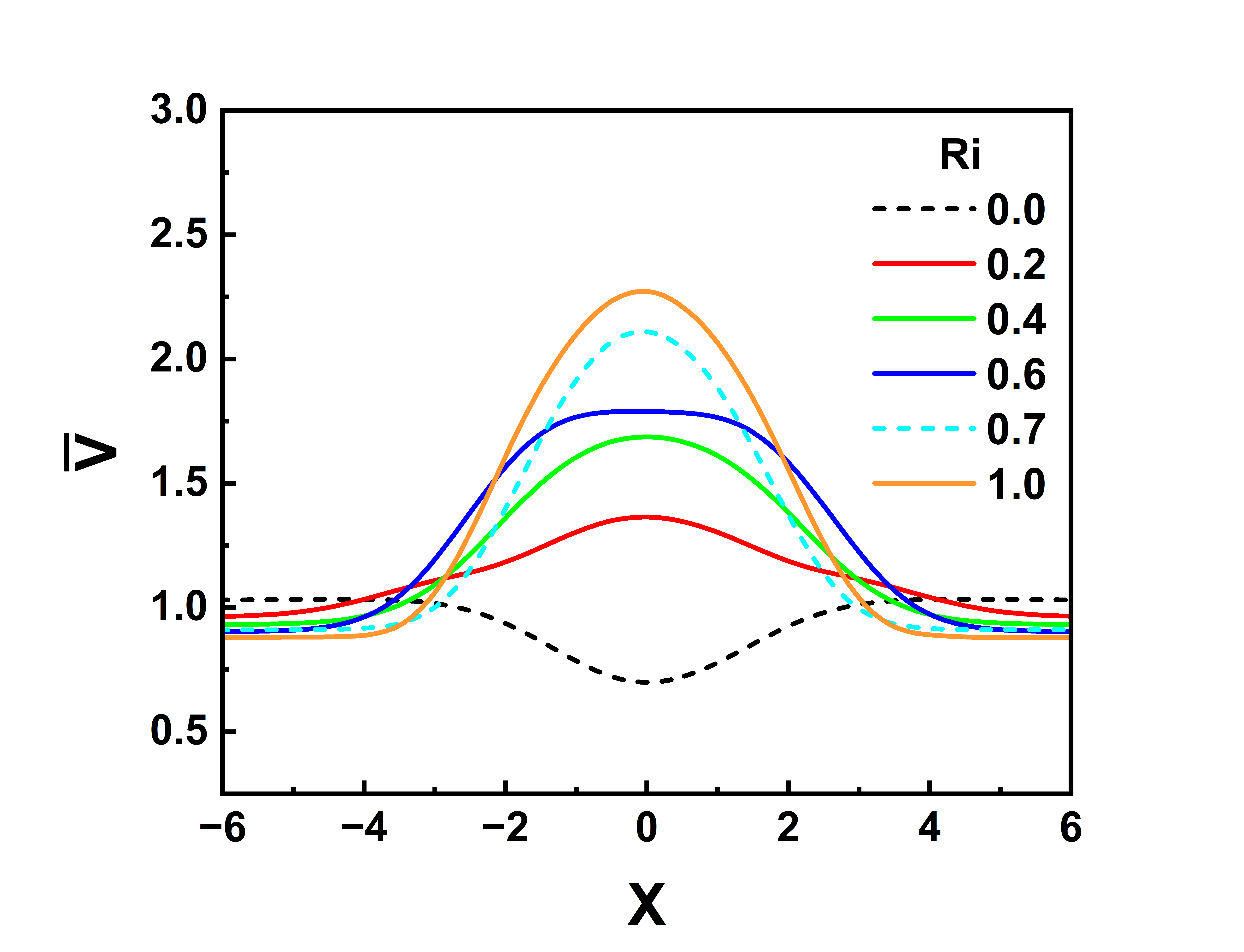}
			\caption{Y $=80$}
			\label{vprof6}
			\end{subfigure}
			
			\caption{Streamwise velocity profile at (a) Y $=1$, (b) Y $=5$, (c) Y $=10$, (d) Y $=30$, (e) Y $=50$, and (f) Y $=80$.}
			\label{vprofs}
		\end{figure}

\begin{figure}[htbp]
			\centering
			\begin{subfigure}[t]{0.45\textwidth}
				\includegraphics[width=\linewidth]{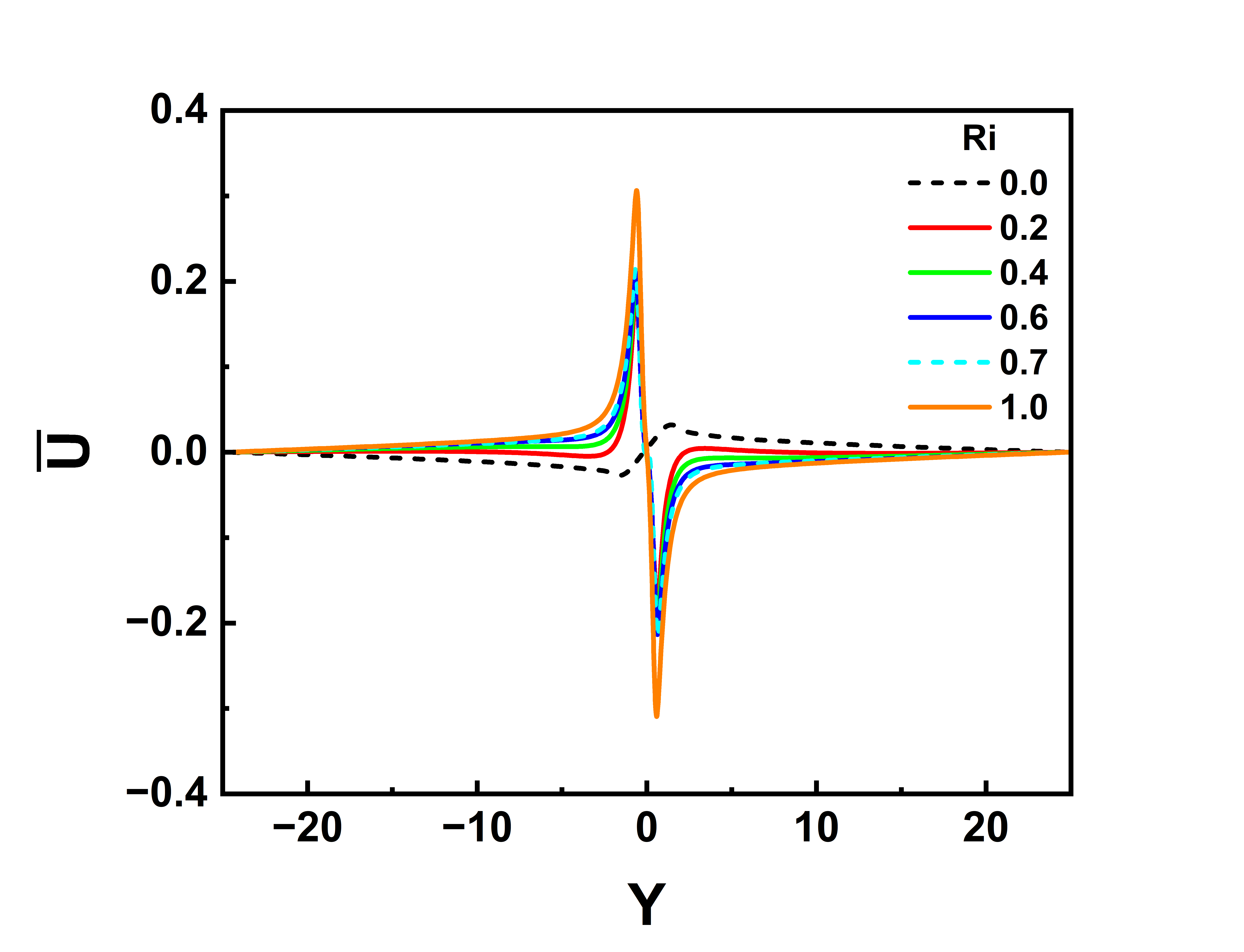}
				\caption{Y $=1$}
				\label{uprof1}
			\end{subfigure}\hspace{30pt}
			\begin{subfigure}[t]{0.45\textwidth}
				\includegraphics[width=\linewidth]{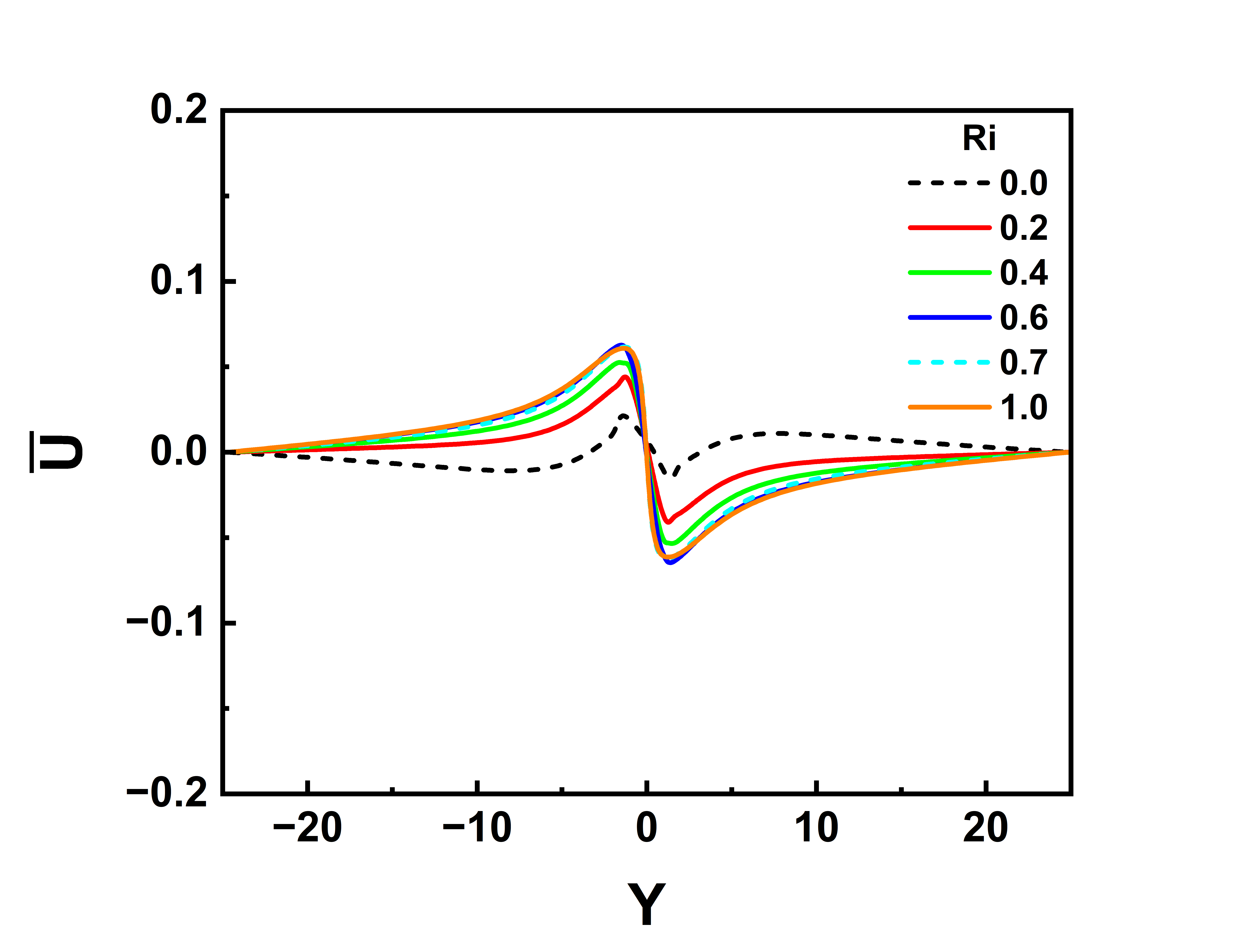}
				\caption{Y $=5$}
				\label{uprof2}
			\end{subfigure}
			
			\begin{subfigure}[t]{0.45\textwidth}
				\includegraphics[width=\linewidth]{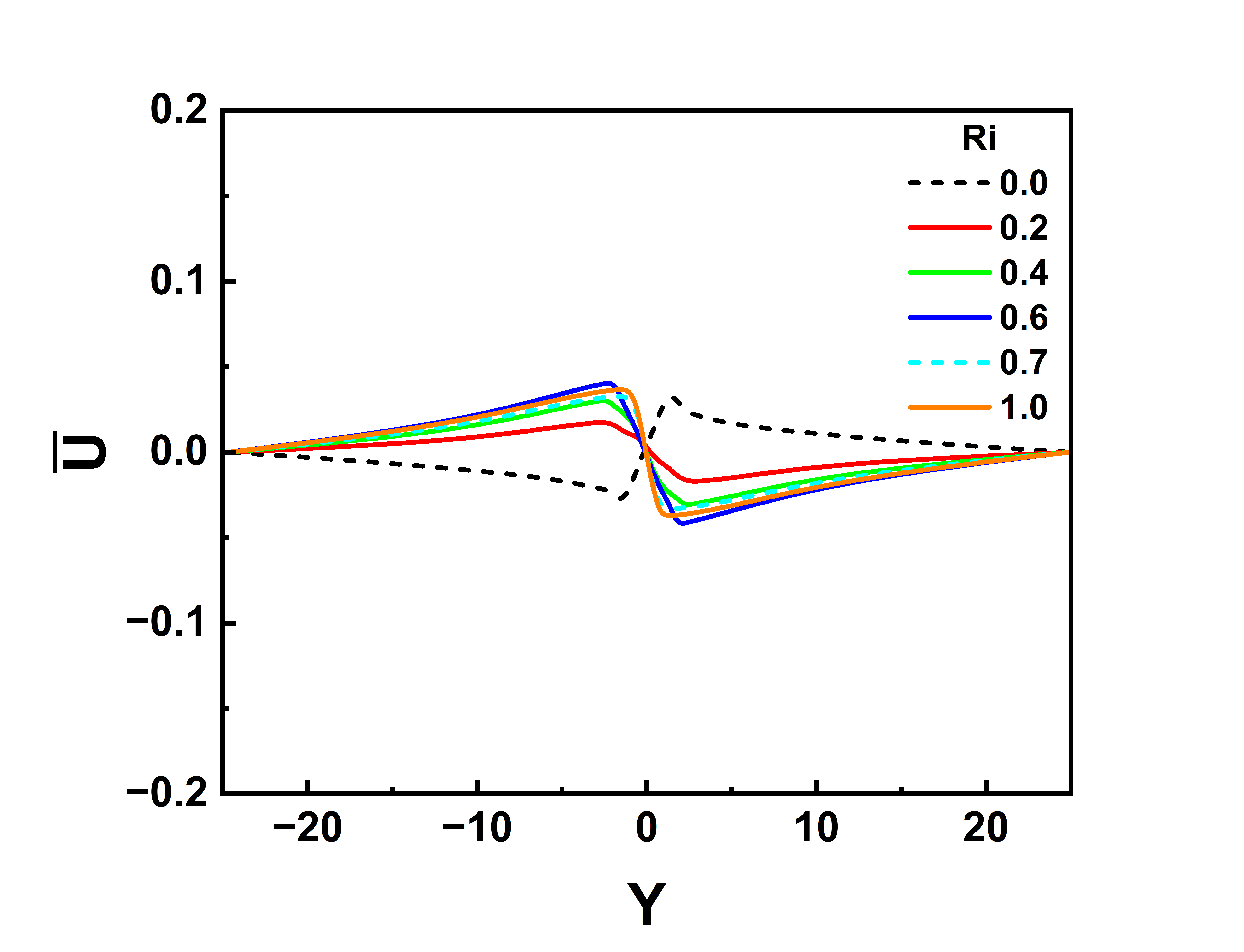}
				\caption{Y $=10$}
				\label{uprof3}
			\end{subfigure}\hspace{30pt}
			\begin{subfigure}[t]{0.45\textwidth}
				\includegraphics[width=\linewidth]{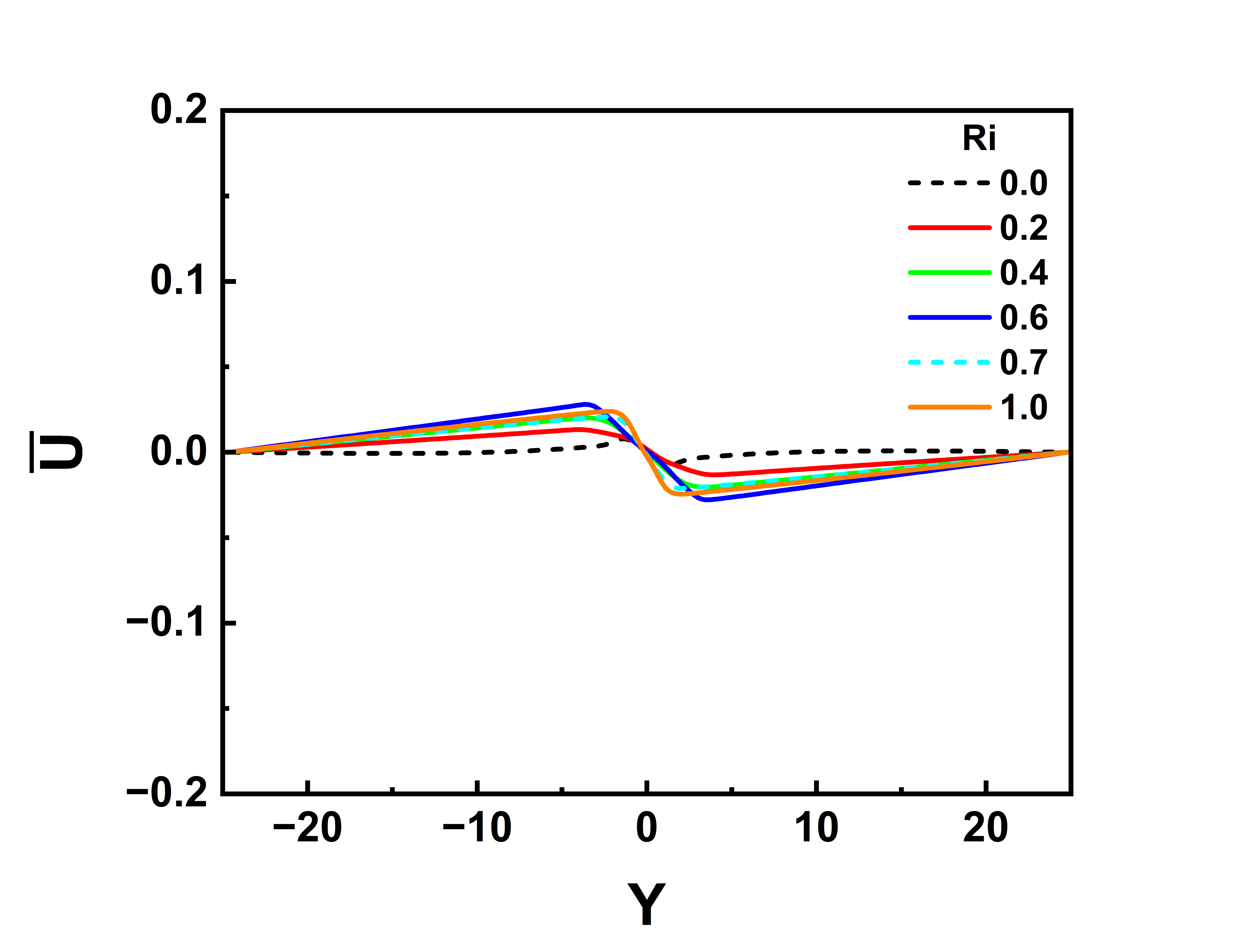}
				\caption{Y $=30$}
				\label{uprof4}
			\end{subfigure}\hspace{30pt}
			
			\begin{subfigure}[t]{0.45\textwidth}
			\includegraphics[width=\linewidth]{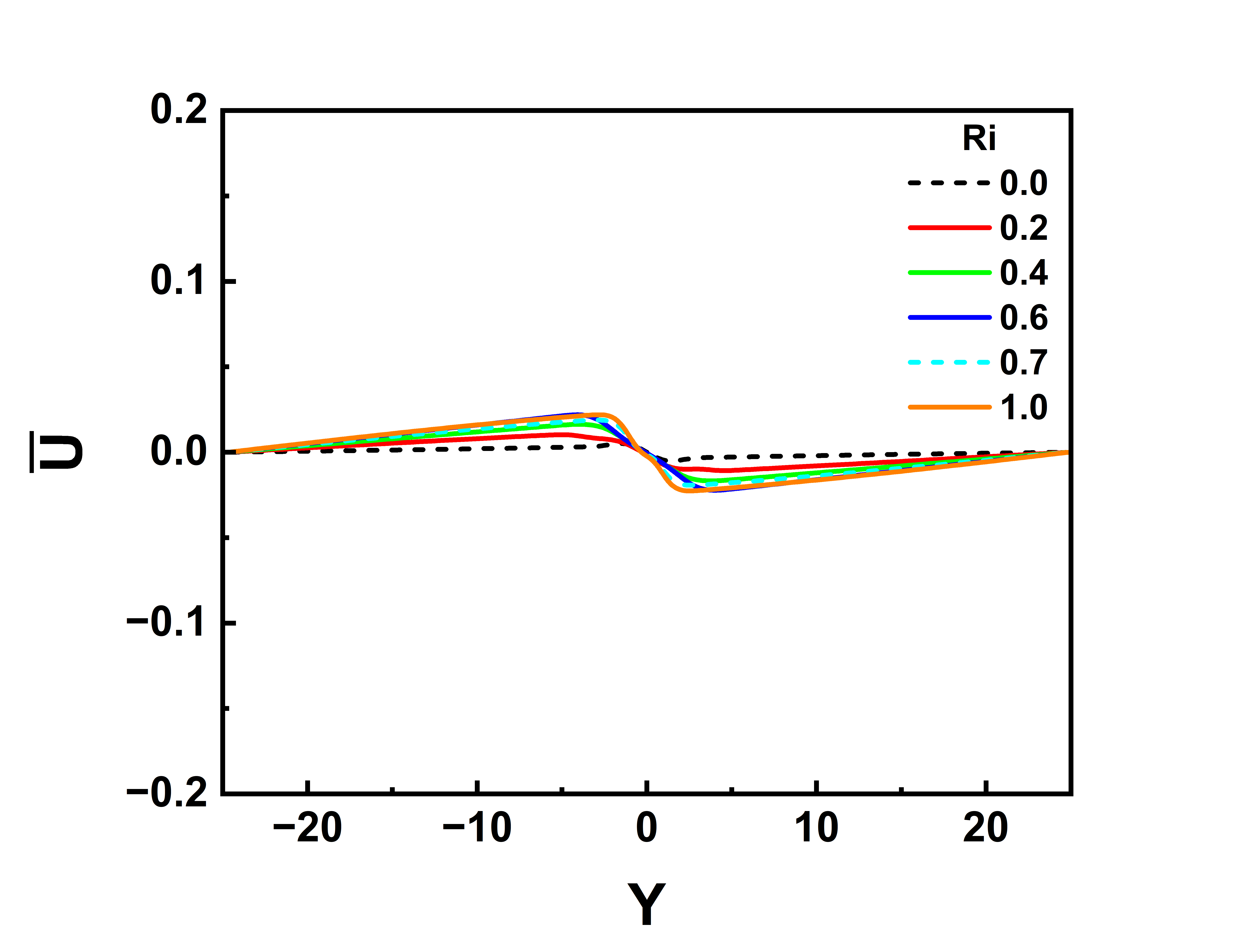}
			\caption{Y $=50$}
			\label{uprof5}
			\end{subfigure}\hspace{30pt}
			\begin{subfigure}[t]{0.45\textwidth}
			\includegraphics[width=\linewidth]{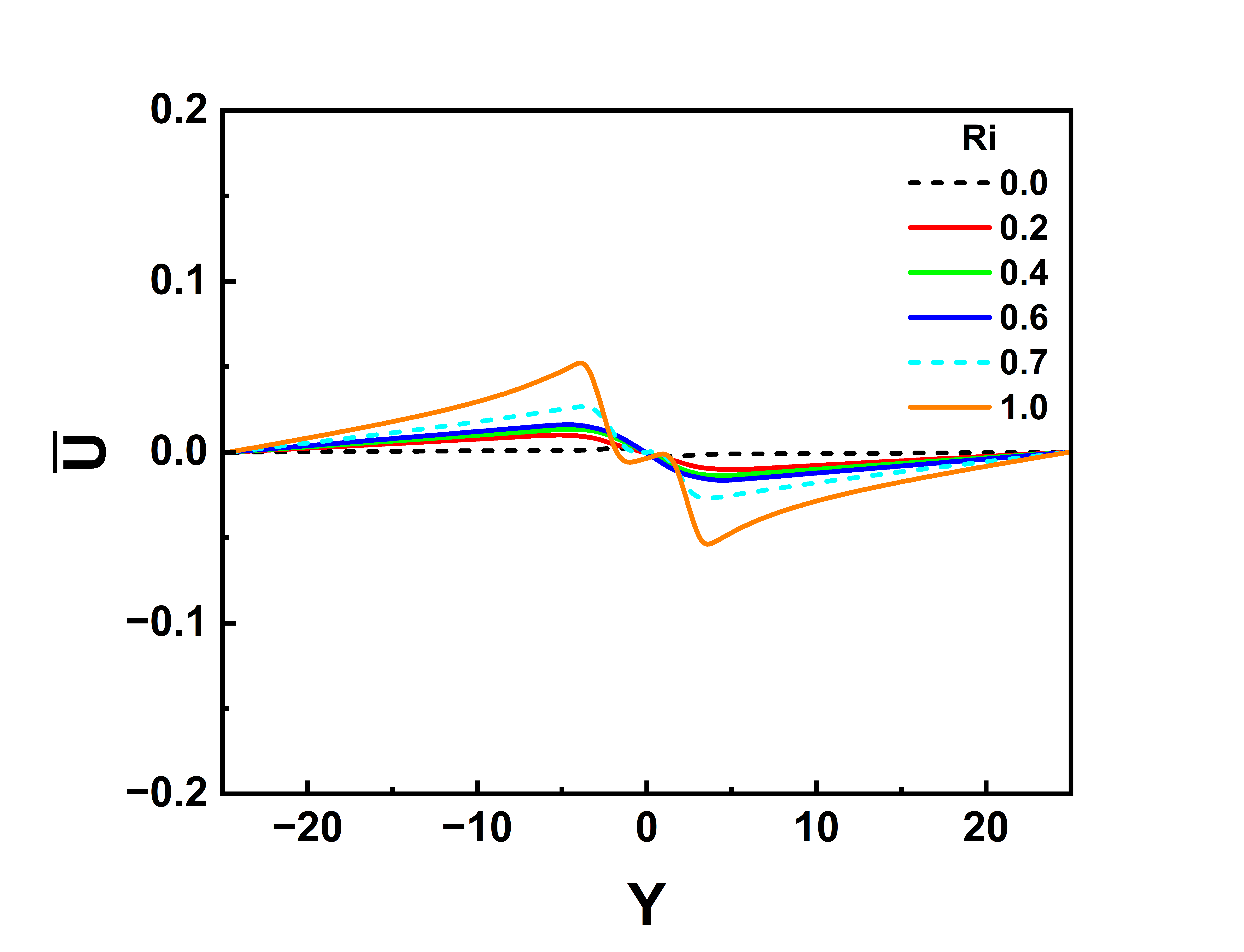}
			\caption{Y $=80$}
			\label{uprof6}
			\end{subfigure}
			
			\caption{Transverse velocity profile at (a) Y $=1$, (b) Y $=5$, (c) Y $=10$, (d) Y $=30$, (e) Y $=50$, and (f) Y $=80$.}
			\label{uprofs}
		\end{figure}

\begin{figure}[htbp]
			\centering
			\begin{subfigure}[t]{0.45\textwidth}
				\includegraphics[width=\linewidth]{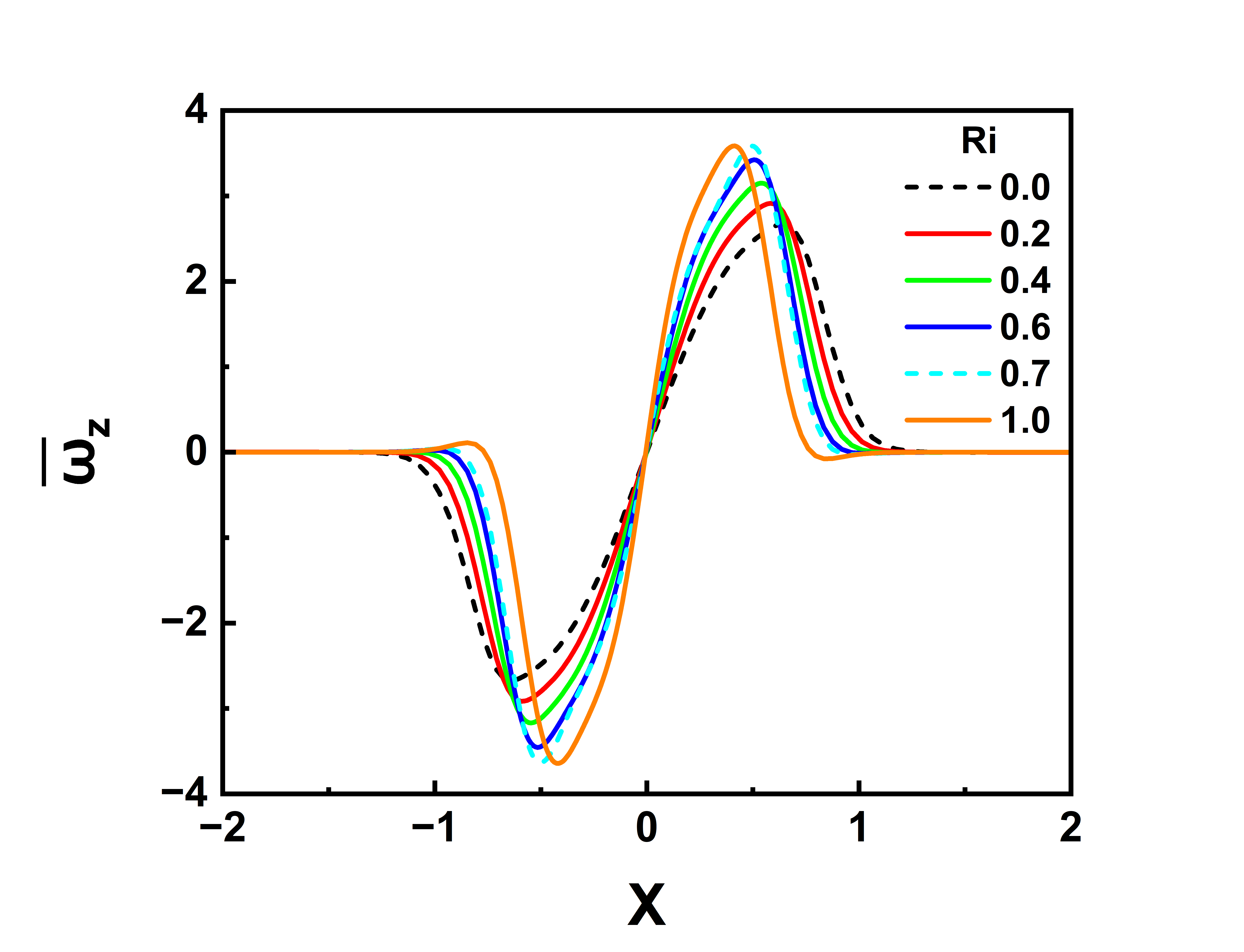}
				\caption{Y $=1$}
				\label{vortprof1}
			\end{subfigure}\hspace{30pt}
			\begin{subfigure}[t]{0.45\textwidth}
				\includegraphics[width=\linewidth]{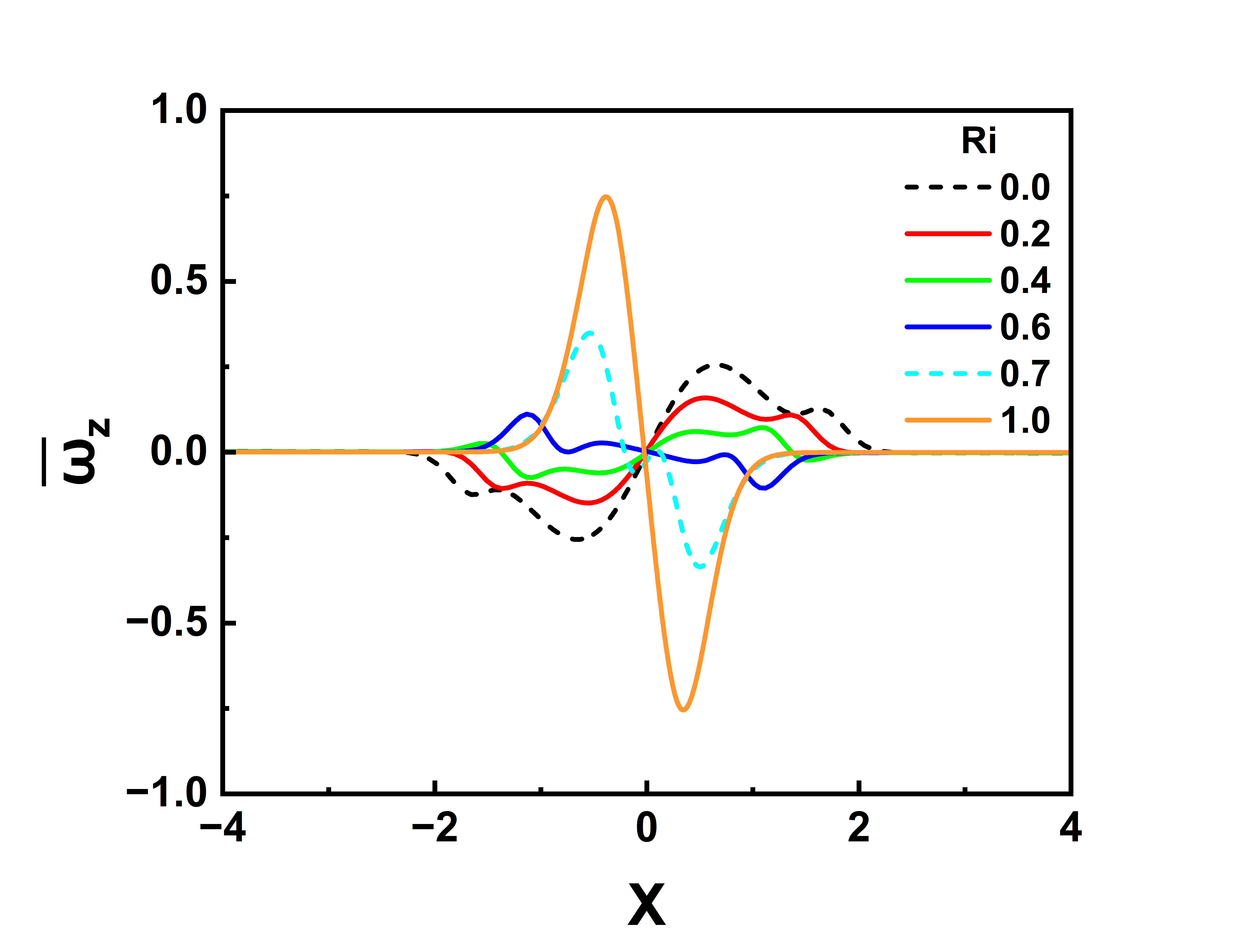}
				\caption{Y $=5$}
				\label{vortprof2}
			\end{subfigure}
			
			\begin{subfigure}[t]{0.45\textwidth}
				\includegraphics[width=\linewidth]{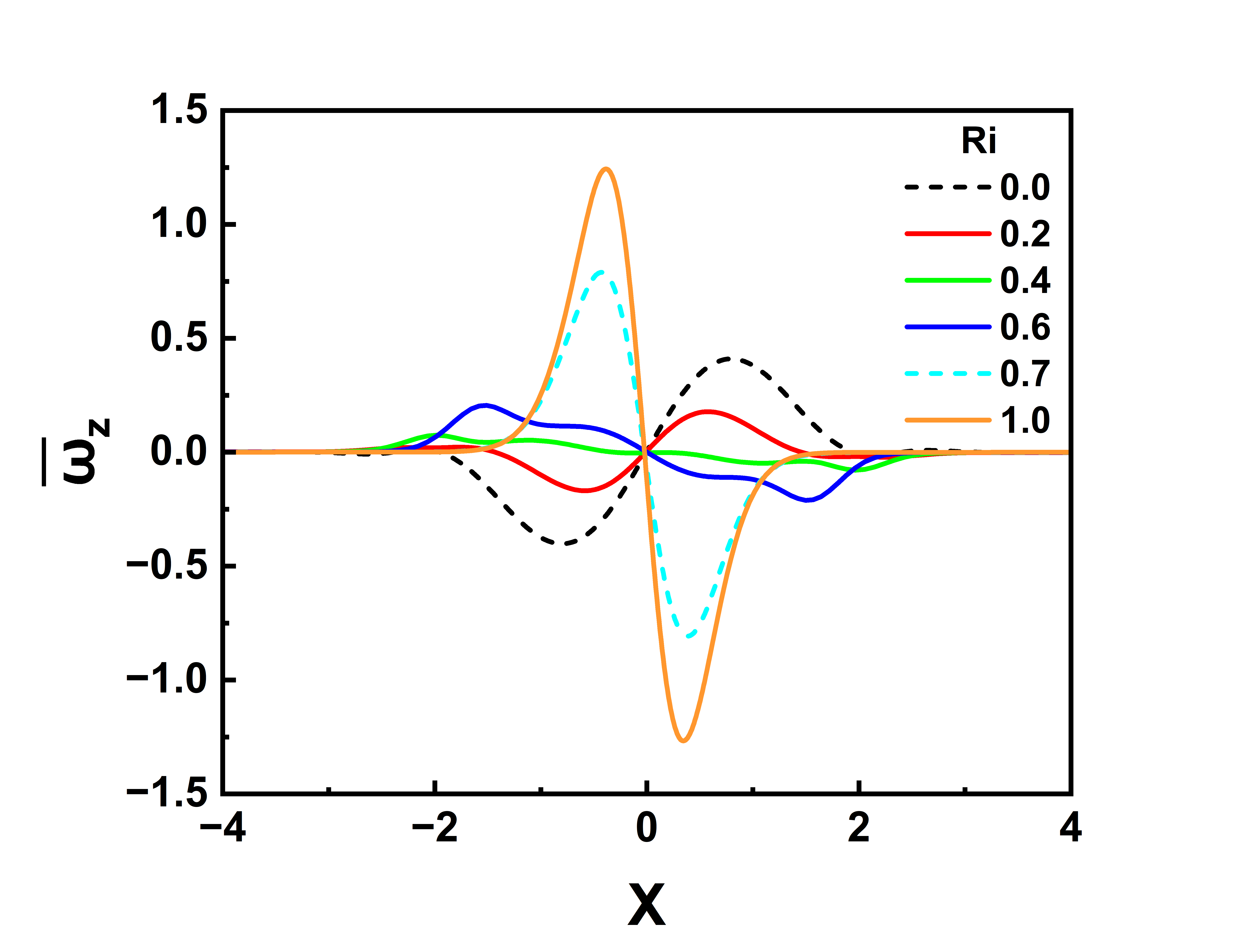}
				\caption{Y $=10$}
				\label{vortprof3}
			\end{subfigure}\hspace{30pt}
			\begin{subfigure}[t]{0.45\textwidth}
				\includegraphics[width=\linewidth]{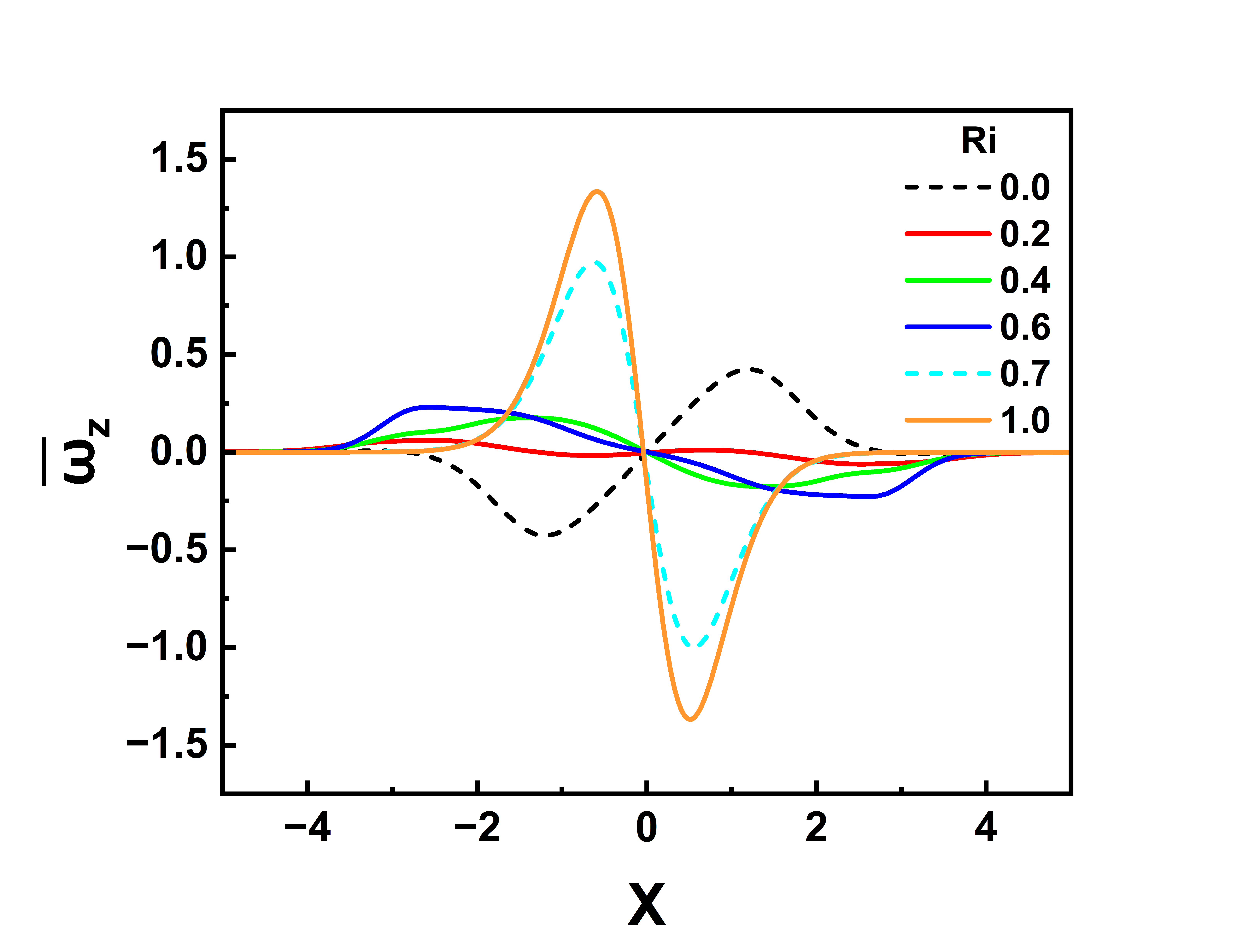}
				\caption{Y $=30$}
				\label{vortprof4}
			\end{subfigure}\hspace{30pt}
			
			\begin{subfigure}[t]{0.45\textwidth}
				\includegraphics[width=\linewidth]{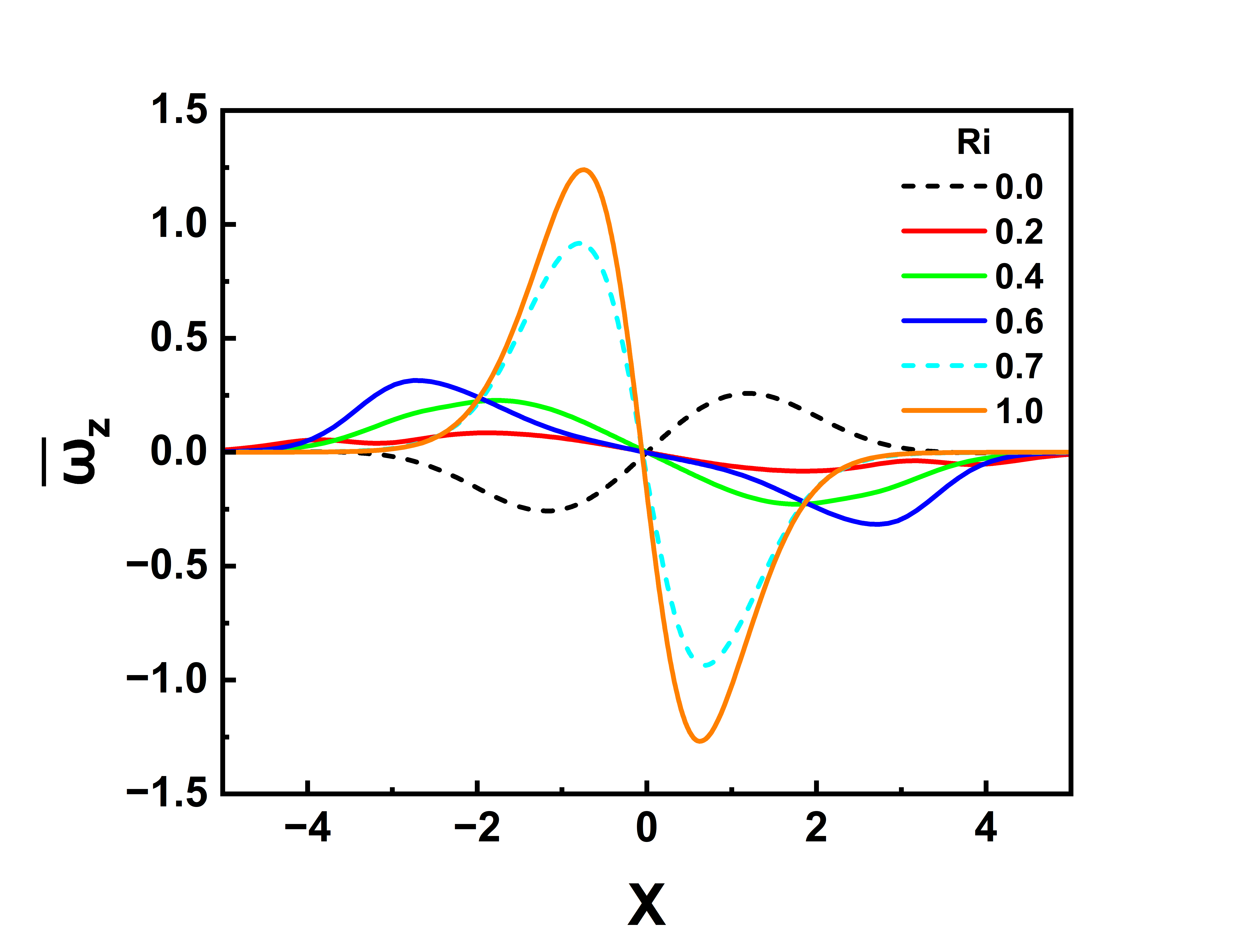}
				\caption{Y $=50$}
				\label{vortprof5}
			\end{subfigure}\hspace{30pt}
			\begin{subfigure}[t]{0.45\textwidth}
				\includegraphics[width=\linewidth]{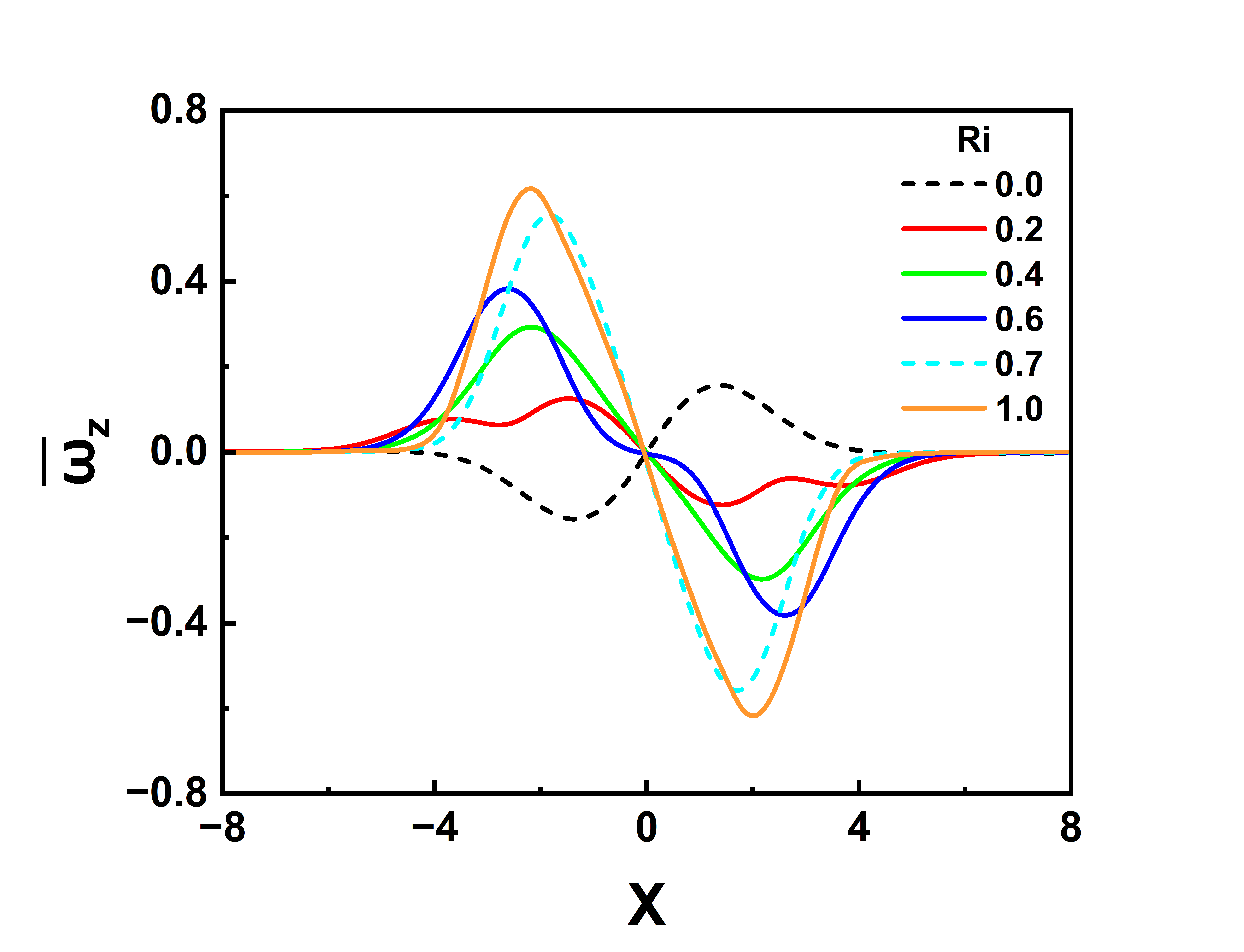}
				\caption{Y $=80$}
				\label{vortprof6}
			\end{subfigure}
			\caption{Vorticity profile at (a) Y $=1$, (b) Y $=5$, (c) Y $=10$, (d) Y $=30$, (e) Y $=50$, and (f) Y $=80$.}
			\label{vortprofs}
		\end{figure}

\begin{figure}[htbp]
			\centering
			\begin{subfigure}[t]{0.45\textwidth}
				\includegraphics[width=\linewidth]{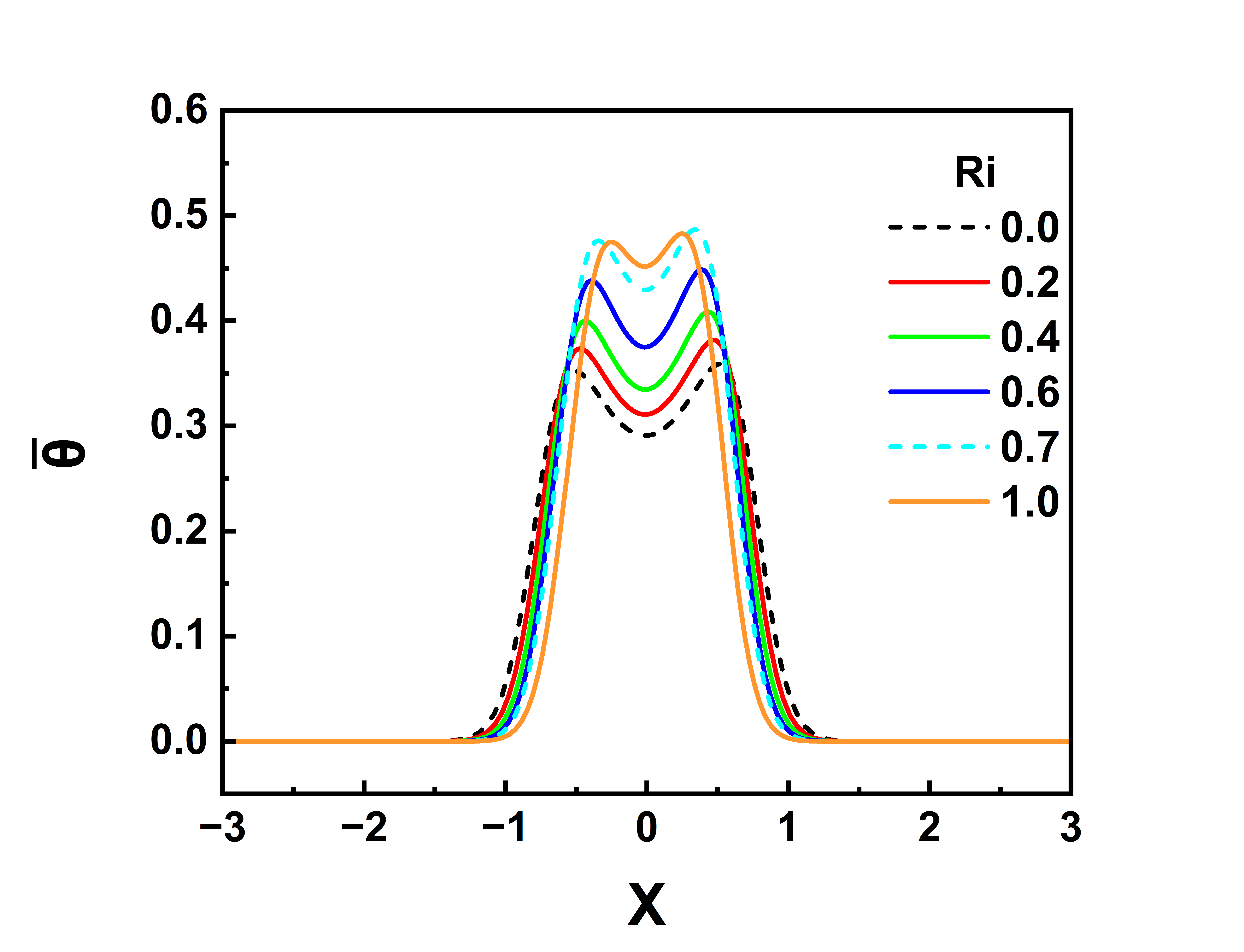}
				\caption{Y $=1$}
				\label{tprof1}
			\end{subfigure}\hspace{30pt}
			\begin{subfigure}[t]{0.45\textwidth}
				\includegraphics[width=\linewidth]{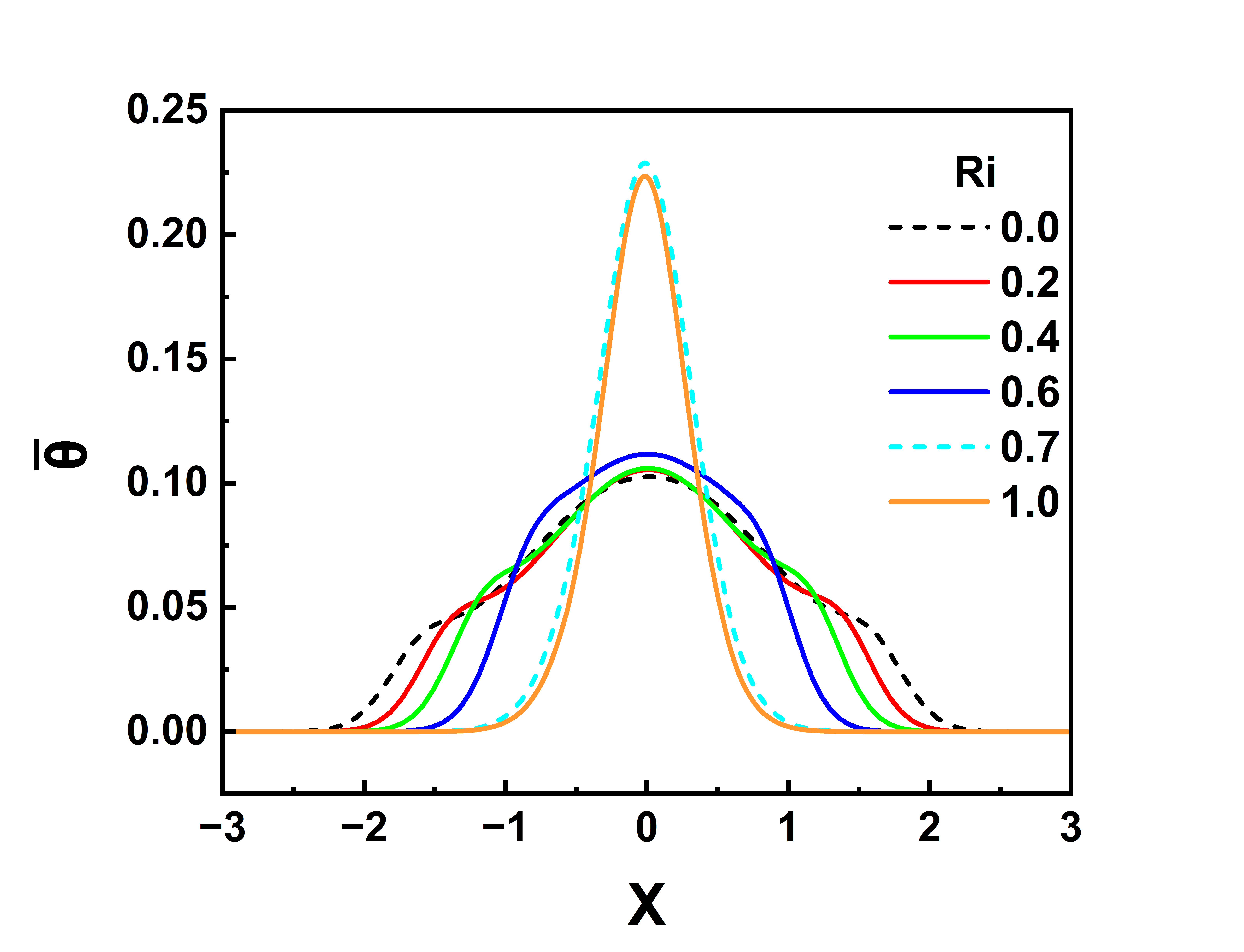}
				\caption{Y $=5$}
				\label{tprof2}
			\end{subfigure}
			
			\begin{subfigure}[t]{0.45\textwidth}
				\includegraphics[width=\linewidth]{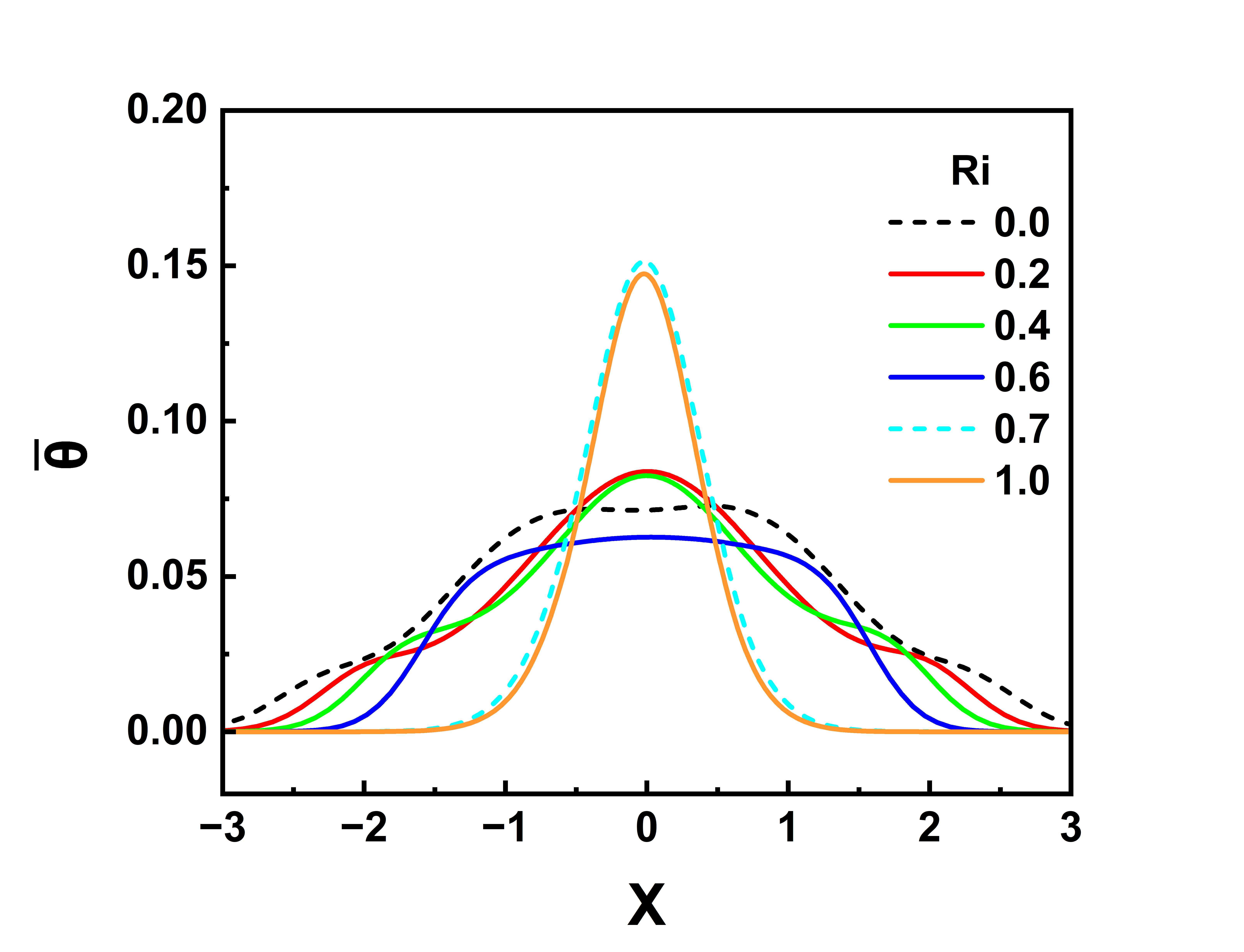}
				\caption{Y $=10$}
				\label{tprof3}
			\end{subfigure}\hspace{30pt}
			\begin{subfigure}[t]{0.45\textwidth}
				\includegraphics[width=\linewidth]{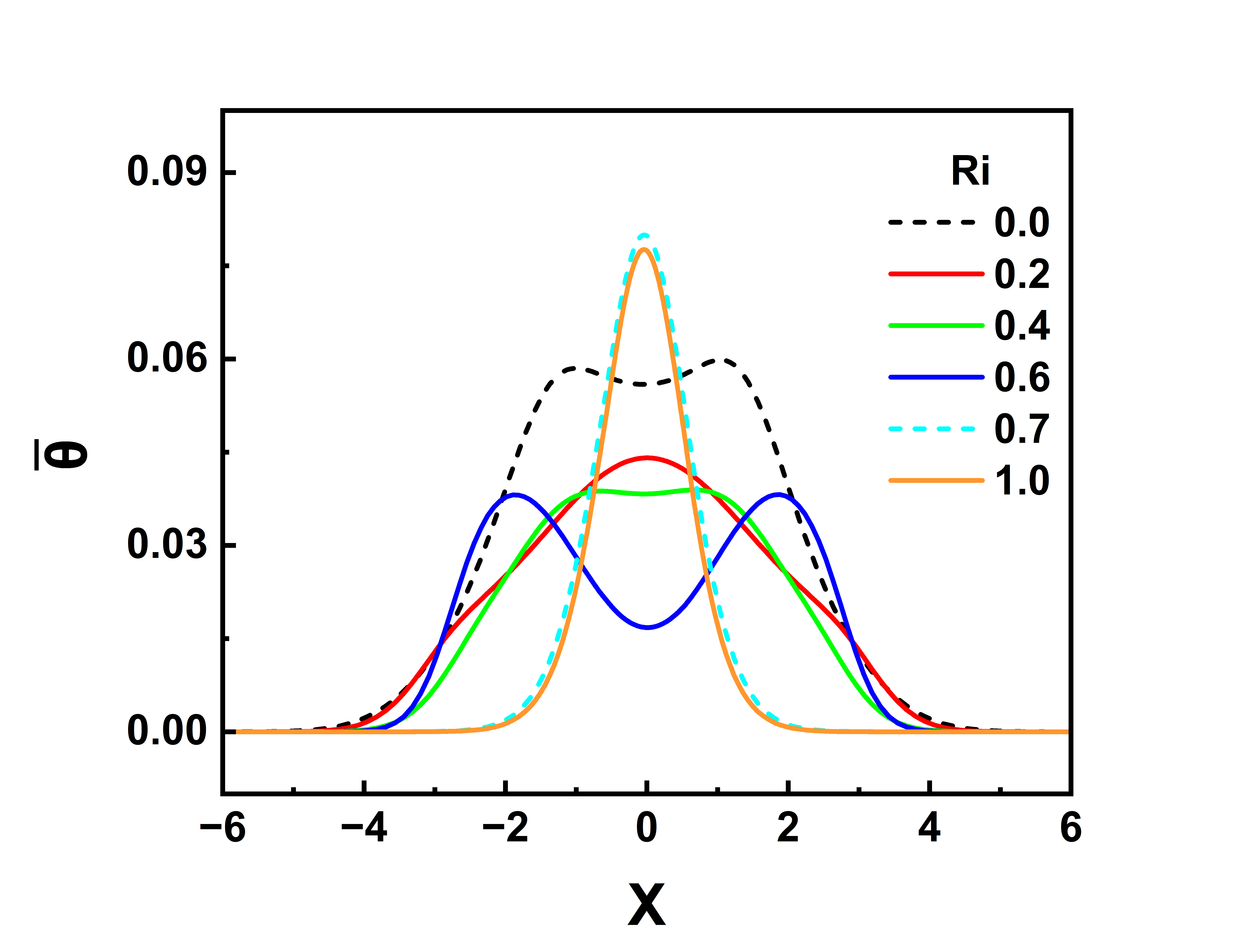}
				\caption{Y $=30$}
				\label{tprof4}
			\end{subfigure}\hspace{30pt}
			
			\begin{subfigure}[t]{0.45\textwidth}
				\includegraphics[width=\linewidth]{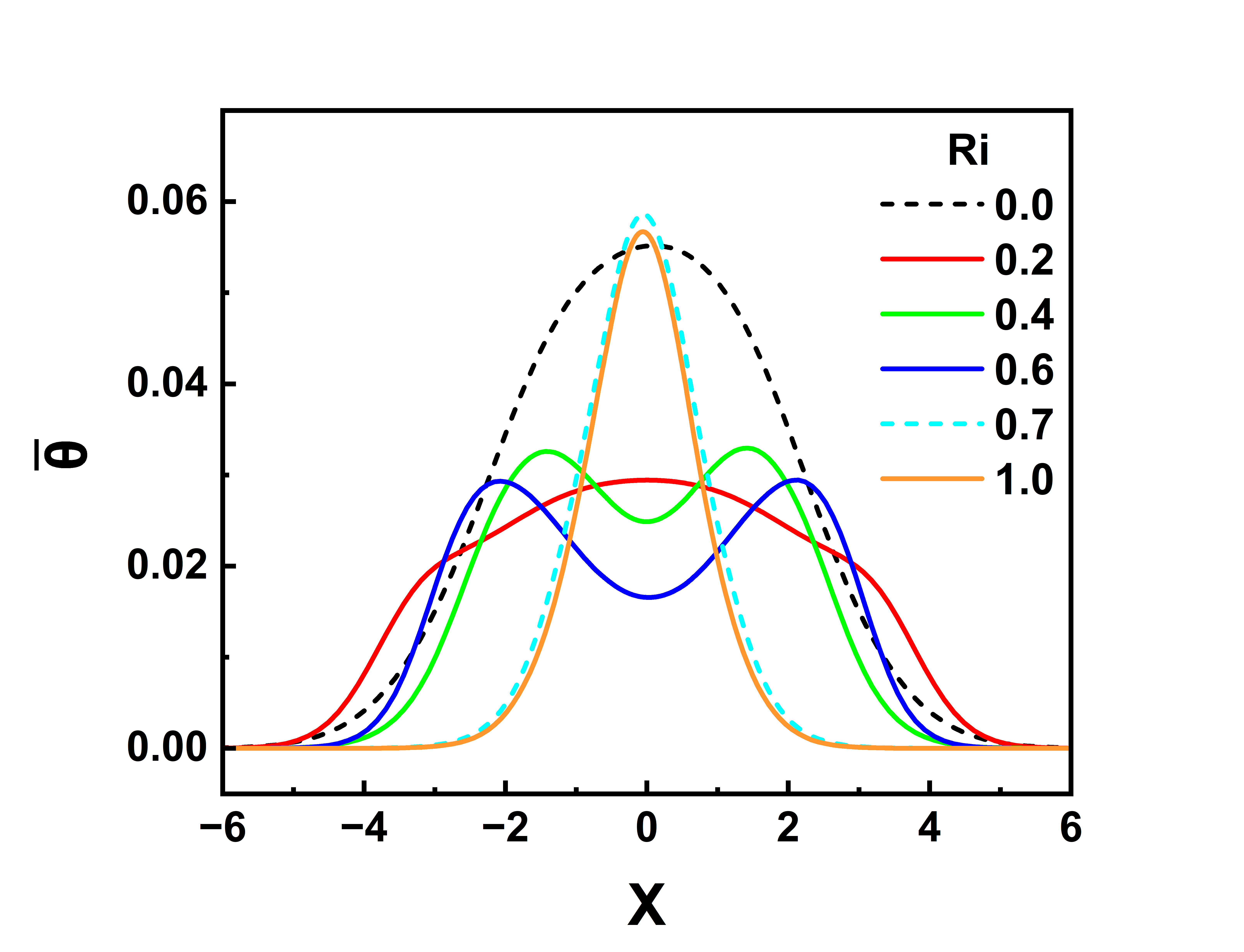}
				\caption{Y $=50$}
				\label{tprof5}
			\end{subfigure}\hspace{30pt}
			\begin{subfigure}[t]{0.45\textwidth}
				\includegraphics[width=\linewidth]{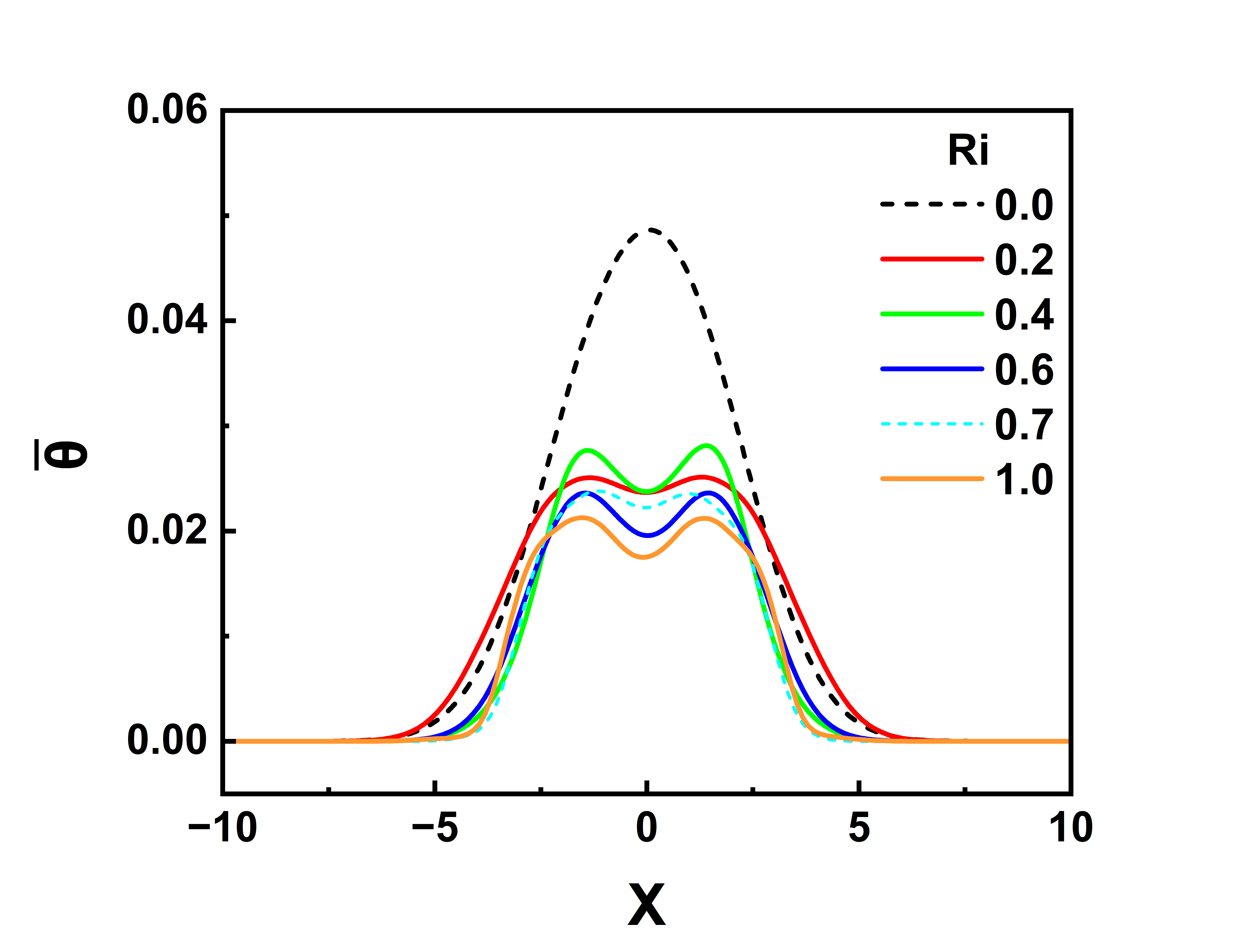}
				\caption{Y $=80$}
				\label{tprof6}
			\end{subfigure}
			\caption{Temperature profile at (a) Y $=1$, (b) Y $=5$, (c) Y $=10$, (d) Y $=30$, (e) Y $=50$, and (f) Y $=80$.}
			\label{tprofs}
		\end{figure}

The streamwise velocity, transverse velocity, vorticity, and temperature profiles at different streamwise locations for different Ri are shown in \Cref{vprofs,uprofs,vortprofs,tprofs}, respectively. \Cref{vprofs} shows the deficit of momentum in the near-field associated with wake-like behavior and a recovery and subsequent excess of momentum in the intermediate and far-fields. In the near-field, the momentum deficit is seen for all Ri, resembling the profile seen in wakes. With increased downstream distance, the recovery of momentum occurs, starting earlier downstream for higher Ri, followed by lower Ri later downstream. At Y $=30$, the recovery of momentum is complete for all positive Ri, and there is an excess of momentum seen for higher Ri, which grows more pronounced further downstream, resembling the profile seen in jets. Therefore, the flow field downstream starts resembling a wake at low Ri and slowly transforms into that of a plume as Ri is increased, first far downstream, then near the cylinder. Such behavior is attributed to buoyancy, which tends to cause a jetting effect in the entire flow field, starting from the far downstream region and slowly moving towards the cylinder. \Cref{uprofs} shows the variation of transverse velocity, which is positive in the left half and negative in the right half of the flow for all positive Ri at all downstream locations. \Cref{vortprofs} reveals the inversion of vorticity in the near and intermediate fields. For higher Ri, the inversion occurs earlier downstream, while for lower Ri, it occurs further downstream as the strength of the natural convective vorticity is lesser at lower Ri. The peak vorticity is seen to be higher for higher Ri. \Cref{tprofs} shows the temperature profiles for different streamwise locations. Overall, the temperature shows a dissipating trend with increased streamwise distance. We have also observed that prior to the onset distance of the far-field unsteadiness, the temperature profile width for the cases above the critical Ri is much lower than that of the cases below the critical Ri (see \Cref{tprof5}). However, after the onset distance of the far-field unsteadiness, the profile widens drastically, and it is comparable with the cases below the critical Ri (see \Cref{tprof6}).

\subsection{Momentum Deficit and Vortex Shedding Suppression in the Near-Field} \label{deficit}
As discussed earlier, shear layer separation occurs at the lateral corners/edges of the cylinder, leading to a deficit of velocity in the cylinder wake. Beyond a critical Re, the flow transitions to unsteadiness, and the B\'{e}nard-von K\'{a}rm\'{a}n vortex street is seen. However, when the cylinder is heated, density differences cause momentum addition, leading to flow acceleration in the wake, particularly around the centerline. This leads to the suppression of vortex shedding in a mechanism similar to other flow control techniques, which prevents the interaction of the two shear layers originating at the two lateral corners.

The increased momentum in the wake at higher Ri suggests that the effect of buoyancy is to add a source that effectively behaves similarly to a jet, thus adding equivalent momentum to the wake, as explained below. At lower Ri, the effective jet caused by buoyancy does not possess sufficient momentum to affect the interaction between the two shear layers. As Ri is increased, a larger velocity of the effective jet causes the undulation of the oppositely signed shear layers to decrease. The interaction between the shear layers is affected by the effective jet, which behaves in a similar way as a splitter plate \cite{sahajaiswal} or a blowing jet \cite{sahashrivastava}. Below the critical Ri, this manifests as an increase in the recirculation length (see \Cref{recircinversion}) due to the increased shear layer length prior to roll-up and detachment. Additionally, the cancellation of the forced convective vorticity by the opposite natural convective vorticity inhibits the interaction between the shear layers. The hindered interaction between the shear layers leads to a loss of strength due to the inability to gain further circulation mutually, leading to a situation where detachment of the shear layers is not possible \cite{gerrardmech} and the vortices remain as a steady separation bubble with no roll-up beyond the critical Ri (see \Cref{insnearvort}).

\subsection{Unsteadiness of the Far-Field} \label{ffinstability}

The effect of the chosen domain in mixed convective flows is significant and changes the observed physics completely. In previous works by Sharma and Eswaran \cite{sharmaeswaran}, Hasan and Ali \cite{baroclinichasan}, and others, the streamwise length of the domain has not exceeded 50 times the projected length of the cylinder. Due to this, transitions to unsteadiness in the far field much beyond the cylinder have not been captured, and the flow field has been remarked as steady, although it may not be the case far downstream of the cylinder. To fully capture the physics of the flow field, we have considered a downstream length ($L_d$) of $110d$. It must be noted that the results contained in the present work pertain only to the domain size chosen here. Additional physics might be seen with the choice of an even larger streamwise domain. 

Instantaneous and time-averaged contours in \Cref{instemp,insvort,meantemp,meanvort} reveal that at Ri $=0.0$, the far-field is steady, while for Ri $>0.0$, the far-field is unsteady, which can be attributed to the increasing strength of the natural convective flow which makes the far-field behave like a plume. At the critical Ri and beyond, the far-field develops large-scale unsteadiness that is very similar to plume-like unsteadiness reported in the literature \cite{qiao1, qiao2, jiang, kimurabejan}. This occurs in conjunction with the suppression of near-field vortex shedding. There is no range of Ri for which the flow field is entirely steady. As Ri is increased beyond the critical Ri, the point of onset of the far-field instability appears to come closer to the cylinder, which further indicates that the plume strength increases closer to the cylinder with increasing Ri. For Ri $\leq 0.6$, there is a widening of the plume in the intermediate-field itself, and in the far-field, the plume narrows (see \Cref{meantemp,meanvort}). For Ri $\geq 0.7$, the plume is narrow in the near and intermediate fields, and it sharply widens in the far-field with the onset of the far-field unsteadiness. The natural convective component grows stronger with increased downstream distance, and the opposite phenomenon occurs with the forced convective component. Due to this, there is a large momentum excess far downstream (Y $>$ 60), which causes strong jet-like behavior in the far-field. This can be seen in \Cref{vprofs,axialvel}. The jetting effect in the far-field is responsible for the plume-like unsteadiness of the far-field, which shows similar characteristics as the instabilities observed in natural convection plumes \cite{qiao1,qiao2,jiang,kimurabejan}. The flow structures observed in the far-field require further detailed investigation to gain a complete understanding of flow physics.

To add to the understanding of the transition from steady to unsteady flow in the far-field, we have analyzed the velocity profiles in the far-field and utilized Fjortoft's theorem, discussed in detail by Arnal \textit{et al.} \cite{arnal}, in order to understand the stability of the far-field flow in a clearer way. Fjortoft's theorem utilizes points of inflexion in the time-averaged velocity profile to give a criterion for the stability of nearly parallel flows such as jets, wakes, and plumes, where the gradients of velocity in the transverse direction are significantly higher than those in the streamwise direction: 

    The point $X_0$ at which the velocity profile $V(X)$ satisfies the following condition is called the point of inflexion (PI):

    \begin{equation*}
    \left. \frac{\partial^2 V}{\partial X^2} \right|_{X = X_0} = 0
    \end{equation*}

\begin{figure}[H]
    \centering
    \includegraphics[width=0.65\linewidth]{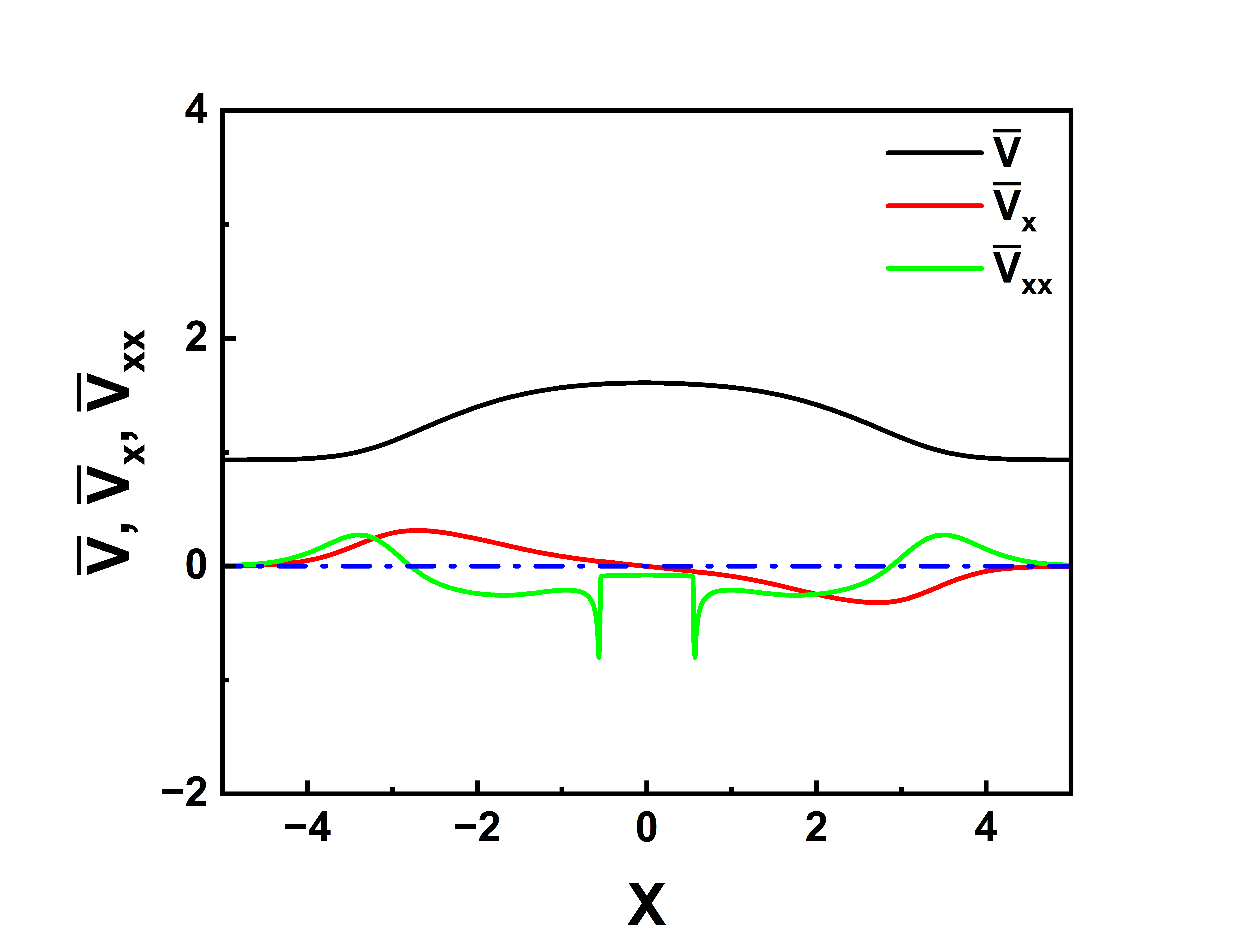}
    \caption{Profiles of time-averaged streamwise velocity, first derivative, and second derivative at Y $= 80$ for Ri $= 0.6$. Points of inflexion in the velocity profile exist at X $=\pm2.75$.}
    \label{inflexion}
    \end{figure}

    According to Fjortoft's theorem, a necessary (but not sufficient) condition for inviscid instability is that

    \begin{equation*}
    \frac{\partial^2 V}{\partial X^2}(V-V_0)  < 0
    \end{equation*}

    must be true somewhere in the flow, where $V_0$ is the velocity at the inflexion point $X=X_0$. We have plotted the streamwise velocity profile against the transverse coordinate along with its first and second derivatives in \Cref{inflexion} at Ri $=0.6$. The transverse locations corresponding to $\overline{V}_{XX} = 0$ intersect at X $=\pm2.75$, which are the two points of inflexion in the velocity profile, and the corresponding velocity at these points is $V_0$. It is clear from the graph that for $-2.75 \leq X \leq 2.75$, the quantity ($V-V_0$) is positive, while the quantity $\dfrac{\partial^2{V}}{\partial X^2}$ is negative. Consequently, one can say that the flow is unstable in the far field when subject to a suitable inviscid perturbation. The far field is stable at Ri $=0.6$, while it becomes unstable (unsteady) at Ri $=0.7$ because of the viscous flow in the present study.

\begin{figure}[htbp]
	\centering
        \begin{subfigure}[t]{0.5\textwidth}
		\includegraphics[width=\linewidth]{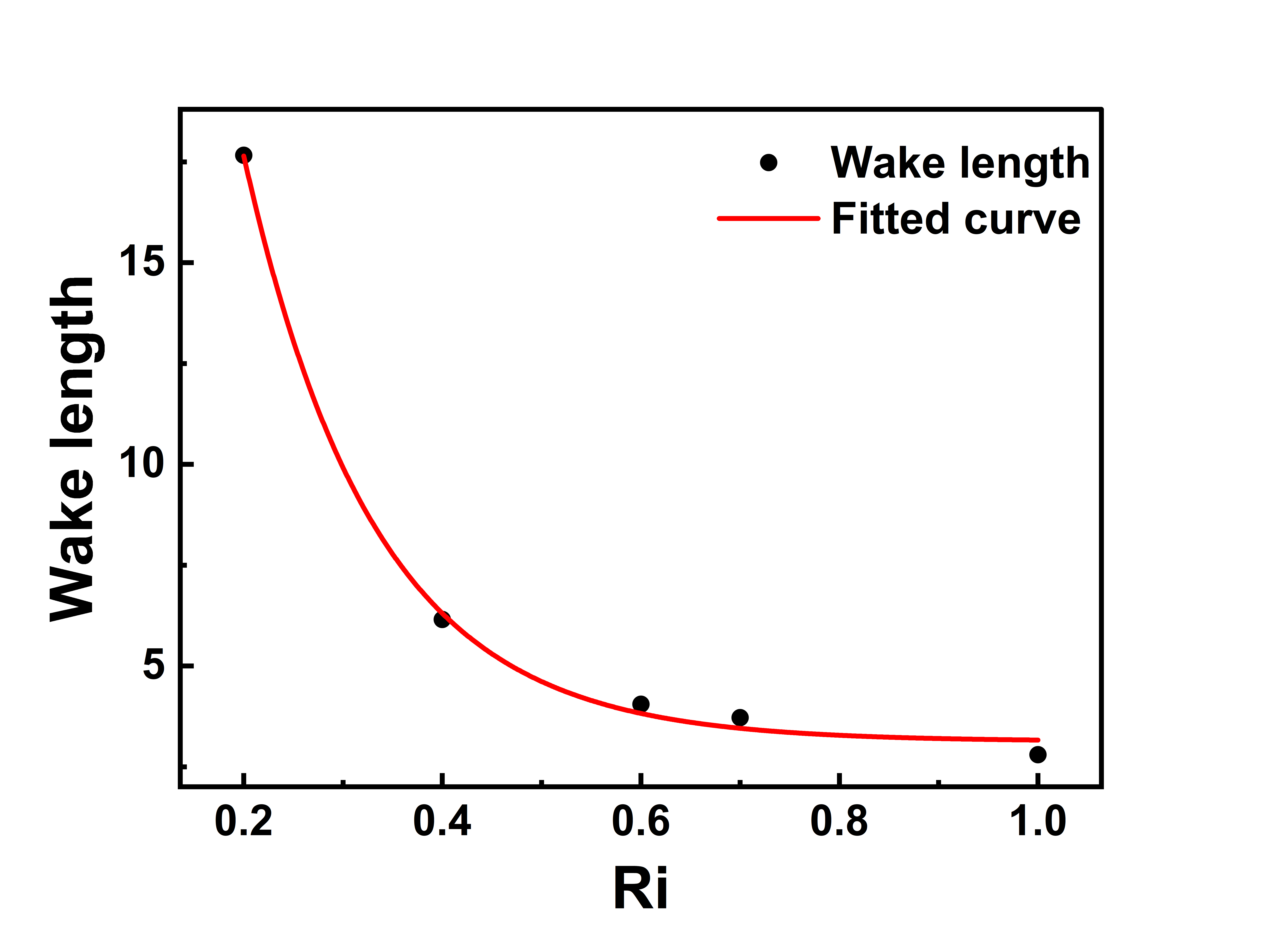}
		\caption{}
		\label{wakelength}
	\end{subfigure}\hfill
	\begin{subfigure}[t]{0.5\textwidth}
		\includegraphics[width=\linewidth]{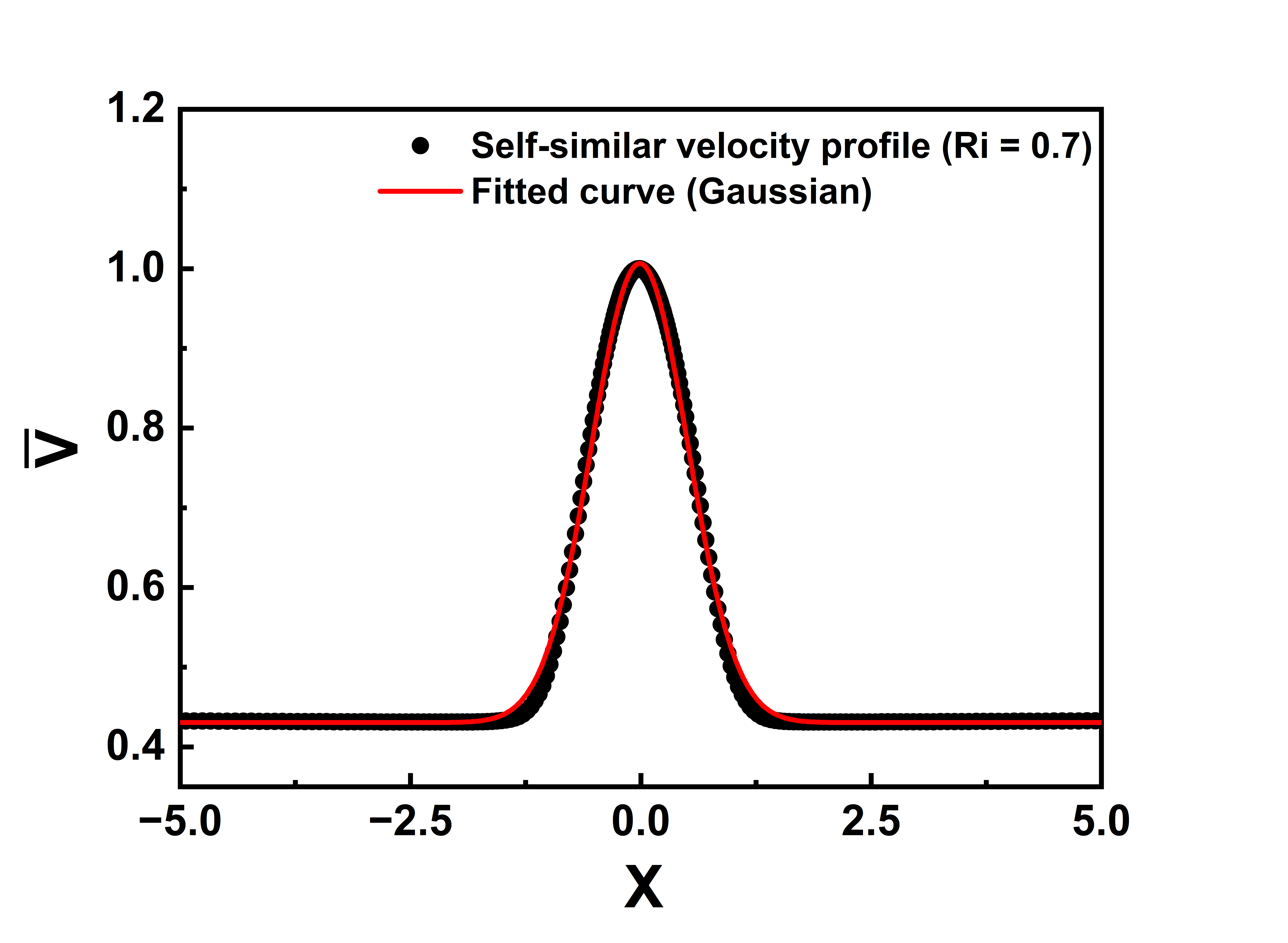}
		\caption{}
		\label{selfsimilar07}
	\end{subfigure}\hfill

	\caption{(a) Variation of wake length with Ri and (b) self-similar profile of streamwise velocity in the far-field for Ri $=0.7$. The remainder of the downstream domain beyond the wake length is the natural convection-dominated zone. The self-similar velocity profile is best described by a Gaussian curve, which has been fitted to the velocity data. }
	\label{selfsimilar}
\end{figure}

The understanding of the transitional nature of the present mixed convective flow may become easier by dividing the flow downstream of the cylinder into two zones using the magnitude of the streamwise velocity along the centerline, as discussed below. Often, it has been discussed in the literature that flows of this nature are strictly forced convective below a certain Ri, and natural convective above a certain Ri. However, at each Ri, the flow in the present study is a spatially evolving one, similar to other transitional forced flows such as laminar-turbulent transitions in boundary layers, jets, etc., where there is no exact value of the control parameter(s) at which the transition of the entire flow occurs. In other words, one may expect, even at a particular Ri, that the flow has both forced and natural convective regimes in different zones of the flow. We have discussed earlier that the downstream flow near the cylinder is dominated by forced convection, while the flow far from the cylinder is dominated by natural convection. One may interpret the deficit in streamwise velocity along the centerline ($\overline{V} < 1$) to be the forced convection-dominated zone (wake), while the excess in streamwise velocity ($\overline{V} > 1$) may be understood as the natural convection-dominated zone (plume) for the present problem. For this purpose, we have plotted the wake length against Ri, where the wake length is defined as the downstream distance where the streamwise velocity along the centerline crosses unity (see \Cref{wakelength}). Therefore, the flow starts out as forced convective in the entire domain at Ri $=0.0$. Subsequently, with increasing positive Ri, the wake length reduces, increasing the natural convection zone length. So, the transition between forced and natural convection is not only governed by Ri, but it is also a spatial phenomenon, as at each Ri, there is a region of the flow behaving like a wake, and the rest seems like a natural convective plume.

We have further analyzed the behaviour of the free convection-dominated far-field region. We have employed scaling analysis to show that the flow in the far-field is similar to that of a plume, as it shows self-similar behaviour. It is well known that plumes in the far-field of heat sources show self-similar behaviour in the time-averaged sense, similar to jets and wakes \cite{huntbremer}. In order to scale the streamwise velocity, we have employed the local maximum velocity $V_s$, and to scale the transverse length X, we have employed the plume half-width ($L_{1/2}$), which is defined as the length measured from the centerline where the streamwise velocity is half of the local maximum velocity $V_s$. On plotting the scaled velocity profiles at different streamwise positions, it is found that they collapse onto a single dimensionless profile, proving a self-similarity behavior in the far-field region. The self-similar velocity follows a Gaussian distribution as given in \Cref{gauss}. The distribution is characterized by four constants whose values are listed in \Cref{params}. The plot of the self-similar velocity along with the fitted Gaussian curve is provided in \Cref{selfsimilar07} for Ri $= 0.7$.  The far field at a higher Ri $= 1.0$ reveals similar self-similar behavior. Hunt and van den Bremer \cite{huntbremer} also reported an identical Gaussian distribution.

    \begin{equation} \label{gauss}
    V= V_0 + \dfrac{A}{w\sqrt{\pi/2}} e^{\dfrac{-2(X-X_c)^2}{w^2}}
    \end{equation}

    The parameters of the equation that were obtained upon fitting the data of Ri $=0.7$ and 1.0 are given below in \Cref{params}.

 \begin{table}[htbp]
    \centering
    \caption{Parameters of fitted curve for $Ri = 0.7$ and $Ri = 1.0$.}
    \label{params}
    \setlength{\tabcolsep}{1em}
    \begin{tabular}{c c c}
    \toprule
    \toprule
    Parameter & $Ri = 0.7$ & $Ri = 1.0$ \\
    \midrule
     $V_0$ & 0.431 & 0.388   \\
     $X_c$ & -0.012 & -0.015  \\
     $w$   & 1.042 & 1.147   \\
     $A$   & 0.752 & 0.892  \\
    \bottomrule
    \bottomrule
    \end{tabular}
    \end{table}

\begin{figure}[htbp]
	\centering
	\begin{subfigure}[t]{0.45\textwidth}
		\includegraphics[width=\linewidth]{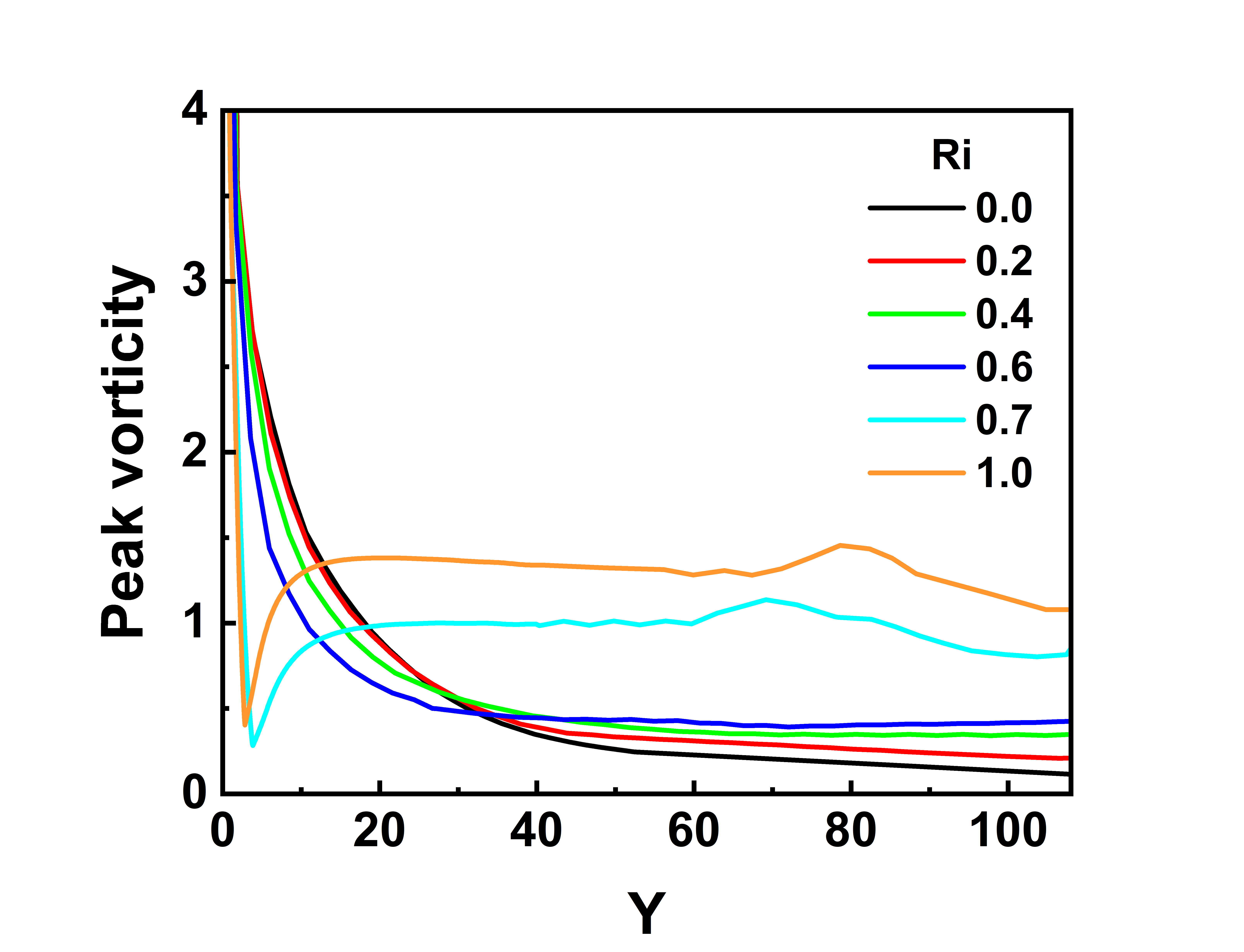}
		\caption{}
		\label{peakvort}
	\end{subfigure}\hfill
	\begin{subfigure}[t]{0.45\textwidth}
		\includegraphics[width=\linewidth]{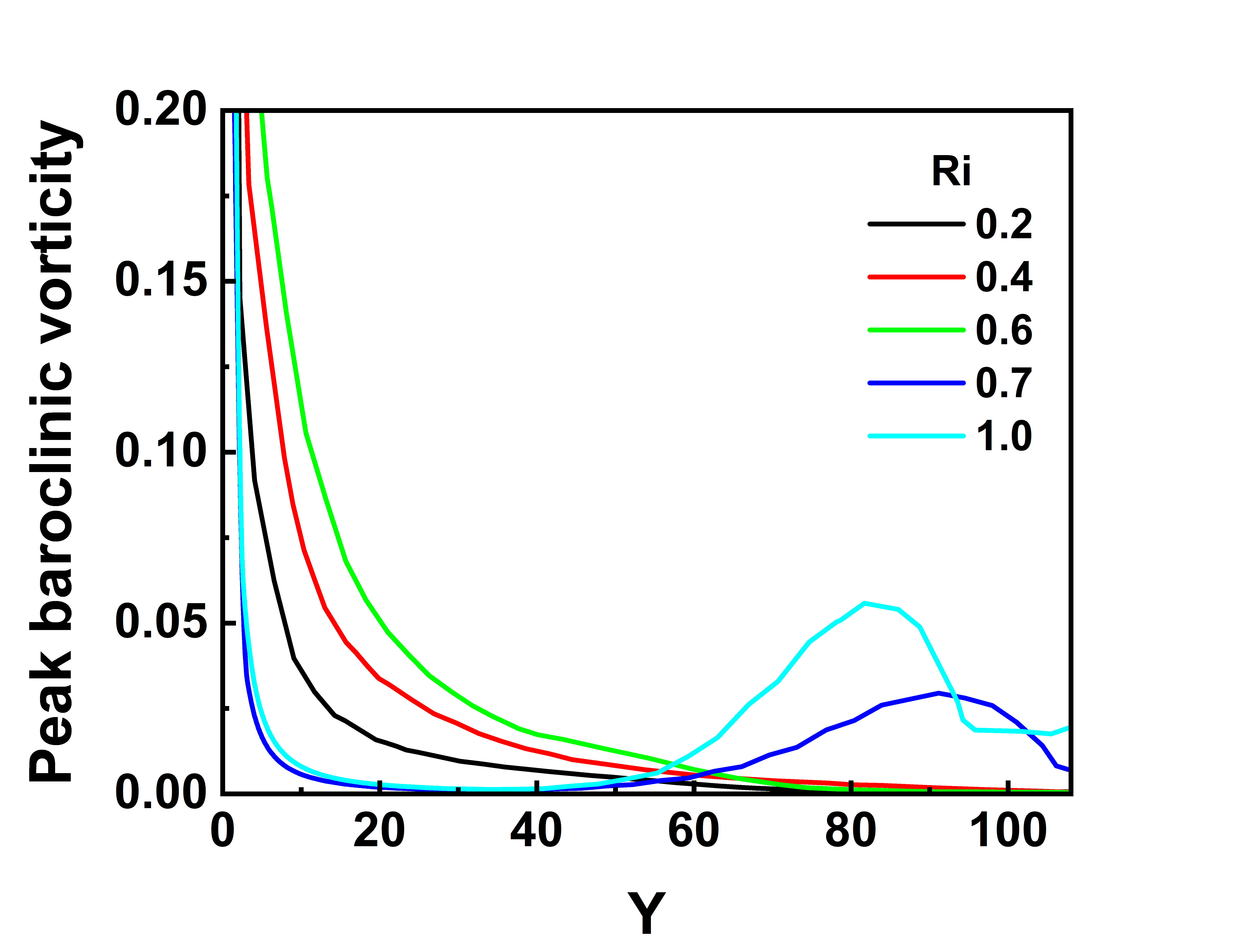}
		\caption{}
		\label{baropeak}
	\end{subfigure}
	\caption{Variation of (a) peak total vorticity, and (b) peak baroclinic vorticity in the streamwise direction, for various Ri. The peak vorticity is the magnitude of the maximum vorticity at any streamwise location.}
	\label{vortvar}
\end{figure}

The curl of the momentum equation (refer \Cref{eq2}) results in the vorticity transport equation and is obtained as:

\begin{equation} \label{vte}
	\frac{\partial \omega}{\partial \tau} + \frac{\partial (\omega u_i)}{\partial x_j} = \frac{1}{Re} \left[ \frac{\partial^2 \omega}{\partial x_i^2} \right] + Ri \frac{\partial \theta}{\partial x_i} \delta_{i2}
\end{equation}

The second term on the right side of \Cref{vte} is the baroclinic vorticity production term due to the buoyancy effect. Buoyancy causes density gradients to arise in the fluid. The baroclinic vorticity appears as a source term in the vorticity transport equation when the density and pressure gradients are not aligned in a fluid. It is seen that this vorticity plays a key role in the suppression of vortex shedding as well as the far-field unsteadiness. 

The baroclinic vorticity (lines) superimposed on the total vorticity (flood) in the near-field region has been depicted in \Cref{insnearvort}. At Ri $=0.0$, there is no baroclinic vorticity due to the absence of thermal gradients. However, for positive Ri, baroclinic vorticity affects the near-field flow structures. The shed vortices appear to be stretched in the transverse direction, matching the shape of the baroclinic vortices. This effect is intensified at higher Ri. On comparing the shed vortices at Ri $=0.7$ with those at Ri $=0.0$, it is clear that the vortices at higher Ri are thinner and elongated compared to those at lower Ri. For Ri $< 0.7$, the baroclinic vorticity follows an alternating pattern, with vortices of opposite signs being shed one after the other. The regions with a high magnitude of total vorticity are accompanied by positive baroclinic vorticity, while regions with a low magnitude of total vorticity are accompanied by negative baroclinic vorticity. This reveals that baroclinic vorticity increases the vorticity content in high-vorticity regions while cancelling the vorticity in the low-vorticity regions. 

\Cref{insfarvort} presents the baroclinic vorticity (lines) superimposed on the total vorticity (flood) in the far-field region. It is seen that the baroclinic vorticity term plays a significant role in the development of the far-field unsteadiness. The baroclinic vorticity at Ri $= 0.2$ creates unsteady flow structures in the far-field, unlike the forced convective flow case. It can also be seen that at this Ri, the effect of the baroclinic vorticity diminishes with increased downstream distance, as does the total vorticity. At Ri $= 0.4$, the baroclinic vorticity is more pronounced. Regions of positive baroclinic vorticity increase the magnitude of total vorticity (in both positive and negative vorticity streams), while regions of negative baroclinic vorticity decrease the magnitude of total vorticity in both vorticity streams. A similar trend is observed for Ri $= 0.6$. However, the strength of baroclinic vorticity increases for Ri beyond 0.7, which causes large-scale undulation of the shear layer, leading to unsteady flow structures. This effect is intensified at higher Ri. Similar to Ri $\leq$ 0.7, the regions of positive baroclinic vorticity enhance the magnitude of total vorticity, while regions of negative baroclinic vorticity minimize or cancel the magnitude of total vorticity. This behaviour is similar to that seen in the near-field.

\Cref{vortvar} shows the streamwise variation of peak total vorticity and peak baroclinic vorticity for different Ri. For Ri $<0.7$, the total vorticity shows a decaying trend in the streamwise direction due to the diffusion of vorticity as it is convected away from the cylinder. Above Ri $=0.7$, the vorticity decays sharply after the cylinder and increases immediately thereafter, remaining constant until the onset of the far-field unsteadiness around Y $=65$. Both Ri of 0.7 and 1.0 show a maxima in the total vorticity with the onset of the far-field unsteadiness. This reveals that the shedding of vortices in the far-field occurs when the shear layer is unable to contain further vorticity beyond a prescribed limit. After the shedding begins, the vorticity once again decays with space due to the diffusion effect. On examining the plot of peak baroclinic vorticity, it is evident that a behavior similar to total vorticity is shown, with maxima in baroclinic vorticity visible in the far-field region.

\section{\label{conclusion}Conclusions}
In this study, we conducted a detailed examination of buoyancy-aided mixed convective flow past a square cylinder at an angle of incidence, $\alpha = 45^{\circ}$ for Re $=100$, $0.0 \leq Ri \leq 1.0$ and Pr $=0.7$ by performing direct numerical simulations (DNS) using the finite-difference based Marker and Cell (MAC) method. 

At a critical Ri between 0.6 and 0.7, vortex shedding suppression takes place, leading to a steady flow in the near-field. Simultaneously, the far-field develops large-scale unsteadiness similar to a plume. There is no range of Ri for which the entire flow field is seen to be steady. At low Ri, the downstream flow behavior resembles that of a wake, having a momentum deficit in the time-averaged velocity. As Ri is increased, the flow behavior gradually transforms into a jet-like plume, having a momentum excess in the time-averaged, which reveals a self-similar behavior. The flow retains the characteristics of both a wake (forced convective flow) and a plume (free convective flow) in this intermediate Ri range. An inversion of vorticity takes place behind the cylinder for all positive Ri, because of the interplay between the forced and natural convective components of the flow. These findings provide valuable insight into the near-field and far-field dynamics of the buoyancy-aided mixed convective flows past cylinders. Also, the point of inflexion found in the far-field flow is believed to be the reason for the unsteady plume at higher Ri.   

We evaluated different integral parameters, such as drag and lift coefficients and Nusselt number. The time-averaged parameters ($\overline{C}_D$ and $\overline{Nu}$) show an increase with Ri, while the corresponding RMS parameters show a decrease with Ri. Additionally, the inversion length decreases with Ri, while the recirculation length increases with increasing Ri till the critical Ri and decreases subsequently.

\begin{acknowledgements}
Two of the authors (Kavin Kabilan and Swapnil Sen) would like to thank SURGE, Indian Institute of Technology Kanpur, for providing financial support to carry out this work. The authors thank the Indian Institute of Technology Kanpur for providing the computational facilities required to carry out this work.

\end{acknowledgements}

\bibliographystyle{ieeetr}
\bibliography{bibliography.bib}

\end{document}